\documentclass[fleqn,10pt]{article}
\usepackage{latexsym, graphicx, epsfig, amsmath, amssymb,amsfonts}
\usepackage{natbib,amsthm,version}
\usepackage{amsbsy,bm,multirow,enumerate}
\usepackage[titletoc,page]{appendix}
\usepackage[mathscr]{eucal}
\usepackage{mathtools}
\usepackage{color}
\usepackage{subfigure}
\usepackage[utf8]{inputenc}
\usepackage[english]{babel}
\usepackage{float}
\usepackage{caption}
\usepackage{systeme}
\usepackage[affil-it]{authblk}
\usepackage{hyperref}

\newtheorem{theorem}{Theorem}[section]
\newtheorem{corollary}{Corollary}[theorem]
\newtheorem{lemma}[theorem]{Lemma}

\begin{document}

\title{ 
	A consistent and conservative Phase-Field method for multiphase incompressible flows
	\footnote{\copyright $<$2022$>$. This manuscript version is made available under the CC-BY-NC-ND 4.0 license \url{http://creativecommons.org/licenses/by-nc-nd/4.0/}.}
	\footnote{This manuscript was accepted for publication in Journal of Computational and Applied Mathematics, Vol 408, Ziyang Huang, Guang Lin, Arezoo M. Ardekani, A consistent and conservative Phase-Field method for multiphase incompressible flows, Page 114116, Copyright Elsevier (2022).}
}

\author[1]{
	Ziyang Huang%
	\thanks{Email: \texttt{huan1020@purdue.edu}; \texttt{ziyangh@umich.edu}. Present address: Mechanical Engineering, University of Michigan, Ann Arbor, MI 48109, USA}}

\author[1,2]{
	Guang Lin%
	\thanks{Email: \texttt{guanglin@purdue.edu}; Corresponding author}}

\author{
	Arezoo M. Ardekani%
	\thanks{Email: \texttt{ardekani@purdue.edu}; Corresponding author}}

\affil[1]{
	School of Mechanical Engineering, Purdue University, West Lafayette, IN 47907, USA}
\affil[2]{
	Department of Mathematics, Purdue University, West Lafayette, IN 47907, USA}

\date{}

\maketitle


\begin{abstract}
In the present study, a consistent and conservative Phase-Field method, including both the model and scheme, is developed for multiphase flows with an arbitrary number of immiscible and incompressible fluid phases.
The \textit{consistency of mass conservation} and the \textit{consistency of mass and momentum transport} are implemented to address the issue of physically coupling the Phase-Field equation, which locates different phases, to the hydrodynamics.
These two consistency conditions, as illustrated, provide the ``optimal'' coupling because (i) the new momentum equation resulting from them is Galilean invariant and implies the kinetic energy conservation, regardless of the details of the Phase-Field equation, and (ii) failures of satisfying the second law of thermodynamics or the \textit{consistency of reduction} of the multiphase flow model only result from the same failures of the Phase-Field equation but are not due to the new momentum equation. 
Physical interpretation of the consistency conditions and their formulations are first provided, and general formulations that are obtained from the consistency conditions and independent of the interpretation of the velocity are summarized.
Then, the present consistent and conservative multiphase flow model is completed by selecting a reduction consistent Phase-Field equation.
Several novel techniques are developed to inherit the physical properties of the multiphase flows after discretization, including the gradient-based phase selection procedure, the momentum conservative method for the surface force, and the general theorems to preserve the consistency conditions on the discrete level. Equipped with those novel techniques, a consistent and conservative scheme for the present multiphase flow model is developed and analyzed. The scheme satisfies the consistency conditions, conserves the mass and momentum, and assures the summation of the volume fractions to be unity, on the fully discrete level and for an arbitrary number of phases.
All those properties are numerically validated. Numerical applications demonstrate that the present model and scheme are robust and effective in studying complicated multiphase dynamics, especially for those with large-density ratios.
\end{abstract}

\vspace{0.05cm}
Keywords: {\em
  Multiphase incompressible flows;
  Phase-Field model; 
  Consistent scheme;
  Mass conservation ;
  Momentum conservation;
  large-density ratio
}

\section{Introduction}\label{Sec Introduction}
Multiphase flows are ubiquitous in real-world problems. Interactions among different fluid phases and evolution of their interfaces are strongly coupled and, as a result, introduce complicated dynamics that casts challenges of solving the problems.
In the present study, we consider that fluid phases have constant (but not necessarily identical) densities, viscosities, and interfacial tensions. Many efforts have been paid on cases including two immiscible fluid phases, and these cases are commonly referred to as the two-phase flows. Under the “one-fluid formulation” framework \citep{Tryggvasonetal2011,ProsperettiTryggvason2007}, where the dynamics of the two fluid phases is governed by a single set of transport equations in the entire domain, many numerical models or methods for the two-phase flows have been developed, greatly improved, and successfully implemented. Examples of those numerical models or methods are the Front-Tracking method \citep{UnverdiTryggvason1992,Tryggvasonetal2001}, the Level-Set method \citep{OsherSethian1988,Sussmanetal1994,SethianSmereka2003,Gibouetal2018}, the Conservative Level-Set method \citep{OlssonKreiss2005,Olssonetal2007,ChiodiDesjardins2017}, the Volume-of-Fluid (VOF) method \citep{HirtNichols1981, ScardovelliZaleski1999, Popinet2009, OwkesDesjardins2017}, the “THINC" method \citep{Xiaoetal2005,Iietal2012,XieXiao2017, Qian2018}, the Phase-Field (or Diffuse-Interface) method \citep{Andersonetal1998,Jacqmin1999,Shen2011,Huangetal2020}, the continuous surface force (CSF) \citep{Brackbilletal1992}, the ghost fluid method (GFM) \citep{Fedkiwetal1999,Lalanneetal2015}, and the balanced-force algorithm \citep{Francoisetal2006,Popinet2018}.

Recent studies are moving towards cases including three fluid phases or, more generally, an arbitrary number of fluid phases. Compared to the two-phase problems, the three-phase or the general $N$-phase ($N \geqslant 1$) flows cast some additional challenges in locating phases and interfaces and in modeling interfacial tensions and moving contact lines. The interface reconstruction schemes in the Volume of Fluid (VOF) and Moment of Fluid (MOF) methods for the two-phase flows are extended to the three-phase cases, e.g., in \citep{Schofieldetal2009,Schofieldetal2010,Francois2015,PathakRaessi2016}, and become more involved. One can expect that those schemes become more and more complicated as the number of phases increases. The Level-Set method is also extended to the three- or $N$-phase problems, e.g., in \citep{Smithetal2002,Losassoetal2006, Starinshaketal2014}, by simply adding more Level-Set functions. However, special cares have to be paid on overlaps or voids introduced by independently advecting the Level-Set functions, and on the issue of mass conservation. 

The Phase-Field method is popularly used to model multiphase dynamics because of its simplicity and effectiveness. It is characterized by introducing a small but finite interface thickness. The Phase-Field method includes thermodynamical compression and diffusion related to a free energy functional, and their competition helps to preserve the interface thickness. Phases are labeled by a set of order parameters governed by the Phase-Field equation.
Many Cahn-Hilliard Phase-Field equations for the three-phase problems are developed, e.g., in \citep{BoyerLapuerta2006,Boyeretal2010,Kim2007}, and some of them have been extended to the $N$-phase cases, e.g., in \citep{BoyerMinjeaud2014,Kim2009,LeeKim2015,Kim2012}. The three-phase moving contact lines on a solid wall are modeled in \citep{ZhangWang2016} and \citep{Zhangetal2016}, using the wall energy functional and the geometrical formulation, respectively. Multiphase Allen-Cahn and Cahn-Hilliard Phase-Field equations including effects of pairwise interfacial tensions are developed in \citep{WuXu2017}. A $N$-phase conservative Allen-Cahn Phase-Field equation is developed in \citep{KimLee2017}, extended from its two-phase counterpart in \citep{BrasselBretin2011}. A series of studies on the Phase-Field modeling of $N$-phase incompressible flows has been performed by Dong \citep{Dong2014,Dong2015,Dong2017,Dong2018}, where not only the Cahn-Hilliard Phase-Field equation but also the contact angle boundary condition is considered.
Early studies only require the Phase-Field equation to follow the mass conservation of each phase and to ensure the summation of the volume fractions of the phases to be unity. Recent studies have realized the significance of the so-called \textit{consistency of reduction}\citep{BoyerLapuerta2006,BoyerMinjeaud2014,Dong2017,Dong2018}, because this principle eliminates the possibility of producing fictitious phases. General theories and discussions about developing a reduction consistent Phase-Field equation are available in \citep{BoyerLapuerta2006,Boyeretal2010,BoyerMinjeaud2014,LeeKim2015,Dong2017,Dong2018,WuXu2017} and the references therein.

The first issue considered in the present study is developing a physical momentum equation that governs the motion of the multiphase fluid mixture, provided a mass conservative Phase-Field equation. In previous studies, most attention was paid to developing the Phase-Field equation that captures phase locations, while physical coupling between the Phase-Field equation and the hydrodynamics was less studied. Once the Phase-Field equation is obtained, the hydrodynamics was included by solving the Navier-Stokes equation:
\begin{equation}\label{Eq NS Intro}
\frac{\partial (\rho \mathbf{u})}{\partial t}
+
\nabla \cdot (\rho \mathbf{u} \otimes \mathbf{u} )
=
-\nabla P
+
\mathbf{f},
\end{equation}
or its equivalent form. Here, $\rho$ is the density of the fluid mixture, $\mathbf{u}$ is the velocity and is divergence-free, i.e., $\nabla \cdot \mathbf{u}=0$, $P$ is the pressure, and $\mathbf{f}$ denotes any forces other than the pressure gradient. Such a strategy of coupling the Phase-Field equation to the hydrodynamics is extended from the so-called modified model H using the Cahn-Hilliard Phase-Field equation in the two-phase cases, and has been popularly used in the two-phase flows, e.g., \citep{GuermondQuartapelle2000,Kim2005,Dingetal2007,DongShen2012,Huangetal2019}, and more recently in the three-phase or $N$-phase flows, e.g., \citep{BoyerLapuerta2006,Boyeretal2010,BoyerMinjeaud2014,Kim2007,Kim2009,Kim2012,LeeKim2015,KimLee2017,ZhangWang2016,Zhangetal2016,Abadi2018}. 
It should be noted that validity of the above coupling strategy is rooted in the presumption that the density of the fluid mixture is governed by
\begin{equation}\label{Eq Mass Intro}
\frac{\partial \rho}{\partial t}
+
\nabla \cdot (\rho \mathbf{u})
=
0,
\end{equation}
and we call Eq.(\ref{Eq Mass Intro}) the sharp-interface mass conservation equation in the present study. However, $\rho$ is never solved from Eq.(\ref{Eq Mass Intro}) but instead computed algebraically from $\sum_{p=1}^N \rho_p C_p$ where $\rho_p$ and $C_p$ are the density and volume fraction of Phase $p$. We show in the present study that the algebraically computed mixture density can be contradicting to the sharp-interface mass conservation equation Eq.(\ref{Eq Mass Intro}), in particular when the velocity is divergence-free. As a result, the Navier-Stokes equation Eq.(\ref{Eq NS Intro}) becomes inconsistent with the ``actual'' mass conservation equation of the mulitphase flow model, and can produce unphysical results. Such an issue in the two-phase flows was realized and addressed in our recent studies \citep{Huangetal2020,Huangetal2020CAC}, by proposing and successfully applying the \textit{consistency of mass conservation} and the \textit{consistency of mass and momentum transport} to the Cahn-Hilliard and conservative Allen-Cahn Phase-Field equations, respectively. Analyses in \citep{Huangetal2020} show that violating those consistency conditions, such as using the modified model H, introduces sources that are proportional to the density difference of the two fluid phases in both the momentum and kinetic energy transports, and therefore results in unphysical velocity fluctuations and interface deformations in large-density-ratio problems. This emphasizes the significance of carefully considering the coupling between the Phase-Field equation and the hydrodynamics.
Compared to the exiting Phase-Field based multiphase flow models using the divergence-free velocity and to our previous studies on the two-phase flows \citep{Huangetal2020,Huangetal2020CAC}, the present study lifts restrictions on the number of phases and on a specific Phase-Field equation, and has the following contributions to the aspect of modeling:
\begin{itemize}
    \item 
    The \textit{consistency of mass conservation} and the \textit{consistency of mass and momentum transport} are first applied to the general $N$-phase ($N \geqslant 1$) cases and to any Phase-Field equation that satisfies the mass conservation. The resulting formulations are therefore generally valid.
    \item 
    Effects of the \textit{consistency of mass conservation} and the \textit{consistency of mass and momentum transport} on the kinetic energy conservation and the Galilean invariance of the momentum equation are analyzed in a formal way, regardless of the number of phases or details of the Phase-Field equation. 
    \item 
    The \textit{consistency of mass conservation} and the \textit{consistency of mass and momentum transport} provide the ``optimal'' coupling to the hydrodynamics in the sense that the second law of thermodynamics or the \textit{consistency of reduction} of the multiphase flow model only relies on the corresponding properties of the Phase-Field equation.
    \item 
    The physical interpretation of the consistency conditions and their formulations are first provided using the control volume analysis and mixture theory. 
\end{itemize}
Thus, the consistency conditions are general, simple, and effective modeling principles for multiphase flows, and the rest is to develop or select a physically admissible Phase-Field equation. Here, we are not attempting to propose a new Phase-Field equation but choose the one in \citep{Dong2018} to finalize the present consistent and conservative multiphase flow model, because that Phase-Field equation is fully reduction consistent, conserves the mass of individual phases, and eliminates local voids or overfilling.
Although we limit our focus in the present study on models using the divergence-free velocity which can be interpreted as the ``volume-averaged'' velocity, there is another category of Phase-Field based multiphase flow models, like those in \citep{LowengrubTruskinovsky1998,Shenetal2013,GuoLin2015,Shenetal2020,KimLowengrub2005,LiWang2014,Odenetal2010}, using the ``mass-averaged'' velocity which is not divergence-free. The consistency conditions can also be applied to that circumstance. We further illustrate that, after implementing the consistency conditions, both categories of models produce the mass conservation equation equivalent to the one from the mixture theory, and share the same form of the momentum equation.

Another issue to be addressed in the present study is to preserve as many physical properties of a multiphase flow model as possible on the discrete level, in order to eliminate unphysical behaviors of numerical solutions. However, this part is still far from complete, particularly in the three- or $N$-phase cases, although some of those schemes are shown to satisfy a discrete energy law, e.g., in \citep{Boyeretal2010,ZhangWang2016, WuXu2017}, and to achieve mass conservation, e.g., in \citep{Boyeretal2010,KimLee2017}. For large-density-ratio problems, the so-called consistent schemes were developed for the two-phase flows, see, e.g., \citep{Rudman1998,Bussmannetal2002,ChenadecPitsch2013,OwkesDesjardins2017} for the Volume-of-Fluid (VOF) method, \citep{RaessiPitsch2012,Nangiaetal2019} for the Level-Set method, and \citep{Huangetal2020,Huangetal2020CAC} for the Cahn-Hilliard and conservative Allen-Cahn Phase-Field equations, respectively. Since there are various Phase-Field equations developed for the multiphase flows and therefore numerical schemes that solve them, it is desirable to develop a unified framework that is applicable to an arbitrary number of phases, various Phase-Field equations for the multiphase flows, and different numerical methods that solve those Phase-Field equations, to obtain the consistent schemes.
In the present study, we develop several novel techniques to satisfy the aforementioned consistency conditions, conserve the mass and momentum, and additionally assure the summation of the volume fractions to be unity. Those novel techniques that are first reported in the present study include:
\begin{itemize}
    \item 
    The gradient-based phase selection procedure is developed to remove mass change, fictitious phases, local voids, or overfilling, produced numerically by the convection term in the Phase-Field equation.
    \item
    The conservative method is developed to discretize the surface force that models interfacial tensions in the multiphase flows, which contributes to the conservation of momentum.
    \item 
    General theorems, applicable to various Phase-Field equations and their discretization, are proposed to preserve the consistency conditions on the discrete level, which highlight correspondences of numerical operations in solving the Phase-Field and momentum equations.
\end{itemize}
Incorporating the above novel techniques to the second-order decoupled scheme in \citep{Dong2018} for the Phase-Field equation and the second-order projection scheme in \citep{Huangetal2020} for the momentum equation, we obtain the consistent and conservative scheme for the present mulitphase flow model. Moreover, we supplement the following analyses and proofs on the fully discrete level, which have not been reported in \citep{Dong2018,Huangetal2020}:
\begin{itemize}
    \item 
    The present scheme for the Phase-Field equation satisfies the \textit{consistency of reduction}, conserves the mass of each phase, and also guarantees the summation of the volume fractions to be unity at every discrete location.
    \item
    The present scheme for the momentum equation conserves the momentum when the surface force is either neglected or computed by the conservative method, and recovers the single-phase dynamics discretely inside each bulk-phase region.
    \item 
    The present scheme physically connects the discrete Phase-Field and momentum equations, following the \textit{consistency of mass conservation} and the \textit{consistency of mass and momentum transport}, and therefore solves advection (or translation) problems exactly with no restrictions on material properties or interface shapes.
\end{itemize}
The properties above are carefully validated by numerical experiments. Comparison studies illustrate productions of fictitious phases, local voids or overfilling, non-conservation of momentum, and unphysical velocity fluctuations and interface deformations, without implementing the novel techniques. It is worth mentioning that all those novel techniques, analyses, and proofs are generally valid for an arbitrary number of phases and dimensions. Applications of the present consistent and conservative model and scheme to the realistic multiphase flows demonstrate their robustness and capability of simulating multiphase dynamics even when there are significant density and/or viscosity ratios.

This article is organized as the following.
In Section \ref{Sec Definitions and governing equations},
after defining the multiphase problems, we implement the consistency conditions to obtain a physical momentum equation, illustrate the significance of satisfying the consistency conditions, and finally provide their physical interpretation.
In Section \ref{Sec Discretization}, 
the novel techniques that help to preserve the physical properties of the multiphase flows are developed.
In Section \ref{Sec Scheme},
the present consistent and conservative scheme, incorporating the novel techniques, is proposed and analyzed.
In Section \ref{Sec Validations and applications},
the present multiphase flow model and scheme are validated by a series of numerical experiments, and then applied to various realistic multiphase flow problems.
Finally,
we conclude the present study in Section \ref{Sec Conclusions and future works}.

\section{Definitions and governing equations} \label{Sec Definitions and governing equations}
In this section, we first define the basic variables in the multiphase flow problems. Then, we implement the \textit{consistency of mass conservation} to a generic form of the Phase-Field equation and derive the ``actual'' mass conservation equation and the consistent mass flux of the multiphase flow model. The momentum equation compatible with the ``actual'' mass conservation equation is obtained following the \textit{consistency of mass and momentum transport}, and the significance of implementing these two consistency conditions are discussed. After that, we proceed to investigate the second law of thermodynamics and the \textit{consistency of reduction} of the entire multiphase flow model. Finally, we select the fully reduction consistent Phase-Field equation in \citep{Dong2018} to specify the present consistent and conservative multiphase flow model, and summarize the physical properties of the present model on the continuous level. At the end of this section, we provide physical interpretation to the consistency conditions, and illustrate that the consistency conditions and their resulting formulations fit in alternative multiphase flow models whose velocity is not necessarily divergence-free.

\subsection{Basic definitions} \label{Sec Basic definitions}
Inside the domain of interest, there are $N$ ($N \geqslant 1$) immiscible and incompressible fluid phases. 
``Immiscible'' means the phases are unable to form a homogeneous mixture, and ``incompressible'' means the densities of the phases are constant. 
The volume fraction of Phase $p$, denoted by $C_p$ here, is defined as the portion of Phase $p$ in a differential volume ($d(vol)_p$) over the same differential volume ($d(vol)$), i.e., $C_p=d(vol)_p/d(vol)$.
Since no void or overfilling is allowed to be generated, the summation of the volume fractions over all the phases should be unity.
Another popular choice to locate the phases is the volume fraction contrast, whose definition is
\begin{equation} \label{Eq Volume fraction contrast}
\phi_p=C_p-\sum_{q=1,q \neq p}^{N} C_q =2C_p-1, 
\quad
1 \leqslant p \leqslant N,
\end{equation}
and its range is from $-1$ to $1$. Correspondingly, the summation of the volume fraction contrasts is
\begin{equation} \label{Eq Summation of volume fraction contrasts}
\sum_{p=1}^N C_p=\sum_{p=1}^N \frac{1+\phi_p}{2}=1
\quad \mathrm{or} \quad
\sum_{p=1}^{N} \phi_p =(2-N).
\end{equation}
Either the volume fractions $\left\{ C_p \right\}_{p=1}^{N}$ or the volume fraction contrasts $\left\{ \phi_p \right\}_{p=1}^{N}$ can be used as the order parameters in a Phase-Field method, and the latter case is considered in the present study unless otherwise specified.

The density and viscosity of Phase $p$ are $\rho_p$ ($\rho_p \geqslant 0$) and $\mu_p$ ($\mu_p \geqslant 0$), respectively. The density and viscosity of the fluid mixture are defined by
\begin{equation} \label{Eq Mixture density}
\rho=\sum_{p=1}^{N} \rho_p C_p=\sum_{p=1}^{N} \rho_p \frac{\phi_p+1}{2},
\end{equation}
\begin{equation} \label{Eq Mixture viscosity}
\mu=\sum_{p=1}^{N} \mu_p C_p=\sum_{p=1}^{N} \mu_p \frac{\phi_p+1}{2}.
\end{equation}
The interfacial tension between Phases $p$ and $q$ is $\sigma_{p,q}$ ($1 \leqslant p,q \leqslant N$), and there is no interfacial tension inside each phase. Thus, we have $\sigma_{q,p}=\sigma_{p,q}$ and $\sigma_{p,p}=0$.
The contact angle of Phases $p$ and $q$ at a wall, measured inside Phase $p$, is $\theta_{p,q}^W$ ($1 \leqslant p,q \leqslant N$). Consequently, we have $\theta_{p,q}^W+\theta_{q,p}^W=\pi$.

Without loss of generality, the Phase-Field equation is written as 
\begin{equation} \label{Eq Volume fraction contrast equation conservative}
\frac{\partial \phi_p}{\partial t}
+
\nabla \cdot ( \mathbf{u} \phi_p )
=
\nabla \cdot \mathbf{J}_p,
\quad
1 \leqslant p \leqslant N,
\end{equation}
where $\mathbf{u}$ is the velocity, and $\{\mathbf{J}_p\}_{p=1}^N$ are the diffusion fluxes of the order parameters. For those Phase-Field equations that include terms in a non-conservative form, like the conservative Allen-Can equation \citep{RubinsteinSternberg1992,BrasselBretin2011}, one can apply the consistent formulation \citep{Huangetal2020CAC} and turn those terms into a conservative form. Therefore, Eq.(\ref{Eq Volume fraction contrast equation conservative}) is adequate to represent all the equations that conserve the order parameters and therefore the mass of the phases. 
At the moment, explicit definition of $\{\mathbf{J}_p\}_{p=1}^N$ is not necessary, but we require $\sum_{p=1}^N \mathbf{J}_p=\mathbf{0}$. As a result, after summing Eq.(\ref{Eq Volume fraction contrast equation conservative}) over all the phases and considering Eq.(\ref{Eq Summation of volume fraction contrasts}), we obtain that the velocity is divergence-free:
\begin{equation} \label{Eq Divergence free}
\nabla \cdot \mathbf{u} = 0.
\end{equation}
Further interpretation of such a divergence-free velocity and the requirement of $\sum_{p=1}^N \mathbf{J}_p=\mathbf{0}$ will be provided in Section \ref{Sec Physical interpretation}.
The divergence-free condition Eq.(\ref{Eq Divergence free}) is enforced by a Lagrange multiplier $P$, called the pressure.

Additionally, we assume that the Phase-Field equation Eq.(\ref{Eq Volume fraction contrast equation conservative}) already satisfies the \textit{consistency of reduction}, whose definition, similar to the one in \citep{BoyerMinjeaud2014,Dong2018}, is
\begin{itemize}
	\item \textit{Consistency of reduction}:
	A $N$-phase system should be able to recover the corresponding $M$-phase system $(1 \leqslant M \leqslant N-1)$ when $(N-M)$ phases don't appear.
\end{itemize}
Violating this consistency condition can easily generate fictitious phases in interfacial regions \citep{BoyerMinjeaud2014,LeeKim2015}. As demonstrated in \citep{BoyerMinjeaud2014}, generation of fictitious phases will significantly change the multiphase dynamics in problems having large density or viscosity ratios, in spite of the fact that amounts of the fictitious phases are small. This is because the density and viscosity of the fluid mixture is extensively modified by the fictitious phases.
To obey this consistency condition, some requirements are posted to the diffusion fluxes $\{\mathbf{J}_p\}_{p=1}^N$ in the Phase-Field equation Eq.(\ref{Eq Volume fraction contrast equation conservative}).
\begin{theorem}\label{Theorem RC Phase-Field}
If and only if the formulation of the diffusion fluxes $\{\mathbf{J}_p\}_{p=1}^N$ in the Phase-Field equation Eq.(\ref{Eq Volume fraction contrast equation conservative}) has the following two properties:
(i) its value becomes zero for the absent phases, and 
(ii) it, for the present phases, recovers the corresponding formulation excluding contributions from the absent phases, 
then the Phase-Field equation Eq.(\ref{Eq Volume fraction contrast equation conservative}) satisfies the consistency of reduction.
\end{theorem}
\begin{proof}
Suppose the number of phases is $N$ ($N \geqslant 2$) and Phase $N$ is absent, i.e., $\phi_N \equiv -1$, the Phase-Field equation Eq.(\ref{Eq Volume fraction contrast equation conservative}) becomes
\begin{eqnarray}\nonumber
\frac{\partial \phi_N}{\partial t}
=
-\underbrace{\nabla \cdot (\mathbf{u} \underbrace{\phi_N}_{-1})}_{-\nabla \cdot \mathbf{u}=0}+\nabla \cdot \underbrace{\mathbf{J}_N}_{0}=0
\quad \mathrm{or} \quad
\phi_N=-1 \quad \forall t>0,\\
\nonumber
\frac{\partial \phi_p}{\partial t}
+
\nabla \cdot ( \mathbf{u} \phi_p )
=
\nabla \cdot \mathbf{J}_p,
\quad
1 \leqslant p \leqslant (N-1).
\end{eqnarray}
Therefore, absent Phase $N$ remains absent. Since $\{\mathbf{J}\}_{p=1}^{N-1}$ recover the corresponding $(N-1)$-phase formulation excluding the contribution from absent Phase $N$, the Phase-Field equation for the present phases, i.e., Phases $1$-$(N-1)$, is identical to the one when there are only $(N-1)$ phases. Any other phase can be chosen as the absent phase and the above procedure can be repeated. As a result, the Phase-Field equation is reduction consistent. On the other hand, if the Phase-Field equation Eq.(\ref{Eq Volume fraction contrast equation conservative}) is reduction consistent already, then, from the definition of the \textit{consistency of reduction}, the diffusion fluxes need to have the two properties.
\end{proof}

From the requirement of $\sum_{p=1}^N \mathbf{J}_p =\mathbf{0}$ and the \textit{consistency of reduction} of the Phase-Field equation, it is straightforward to deduce $\mathbf{J}_p=\mathbf{0}$ when there is only single Phase $p$ ($1 \leqslant p \leqslant N$). Another possible consequence of violating this consistency condition is producing incorrect dynamics inside bulk-phase regions, and this will be discussed in more detail when considering the momentum equation in Section \ref{Sec Momentum reduction}.

\subsection{The mass conservation}\label{Sec Mass}
As the density of the fluid mixture is computed from Eq.(\ref{Eq Mixture density}), instead of presuming that the mixture density is governed by the sharp-interface mass conservation equation Eq.(\ref{Eq Mass Intro}), we follow the \textit{consistency of mass conservation} which was proposed and successfully implemented in the two-phase flows \citep{Huangetal2020}. The definition of this consistency condition is: 
\begin{itemize}
	\item \textit{Consistency of mass conservation}:
	The mass conservation equation should be consistent with the Phase-Field equation and the density of the fluid mixture. In this mass conservation equation, the consistent mass flux should lead to a zero mass source.
\end{itemize}

Following this consistency condition, one can, in general, reach the transport equation below for the density of the fluid mixture:
\begin{equation}\label{Eq Mass Phase-Field}
\frac{\partial \rho}{\partial t}
+
\nabla \cdot \mathbf{m}^*
=
S_{m^*},
\end{equation}
after using Eq.(\ref{Eq Mixture density}) and Eq.(\ref{Eq Volume fraction contrast equation conservative}), and noticing that the velocity is divergence-free, i.e., Eq.(\ref{Eq Divergence free}). Here, $\mathbf{m}^*$ is the mass flux and $S_{m^*}$ is the mass source. Depending on how the mass flux is defined, the mass source is determined correspondingly. For example, if the mass flux is defined as $\mathbf{m}^*=\rho \mathbf{u}$, then the corresponding mass source is $S_{m^*}=\sum_{p=1}^N \frac{\rho_p}{2} \nabla \cdot \mathbf{J}_p$. Therefore, the mixture density, in general, is not necessarily governed by the sharp-interface mass conservation equation Eq.(\ref{Eq Mass Intro}). Instead, Eq.(\ref{Eq Mass Phase-Field}) is the ``actual'' mass conservation equation of the multiphase flow model. 

As the ``actual'' mass conservation equation Eq.(\ref{Eq Mass Phase-Field}) is obtained, the next step is to specify the consistent mass flux $\mathbf{m}$ among various choices. The \textit{consistency of mass conservation} points out that the corresponding mass source of the consistent mass flux is zero, i.e., $S_m=0$. As a result, Eq.(\ref{Eq Mass Phase-Field}) becomes
\begin{equation} \label{Eq Mass conservation}
\frac{\partial \rho}{\partial t}+\nabla \cdot \mathbf{m} =0,
\end{equation}
after using the consistent mass flux $\mathbf{m}$. 
Applying Eq.(\ref{Eq Mass conservation}), Eq.(\ref{Eq Mixture density}), and Eq.(\ref{Eq Volume fraction contrast equation conservative}), the divergence of the consistent mass flux is
\begin{eqnarray} \label{Eq Divergence of mass flux}
\nabla \cdot \mathbf{m}
=
-\frac{\partial \rho}{\partial t}
=
-\frac{\partial}{\partial t} \sum_{p=1}^{N} \rho_p \frac{1+\phi_p}{2}
=
- \sum_{p=1}^{N} \frac{\rho_p}{2} \frac{\partial \phi_p}{\partial t}
=
\nabla \cdot \sum_{p=1}^{N} \frac{\rho_p}{2} ( \mathbf{u} \phi_p - \mathbf{J}_p ),
\end{eqnarray} 
and therefore the consistent mass flux is
\begin{equation} \label{Eq Mass flux Phase-Field}
\mathbf{m}
=
\sum_{p=1}^{N} 
\frac{\rho_p}{2} (\mathbf{u} + \mathbf{u} \phi_p - \mathbf{J}_p).
\end{equation}
It should be noted that one more $\mathbf{u}$ is added inside the parentheses of Eq.(\ref{Eq Mass flux Phase-Field}), due to the \textit{consistency of reduction}. If only Phase $p$ appears, the consistent mass flux $\mathbf{m}$ should become $\rho_p \mathbf{u}$, which is true from Eq.(\ref{Eq Mass flux Phase-Field}) since $\phi_p=1$, $\{\phi_q\}_{q=1,q \neq p}^N=-1$, and $\{\mathbf{J}_p\}_{p=1}^N=\mathbf{0}$. On the other hand, if the consistent mass flux is defined as $\sum_{p=1}^{N} \frac{\rho_p}{2} (\mathbf{u} \phi_p - \mathbf{J}_p)$, directly from Eq.(\ref{Eq Divergence of mass flux}), it becomes $\frac{1}{2} (\rho_p-\sum_{q=1,q \neq p}^N \rho_q) \mathbf{u}$ when only Phase $p$ is present, and consequently violates the \textit{consistency of reduction}. It should also be noted that the definition of the consistent mass flux in Eq.(\ref{Eq Mass flux Phase-Field}) is also consistent with the given example of Eq.(\ref{Eq Mass Phase-Field}) where $\mathbf{m}^*=\rho \mathbf{u}$ and $S_{m^*}=\sum_{p=1}^N \frac{\rho_p}{2} \nabla \cdot \mathbf{J}_p$. Since the velocity is divergence-free, the divergence of the consistent mass flux defined in Eq.(\ref{Eq Mass flux Phase-Field}) follows Eq.(\ref{Eq Divergence of mass flux}), and therefore the \textit{consistency of mass conservation} is held. 

To further simplify the formulations, we introduce the Phase-Field fluxes $\{\mathbf{m}_{\phi_p}=\mathbf{u} \phi_p - \mathbf{J}_p\}_{p=1}^N$ that include both the convection and diffusion fluxes in the Phase-Field equation, and we have the following theorem:
\begin{theorem}\label{Theorem Consistency of mass conservation}
Given the Phase-Field fluxes $\{\mathbf{m}_{\phi_p}\}_{p=1}^N$ that satisfy the Phase-Field equation:
\begin{equation} \label{Eq Volume fraction contrast equation Flux}
\frac{\partial \phi_p}{\partial t}
+
\nabla \cdot \mathbf{m}_{\phi_p}
=
0,
\quad
1 \leqslant p \leqslant N,
\end{equation}
the corresponding consistent mass flux that satisfies the consistency of mass conservation is
\begin{equation} \label{Eq Mass flux}
\mathbf{m}
=
\sum_{p=1}^{N} 
\frac{\rho_p}{2} (\mathbf{u}+\mathbf{m}_{\phi_p}).
\end{equation}
\end{theorem}
Formulations in Theorem~\ref{Theorem Consistency of mass conservation} not only have a simpler form but also are more convenient to preserve the \textit{consistency of mass conservation} in numerical practice, which will be seen in Section \ref{Sec Discretization}. Hereafter, the (actual) mass conservation equation is referred to Eq.(\ref{Eq Mass conservation}) and the consistent mass flux is referred to Eq.(\ref{Eq Mass flux}).
It is worth mentioning that Theorem~\ref{Theorem Consistency of mass conservation} has no restrictions on the Phase-Field fluxes $\{\mathbf{m}_{\phi_p}\}_{p=1}^N$ (or the diffusion fluxes $\{\mathbf{J}_p\}_{p=1}^N$) in the Phase-Field equation. Therefore, it is generally applicable to different Phase-Field equations.
Physical interpretation of the actual mass conservation equation Eq.(\ref{Eq Mass conservation}) and the consistent mass flux Eq.(\ref{Eq Mass flux}) is provided in Section~\ref{Sec Physical interpretation}.

\subsection{The momentum equation}\label{Sec Momentum}
As already indicated in Section \ref{Sec Mass}, it is possible that the density of the fluid mixture is not governed by the sharp-interface mass conservation equation, i.e., Eq.(\ref{Eq Mass Intro}).
As a result, the Navier-Stokes equation Eq.(\ref{Eq NS Intro}) needs to be modified in order to be compatible with the ``actual'' mass conservation equation of the multiphase flow model. The question becomes what the corresponding momentum transport of the fluid mixture is provided the mass conservation equation written in Eq.(\ref{Eq Mass conservation}). This question is answered by the \textit{consistency of mass and momentum transport} which was proposed and successfully implemented in the two-phase flows \citep{Huangetal2020}. The definition of this consistency condition is:
\begin{itemize}
	\item \textit{Consistency of mass and momentum transport}:
	The momentum flux in the momentum equation should be the tensor product between the mass flux and the velocity, where the mass flux should be identical to the one in the mass conservation equation. 
\end{itemize}

Following this consistency condition, the momentum equation governing the motion of the fluid mixture becomes
\begin{equation} \label{Eq Momentum}
\frac{\partial (\rho \mathbf{u} )}{\partial t}
+
\nabla \cdot \left( \mathbf{m} \otimes \mathbf{u} \right)
=
-\nabla P
+\nabla \cdot \left[
\mu (\nabla \mathbf{u}+\nabla \mathbf{u} ^T) 
\right]
+
\rho \mathbf{g} 
+
\mathbf{f}_s,
\end{equation}
where $\otimes$ denotes the tensor product, $\mathbf{g}$ is the gravity, and $\mathbf{f}_s$ is the surface force modeling interfacial tensions. It should be noted that the inertia term in Eq.(\ref{Eq Momentum}) is written in its conservative form, which is essential to achieve momentum conservation on the discrete level, and the consistent mass flux $\mathbf{m}$ obtained in Section~\ref{Sec Mass} has been applied.
The surface force $\mathbf{f}_s$ in the present study will be obtained after considering the energy law in Section \ref{Sec Energy}, and it is shown in that section that the surface force is equivalent to a conservative form. Therefore, without considering the gravity, the momentum, i.e., $\int_\Omega \rho \mathbf{u} d\Omega$, is conserved by Eq.(\ref{Eq Momentum}) even including effects of interfacial tensions.

The significance of implementing the \textit{consistency of mass conservation} and the \textit{consistency of mass and momentum transport} can be illustrated when considering the Galilean invariance of the momentum equation and the conservation of the kinetic energy.

\begin{theorem}\label{Theorem Galilean invariance}
If both the consistency of mass conservation and the consistency of mass and momentum transport are satisfied,
then the momentum equation Eq.(\ref{Eq Momentum}) is Galilean invariant.
\end{theorem}
\begin{proof}\label{Proof Galilean invariance}
The Galilean transformation is
\begin{equation}\label{Eq Galilean transform}\nonumber
\mathbf{x}'=\mathbf{x}-\mathbf{u}_0 t,
\quad
t'=t,
\quad
\mathbf{u}'=\mathbf{u}-\mathbf{u}_0,
\quad
f'=f,
\quad
\nabla' f'=\nabla f,
\quad
\frac{\partial f'}{\partial t'}=\frac{\partial f}{\partial t}+\mathbf{u}_0 \cdot \nabla f.
\end{equation}
Here, $(\mathbf{x},t)$ is the fixed frame and $(\mathbf{x}',t')$ is the moving frame with respect to the fixed frame with a constant velocity $\mathbf{u}_0$. $f$ is a scalar variable measured in the fixed frame, while $f'$ is the same variable measured in the moving frame. Applying the Galilean transformation, the momentum equation in the moving frame has the same form as the one in the fixed frame:
\begin{eqnarray}\label{Eq Galilean invariance}\nonumber
\frac{\partial (\rho' \mathbf{u}' )}{\partial t'}
+
\nabla' \cdot \left( \mathbf{m}' \otimes \mathbf{u}' \right)
+
\nabla' P'
-\nabla' \cdot \left[
\mu' (\nabla' \mathbf{u}'+\nabla' \mathbf{u'} ^T) 
\right]
-
\rho' \mathbf{g}' 
-
\mathbf{f}'_s\\
\nonumber
=
\underbrace{\left(\frac{\partial (\rho \mathbf{u} )}{\partial t}
+
\nabla \cdot \left(  \mathbf{m}\otimes \mathbf{u} \right) 
+
\nabla P
-\nabla \cdot \left[
\mu (\nabla \mathbf{u}+\nabla\mathbf{u} ^T) 
\right]
-
\rho \mathbf{g} 
-
\mathbf{f}_s\right)}_{\mathbf{0}}
-
\mathbf{u}_0 \underbrace{\left(
\frac{\partial \rho  }{\partial t}
+\nabla \cdot \mathbf{m}
\right)}_{0}
=
\mathbf{0}.
\end{eqnarray}
The two parentheses group, respectively, the momentum equation Eq.(\ref{Eq Momentum}) and the mass conservation equation Eq.(\ref{Eq Mass conservation}) in the fixed frame, both of which are zero. Therefore, the momentum equation Eq.(\ref{Eq Momentum}) satisfying the two consistency conditions is Galilean invariant. 
\end{proof}

\begin{corollary}\label{Corollary constent velocity}
If the mechanical equilibrium in the hydrostatic state is reached, i.e.,
\begin{equation}\label{Eq Mechanical Equilibrium}\nonumber
-\nabla P
+
\rho \mathbf{g} 
+
\mathbf{f}_s
=\mathbf{0},
\end{equation}
then any homogeneous velocity is an admissible solution of the momentum equation Eq.(\ref{Eq Momentum}).
\end{corollary}
\begin{proof}\label{Proof constant velocity}
Suppose the velocity in the moving frame, i.e., $\mathbf{u}'$, is zero, then the mechanical equilibrium is true in the moving frame. From the Galilean transformation, the velocity in the fixed frame becomes $\mathbf{u}=\mathbf{u}_0$, and the mechanical equilibrium is still valid. As a result, the left-hand side of the momentum equation Eq.(\ref{Eq Momentum}) becomes $\mathbf{u}_0$ times the mass conservation equation Eq.(\ref{Eq Mass conservation}) and therefore is zero. The right-hand side remains the viscous force which is again zero since there is no velocity gradient. Therefore, any homogeneous velocity $\mathbf{u}_0$ is the solution of the momentum equation Eq.(\ref{Eq Momentum}) if the mechanical equilibrium is reached.
\end{proof}

\begin{theorem}\label{Theorem Kinetic energy}
If both the consistency of mass conservation and the consistency of mass and momentum transport are satisfied, then the momentum equation Eq.(\ref{Eq Momentum}) implies the following kinetic energy equation:
\begin{eqnarray} \label{Eq Kinetic energy equation}\nonumber
\frac{\partial e_K}{\partial t}
+
\nabla \cdot \left( \mathbf{m} \frac{\mathbf{u}\cdot\mathbf{u}}{2} \right)
=
-\nabla \cdot (\mathbf{u} P)
+\nabla \cdot [\mu (\nabla \mathbf{u}+\nabla \mathbf{u}^T) \cdot \mathbf{u}]\\
\nonumber
-\frac{1}{2} \mu (\nabla \mathbf{u}+\nabla \mathbf{u}^T):(\nabla \mathbf{u}+\nabla \mathbf{u}^T)
+\rho \mathbf{u} \cdot \mathbf{g}
+\mathbf{u} \cdot \mathbf{f}_s.
\end{eqnarray}
Here, $e_K= \frac{1}{2} \rho \mathbf{u} \cdot \mathbf{u}$ is the kinetic energy density.
\end{theorem}
\begin{proof}\label{Proof Kinetic energy}
Performing the dot product between $\mathbf{u}$ and Eq.(\ref{Eq Momentum}), and applying the integration by part, the right-hand side (RHS) of the kinetic energy equation is obtained. The left-hand side (LHS) of the kinetic energy equation is obtained from
\begin{equation}\label{Eq Kinetic energy LHS}\nonumber
\mathbf{u} \cdot 
\left( \frac{\partial (\rho \mathbf{u})}{\partial t}+\nabla \cdot ( \mathbf{m} \otimes \mathbf{u} ) \right)
=
\frac{\partial e_K}{\partial t}+\nabla \cdot \left( \mathbf{m} \frac{1}{2}\mathbf{u}\cdot \mathbf{u} \right)
+\frac{1}{2} \mathbf{u} \cdot \mathbf{u} \underbrace{\left( \frac{\partial \rho}{\partial t}+\nabla \cdot \mathbf{m}\right)}_{0}.
\end{equation}
The right-most parentheses group the mass conservation equation Eq.(\ref{Eq Mass conservation}) which therefore is zero.
\end{proof}

\begin{corollary}\label{Corollary Kinetic energy conservation}
If all the forces except the pressure gradient are neglected and all the boundary integrals vanish with some proper boundary conditions, then the momentum equation Eq.(\ref{Eq Momentum}) implies the kinetic energy conservation, i.e.,
\begin{equation}\label{Eq Kinetic energy conservation}\nonumber
\frac{d E_K}{dt}=\frac{d}{dt} \int_{\Omega} e_K d\Omega=0.
\end{equation}
\end{corollary}
\begin{proof}\label{Proof Kinetic energy conservation}
Integrating the kinetic energy equation in Theorem \ref{Theorem Kinetic energy} over domain $\Omega$, keeping only the first term on RHS, applying the integration by part, and dropping all the boundary integrals, then the kinetic energy conservation is obtained.
\end{proof}

Thanks to satisfying the \textit{consistency of mass conservation} and the \textit{consistency of mass and momentum transport}, the momentum equation Eq.(\ref{Eq Momentum}) honors the physical properties in Theorem \ref{Theorem Galilean invariance}, Theorem \ref{Theorem Kinetic energy}, and their corollaries. The critical step in those proofs is to obtain the ``actual'' mass conservation equation which is zero because of the \textit{consistency of mass conservation}. This is achieved by applying the consistent mass flux in the inertia term of the momentum equation, following the \textit{consistency of mass and momentum transport}. Violating either of the consistency conditions results in failures of proving Theorem \ref{Theorem Galilean invariance}, Theorem \ref{Theorem Kinetic energy}, or their corollaries, and such a momentum equation will not only produce a problematic kinetic energy equation but also be Galilean variant. An informative example will be considering the usage of the Navier-Stokes equation Eq.(\ref{Eq NS Intro}) to govern the motion of the fluid mixture, which is equivalent to replacing the consistent mass flux $\mathbf{m}$ in Eq.(\ref{Eq Momentum}) with $\mathbf{m}^*=\rho \mathbf{u}$, and the same in the proofs of Theorem \ref{Theorem Galilean invariance} and Theorem \ref{Theorem Kinetic energy}. As a result, we obtain in the proofs $\frac{\partial \rho}{\partial t}+\nabla \cdot \mathbf{m}^*$, which equals to $S_{m^*}=\sum_{p=1}^N \frac{\rho_p}{2} \nabla \cdot \mathbf{J}_p$ from Eq.(\ref{Eq Mass Phase-Field}) and is not zero in general. Therefore, Theorem \ref{Theorem Galilean invariance}, Theorem \ref{Theorem Kinetic energy}, or their corollaries fails. Due to the failure of Corollary \ref{Corollary constent velocity}, circular interfaces will be deformed even in advection (or translation) problems. Due to the failure of Corollary \ref{Corollary Kinetic energy conservation}, the kinetic energy is not conserved in a periodic domain even though all the forces except the pressure gradient are absent, and therefore velocity fluctuations are produced. These are some unphysical phenomena that probably appear when the Navier-Stokes equation Eq.(\ref{Eq NS Intro}) is applied to govern the motion of the fluid mixture.

It is worth mentioning that, in the proofs of Theorem \ref{Theorem Galilean invariance}, Theorem \ref{Theorem Kinetic energy}, and their corollaries, we do not need to know the detail expression of the consistent mass flux or therefore the Phase-Field equation. Consequently, the momentum equation Eq.(\ref{Eq Momentum}), as well as the theorems and corollaries, in this section is generally valid. The \textit{consistency of mass and momentum transport}, as well as the momentum equation Eq.(\ref{Eq Momentum}) derived from it, will be further justified in Section \ref{Sec Physical interpretation}.

\subsection{The energy law}\label{Sec Energy}
Here, we investigate the second law of thermodynamics of the multiphase flow model using the momentum equation Eq.(\ref{Eq Momentum}) derived from the consistency conditions in Section \ref{Sec Momentum}. The second law of thermodynamics states that the total energy of an isothermal multiphase system, including both the free energy and kinetic energy, is not increasing, without considering any external energy input. Different from in Section \ref{Sec Mass} and Section \ref{Sec Momentum}, more details of the Phase-Field equation are needed to investigate this physical principle. Along with two possible forms of the diffusion flux in the Phase-Field equation, we show in this section that the momentum equation Eq.(\ref{Eq Momentum}) helps to achieve this physical principles. 

Here, we denote the free energy as $E_F=\int_\Omega e_F d\Omega$, where $e_F$ is the free energy density, and denote the chemical potential of Phase $p$ as $\xi_p$, which is the functional derivative of the free energy with respect to the order parameter of Phase $p$, i.e., $\xi_p=\delta E_F/\delta \phi_p$.

\begin{theorem}\label{Theorem energy law}
Provided the diffusion flux in Eq.(\ref{Eq Volume fraction contrast equation conservative}) to be
\begin{equation}\nonumber
\mathbf{J}_p=M_p \nabla \xi_p
\quad \mathrm{or} \quad 
\mathbf{J}_p=\sum_{q=1}^N M_{p,q} \nabla \xi_q,
\quad
1 \leqslant p \leqslant N,
\end{equation}
where $M_p$ is non-negative and $M_{p,q}$ is symmetric positive semi-definite, and provided the surface force in Eq.(\ref{Eq Momentum}) to be
\begin{equation}\nonumber
\mathbf{f}_s=\sum_{p=1}^N \beta \xi_p \nabla \phi_p,
\end{equation}
where $\beta$($>0$) is to match the two-phase formulation, e.g., in \citep{DongShen2012,Huangetal2020}, required by the consistency of reduction,
the multiphase flow model, including the Phase-Field equation Eq.(\ref{Eq Volume fraction contrast equation conservative}) and the momentum equation Eq.(\ref{Eq Momentum}), has the following energy law:
\begin{equation}\nonumber
    \frac{d}{dt}\int_{\Omega} (e_K+\beta e_F) d\Omega=-\frac{1}{2} \int_\Omega \mu (\nabla \mathbf{u}+\nabla \mathbf{u}^T): (\nabla \mathbf{u}+\nabla \mathbf{u}^T) d\Omega
-\mathcal{D}_F,
\end{equation}
if all the boundary integrals vanish with some proper boundary conditions. Here, $\mathcal{D}_F$ is the dissipation from the Phase-Field equation:
\begin{equation}\nonumber
    \mathcal{D}_F=\beta \int_\Omega \sum_{p=1}^N M_p \nabla \xi_p \cdot \nabla \xi_p d\Omega
    \quad \mathrm{or} \quad 
    \mathcal{D}_F=\beta \int_\Omega \sum_{p,q=1}^N M_{p,q} \nabla \xi_p \cdot \nabla \xi_q d\Omega,
\end{equation}
and it is non-negative. 
\end{theorem}
\begin{proof}\label{Proof energy law}
We only consider the case of $\mathbf{J}_p=\sum_{q=1}^N M_{p,q} \nabla \xi_q$, and the case of $\mathbf{J}_p=M_p \nabla \xi_p$ follows the same procedure. Multiplying Eq,(\ref{Eq Volume fraction contrast equation conservative}) by $\beta \xi_p$, summing over $p$, integrating over domain $\Omega$, performing the integration by part, and dropping all the boundary integrals, we obtain
\begin{equation}\nonumber
\frac{d}{dt}\int_{\Omega} \beta e_F d\Omega
+
\int_{\Omega} \mathbf{u} \cdot \sum_{p=1}^N \beta \xi_p \nabla \phi_p d\Omega
=
-
\beta \int_{\Omega} \sum_{p,q=1}^N M_{p,q} \nabla \xi_p \cdot \nabla \xi_q d\Omega.
\end{equation}
Integrating the kinetic energy equation (without the gravity) in Theorem \ref{Theorem Kinetic energy} over domain $\Omega$, performing the integration by part, dropping all the boundary integrals, and summing the above equation, the energy law is obtained.
\end{proof}

The physical explanation of the surface force in Theorem \ref{Theorem energy law} is that the amount of work done by the surface force should compensate for the increase of the free energy due to convection \citep{Jacqmin1999,BoyerLapuerta2006,Boyeretal2010,BoyerMinjeaud2014,Huangetal2020}. As a result, the total energy, which is the summation of the kinetic energy and the free energy, should not change because of the surface force. The first term on the right-hand side (RHS) of the energy law comes from the viscosity of the fluid mixture and the second term is introduced by the non-equilibrium thermodynamical state. It should be noted that there will be an extra term in the kinetic energy in Theorem \ref{Theorem Kinetic energy} and, therefore, in the energy law, if the momentum equation violates the \textit{consistency of mass conservation} and the \textit{consistency of mass and momentum transport}, as discussed in Section \ref{Sec Momentum}. In other words, the total energy of the multiphase system can be changed, even when all the fluids are inviscid and the thermodynamical equilibrium is reached, which is unphysical.

With an appropriately chosen free energy density, the surface force in Theorem \ref{Theorem energy law} is equivalent to a conservative form $\sum_{p=1}^N \nabla \cdot \left( -\beta \frac{\partial e_F}{\partial (\nabla \phi_p)} \otimes \nabla \phi_p\right)$, which contributes to the conservation of momentum.
\begin{theorem}\label{Theorem surface force equivalent}
The surface force in Theorem \ref{Theorem energy law} is equivalent to a conservative form $\sum_{p=1}^N \nabla \cdot \left( -\beta \frac{\partial e_F}{\partial (\nabla \phi_p)} \otimes \nabla \phi_p\right)$, provided that the free energy density only depends on $\{\phi_p\}_{p=1}^N$ and $\{\nabla \phi_p\}_{p=1}^N$, i.e., $e_F=e_F\left(\{\phi_p\}_{p=1}^N,\{\nabla \phi_p\}_{p=1}^N\right)$.
\end{theorem}
\begin{proof}\label{Proof surface force equivalent}
Since $e_F$ only depends on $\{\phi_p\}_{p=1}^N$ and $\{\nabla \phi_p\}_{p=1}^N$, we have
\begin{equation}\nonumber
\xi_p=\frac{\delta E_F}{\delta \phi_p}=\frac{\partial e_F}{\partial \phi_p}-\nabla \cdot \frac{\partial e_F}{\partial (\nabla \phi_p)},
\quad
\nabla e_F=\sum_{p=1}^N \left( \frac{\partial e_F}{\partial \phi_p} \nabla \phi_p + \frac{\partial e_F}{\partial (\nabla \phi_p)} \cdot \nabla \otimes (\nabla \phi_p ) \right).
\end{equation}
Therefore, 
\begin{eqnarray}\label{Eq surface force equivalent}\nonumber
\sum_{p=1}^N \nabla \cdot \left( -\beta \frac{\partial e_F}{\partial (\nabla \phi_p)} \otimes \nabla \phi_p\right)
=
-\beta \sum_{p=1}^N \left( \nabla \cdot \frac{\partial e_F}{\partial (\nabla \phi_p)} \nabla \phi_p + \frac{\partial e_F}{\partial (\nabla \phi_p)} \cdot \nabla \otimes (\nabla \phi_p) \right)\\
\nonumber
=
\beta \sum_{p=1}^N \underbrace{\left(\frac{\partial e_F}{\partial \phi_p} 
-
\nabla \cdot \frac{\partial e_F}{\partial (\nabla \phi_p)} \right)}_{\xi_p} \nabla \phi_p
-
\beta \underbrace{\sum_{p=1}^N \left(
\frac{\partial e_F}{\partial \phi_p} \nabla \phi_p 
+
\frac{\partial e_F}{\partial (\nabla \phi_p)} \cdot \nabla \otimes (\nabla \phi_p) 
\right)}_{\nabla e_F}\\
\nonumber
=
\sum_{p=1}^N \beta \xi_p \nabla \phi_p
-
\beta \nabla e_F
=
\mathbf{f}_s
-
\nabla (\beta e_F).
\end{eqnarray}
The gradient term $\nabla (\beta e_F)$ can be absorbed into the pressure gradient, and, as a result, the surface force in Theorem \ref{Theorem energy law} is equivalent to $\sum_{p=1}^N \nabla \cdot \left( -\beta \frac{\partial e_F}{\partial (\nabla \phi_p)} \otimes \nabla \phi_p\right)$.
\end{proof}

\textit{\textbf{Remark:}
\begin{itemize}
    \item 
    Since the momentum equation Eq.(\ref{Eq Momentum}) from the consistency conditions enjoys Theorem \ref{Theorem Kinetic energy} (for kinetic energy) regardless of the Phase-Field equation, the multiphase flow model satisfying the second law of thermodynamics (or Theorem \ref{Theorem energy law}) only relies on the property of the Phase-Field equation. In other words, as long as the Phase-Field equation on its own, i.e., without convection, follows energy dissipation, Theorem \ref{Theorem energy law} will be true. Failure of Theorem \ref{Theorem energy law} is totally irrelevant to the momentum equation Eq.(\ref{Eq Momentum}) obtained from the consistency of mass conservation and the consistency of mass and momentum transport.
    \item 
    The examples of the diffusion flux in Theorem \ref{Theorem energy law} need to be further refined to meet the requirements of the diffusion fluxes already mentioned below Eq.(\ref{Eq Volume fraction contrast equation conservative}) in Section \ref{Sec Basic definitions}, i.e., $\sum_{p=1}^N \mathbf{J}_p=\mathbf{0}$ and those in Theorem \ref{Theorem RC Phase-Field} for the consistency of reduction. This is not a simple task and requires careful design of the mobility ($M_p$ or $M_{p,q}$) as well as the free energy $E_F$ which determines $\xi_p$ in Theorem \ref{Theorem energy law}. We are not attempting to propose general solutions to this question, and related discussions are referred to \citep{BoyerLapuerta2006,Boyeretal2010,BoyerMinjeaud2014,LeeKim2015,Dong2017,Dong2018,WuXu2017} and the references therein. In the present study, we focus on coupling any given Phase-Field equation to the hydrodynamics, and on illustrating the significance of implementing the consistency conditions during the coupling.
    \item 
    Both Corollary \ref{Corollary Kinetic energy conservation} and Theorem \ref{Theorem energy law} require proper boundary conditions (BCs), such as the periodic BC for all the variables, homogeneous Neumann BC for the order parameters, no-flux BC for the diffusion flux in the Phase-Field equation, and no-slip BC for the velocity, to result in zero boundary integrals.
\end{itemize}
}

\subsection{The consistency of reduction}\label{Sec Momentum reduction}
Here, we investigate the \textit{consistency of reduction} of the multiphase flow model using the momentum equation Eq.(\ref{Eq Momentum}) derived from the \textit{consistency of mass conservation} and the \textit{consistency of mass and momentum transport}. Since we have assumed that the Phase-Field equation Eq.(\ref{Eq Volume fraction contrast equation conservative}) satisfies the \textit{consistency of reduction}, see Theorem \ref{Theorem RC Phase-Field} in Section \ref{Sec Basic definitions}, we particularly focus on the momentum equation Eq.(\ref{Eq Momentum}).
\begin{theorem}\label{Theorem consistency of reduction momentum}
If the density and viscosity of the fluid mixture, the consistent mass flux, and the surface force satisfy the consistency of reduction, then the momentum equation Eq.(\ref{Eq Momentum}) is also reduction consistent.
\end{theorem}
\begin{proof}
Since only the density and viscosity of the fluid mixture, the consistent mass flux, and the surface force are related to the number of phases in the momentum equation Eq.(\ref{Eq Momentum}), if all of them are reduction consistent, then the momentum equation Eq.(\ref{Eq Momentum}) is reduction consistent as well.
\end{proof}
\begin{corollary}\label{Corollary consistency of reduction momentum bulk-phase}
If the momentum equation Eq.(\ref{Eq Momentum}) satisfies the consistency of reduction, then it recovers inside each bulk-phase region the single-phase Navier-Stokes equation with the corresponding density and viscosity of that phase, e.g., inside the bulk-phase region of Phase $p$, the momentum equation Eq.(\ref{Eq Momentum}) becomes
\begin{equation}\nonumber
\frac{\partial (\rho_p \mathbf{u} )}{\partial t}
+
\nabla \cdot \left( \rho_p \mathbf{u} \otimes \mathbf{u} \right)
=
-
\nabla P
+
\mu_p \nabla^2 \mathbf{u}
+
\rho_p \mathbf{g}.
\end{equation}
\end{corollary}

Although the Navier-Stokes equation Eq.(\ref{Eq NS Intro}) is not generally valid to govern the motion of the fluid mixture in the entire domain, as discussed in Section \ref{Sec Momentum}, it still locally governs the motion of the single-phase fluids inside individual bulk-phase regions. It should be noted that in the multiphase flow problems, most of the domain is occupied by bulk-phase regions inside which there is a sing fluid phase. Therefore, it is critical for the momentum equation Eq.(\ref{Eq Momentum}) to recover the correct single-phase dynamics, highlighted in Corollary \ref{Corollary consistency of reduction momentum bulk-phase}, in those regions. This emphasizes the significance of satisfying the \textit{consistency of reduction}.

From Theorem \ref{Theorem consistency of reduction momentum}, we need to further demonstrate the \textit{consistency of reduction} of the density Eq.(\ref{Eq Mixture density}) and viscosity Eq.(\ref{Eq Mixture viscosity}) of the fluid mixture, the consistent mass flux in Theorem \ref{Theorem Consistency of mass conservation}, and the surface force in Theorem \ref{Theorem energy law}. Based on the definition of the \textit{consistency of reduction} in Section \ref{Sec Basic definitions}, we need to investigate how a multiphase formulation behaves when some of the phases are absent. The upcoming analyses are not related to which phase is chosen to be absent, and thus the same conclusion will be drawn by arbitrarily choosing an absent phase among the $N$ $(N \geqslant 2)$ phases. Besides, the analyses are repeatable. If the analyses are performed $(N-M)$ times $(1 \leqslant M \leqslant N-1)$, they produce results of $(N-M)$ absent phases. Therefore, we only need to consider the case where only a single phase is absent. Without loss of generality and for a clear presentation, we consider a $N$-phase case $(N \geqslant 2)$ where the last phase, i.e., Phase $N$, doesn't appear, i.e., $\phi_N \equiv -1$. If a $N$-phase formulation recovers the corresponding $(N-1)$-phase formulation excluding the contribution from absent Phase $N$, then that multiphase formulation is reduction consistent.

It is obvious that the density and viscosity of the fluid mixture in Eq.(\ref{Eq Mixture density}) and Eq.(\ref{Eq Mixture viscosity}), respectively, are reduction consistent. We pay attention to the consistent mass flux in Theorem \ref{Theorem Consistency of mass conservation} and the surface force in Theorem \ref{Theorem energy law} in the following, although some of their derivations have considered the \textit{consistency of reduction}.
 
\begin{theorem}\label{Theorem consistency of reduction mass flux}
If the Phase-Field equation satisfies the consistency of reduction, then the consistent mass flux in Theorem \ref{Theorem Consistency of mass conservation} is also reduction consistent.
\end{theorem}
\begin{proof}\label{Proof consistency of reduction mass flux}
Provided that Phase $N$ is absent, i.e., $\phi_N \equiv -1$, and recall that the Phase-Field flux includes both the convection and diffusion fluxes in the Phase-Field equation, i.e., $\mathbf{m}_{\phi_p}=\mathbf{u}\phi_p-\mathbf{J}_p$, the consistent mass flux in Theorem \ref{Theorem Consistency of mass conservation} becomes:
\begin{eqnarray}\nonumber
\mathbf{m}
=\sum_{p=1}^N \frac{\rho_p}{2} (\mathbf{u}+\mathbf{m}_{\phi_p})
=
\sum_{p=1}^{N-1} \frac{\rho_p}{2} (\mathbf{u}+\mathbf{u}\phi_p-\mathbf{J}_p)
+
\frac{\rho_N}{2} (\mathbf{u}+\mathbf{u}\underbrace{\phi_N}_{-1}-\underbrace{\mathbf{J}_N}_{\mathbf{0}})\\
\nonumber
=\sum_{p=1}^{N-1} \frac{\rho_p}{2} (\mathbf{u}+\mathbf{u}\phi_p-\mathbf{J}_p)
=\sum_{p=1}^{N-1} \frac{\rho_p}{2} (\mathbf{u}+\mathbf{m}_{\phi_p}).
\end{eqnarray}
Since the Phase-Field equation is reduction consistent (Theorem \ref{Theorem RC Phase-Field}), the contribution of Phase $N$ to $\{\mathbf{J}_p\}_{p=1}^{N-1}$ is excluded. As a result, the consistent mass flux in Theorem \ref{Theorem Consistency of mass conservation} recovers its $(N-1)$-phase formulation excluding the contribution from absent Phase $N$, and therefore it is reduction consistent.
\end{proof}

\begin{theorem}\label{Theorem consistency of reduction surface force}
If the Phase-Field equation with the diffusion flux in Theorem \ref{Theorem energy law} satisfies the consistency of reduction, then the surface force in Theorem \ref{Theorem energy law} is also reduction consistent.
\end{theorem}
\begin{proof}\label{Proof consistency of reduction surface force}
Provided that Phase $N$ is absent, i.e., $\phi_N \equiv -1$, the contribution of Phase $N$ to $\{\mathbf{J}_p\}_{p=1}^{N-1}$ in Theorem~\ref{Theorem energy law} and therefore $\{\xi_p\}_{p=1}^{N-1}$ is excluded, since the Phase-Field equation is reduction consistent (Theorem~\ref{Theorem RC Phase-Field}). The surface force in Theorem \ref{Theorem energy law} becomes
\begin{equation}\nonumber
\mathbf{f}_s
=\sum_{p=1}^N \beta \xi_p \nabla \phi_p
=\sum_{p=1}^{N-1} \beta \xi_p \nabla \phi_p + \beta \xi_N \nabla \underbrace{\phi_N}_{-1}
=\sum_{p=1}^{N-1} \beta \xi_p \nabla \phi_p.
\end{equation}
 As a result, the surface force in Theorem \ref{Theorem energy law} recovers its $(N-1)$-phase formulation excluding the contribution from absent Phase $N$, and therefore it is reduction consistent.
\end{proof}

\textit{\textbf{Remark}:
\begin{itemize}
    \item 
    As illustrated in Theorem \ref{Theorem consistency of reduction momentum}, Theorem \ref{Theorem consistency of reduction mass flux}, and Theorem \ref{Theorem consistency of reduction surface force}, after implementing the consistency of mass conservation and the consistency of mass and momentum transport to obtain the momentum equation Eq.(\ref{Eq Momentum}), the consistency of reduction of the multiphase flow model only relies on the property of the Phase-Field equation. In other words, as long as the Phase-Field equation follows the consistency of reduction, the same will be true for the momentum equation Eq.(\ref{Eq Momentum}) and for the multiphase flow model. Using Eq.(\ref{Eq Momentum}) produces a stronger reliance than using the Navier-Stokes equation Eq.(\ref{Eq NS Intro}) in the sense that the consistent mass flux includes the diffusion flux of the Phase-Field equation.
    \item
    It should be noted that the viscous force, i.e., $\nabla \cdot \left[\mu (\nabla \mathbf{u}+\nabla \mathbf{u} ^T) \right]$, becomes $\mu \nabla^2 \mathbf{u}$ when it reduces from the $N$-phase ($N \geqslant 2$) to single-phase case, as indicated in Corollary \ref{Corollary consistency of reduction momentum bulk-phase}, due to $\nabla \cdot (\nabla \mathbf{u}^T)=\nabla (\nabla \cdot \mathbf{u})=\mathbf{0}$ in the single-phase case.
\end{itemize}
}

\subsection{The present consistent and conservative multiphase flow model}\label{Sec Model summary}
So far, we have implemented the \textit{consistency of mass conservation} and the \textit{consistency of mass and momentum transport} to obtain the consistent mass flux Eq.(\ref{Eq Mass flux}) and the momentum equation Eq.(\ref{Eq Momentum}) in Section~\ref{Sec Mass} and Section \ref{Sec Momentum}, respectively, based on a generic Phase-Field equation Eq.(\ref{Eq Volume fraction contrast equation conservative}), and have illustrated the critical role played by these two consistency conditions on the Galilean invariance of the momentum equation and the conservation of the kinetic energy. We have also indicated in Section \ref{Sec Energy} and Section \ref{Sec Momentum reduction} that these two consistency conditions provide the optimal coupling between the Phase-Field equation and the hydrodynamics such that the second law of thermodynamics and the \textit{consistency of reduction} of the multiphase flow model only depend on the corresponding properties of the Phase-Field equation. 

To complete the present consistent and conservative multiphase flow model, we need to specify the Phase-Field equation. Here, we select the Phase-Field equation developed in \citep{Dong2018} because it is reduction consistent and additionally follows energy dissipation. The diffusion flux of this Phase-Field equation is
\begin{equation}\label{Eq Volume fraction contrast equation}
\mathbf{J}_p
=
\sum_{q=1}^{N} M_{p,q} \nabla \xi_q,
\quad
1 \leqslant p \leqslant N,
\end{equation}
where
\begin{equation} \label{Eq Mobility Phi}
M_{p,q}=\left\{
\begin{array}{ll}
- M_0 (1+\phi_p) (1+\phi_q), \quad p \neq q\\
M_0 (1+\phi_p) (1-\phi_q), \quad p=q\\
\end{array}
\right.,
\quad
1 \leqslant p,q \leqslant N,
\end{equation}
is the mobility between Phases $p$ and $q$, and 
\begin{equation}\label{Eq Chemical potential Phi}
\xi_{p}
=
\sum_{q=1}^{N}
\lambda_{p,q} \left[
\frac{1}{\eta^2} \left( g'_1(\phi_p)-g'_2(\phi_p+\phi_q) \right)
+
\nabla^2 \phi_q
\right],
\quad 
1 \leqslant p \leqslant N,
\end{equation}
is the chemical potential of Phase $p$. $M_0$ in Eq.(\ref{Eq Mobility Phi}) is a positive constant. Thanks to Eq.(\ref{Eq Summation of volume fraction contrasts}), we have $\sum_{p=1}^N M_{p,q}=0$ and therefore $\sum_{p=1}^N \mathbf{J}_p=\mathbf{0}$. 
The mixing energy densities $\{\lambda_{p,q}\}_{p,q=1}^N$ in Eq.(\ref{Eq Chemical potential Phi}) are
\begin{equation} \label{Eq Mixing energy density Phi}
\lambda_{p,q}=\frac{3}{2\sqrt{2}} \sigma_{p,q} \eta,
\quad
1 \leqslant p,q \leqslant N,
\end{equation}
depending on the pairwise interfacial tension $\sigma_{p,q}$ and on the interface thickness $\eta$. $\{\lambda_{p,q}\}_{p,q=1}^N$ are symmetric and have a zero diagonal. The potential functions in Eq.(\ref{Eq Chemical potential Phi}) are
\begin{equation} \label{Eq g1 and g2}
g_1(\phi)=\frac{1}{4} (1-\phi^2)^2,
\quad
g_2(\phi)=\frac{1}{4} \phi^2 (\phi+2)^2,
\end{equation}
and $g'_1(\phi)$ and $g'_2(\phi)$ are the derivatives of $g_1(\phi)$ and $g_2(\phi)$ with respect to $\phi$, respectively. Properties of $g'_2(\phi-1)=g'_1(\phi)$ and $M_{p,q}|_{\phi_p\ \mathrm{or}\ \phi_q=-1}=0$ contribute to the \textit{consistency of reduction} of the Phase-Field equation, and more details should refer to \citep{Dong2018}.

The free energy density of this Phase-Field equation is
\begin{equation} \label{Eq Free energy density Phi}
e_F=\sum_{p,q=1}^{N} \frac{\lambda_{p,q}}{2} \left[
\frac{1}{\eta^2} ( g_1(\phi_p)+g_1(\phi_q)-g_2(\phi_p+\phi_q) 
)
-\nabla \phi_p \cdot \nabla \phi_q
\right],
\end{equation}
which relates the chemical potentials to $\{\xi_p=\delta \int_\Omega e_F d\Omega/ \delta \phi_p\}_{p=1}^N$.
Implementing Theorem \ref{Theorem energy law}, we obtain the energy law of the present multiphase flow model
\begin{equation}\label{Eq Energy law Phi}
\frac{d}{d t} \int_\Omega (e_K+\frac{1}{2}e_F) d\Omega
=
-\frac{1}{2} \int_\Omega \mu (\nabla \mathbf{u}+\nabla \mathbf{u}^T): (\nabla \mathbf{u}+\nabla \mathbf{u}^T) d\Omega
-\frac{1}{2} \int_\Omega \sum_{p,q=1}^N M_{p,q} \nabla \xi_p \cdot \nabla \xi_q d\Omega.
\end{equation}
Here, $M_{p,q}$, $\xi_p$, and $e_F$ are defined in Eq.(\ref{Eq Mobility Phi}), Eq.(\ref{Eq Chemical potential Phi}), and Eq.(\ref{Eq Free energy density Phi}), respectively. Since $M_{p,q}$ is symmetric positive semi-definite, the dissipation from the Phase-Field equation, i.e., the second term on the right-hand side of Eq.(\ref{Eq Energy law Phi}), is non-negative. The corresponding surface force is  
\begin{equation} \label{Eq Surface force Phi}
\mathbf{f}_s=\frac{1}{2} \sum_{p=1}^{N}  \xi_p \nabla  \phi_p.
\end{equation}
This surface force is equivalent to a conservative form $\frac{1}{2}\sum_{p,q=1}^N\nabla \cdot (\lambda_{p,q} \nabla \phi_p \otimes \nabla \phi_q)$ from Theorem \ref{Theorem surface force equivalent}, since the free energy density in Eq.(\ref{Eq Free energy density Phi}) only depends on the order parameters and the gradients of the order parameters. Compared to the conservative form, the surface force in Eq.(\ref{Eq Surface force Phi}) is more convenient to be numerically implemented. First, the number of terms in Eq.(\ref{Eq Surface force Phi}) doesn't change with problem dimensions. Second, it doesn't need to evaluate the mixed derivatives. The only second derivative comes from the Laplace operator in $\xi_p$, which can be conveniently computed in any dimensions. Lastly, writing the surface force in a gradient form, i.e., Eq.(\ref{Eq Surface force Phi}), the balanced-force algorithm \citep{Francoisetal2006,Huangetal2020} can be applied to reduce the spurious current caused by the numerical force imbalance.

The contact angle boundary condition for the $N$-phase flows, developed in \citep{Dong2017}, i.e.,
\begin{equation} \label{Eq Contact angle boundary condition Phi}
\mathbf{n} \cdot \nabla \phi_p
=
\sum_{q=1}^{N} \zeta_{p,q} \frac{1+\phi_p}{2}\frac{1+\phi_q}{2},
\quad
1 \leqslant p \leqslant N,
\end{equation}
where
\begin{equation} \label{Eq Zeta Phi}
\zeta_{p,q}=\frac{2\sqrt{2}}{\eta} \cos(\theta_{p,q}^W),
\quad
1 \leqslant p,q \leqslant N,
\end{equation}
is implemented when effects of pairwise contact angles $\{\theta_{p,q}^W\}_{p,q=1}^N$ at a wall boundary are considered.

In summary, the present multiphase flow model includes the Phase-Field equation Eq.(\ref{Eq Volume fraction contrast equation conservative}) with the diffusion flux in Eq.(\ref{Eq Volume fraction contrast equation}) from \citep{Dong2018} and the momentum equation Eq.(\ref{Eq Momentum}), along with the density and viscosity of the fluid mixture in Eq.(\ref{Eq Mixture density}) and Eq.(\ref{Eq Mixture viscosity}), respectively, the consistent mass flux Eq.(\ref{Eq Mass flux}), the surface force Eq.(\ref{Eq Surface force Phi}), and the divergence-free condition Eq.(\ref{Eq Divergence free}). The present multiphase flow model is consistent because it satisfies the \textit{consistency of mass conservation} and the \textit{consistency of mass and momentum transport}, see Section \ref{Sec Mass} and Section \ref{Sec Momentum}. It additionally satisfies the \textit{consistency of reduction} because this is the case for the selected Phase-Field equation, see Section \ref{Sec Momentum reduction}. The present multiphase flow model is conservative because the Phase-Field equation Eq.(\ref{Eq Volume fraction contrast equation conservative}) conserves the order parameters and therefore the mass of individual phases. The momentum is conserved, even including effects of interfacial tensions, see Eq.(\ref{Eq Momentum}) and thanks to Theorem \ref{Theorem surface force equivalent}. Moreover, the present model guarantees that the summation of the order parameters follows Eq.(\ref{Eq Summation of volume fraction contrasts}) due to $\sum_{p=1}^N \mathbf{J}_p=\mathbf{0}$ and $\nabla \cdot \mathbf{u}=0$. Besides, the present model is Galilean invariant, see Theorem~\ref{Theorem Galilean invariance}, and enjoys the energy law in Eq.(\ref{Eq Energy law Phi}).

One may choose another Phase-Field equation, such as those in \citep{BoyerLapuerta2006,Boyeretal2010,Kim2007,BoyerMinjeaud2014,Kim2009,LeeKim2015,Kim2012,ZhangWang2016,Zhangetal2016,Dong2014,Dong2015,Dong2017}, and the only change is the diffusion flux or the Phase-Field flux of the Phase-Field equation. Notice that those Phase-Field equations may not necessarily satisfy the \textit{consistency of reduction}. Implementing the present theory and formulations to the conservative Allen-Cahn equation is reported in our other works \citep{Huangetal2020CAC,Huangetal2020B}.

\subsection{Physical interpretation of the consistency conditions and alternative models}\label{Sec Physical interpretation}
Here, we provide the physical interpretation to the consistency conditions and their formulations, and illustrate their validity in alternative multiphase flow models that have a non-divergence-free velocity. We first justify the \textit{consistency of mass conservation} and the \textit{consistency of mass and momentum transport} using the control volume analysis. The Reynolds transport theorem (Lemma \ref{Lemma Reynolds}) and Divergence (Gauss's) theorem (Lemma \ref{Lemma Divergence}) \citep{Leal2007} are to be implemented. Given an arbitrary control volume $\Omega(t)$ whose boundary is $\partial \Omega(t)$. The velocity and outward unit normal on $\partial \Omega(t)$ are $\mathbf{u}_{\partial \Omega(t)}$ and $\mathbf{n}$, respectively. Then, the Reynolds transport theorem and Divergence theorem on a scalar (vector or tensor) function $\mathcal{F}$ are written as
\begin{lemma}[Reynolds transport theorem]\label{Lemma Reynolds}
\begin{equation}\label{Eq Reynolds transport theorem}\nonumber
\frac{d}{dt} \int_{\Omega(t)} \mathcal{F} d\Omega
=
\int_{\Omega(t)} \frac{\partial \mathcal{F}}{\partial t} d\Omega
+
\int_{\partial \Omega(t)} \mathcal{F} (\mathbf{n} \cdot \mathbf{u}_{\partial \Omega(t)}) d\Gamma.
\end{equation}
\end{lemma}
\begin{lemma}[Divergence (Gauss's) theorem]\label{Lemma Divergence}
\begin{equation}\label{Eq Divergence theorem}\nonumber
\int_{\Omega(t)} \nabla \cdot \mathcal{F} d\Omega
=
\int_{\partial \Omega(t)} \mathbf{n} \cdot \mathcal{F} d\Gamma.
\end{equation}
\end{lemma}

The mass, momentum, and force in $\Omega(t)$ are
\begin{equation}\label{Eq CV}
\int_{\Omega(t)} \rho d\Omega,
\quad
\int_{\Omega(t)} \rho \mathbf{u} d\Omega,
\quad
\int_{\Omega(t)} \mathbf{f} d\Omega.
\end{equation}
Notice that forces acting on $\partial \Omega(t)$ have been included in $\mathbf{f}$ after using the Divergence theorem (Lemma \ref{Lemma Divergence}). Suppose the mass flux on $\partial \Omega(t)$ is $\mathbf{m}$, then the mass and momentum leaving the control volume boundary per unit time are
\begin{eqnarray}\label{Eq CVB}
\int_{\partial \Omega(t)} \mathbf{n} \cdot \mathbf{m} d\Gamma
-
\int_{\partial \Omega(t)} \rho \mathbf{n} \cdot \mathbf{u}_{\partial \Omega(t)} d\Gamma,\\
\nonumber
\int_{\partial \Omega(t)} (\mathbf{n} \cdot \mathbf{m}) \mathbf{u} d\Gamma
-
\int_{\partial \Omega(t)} (\rho \mathbf{n} \cdot \mathbf{u}_{\partial \Omega(t)}) \mathbf{u} d\Gamma.
\end{eqnarray}
Provided Eq.(\ref{Eq CV}) and Eq.(\ref{Eq CVB}), we obtain the mass and momentum balances of $\Omega(t)$:
\begin{eqnarray}\label{Eq CV Mass Momentum}
\frac{d}{dt} \int_{\Omega(t)} \rho d\Omega
=
-\left[
\int_{\partial \Omega(t)} \mathbf{n} \cdot \mathbf{m} d\Gamma
-
\int_{\partial \Omega(t)} \rho \mathbf{n} \cdot \mathbf{u}_{\partial \Omega(t)} d\Gamma
\right],\\
\nonumber
\frac{d}{dt} \int_{\Omega(t)} \rho \mathbf{u} d\Omega
=
-\left[
\int_{\partial \Omega(t)} (\mathbf{n} \cdot \mathbf{m}) \mathbf{u} d\Gamma
-
\int_{\partial \Omega(t)} (\rho \mathbf{n} \cdot \mathbf{u}_{\partial \Omega(t)}) \mathbf{u} d\Gamma
\right]
+
\int_{\Omega(t)} \mathbf{f} d\Omega.
\end{eqnarray}
After implementing the Reynolds transport theorem (Lemma \ref{Lemma Reynolds}) and then the Divergence theorem (Lemma \ref{Lemma Divergence}) to Eq.(\ref{Eq CV Mass Momentum}), we obtain:
\begin{eqnarray}\label{Eq CV Mass Momentum diff}
\int_{\Omega(t)} \left(
\frac{\partial \rho}{\partial t} 
+
\nabla \cdot \mathbf{m} 
\right) d\Omega 
=
0,\\
\nonumber
\int_{\Omega(t)} \left(
\frac{\partial (\rho \mathbf{u})}{\partial t} 
+
\nabla \cdot (\mathbf{m} \otimes \mathbf{u})
-
\mathbf{f} 
\right) d\Omega
=
0.
\end{eqnarray}
Notice that the control volume is arbitrary. Consequently, Eq.(\ref{Eq CV Mass Momentum diff}) is true everywhere, and we obtain the mass and momentum equations identical to Eq.(\ref{Eq Mass conservation}) and Eq.(\ref{Eq Momentum}), respectively, derived from the \textit{consistency of mass conservation} and \textit{consistency of mass and momentum transport}. So far, we only use the Reynolds transport theorem (Lemma \ref{Lemma Reynolds}) and Divergence theorem (Lemma \ref{Lemma Divergence}) without any further assumptions, like on the divergence of the velocity. 

Then, we need to determine the mass flux $\mathbf{m}$ in Eq.(\ref{Eq CV Mass Momentum diff}). For a clear presentation, we use the volume fractions $\{C_p\}_{p=1}^N$ as the order parameters in the following discussions. As pointed out in \citep{Bowen1976,GrayHassanizadeh1991,Achantaetal1994,BennethumCushman2002}, one has the flexibility to choose the governing equation for the volume fractions. Here, we consider the following two cases:
\begin{equation}\label{Eq Volume fraction 1}
\frac{\partial C_p}{\partial t}
+
\nabla \cdot (\mathbf{u} C_p)
=
\nabla \cdot \mathbf{J}_p^C,
\quad 1 \leqslant p \leqslant N,
\end{equation}
\begin{equation}\label{Eq Volume fraction 2}
\frac{\partial C_p}{\partial t}
+
\mathbf{u} \cdot \nabla C_p
=
\nabla \cdot \mathbf{J}_p^C,
\quad 1 \leqslant p \leqslant N.
\end{equation}
where $\{\mathbf{J}_p^C\}_{p=1}^N$ are the diffusion fluxes of the volume fractions. The only difference in Eq.(\ref{Eq Volume fraction 1}) and Eq.(\ref{Eq Volume fraction 2}) is in the convection term since we don't need a divergence-free velocity at the moment.
Next, we supplement the assumption that the densities of the fluid phases are constant, which is the case we consider in the present study as pointed out in Section \ref{Sec Basic definitions}. Under this assumption and recalling that $\rho$ is defined in Eq.(\ref{Eq Mixture density}), we can obtain the mass conservation equation algebraically, starting from Eq.(\ref{Eq Volume fraction 1}) or Eq.(\ref{Eq Volume fraction 2}). After appropriately arranging the terms, we can finalize the mass flux $\mathbf{m}$ in Eq.(\ref{Eq CV Mass Momentum diff}). This is exactly the procedure stated in the \textit{consistency of mass conservation}. After multiplying Eq.(\ref{Eq Volume fraction 1}) or Eq.(\ref{Eq Volume fraction 2}) by $\rho_p$, summing over $p$, and comparing to the first equation of Eq.(\ref{Eq CV Mass Momentum diff}) (or Eq.(\ref{Eq Mass conservation})), we have the following four options.
\begin{itemize}
    \item \textit{Option 1}: Provided Eq.(\ref{Eq Volume fraction 1}) and $\sum_{p=1}^N \mathbf{J}_p^C=\mathbf{0}$,
        \begin{equation}\nonumber
            \frac{\partial \rho}{\partial t}
            +
            \nabla \cdot \left( \sum_{p=1}^N \rho_p (\mathbf{u} C_p-\mathbf{J}_p^C) \right)
            =
            0
            \quad \mathrm{or} \quad
            \nabla \cdot \mathbf{u}=0 
            \quad \mathrm{and} \quad
            \mathbf{m}=\sum_{p=1}^N \rho_p (\mathbf{u} C_p-\mathbf{J}_p^C).
        \end{equation}
        To satisfy the second law of thermodynamics, the diffusion flux can be
        \begin{equation}\nonumber
            \mathbf{J}_p^C=M_p^C \nabla \xi_p^C
            \quad \mathrm{or} \quad
            \mathbf{J}_p^C= \sum_{q=1}^N M_{p,q}^C \nabla \xi_q^C,
            \quad
            1\leqslant p \leqslant N.
        \end{equation}
    \item \textit{Option 2}: Provided Eq.(\ref{Eq Volume fraction 1}) and $\sum_{p=1}^N \rho_p \mathbf{J}_p^C=\mathbf{0}$,
        \begin{equation}\nonumber
            \frac{\partial \rho}{\partial t}
            +
            \nabla \cdot (\rho \mathbf{u})
            =
            0
            \quad \mathrm{or} \quad 
            \nabla \cdot \mathbf{u}=\nabla \cdot \left(\sum_{p=1}^N \mathbf{J}_p^C \right)
            \quad \mathrm{and} \quad
            \mathbf{m}=\rho \mathbf{u}.
        \end{equation}
        To satisfy the second law of thermodynamics, the diffusion flux can be
        \begin{equation}\nonumber
            \mathbf{J}_p^C= M_p^C \nabla \left( \beta_C \xi_p^C + P \right)
            \quad \mathrm{or} \quad
            \mathbf{J}_p^C= \sum_{q=1}^N M_{p,q}^C \nabla \left( \beta_C \xi_q^C+P \right),
            \quad
            1\leqslant p \leqslant N.
        \end{equation}
    \item \textit{Option 3}: Provided Eq.(\ref{Eq Volume fraction 2}) and $\sum_{p=1}^N \mathbf{J}_p^C=\mathbf{0}$,
        \begin{equation}\nonumber
            \frac{\partial \rho}{\partial t}
            +
            \nabla \cdot \left( \sum_{p=1}^N \rho_p ( \mathbf{u} C_p -\mathbf{J}_p^C ) \right)
            =
            \rho \nabla \cdot \mathbf{u}
            \quad \mathrm{or} \quad
            \nabla \cdot \mathbf{u}=0
            \quad \mathrm{and} \quad
            \mathbf{m}=\sum_{p=1}^N \rho_p ( \mathbf{u} C_p - \mathbf{J}_p^C ).
        \end{equation}
        To satisfy the second law of thermodynamics, the diffusion flux can be
        \begin{equation}\nonumber
            \mathbf{J}_p^C=M_p^C \nabla \xi_p^C
            \quad \mathrm{or} \quad
            \mathbf{J}_p^C= \sum_{q=1}^N M_{p,q}^C \nabla \xi_q^C,
            \quad
            1\leqslant p \leqslant N.
        \end{equation}
    \item \textit{Option 4}: Provided Eq.(\ref{Eq Volume fraction 2}) and $\sum_{p=1}^N \mathbf{J}_p^C=\mathbf{0}$,
        \begin{equation}\nonumber
            \frac{\partial \rho}{\partial t}
            +
            \nabla \cdot (\rho \mathbf{u})
            =
            \rho \nabla \cdot \mathbf{u}
            +
            \nabla \cdot \sum_{p=1}^N \rho_p \mathbf{J}_p^C
            \quad \mathrm{or} \quad
            \nabla \cdot \mathbf{u}=-\frac{1}{\rho}\nabla \cdot \sum_{p=1}^N \rho_p \mathbf{J}_p^C
            \quad \mathrm{and} \quad
            \mathbf{m}=\rho \mathbf{u}
        \end{equation}
        To satisfy the second law of thermodynamics, the diffusion flux can be
        \begin{equation}\nonumber
            \mathbf{J}_p^C=M_p^C \nabla \left( \beta_C \xi_p^C - \frac{\rho_p}{\rho} P \right)
            \quad \mathrm{or} \quad
            \mathbf{J}_p^C= \sum_{q=1}^N M_{p,q}^C \nabla \left( \beta_C \xi_q^C - \frac{\rho_q}{\rho} P\right),
            \quad
            1\leqslant p \leqslant N.
        \end{equation}
\end{itemize}
Here, $\xi_p^C$($=\delta E_F/\delta C_p$) and $M_p^C$ (or $M_{p,q}^C$) are the chemical potential and mobility, respectively, based on the volume fractions, and $\beta_C$($>0$) is the coefficient for matching the two-phase formulation, like $\beta$ in Theorem~\ref{Theorem energy law}.

\textit{Option~1} is equivalent to the results in Section \ref{Sec Mass}. The only difference is that $\{\phi_p\}_{p=1}^N$ are the order parameters in Section \ref{Sec Mass} but they are $\{C_p\}_{p=1}^N$ here. We obtain a divergence-free velocity but the mass conservation equation is different from the sharp-interface one Eq.(\ref{Eq Mass Intro}), as pointed out in Introduction (Section~\ref{Sec Introduction}) and further demonstrated in Section~\ref{Sec Mass}.
\textit{Option~2}, on the other hand, recovers the form of the sharp-interface mass conservation equation Eq.(\ref{Eq Mass Intro}) but results in a non-divergence-free velocity, although all the densities of the fluid phases are constant. Similar formulations are presented in \citep{Shenetal2013} for the two-phase case and in \citep{LiWang2014} for the $N$-phase case. Notice that the mass flux in \textit{Option~2} can also be written as $\mathbf{m}=\sum_{p=1}^N \rho_p (\mathbf{u} C_p -\mathbf{J}_p^C)$, identical to the one in \textit{Option~1}, since \textit{Option~2} requires $\sum_{p=1}^N \rho_p \mathbf{J}_p^C=\mathbf{0}$.
\textit{Option~3} is identical to \textit{Option~1}, both of which have a divergence-free velocity. As a result, Eq.(\ref{Eq Volume fraction 1}) and Eq.(\ref{Eq Volume fraction 2}) have no difference since the convection terms of them are equal.
\textit{Option 4} is similar to \textit{Option 2} in the sense that it reproduces the form of the sharp-interface mass conservation equation Eq.(\ref{Eq Mass Intro}) with a non-divergence-free velocity. However, it does not conserve $(\rho_p C_p)$, or the mass of Phase $p$ ($1 \leqslant p \leqslant N$), due to $\nabla \cdot \mathbf{u} \neq 0$ in Eq.(\ref{Eq Volume fraction 2}), although it conserves $\rho$, or the mass of the fluid mixture. Therefore, \textit{Option 4} is not suitable for the present problem. 
The divergence of the velocity in \textit{Options 1 and 2} is obtained by summing Eq.(\ref{Eq Volume fraction 1}) over $p$ and notice that $\sum_{p=1}^N C_p=1$ as in Eq.(\ref{Eq Summation of volume fraction contrasts}). In order to satisfy the second law of thermodynamics, the diffusion fluxes in \textit{Options 2 and 4} include the pressure, which is used to compensate the work done by the pressure due to volume change, i.e., $P \nabla \cdot \mathbf{u}$, from the momentum equation Eq.(\ref{Eq Momentum}) (or the second equation of Eq.(\ref{Eq CV Mass Momentum diff})). 
Since \textit{Options 1 and 3} are identical and \textit{Option 4} is unphysical, we only discuss \textit{Options 1 and 2} in the following.

To further illustrate the physical meaning of the velocity, the mass flux, and the conditions of summing $\{\mathbf{J}_p^C\}_{p=1}^N$ in \textit{Options 1 and 2}, we use the mixture theory \citep{Drew1983,SunBeckermann2004}. The volume fractions is advected by the corresponding phase velocity:
\begin{equation}\label{Eq Volume fraction Mixture}
\frac{\partial C_p}{\partial t}
+
\mathbf{u}_p \cdot \nabla C_p
=
0,\quad
\nabla \cdot \mathbf{u}_p=0,\quad
1 \leqslant p \leqslant N.
\end{equation}
Here, $\{\mathbf{u}_p\}_{p=1}^N$ are the phase velocities and they are divergence-free since the density of each phase is a constant. From Eq.(\ref{Eq Volume fraction Mixture}), one can easily derive the mass conservation equation of the mixture:
\begin{equation}\label{Eq Mass Mixture}
\frac{\partial \rho}{\partial t}
+
\nabla \cdot \left(\sum_{p=1}^N \rho_p C_p \mathbf{u}_p\right)
=
0.
\end{equation}
After comparing Eq.(\ref{Eq Volume fraction Mixture}) to Eq.(\ref{Eq Volume fraction 1}), the diffusion flux should be interpreted as $\mathbf{J}_p^C=(\mathbf{u}-\mathbf{u}_p)C_p$, the volume transport by the relative velocity. Provided the condition in \textit{Option 1}, i.e., $\sum_{p=1}^N \mathbf{J}_p^C=\mathbf{0}$, we obtain $\mathbf{u}=\sum_{p=1}^N C_p \mathbf{u}_p$, which is the volume-averaged velocity. \textit{Option 1} tells that the volume-averaged velocity is divergence-free, which has been used in Sections \ref{Sec Basic definitions}-\ref{Sec Model summary}. An alternative way to show this is from its definition and Eq.(\ref{Eq Volume fraction Mixture}), i.e., $\nabla \cdot \mathbf{u}=\sum_{p=1}^N \nabla \cdot (\mathbf{u}_p C_p)=\sum_{p=1}^N \mathbf{u}_p \cdot \nabla C_p=-\sum_{p=1}^N \frac{\partial C_p}{\partial t}=0$, presented in \citep{Abelsetal2012} for the two-phase case and in \citep{Dong2018} for the $N$-phase case. The mass flux in \textit{Option 1} now is $\mathbf{m}=\sum_{p=1}^N \rho_p(\mathbf{u}C_p-\mathbf{J}_p^C)=\sum_{p=1}^N \rho_p C_p \mathbf{u}_p$, the same as the one in Eq.(\ref{Eq Mass Mixture}) from the mixture theory. Since \textit{Option 1} is equivalent to the case in Section \ref{Sec Mass}, the consistent mass flux in Theorem \ref{Theorem Consistency of mass conservation} is actually the mass flux from the mixture theory. The failure of using $\rho \mathbf{u}$ as the mass flux in the volume-averaged velocity case is because it only considers the mass transport by the averaged velocity but misses those by the relative velocity (or motion) of the phases.
\textit{Option 2} requires $\sum_{p=1}^N \rho_p \mathbf{J}_p^C=\mathbf{0}$, and therefore we have $\mathbf{u}=(\sum_{p=1}^N \rho_p C_p \mathbf{u}_p)/\rho$, which is the mass-averaged velocity. Therefore, the mass-averaged velocity is non-divergence-free although each phase is incompressible, which has also been realized in \citep{LowengrubTruskinovsky1998,Shenetal2013} for the two-phase case, \citep{KimLowengrub2005} for the three-phase case, and \citep{LiWang2014,Odenetal2010} for the $N$-phase case. The mass flux in \textit{Option 2} is $\mathbf{m}=\rho \mathbf{u}=\sum_{p=1}^N \rho_p C_p \mathbf{u}_p$, correspondingly, which is again identical to the mass flux in Eq.(\ref{Eq Mass Mixture}) from the mixture theory. As the mass flux in \textit{Option 2} can also be written as the one in \textit{Option 1}, i.e., $\mathbf{m}=\sum_{p=1}^N \rho_p(\mathbf{u}C_p-\mathbf{J}_p^C)$, there is no surprise that both \textit{Options 1 and 2} reproduce the mass flux of the mixture theory although the physical interpretation of the velocity in them is different.

Finally, we consider the \textit{consistency of reduction}. Provided Eq.(\ref{Eq Volume fraction 1}), the \textit{consistency of reduction} requires $\mathbf{J}_p^C|_{C_p \equiv 0}=\mathbf{0}$, see Theorem \ref{Theorem RC Phase-Field}. Due to $\sum_{p=1}^N \mathbf{J}_p^C=\mathbf{0}$ in \textit{Option 1} or to $\sum_{p=1}^N \rho_p \mathbf{J}_p^C=\mathbf{0}$ in \textit{Option 2}, it can be inferred that $\mathbf{J}_p^C|_{C_p \equiv 1}=\mathbf{0}$. In other words, $\mathbf{J}_p^C$ is zero inside bulk phase regions. This can be interpreted again from the mixture theory where the diffusion flux is interpreted as $\mathbf{J}_p^C=(\mathbf{u}-\mathbf{u}_p)C_p$. Obviously, $\mathbf{J}_p^C$ vanishes when Phase $p$ is absent (or $C_p=0$). On the other hand, both the volume-averaged velocity in \textit{Option 1} and the mass-averaged velocity in \textit{Option 2} have the property: $\mathbf{u}=\mathbf{u_p}$ inside Phase $p$ (or $C_p=1$) from their definitions. As a result, $\mathbf{J}_p^C$ vanishes inside bulk-phase regions, deduced from the mixture theory. 

Below we summarize in this section the general formulations that are independent of the interpretation of the velocity:
\begin{eqnarray}\label{Eq General formulations}
\nabla \cdot \mathbf{u}=\nabla \cdot \left( \sum_{p=1}^N \mathbf{J}_p^C \right),
\quad
\mathbf{m}=\sum_{p=1}^N  \rho_p (C_p \mathbf{u} -\mathbf{J}_p^C),
\quad
\frac{\partial  \rho}{\partial t}
+
\nabla \cdot \mathbf{m}
=
0,
\quad
\frac{\partial (\rho \mathbf{u})}{\partial t} 
+
\nabla \cdot (\mathbf{m} \otimes \mathbf{u})
=
\mathbf{f}. 
\end{eqnarray}
The above formulations can be directly derived from the \textit{consistency of mass conservation} and the \textit{consistency of mass and momentum transport}, except that the divergence of the velocity is obtained by summing Eq.(\ref{Eq Volume fraction 1}) (or the Phase-Field equation specifically in this section). Once the diffusion flux is provided, the multiphase flow model is completed. It should be noted that the suggested diffusion fluxes in \textit{Options 1-4} only consider the second law of thermodynamics. Additional care has to be paid to satisfy the requirements, i.e., $\sum_{p=1}^N \mathbf{J}_p^C=\mathbf{0}$ in \textit{Option 1} or $\sum_{p=1}^N \rho_p \mathbf{J}_p^C=\mathbf{0}$ in \textit{Option 2}, and those in Theorem \ref{Theorem RC Phase-Field} for the \textit{consistency of reduction}. In the present study, we focus on the divergence-free volume-averaged velocity, i.e., \textit{Option 1}, and the multiphase flow model developed and analyzed in Sections \ref{Sec Mass}-\ref{Sec Model summary} is equivalent to the special case of Eq.(\ref{Eq General formulations}) under $\sum_{p=1}^N \mathbf{J}_p^C=\mathbf{0}$. We implement the diffusion flux in Eq.(\ref{Eq Volume fraction contrast equation}) from \citep{Dong2018} because it satisfies all the aforementioned requirements, see Section \ref{Sec Model summary} and more details in \citep{Dong2018}. There are several two-phase \citep{LowengrubTruskinovsky1998,Shenetal2013,GuoLin2015,Shenetal2020}, three-phase \citep{KimLowengrub2005}, and $N$-phase \citep{LiWang2014,Odenetal2010} models considering the mass-averaged velocity as presented in \textit{Option 2}, and these models are also called the ``quasi-incompressible'' models. Although part of the aforementioned requirements is considered, we did not find explicit discussions about whether the diffusion flux in those models meets all the required aspects simultaneously. Such discussions are outside the scope of the present study. Since the interpretation of the velocity only depends on the condition of summing the diffusion fluxes, one may expect Eq.(\ref{Eq General formulations}) works in other models using the velocity other than the volume-averaged or mass-averaged one. 

In summary, the physical insights of the consistency conditions and their formulations in the present study are provided using the control volume analysis and the mixture theory. The \textit{consistency of mass and momentum transport} is a direct consequence of the control volume analysis. Without priorly interpreting the velocity, one can straightforwardly obtain the mass flux as well as the divergence of the velocity following the \textit{consistency of mass conservation} and the condition of summing the diffusion fluxes. These formulations can be further physically explained with the mixture theory, and the obtained mass flux is always identical to the one from the mixture theory regardless of the interpretation of the velocity. The \textit{consistency of reduction} is supported by the mixture theory as well. We illustrate that these consistency conditions are general for both models using the volume-averaged and mass-averaged velocities, respectively, and probably for models using other averaged velocity, although we focus on the first case in the present study. In practice, these consistency conditions provide a ``shortcut'' of developing physical models that couple the Phase-Field equation to the hydrodynamics, without explicitly exploring their physical background. 

\section{Novel numerical techniques for the multiphase flows} \label{Sec Discretization}
It is important to preserve the physical properties of the multiphase flow model developed and analyzed in Section \ref{Sec Definitions and governing equations}, so that unphysical behaviors are not produced in numerical results. In this section, we first introduce the notations of discretization and then several novel numerical techniques that help to develop a consistent and conservative scheme.

\subsection{Notations of discretization} \label{Sec Notations of discritization}
\begin{figure}[!t]
	\centering
	\includegraphics[scale=0.4]{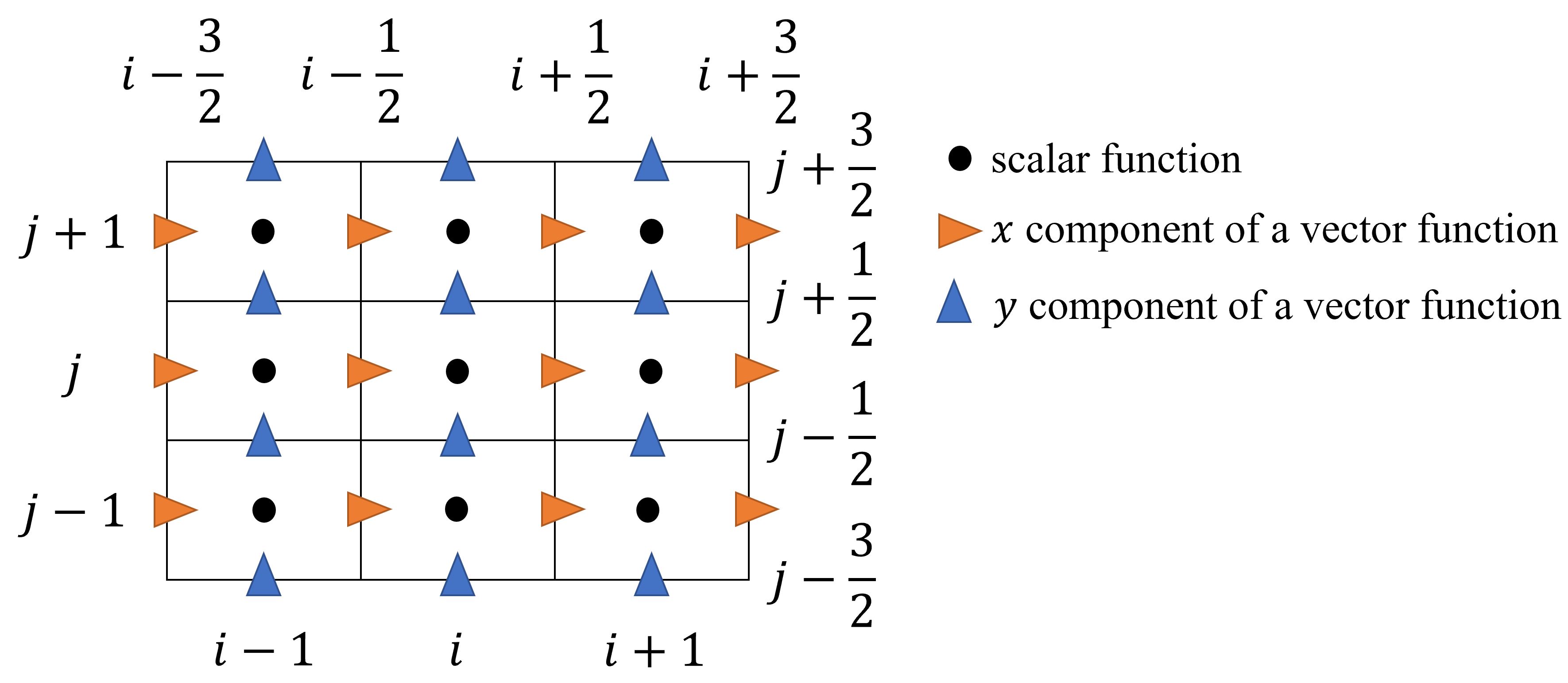}
	\caption{Schematic of the collocated grid. Scalar functions are stored at cell centers. Vector functions are stored at cell faces} \label{Fig Collocated grid}
\end{figure}

In the present study, we mostly follow the notations in \citep{Huangetal2020} to denote numerical operations. For a clear presentation, we consider two-dimensional cases, while extension to three dimensional problems is straightforward. We use the collocated grid arrangement, as shown in Fig.\ref{Fig Collocated grid}. Cell centers are denoted by $(x_i,y_j)$, while cell faces are denoted by $(x_{i+1/2},y_j)$ and $(x_i,y_{j+1/2})$. 
Nodal values of a scalar function $f$ are stored at the cell centers and are denoted by $f_{i,j}$ at $(x_i,y_j)$. 
Nodal values of a vector function $\mathbf{f}$ are stored at the cell faces, and are denoted by $f_{i+1/2,j}^x$ at $(x_{i+1/2},y_j)$ and $f_{i,j+1/2}^y$ at $(x_i,y_{j+1/2})$, where $f^x$ and $f^y$ are the $x$- and $y$-components of the vector function. It should be noted that gradient of a scalar function is also a vector function.
In the collocated grid arrangement, we have a cell-center velocity, whose individual components are considered as a scalar function at the cell centers. In addition, a cell-face velocity is stored at the cell faces, as a vector function. To distinguish from their continuous counterparts, discrete differential operators are denoted with superscript $\tilde{(\cdot)}$, e.g., $\tilde{\nabla}$ represents the discrete gradient operator. If $\tilde{(\cdot)}$ applies to a variable, e.g., $\tilde{f}$, it represents an interpolation from nodal points of the variable, i.e., $f$ in the example, to a desired location.
We use $f^n$ to denote the function value of $f$ at time $t^n$, and $f^{*,n}$ to denote an approximation of $f^n$ from the function values of $f$ at previous time levels. 
We use $\Delta x$ and $\Delta y$ to denote the cell/grid size, along the $x$ and $y$ axes, respectively, $\Delta \Omega=\Delta x \Delta y$ to denote the volume of the discrete cells, and $\Delta t$ to denote the time step. 

The time derivative is approximated by a time discretization scheme denoted by 
\begin{equation} \label{Eq Time discretization scheme}
\frac{\partial f}{\partial t}
\approx
\frac{\gamma_t f^{n+1}-\hat{f}}{\Delta t},
\end{equation} 
where $\gamma_t$ and $\hat{(\cdot)}$ are scheme-dependent coefficient and linear operator, respectively, with $\gamma_t=\hat{1}$. Notice that Eq.(\ref{Eq Time discretization scheme}) can conceptually represent various time discretization schemes, such as the Runge–Kutta and linear multistep methods.

We denote the linear interpolation along the $x$ axis at $(x_i,y_j)$ as
\begin{equation} \label{Eq Linear Interpolation}
\left[ \overline{f}^x \right]_{i,j}
=
\frac{1}{2}f_{i-1/2,j}+\frac{1}{2}f_{i+1/2,j}.
\end{equation}
The linear interpolation along the other axis can be defined in the same manner. Notice that $\tilde{f}$ can also denote but does not limit to the linear interpolation.

The cell-face gradient is defined at the cell faces and is approximated by the second-order central difference:
\begin{equation} \label{Eq Discrete Gradient Face}
[\tilde{\nabla}^x f]_{i+1/2,j}=\frac{f_{i+1,j}-f_{i,j}}{\Delta x},
\quad
[\tilde{\nabla}^y f]_{i,j+1/2}=\frac{f_{i,j+1}-f_{i,j}}{\Delta y},
\end{equation}
where $f$ is a scalar function defined at the cell centers. As a result, $\tilde{\nabla} f$ is a vector function defined at the cell faces. The cell-center gradient is linearly interpolated from the cell-face gradient:
\begin{equation} \label{Eq Discrete Gradient Center}
\left[\overline{\tilde{\nabla}f} \right]_{i,j}
=
\left( \left[\overline{\tilde{\nabla}^x f}^x \right]_{i,j},
\left[\overline{\tilde{\nabla}^y f}^y\right]_{i,j} \right)
\end{equation}

The discrete divergence operator is defined as
\begin{equation} \label{Eq Discrete Divergence}
[ \tilde{\nabla} \cdot \mathbf{f} ]_{i,j}
=
\frac{ 
	f_{i+1/2,j}^x
	-
	f_{i-1/2,j}^x
}{\Delta x}
+
\frac{
	f_{i,j+1/2}^y
	-
	f_{i,j-1/2}^y
}{\Delta y},
\end{equation} 
where $\mathbf{f}$ is a vector function defined at the cell faces. 
For the convection term, i.e., $\mathbf{f}=\mathbf{f}_1 \widetilde{f_2}$, $\mathbf{f}_1$ is a vector function defined at the cell faces, and $f_2$ is a scalar function defined at the cell centers. 
For the diffusion term, i.e., $\mathbf{f}= \widetilde{f_2} \tilde{\nabla} f_1$, both $f_1$ and $f_2$ are scalar functions defined at the cell centers.
For the so-called DGT term, i.e., $\mathbf{f}=\widetilde{f_2} (\tilde{\nabla}\widetilde{\mathbf{f}_1})^T$, $\mathbf{f}_1$ is a vector function defined at the cell faces and $f_2$ is a scalar function defined at the cell centers. A special interpolation to compute $\widetilde{\mathbf{f}_1}$ in the DGT term is developed in \citep{Huangetal2020} so that 
$\tilde{\nabla} \cdot (\tilde{\nabla} \widetilde{\mathbf{f}_1})^T
=
\overline{\tilde{\nabla} (\tilde{\nabla} \cdot \mathbf{f}_1)}$ is true. 
We refer interested readers to \citep{Huangetal2020} for detailed definitions of all the discrete operators and for treatments of boundary conditions.

In the present study, a discrete differential operator is called conservative if its summation over all its nodal points is zero in a periodic domain. Conservative operators help to achieve mass and momentum conservation.
\begin{lemma}\label{Lemma Conservation discrete gradient face}
The discrete gradient and divergence operators defined in Eq.(\ref{Eq Discrete Gradient Face}), Eq.(\ref{Eq Discrete Gradient Center}), and Eq.(\ref{Eq Discrete Divergence}) are conservative, i.e.,
\begin{eqnarray}\nonumber
\sum_{i,j} [\tilde{\nabla}^x f]_{i+1/2,j}=\sum_{i,j} [\tilde{\nabla}^y f]_{i,j+1/2}=0,\\
\nonumber
\sum_{i,j} \left[f_2 \overline{\left(\frac{1}{\overline{f_2}}\tilde{\nabla}f_1 \right)} \right]_{i,j}=\mathbf{0},\\
\nonumber
\sum_{i,j} [ \tilde{\nabla} \cdot \mathbf{f} ]_{i,j}=0.
\end{eqnarray}
\end{lemma}
\begin{proof}\label{Proof Conservation discrete gradient face}
For the gradient operators in Eq.(\ref{Eq Discrete Gradient Face}) and Eq.(\ref{Eq Discrete Gradient Center}), we only show their $x$-component, and the procedure is the same for the other components. 

For the cell-face gradient operator in Eq.(\ref{Eq Discrete Gradient Face}), we have
\begin{equation} \label{Eq Conservation discrete gradient face}\nonumber
\sum_{i,j} [\tilde{\nabla}^x f]_{i+1/2,j}
=
\sum_{j} \frac{\sum_{i} f_{i+1,j}-\sum_{i} f_{i,j}}{\Delta x}
=
\sum_{j} \frac{\sum_{i} f_{i,j}-\sum_{i} f_{i,j}}{\Delta x}
=
\sum_{j} 0
=
0.
\end{equation}

For the cell-center gradient operator in Eq.(\ref{Eq Discrete Gradient Center}), using the definition of the linear interpolation in Eq.(\ref{Eq Linear Interpolation}), we have
\begin{eqnarray}\nonumber
\sum_{i,j} \left[f_2 \overline{\left(\frac{1}{\overline{f_2}^x}\tilde{\nabla}^x f_1 \right)}^x \right]_{i,j}
=
\sum_{i,j}  
\left( 
\frac{[f_2]_{i,j}}{2} \left[\frac{1}{\overline{f_2}^x}\tilde{\nabla}^x f_1\right]_{i-1/2,j} 
+ 
\frac{[f_2]_{i,j}}{2} \left[\frac{1}{\overline{f_2}^x}\tilde{\nabla}^x f_1\right]_{i+1/2,j} 
\right)\\
\nonumber
=
\sum_{i,j}  
\frac{[f_2]_{i,j}+[f_2]_{i+1,j}}{2} \left[\frac{1}{\overline{f_2}^x}\tilde{\nabla}^x f_1\right]_{i+1/2,j}
=
\sum_{i,j}  
\left[\overline{f_2}^x\right]_{i+1/2,j} \left[\frac{1}{\overline{f_2}^x}\tilde{\nabla}^x f_1\right]_{i+1/2,j}
=
\sum_{i,j} \left[\tilde{\nabla}^x f_1\right]_{i+1/2,j}
=0.
\end{eqnarray}

For the divergence operator in Eq.(\ref{Eq Discrete Divergence}), we have
\begin{eqnarray} \label{Eq Conservation discrete divergence}\nonumber
\sum_{i,j} [\tilde{\nabla} \cdot \mathbf{f}]_{i,j}
=
\sum_{j} 
\frac{ 
	\sum_{i} f_{i+1/2,j}^x
	-
	\sum_{i} f_{i-1/2,j}^x
}{\Delta x}
+
\sum_{i} 
\frac{
	\sum_{j} f_{i,j+1/2}^y
	-
	\sum_{j} f_{i,j-1/2}^y
}{\Delta y} \\
\nonumber
=
\sum_{j} 
\frac{ 
	\sum_{i} f_{i+1/2,j}^x
	-
	\sum_{i} f_{i+1/2,j}^x
}{\Delta x}
+
\sum_{i} 
\frac{
	\sum_{j} f_{i,j+1/2}^y
	-
	\sum_{j} f_{i,j+1/2}^y
}{\Delta y}
=
\sum_{j} 0
+
\sum_{i} 0
=
0.
\end{eqnarray}
\end{proof}
\textit{\textbf{Remark}:
Lemma \ref{Lemma Conservation discrete gradient face} is true with a zero-gradient boundary condition for the gradient operators and with a zero-flux boundary condition for the divergence operator.
}

\subsection{Gradient-based phase selection procedure}\label{Sec Phase selection}
On the continuous level, the convection term in the Phase-Field equation Eq.(\ref{Eq Volume fraction contrast equation conservative}) has the following two properties: (i) $\sum_{p=1}^N \nabla \cdot (\mathbf{u} \phi_p)=(2-N)\nabla \cdot \mathbf{u}$ and (ii) $\nabla \cdot (\mathbf{u} \phi_p)=-\nabla \cdot \mathbf{u}$ provided $\phi_p=-1$ and $\nabla \phi_p=\mathbf{0}$. We discover that the discrete convection term can produce fictitious phases, local voids, or overfilling, due to violating the two properties above, which has not been realized in previous studies. The gradient-based phase selection procedure is developed to address those unphsyical behaviors.

The discrete convection term is denoted by $\{\tilde{\nabla} \cdot (\mathbf{u} \widetilde{\phi_p})\}_{p=1}^N$, where $\mathbf{u}$ is the cell-face velocity and $\tilde{\phi}$ represents interpolated values at the cell-faces from the nodal values of $\phi$ at the cell centers. The gradient-based phase selection procedure has the following steps at each cell face:
\begin{itemize}
    \item \textit{Step 1:} compute the convection fluxes $\{\mathbf{u} \widetilde{\phi_p}\}_{p=1}^N$ with any interpolation method, linear or non-linear,
    \item \textit{Step 2:} compute the normal gradients of the order parameters, i.e., $\{\tilde{\nabla}^n \phi_p\}_{p=1}^N=\{\tilde{\nabla}^x \phi_p\}_{p=1}^N$ at $(x_{i+1/2},y_j)$ and $\{\tilde{\nabla}^n \phi_p\}_{p=1}^N=\{\tilde{\nabla}^y \phi_p\}_{p=1}^N$ at $(x_i,y_{j+1/2})$,
    \item \textit{Step 3:} specify Phase $q$ such that $|\tilde{\nabla}^n \phi_q|$ is maximum among all the phases,
    \item \textit{Step 4:} correct the convection flux of Phase $q$ using
            \begin{equation} \label{Eq Convective flux correction}\nonumber
                \mathbf{u} \widetilde{\phi_q}= (2-N) \mathbf{u}  - \sum_{p=1,p\neq q}^{N} \mathbf{u} \widetilde{\phi_p}.
            \end{equation}
\end{itemize}
Then, one can compute the discrete convection term $\{\tilde{\nabla} \cdot (\mathbf{u} \widetilde{\phi_p})\}_{p=1}^N$ at all the cell centers. After implementing the gradient-based phase selection procedure, it is obvious that the discrete convection term satisfies property (i), i.e., $\sum_{p=1}^N \{\tilde{\nabla} \cdot (\mathbf{u} \widetilde{\phi_p})\}_{p=1}^N=(2-N) \tilde{\nabla} \cdot \mathbf{u}$. Suppose $\phi_p=-1$ in the stencil of the interpolation, therefore, we have $\widetilde{\phi_p}=-1$ and $\tilde{\nabla}^n \phi_p=0$. As a result, Phase $p$ will not be selected in \textit{Step~3} to perform the correction in \textit{Step~4}, and we achieve $\tilde{\nabla} \cdot (\mathbf{u} \widetilde{\phi_p})=-\tilde{\nabla} \cdot \mathbf{u}$, which is property (ii). The gradient-based phase selection procedure may also improve the robustness and accuracy of the discrete convection term because it corrects the interpolation of the order parameter having the largest local gradient. Usually, interpolation methods are less accurate and may produce oscillations where the gradient is large. It is worth mentioning that the gradient-based phase selection procedure works in any interpolation method to obtain $\{\widetilde{\phi_p}\}_{p=1}^N$ at the cell faces.
Since the convection term is treated explicitly in the present study, the CFL condition is a suitable guidance for numerical stability. Rigorous analyses of the effect of the gradient-based phase selection procedure on numerical stability are difficult, but our numerical test in Section \ref{Sec Advection} indicates that using a CFL number of $0.1$ is able to successfully solve an advection problem with a maximum density ratio of $10^9$. Implicit treatments to the convection term with the gradient-based phase selection procedure are possible but involved in practice.

Without implementing the gradient-based phase selection procedure, there is nothing to guarantee that $\sum_{p=1}^N \widetilde{\phi_p}=(2-N)$, see Eq.(\ref{Eq Summation of volume fraction contrasts}), even though $\sum_{p=1}^N \phi_p=(2-N)$ is true at every cell center, because the summation and interpolation may not be interchangeable, especially when the interpolation is non-linear. Failure of satisfying $\sum_{p=1}^N \tilde{\nabla} \cdot (\mathbf{u} \widetilde{\phi_p})=(2-N) \tilde{\nabla} \cdot \mathbf{u}$ results in the same failure of $\sum_{p=1}^N \phi_p=(2-N)$, see the proof of Theorem~\ref{Theorem Summation Phi} and the numerical studies in Section \ref{Sec Horizontal shear layer}. 

In many previous studies, e.g., in \citep{LeeKim2015,KimLee2017,Dong2014,Dong2018}, only the first $(N-1)$ phases are numerically solved from their Phase-Field equation, and the last phase is obtained algebraically from their summation, i.e., Eq.(\ref{Eq Summation of volume fraction contrasts}). This is equivalent to always correcting the convection flux of Phase $N$. Such a strategy can violate the \textit{consistency of reduction} and, as a result, generate fictitious phases. Suppose Phase $N$ is absent, i.e., $\phi_N \equiv -1$, we have $\sum_{p=1}^{N-1} \phi_p=(3-N)$ from Eq.(\ref{Eq Summation of volume fraction contrasts}). However, this is not guaranteed after the interpolation since $\{\widetilde{\phi_p}\}_{p=1}^{N-1}$ are computed independently. The discrete convection flux of Phase $N$ now is $\left[(2-N) \tilde{\nabla} \cdot \mathbf{u}-\sum_{p=1}^{N-1} \tilde{\nabla} \cdot (\mathbf{u} \widetilde{\phi_p})\right]=\tilde{\nabla} \cdot \left( (2-N - \sum_{p=1}^{N-1} \widetilde{\phi_p} ) \mathbf{u} \right)$ and may not be $-\tilde{\nabla} \cdot \mathbf{u}$. As a result, the convection term becomes a numerical source to produce Phase $N$. Therefore, the phase selection in \textit{Step~3} is essential to preserve the \textit{consistency of reduction}, see the proof of Theorem~\ref{Theorem consistency of reduction Phase-Field discrete} and the numerical studies in Section \ref{Sec Horizontal shear layer}.

\subsection{Discretization of the surface force} \label{Sec Discreteization of the surface force}
In this section, we consider the discretization of the surface force $\mathbf{f}_s$ defined in Eq.(\ref{Eq Surface force Phi}). Under the hydrostatic case, i.e., $\mathbf{u} \equiv \mathbf{0}$, the net force acting on the fluid mixture, i.e., the summation of the pressure gradient, the gravity, and the surface force, should be zero, from the momentum equation Eq.(\ref{Eq Momentum}).
The discretization of the surface force should reproduce this property on the discrete level as much as possible to avoid the spurious currents due to numerical force imbalance. We use $\mathbf{G}$ to denote the discretized net force per unit mass and $\mathbf{G}_s$ is the net force per unit mass excluding the pressure gradient, i.e.,
\begin{equation} \label{Eq G}
\mathbf{G}=-\frac{1}{\overline{\rho}} \tilde{\nabla} P + \mathbf{G}_s,
\quad
\mathbf{G}_s=\frac{1}{\overline{\rho}} \mathbf{f}_s + \mathbf{g},
\end{equation}
As defined in Section \ref{Sec Notations of discritization}, the discrete gradient operator is first computed at the cell faces and then interpolated to the cell centers, see Eq.(\ref{Eq Discrete Gradient Face}) and Eq.(\ref{Eq Discrete Gradient Center}). In order to achieve a good numerical balance between the surface force and the pressure gradient, the surface force should follow the same procedure, as pointed out in \citep{Francoisetal2006}.

In the following, we first introduce the balanced-force method, which is directly extended from its two-phase counterpart, and then develop a novel method, called the conservative method, which helps to conserve the momentum on the discrete level. To avoid repeated algebra, we focus on only the $x$-component of the surface force, and the other components are computed following the same manner. We will further numerically study the performances of these two methods on discrete force balance, see Section \ref{Sec Steady drops}, and on discrete momentum conservation, see Section \ref{Sec Horizontal shear layer}, in the multiphase flow problems.

\subsubsection{The balanced-force method} \label{Sec The balanced-force method}
The balanced-force method for the surface force in Eq.(\ref{Eq Surface force Phi}) reads
\begin{equation} \label{Eq The balanced-force method}
[f_s^x]_{i+1/2,j}=
\frac{1}{2} \sum_{p=1}^N \left[\overline{\xi_p}^x\right]_{i+1/2,j} [\tilde{\nabla}^x \phi_p]_{i+1/2,j}.
\end{equation}
The balanced-force method is directly extended from the one for the two-phase flows, e.g., in \citep{Francoisetal2006,Huangetal2020}, because the surface force in Eq.(\ref{Eq Surface force Phi}) has a form of a scalar times the gradient of another scalar, the same structure as the two-phase surface tension models, e.g., in \citep{Brackbilletal1992,Sussmanetal1994,Jacqmin1999}. The two-phase balanced-force method has been extensively studied and popularly used, e.g., in \citep{Francoisetal2006,Huangetal2020,Huangetal2020CAC,Mirjalili2019}.

\subsubsection{The conservative method} \label{Sec The conservative method}
The newly developed conservative method discretizes the following equivalent form of the surface force in Eq.(\ref{Eq Surface force Phi}), i.e.,
\begin{eqnarray}\nonumber
\mathbf{f}_s
=
\frac{1}{2} \sum_{p=1}^N \xi_p \nabla \phi_p
=
\frac{1}{2} \sum_{p,q=1}^N \lambda_{p,q} \left[ \frac{1}{\eta^2} \left(g'_1(\phi_p)-g'_2(\phi_p+\phi_q) \right)+\nabla^2 \phi_q \right] \nabla \phi_p \\
\nonumber
=
\frac{1}{2} \sum_{p,q=1}^N \lambda_{p,q} \left[ 
\frac{1}{\eta^2} \left(  
\nabla g_1(\phi_p) 
-
\frac{1}{2} \nabla g_2(\phi_p+\phi_q)  \right)
+  
\nabla^2 \phi_q \nabla \phi_p 
\right],
\end{eqnarray}
and the conservative method reads
\begin{eqnarray} \label{Eq The conservative method}
[f_s^x]_{i+1/2,j}
=
\frac{1}{2} \sum_{p,q=1}^N \lambda_{p,q} \left[ 
\frac{1}{\eta^2} \left(  
[\tilde{\nabla}^x g_1(\phi_p) ]_{i+1/2,j} 
-
\frac{1}{2} [\tilde{\nabla}^x g_2(\phi_p+\phi_q)]_{i+1/2,j}  \right) \right.\\
\nonumber
\left.
+  
\left[\overline{\tilde{\nabla} \cdot (\tilde{\nabla} \phi_q)}^x\right]_{i+1/2,j} [\tilde{\nabla}^x \phi_p]_{i+1/2,j} 
\right].
\end{eqnarray}

The conservative method is a general momentum conservative numerical model for interfaceial tensions that can include an arbitrary number of phases. Such a general model has not been developed in previous studies.
\begin{theorem}\label{Theorem conservative surface force}
The conservative method Eq.(\ref{Eq The conservative method}) for the surface force in Eq.(\ref{Eq Surface force Phi}) is conservative, i.e.,
\begin{equation}\label{Eq conservative surface force}\nonumber
\sum_{i,j} [f_s^x]_{i+1/2,j}=0,
\quad
\sum_{i,j} [f_s^y]_{i,j+1/2}=0.
\end{equation}
\end{theorem}
\begin{proof}\label{Proof conservative surface force}
Following the definition of the gradient operator in Eq.(\ref{Eq Discrete Gradient Face}), the mixed derivative evaluated at the cell corners is commutable, i.e.,
\begin{eqnarray} \label{Eq conservative surface force Eq1} \nonumber
\left[ \tilde{\nabla}^y (\tilde{\nabla}^x \phi) \right]_{i+1/2,j+1/2}
=
\left[ \tilde{\nabla}^x (\tilde{\nabla}^y \phi) \right]_{i+1/2,j+1/2}.
\end{eqnarray}
After defining
\begin{eqnarray} \label{Eq conservative surface force Eq2} \nonumber
\langle \nabla^x \phi_p \nabla^x \phi_q \rangle_{i,j}
=
\left[\overline{\tilde{\nabla}^x \phi_p}^x\right]_{i,j}
\left[\overline{\tilde{\nabla}^x \phi_q}^x\right]_{i,j},
\end{eqnarray}

\begin{eqnarray} \label{Eq conservative surface force Eq3} \nonumber
\langle \nabla^y \phi_p \nabla^x \phi_q \rangle_{i+1/2,j+1/2}
=
\left[\overline{\tilde{\nabla}^y \phi_p}^x\right]_{i+1/2,j+1/2}
\left[\overline{\tilde{\nabla}^x \phi_q}^y\right]_{i+1/2,j+1/2},
\end{eqnarray}

\begin{eqnarray} \label{Eq conservative surface force Eq4} \nonumber
\langle \nabla \phi_p \cdot \nabla \phi_q \rangle_{i,j}
=
\left[\overline{\tilde{\nabla}^x \phi_p \tilde{\nabla}^x \phi_q}^x\right]_{i,j}
+
\left[\overline{\tilde{\nabla}^y \phi_p \tilde{\nabla}^y \phi_q}^y\right]_{i,j},
\end{eqnarray}

\begin{eqnarray} \label{Eq conservative surface force Eq5} \nonumber
\langle \nabla^x \phi_p \nabla^x (\nabla^x \phi_q) \rangle_{i,j}
=
\left[\overline{\tilde{\nabla}^x \phi_p}^x \right]_{i,j} [\tilde{\nabla}^x(\tilde{\nabla}^x \phi_q)]_{i,j},
\end{eqnarray}

\begin{eqnarray}  \label{Eq conservative surface force Eq6} \nonumber
\langle \nabla^y \phi_p \nabla^x (\nabla^y \phi_q) \rangle_{i+1/2,j+1/2}
=
\left[\overline{\tilde{\nabla}^y \phi_p}^x\right]_{i+1/2,j+1/2} [\tilde{\nabla}^x (\tilde{\nabla}^y \phi_q)]_{i+1/2,j+1/2},
\end{eqnarray}
we obtain the following two identities
\begin{eqnarray} \label{Eq conservative surface force 1}\nonumber
\left[\overline{\tilde{\nabla}\cdot (\tilde{\nabla}\phi_p)}^x\right]_{i+1/2,j} [\tilde{\nabla}^x \phi_q]_{i+1/2,j}
=
\left[ \tilde{\nabla}^x 
\langle \nabla^x \phi_p \nabla^x \phi_q \rangle \right]_{i+1/2,j}
+
\left[ \tilde{\nabla}^y 
\langle \nabla^y \phi_p \nabla^x \phi_q \rangle
\right]_{i+1/2,j} \\
\nonumber
-
\left[\overline{\langle \nabla^x \phi_p \nabla^x (\nabla^x \phi_q) \rangle}^x\right]_{i+1/2,j}
-
\left[\overline{\langle \nabla^y \phi_p \nabla^x (\nabla^y \phi_q) \rangle}^y \right]_{i+1/2,j},
\end{eqnarray}
\begin{eqnarray} \label{Eq conservative surface force 2}\nonumber
[\tilde{\nabla}^x \langle \nabla \phi_p \cdot \nabla \phi_q \rangle]_{i+1/2,j}
=
\left[\overline{\langle \nabla^x \phi_p \nabla^x (\nabla^x \phi_q) \rangle}^x\right]_{i+1/2,j}
+
\left[\overline{\langle \nabla^y \phi_p \nabla^x (\nabla^y \phi_q) \rangle}^y\right]_{i+1/2,j}\\
\nonumber
+
\left[\overline{\langle \nabla^x \phi_q \nabla^x (\nabla^x \phi_p)\rangle}^x\right]_{i+1/2,j}
+
\left[\overline{\langle \nabla^y \phi_q \nabla^x (\nabla^y \phi_p)\rangle}^y\right]_{i+1/2,j}. 
\end{eqnarray}
Multiplying them with $\lambda_{p,q}$ and summing over $p$ and $q$, we obtain
\begin{eqnarray} \label{Eq conservative surface force 3}\nonumber
\sum_{p,q=1}^N \lambda_{p,q} \left[\overline{\tilde{\nabla}\cdot (\tilde{\nabla}\phi_q)}^x\right]_{i+1/2,j} [\tilde{\nabla}^x \phi_p]_{i+1/2,j} 
=
\sum_{p,q=1}^N \lambda_{p,q} \left[ \tilde{\nabla}^x 
\langle \nabla^x \phi_p \nabla^x \phi_q \rangle \right]_{i+1/2,j}\\
\nonumber
+
\sum_{p,q=1}^N \lambda_{p,q} \left[ \tilde{\nabla}^y 
\langle \nabla^y \phi_p \nabla^x \phi_q \rangle
\right]_{i+1/2,j} 
-
\frac{1}{2}\sum_{p,q=1}^N \lambda_{p,q} [\tilde{\nabla}^x \langle \nabla \phi_p \cdot \nabla \phi_q \rangle]_{i+1/2,j}.
\end{eqnarray}
Finally, applying Lemma \ref{Lemma Conservation discrete gradient face}, we have
\begin{equation} \label{Eq Summation of surface force conservative method}\nonumber
\sum_{i,j} [f_s^x]_{i+1/2,j}=0.
\end{equation}
With the same procedure, we have $\sum_{i,j} [f_s^y]_{i,j+1/2}=0$.
\end{proof}

The difference between the balanced-force method Eq.(\ref{Eq The balanced-force method}) and the conservative method Eq.(\ref{Eq The conservative method}) is rooted in discretizing $\nabla g_1$ and $\nabla g_2$. In the conservative method Eq.(\ref{Eq The conservative method}), these two terms are discretized directly using the gradient operators defined in Eq.(\ref{Eq Discrete Gradient Face}), while in the balanced-force method Eq.(\ref{Eq The balanced-force method}), the chain rule was applied first and then the gradient operator is applied to $\phi_p$ instead. Such a difference between the two methods changes their discrete conservation properties, although the difference is in the order of the truncation error.

\subsection{Consistency on the discrete level} \label{Sec Consistency on the discrete level}
In this section, general theorems are newly developed to preserve the consistency conditions on the discrete level, so that the physical coupling between the Phase-Field and momentum equations, discussed in Section~\ref{Sec Definitions and governing equations}, is inherited after discretization. Those theorems are on the fully discrete level and independent of the number of phases, the Phase-Field equation, or the scheme to solve the Phase-Field equation, which have not been proposed in previous studies. 

Without loss of generality, the fully discrete Phase-Field equation is represented by
\begin{equation}\label{Eq Phase-Field discrete}
\frac{\gamma_t \phi_p^{n+1}-\widehat{\phi_p}}{\Delta t}
+
\tilde{\nabla} \cdot (\mathbf{u} \widetilde{\phi_p}) 
=
\tilde{\nabla} \cdot \langle\mathbf{J}_p\rangle,
\quad 1 \leqslant p \leqslant N,
\end{equation}
where $(\mathbf{u} \widetilde{\phi_p})$ and $\langle\mathbf{J}_p\rangle$ are the discrete convection and diffusion fluxes, respectively, of Phase $p$. Here, we don't need to know how $\langle\mathbf{J}_p\rangle$ is defined or computed. The time discretization scheme (denoted by $\gamma_t$ and $\hat{(\cdot)}$), the discrete divergence operator (denoted by $\tilde{\nabla} \cdot (\cdot)$), and the interpolation (denoted by $\tilde{\phi}$) can be arbitrary. We only assume that Eq.(\ref{Eq Phase-Field discrete}) follows the \textit{consistency of reduction}, and therefore $\{\langle\mathbf{J}_p\rangle\}_{p=1}^N$ meet the requirements in Theorem \ref{Theorem RC Phase-Field}. To simplify the notation, we define the discrete Phase-Field flux of Phase~$p$ as $\langle \mathbf{m}_{\phi_p} \rangle=\mathbf{u} \widetilde{\phi_p}-\langle\mathbf{J}_p\rangle$.

\subsubsection{The consistency of mass conservation on the discrete level} \label{Sec Consistency of mass conservation}
The following theorem guarantees the \textit{consistency of mass conservation} on the discrete level.
\begin{theorem}\label{Theorem Consistency of mass conservation discrete}
Given the discrete Phase-Field fluxes $\{\langle \mathbf{m}_{\phi_p} \rangle\}_{p=1}^N$ that satisfy the fully discrete Phase-Field equation, i.e.,
\begin{equation}\nonumber
\frac{\gamma_t \phi_p^{n+1}-\widehat{\phi_p}}{\Delta t}
+
\tilde{\nabla} \cdot \langle \mathbf{m}_{\phi_p} \rangle=0,
\quad
1 \leqslant p \leqslant N,
\end{equation}
and provided a cell-face velocity $\mathbf{u}$ that is discretely divergence-free, i.e.,
\begin{equation}\nonumber
\tilde{\nabla} \cdot \mathbf{u}=0,
\end{equation}
the corresponding discrete consistent mass flux that satisfies the consistency of mass conservation on the discrete level is
\begin{equation}\label{Eq Mass flux discrete}
\langle \mathbf{m} \rangle=\sum_{p=1}^N \frac{\rho_p}{2} (\mathbf{u}+\langle \mathbf{m}_{\phi_p} \rangle ),
\end{equation}
which results in the fully discrete mass conservation equation
\begin{equation}\label{Eq Mass discrete}
\frac{\gamma_t \rho^{n+1}-\hat{\rho}}{\Delta t}
+
\tilde{\nabla} \cdot \langle \mathbf{m} \rangle
=
0.
\end{equation}
It should be noted that $\gamma_t$, $\hat{(\cdot)}$, and $\tilde{\nabla} \cdot (\cdot)$ in the fully discrete mass conservation equation are identical to those in the fully discrete Phase-Field equation.
\end{theorem}
\begin{proof}\label{Proof Consistency of mass conservation discrete}
Using Eq.(\ref{Eq Mixture density}) for $\rho$, Eq.(\ref{Eq Mass flux discrete}) for $\langle \mathbf{m} \rangle$, and $\gamma_t$, $\hat{(\cdot)}$, and $\tilde{\nabla} \cdot (\cdot)$ in the fully discrete Phase-Field equation, we have
\begin{eqnarray} \label{Eq Consistency of mass conservation Discrete}\nonumber
\frac{\gamma_t \rho^{n+1} - \hat{\rho}}{\Delta t}
+
\tilde{\nabla} \cdot \langle \mathbf{m} \rangle
=
\sum_{p=1}^N \frac{\rho_p}{2} \frac{\gamma_t (1+\phi_p^{n+1}) - \widehat{ (1+\phi_p) }}{\Delta t}
+
\tilde{\nabla} \cdot \sum_{p=1}^{N} \frac{\rho_p}{2}   (\mathbf{u}+\langle\mathbf{m}_{\phi_p}\rangle) \\
\nonumber
=
\sum_{p=1}^N \frac{\rho_p}{2} \left( \frac{ \gamma_t - \overbrace{\hat{1}}^{\gamma_t}}{\Delta t} + \underbrace{\tilde{\nabla} \cdot  \mathbf{u}}_{0} \right)
+
\sum_{p=1}^{N} \frac{\rho_p}{2} \underbrace{\left( \frac{\gamma_t \phi_p^{n+1}- \widehat{\phi_p}}{\Delta t}+\tilde{\nabla} \cdot  \langle \mathbf{m}_{\phi_p} \rangle\right)}_{0}
=
0.
\end{eqnarray}
Therefore, the \textit{consistency of mass conservation} is satisfied on the discrete level, and $\langle \mathbf{m} \rangle$ in Eq.(\ref{Eq Mass flux discrete}) is the discrete consistent mass flux. 
\end{proof}

\subsubsection{The consistency of mass and momentum transport on the discrete level} \label{Sec Consistency of mass and momentum transport}
The \textit{consistency of mass and momentum transport} on the discrete level can be formulated in the following theorem:
\begin{theorem}\label{Theorem consistency of mass and momentum transport discrete}
Given the discrete consistent mass flux $\langle \mathbf{m} \rangle$ that satisfies the consistency of mass conservation, i.e., the fully discrete mass conservation equation is
\begin{equation}\nonumber
\frac{\gamma_t \rho^{n+1}-\hat{\rho}}{\Delta t}
+
\tilde{\nabla} \cdot \langle \mathbf{m} \rangle
=0,
\end{equation}
to satisfy the consistency of mass and momentum transport on the discrete level, the momentum equation Eq.(\ref{Eq Momentum}) is discretized as
\begin{equation}\nonumber
\frac{\gamma_t \rho^{n+1} \mathbf{u}^{n+1} - \widehat{(\rho \mathbf{u})}}{\Delta t}
+
\tilde{\nabla} \cdot (\langle \mathbf{m} \rangle \otimes \tilde{\mathbf{u}} )=\langle \mathbf{f} \rangle,
\end{equation}
where $\tilde{\mathbf{u}}$ represents an interpolation of $\mathbf{u}$ to the cell faces, and for simplicity $\langle \mathbf{f} \rangle$ denotes the discretized right-hand side of the momentum equation Eq.(\ref{Eq Momentum}). It should be noted that $\gamma_t$, $\hat{(\cdot)}$, $\tilde{\nabla} \cdot (\cdot)$, and $\langle \mathbf{m} \rangle$ in the discretized momentum equation are identical to those in the fully discrete mass conservation equation.
\end{theorem}
\begin{corollary}\label{Corollary Discrete momentum transport}
The discrete momentum transport, i.e.,
\begin{equation}\nonumber
\frac{\gamma_t \rho^{n+1} \mathbf{u}^{n+1} - \widehat{(\rho \mathbf{u})}}{\Delta t}
+
\tilde{\nabla} \cdot (\langle \mathbf{m} \rangle \otimes \tilde{\mathbf{u}} )=\mathbf{0},
\end{equation}
admits an arbitrary homogeneous velocity as the solution.
\end{corollary}
\begin{proof}
The discrete momentum transport with a homogeneous velocity $\mathbf{u}_0$ becomes
\begin{equation}\nonumber
\mathbf{u}_0 \left( \frac{\gamma_t \rho^{n+1}  - \hat{\rho}}{\Delta t}
+
\tilde{\nabla} \cdot \langle \mathbf{m} \rangle \right)=0.
\end{equation}
The parentheses group the fully discrete mass conservation equation which is zero, and therefore $\mathbf{u}_0$ is arbitrary.
\end{proof}

Theorem \ref{Theorem consistency of mass and momentum transport discrete} follows the definition of the \textit{consistency of mass and momentum transport} in Section \ref{Sec Momentum} but additionally considers the numerical operations in the mass and momentum equations. It is actually the discrete version of the control volume analysis in Section \ref{Sec Physical interpretation} but here the control volume is fixed and the differential operators are replaced by their discrete counterparts. If all the forces acting on the fluid mixture is zero, then the fluid mixture has no acceleration. In other words, the velocity of the fluid mixture will not change. This physical configuration is discretely reproduced, as illustrated in Corollary \ref{Corollary Discrete momentum transport}, thanks to satisfying the two consistency conditions on the discrete level.

\textit{\textbf{Remark:}
From Theorem \ref{Theorem Consistency of mass conservation discrete} and Theorem \ref{Theorem consistency of mass and momentum transport discrete}, to achieve the \textit{consistency of mass conservation} and the \textit{consistency of mass and momentum transport} on the discrete level simultaneously, the Phase-Field and momentum equations need to be discretized using the same time discretization scheme and discrete divergence operator.
}

\subsubsection{The consistency of reduction on the discrete level} \label{Sec Consistency of reduction}
In this section, we consider the \textit{consistency of reduction} on the discrete level, provided that this consistency condition has been satisfied by the fully discrete Phase-Field equation Eq.(\ref{Eq Phase-Field discrete}) as pointed out at the beginning of Section \ref{Sec Consistency on the discrete level}. Therefore, we in particular focus on the discrete version of the momentum equation Eq.(\ref{Eq Momentum}) and its related terms, which has seldom been discussed in previous studies.

Without loss of generality, the fully discrete momentum equation can be denoted as
\begin{equation}\label{Eq Momentum discrete}
\frac{\gamma_t \rho^{n+1} \mathbf{u}^{n+1}-\widehat{(\rho \mathbf{u})}}{\Delta t}
+
\tilde{\nabla} \cdot (\langle \mathbf{m} \rangle \otimes \tilde{\mathbf{u}})
=
-
\langle \nabla P \rangle
+
\langle \nabla \cdot (\mu \nabla \mathbf{u}) \rangle
+
\langle \nabla \cdot (\mu \nabla \mathbf{u}^T) \rangle
+
\langle \rho \mathbf{g} \rangle 
+
\langle \mathbf{f}_s \rangle,
\end{equation}
where $\langle \cdot \rangle$ represents the discrete mass flux and forces. Again, details of the time discretization scheme, discrete differential operators, interpolations, or computation methods for the mass flux and forces, are not required here.
\begin{theorem}\label{Theorem consistency of reduction momentum discrete essential}
If the density and viscosity of the fluid mixture, the discrete consistent mass flux, and the discrete surface force are reduction consistent, then the fully discrete momentum equation Eq.(\ref{Eq Momentum discrete}) is essentially reduction consistent in the sense that the single-phase case may not be recovered.
\end{theorem}
\begin{proof}
There is no guarantee that the discrete viscous force 
$\langle \nabla \cdot (\mu \nabla \mathbf{u}) \rangle
+
\langle \nabla \cdot (\mu \nabla \mathbf{u}^T) \rangle$
for the multiphase case recovers $\langle \nabla \cdot (\mu \nabla \mathbf{u}) \rangle$ for the single-phase case. Therefore, the fully discrete momentum equation Eq.(\ref{Eq Momentum discrete}) may not be able to recover the sing-phase case, although only the density and viscosity of the fluid mixture, the discrete consistent mass flux, and the discrete surface force are related to the number of phases in Eq.(\ref{Eq Momentum discrete}) and they are all reduction consistent.
\end{proof}
\begin{theorem}\label{Theorem consistency of reduction momentum discrete}
If the density and viscosity of the fluid mixture, the discrete consistent mass flux, and the discrete surface force are reduction consistent, and $\langle \nabla \cdot (\mu \nabla \mathbf{u}^T) \rangle$ is exactly zero when $\mu$ is a constant, then the fully discrete momentum equation Eq.(\ref{Eq Momentum discrete}) is reduction consistent.
\end{theorem}
\begin{proof}
Continuing the proof of Theorem \ref{Theorem consistency of reduction momentum discrete essential} and supplementing the condition that $\langle \nabla \cdot (\mu \nabla \mathbf{u}^T) \rangle$ becomes zero when $\mu$ is a constant, the fully discrete momentum equation Eq.(\ref{Eq Momentum discrete}) is able to recover the single-phase case, and therefore is reduction consistent.
\end{proof}
\begin{corollary}\label{Corollary consistency of reduction momentum discrete  bulk-phase}
If the fully discrete momentum equation Eq.(\ref{Eq Momentum discrete}) satisfies the consistency of reduction, then it recovers inside each bulk-phase region the corresponding discrete version of the single-phase Navier-Stokes equation with the density and viscosity of that phase, e.g., inside the bulk-phase region of Phase~$p$, the fully discrete the momentum equation Eq.(\ref{Eq Momentum discrete}) becomes
\begin{equation}\nonumber
\frac{\gamma_t \rho_p \mathbf{u}^{n+1}-\rho_p \widehat{\mathbf{u}}}{\Delta t}
+
\tilde{\nabla} \cdot \left( \langle\rho_p \mathbf{u}\rangle \otimes \tilde{\mathbf{u}} \right)
=
-
\langle \nabla P \rangle
+
\langle \nabla \cdot (\mu_p \nabla \mathbf{u}) \rangle
+
\langle \rho_p \mathbf{g} \rangle,
\end{equation}
which is the fully discrete Navier-Stokes equation with density $\rho_p$ and viscosity $\mu_p$.
\end{corollary}

It is obvious that the density and viscosity of the fluid mixture in Eq.(\ref{Eq Mixture density}) and Eq.(\ref{Eq Mixture viscosity}), respectively, meet the requirement in Theorem \ref{Theorem consistency of reduction momentum discrete essential} and Theorem \ref{Theorem consistency of reduction momentum discrete}. For the viscous term, we implement the ``DGT'' operator in \citep{Huangetal2020} to discretize $\nabla \cdot (\mu \nabla \mathbf{u}^T)$, which has the property $\tilde{\nabla} \cdot (\tilde{\nabla}\tilde{\mathbf{u}}^T)=\overline{\tilde{\nabla} (\tilde{\nabla}\cdot \mathbf{u})}$, as mentioned in Section \ref{Sec Notations of discritization}. Therefore, as long as the cell-face velocity is discretely divergence-free, the requirement in Theorem \ref{Theorem consistency of reduction momentum discrete} for the viscous term is satisfied. The remaining task is to analysis the \textit{consistency of reduction} of the discrete consistent mass flux and the discrete surface force. Following the same argument in Section \ref{Sec Momentum reduction}, we only need to consider the case where there are $N$ phases ($N \geqslant 2$) and Phase $N$ is absent.

\begin{theorem}\label{Theorem consistency of reduction mass flux discrete}
If the fully discrete Phase-Field equation satisfies the consistency of reduction, and the discretely divergence-free cell-face velocity in Theorem \ref{Theorem Consistency of mass conservation discrete} is identical to the one in the convection term of the fully discrete Phase-Field equation, then the discrete consistent mass flux in Theorem \ref{Theorem Consistency of mass conservation discrete} is also reduction consistent.
\end{theorem}
\begin{proof}\label{Proof consistency of reduction mass flux discrete}
Provided that Phase $N$ is absent, i.e, $\phi_N \equiv -1$, and recall that the discrete Phase-Field flux includes both the discrete convection and diffusion fluxes, i.e., $\langle\mathbf{m}_{\phi_p}\rangle=\mathbf{u}^{(1)} \widetilde{\phi_p}-\langle\mathbf{J}_p\rangle$, the discrete consistent mass flux in Theorem \ref{Theorem Consistency of mass conservation discrete} becomes
\begin{eqnarray}\nonumber
\langle \mathbf{m} \rangle=\sum_{p=1}^N \frac{\rho_p}{2} (\mathbf{u}^{(2)}+\langle\mathbf{m}_{\phi_p}\rangle)
=\sum_{p=1}^{N-1} \frac{\rho_p}{2} (\mathbf{u}^{(2)}+\langle\mathbf{m}_{\phi_p}\rangle)
+\frac{\rho_N}{2} (\mathbf{u}^{(2)}+\mathbf{u}^{(1)} \underbrace{\widetilde{\phi_N}}_{-1}-\underbrace{\langle\mathbf{J}_N\rangle}_{\mathbf{0}})\\
\nonumber
=\sum_{p=1}^{N-1} \frac{\rho_p}{2} (\mathbf{u}^{(2)}+\langle\mathbf{m}_{\phi_p}\rangle)
+\frac{\rho_N}{2} (\mathbf{u}^{(2)}-\mathbf{u}^{(1)}).
\end{eqnarray}
Here, $\mathbf{u}^{(1)}$ is the velocity used in the convection term of the fully discrete Phase-Field equation, and $\mathbf{u}^{(2)}$ is the discretely divergence-free cell-face velocity, i.e., $\tilde{\nabla} \cdot \mathbf{u}^{(2)}=0$, in Theorem \ref{Theorem Consistency of mass conservation discrete}. After requiring $\mathbf{u}^{(2)}=\mathbf{u}^{(1)}$, the contribution from Phase $N$ to the discrete consistent mass flux is removed, thanks to the reduction consistent fully discrete Phase-Field equation. As a result, the discrete consistent mass flux in Theorem \ref{Theorem Consistency of mass conservation discrete} recovers its ($N-1$)-phase formulation excluding the contribution from absent Phase $N$, and therefore it is reduction consistent. 
\end{proof}

\begin{theorem}\label{Theorem consistency of reduction surface force discrete}
Both the balanced-force method Eq.(\ref{Eq The balanced-force method}) and the conservative method Eq.(\ref{Eq The conservative method}) for discretizing the surface force in Eq.(\ref{Eq Surface force Phi}) are reduction consistent.
\end{theorem}
\begin{proof}\label{Proof consistency of reduction surface force discrete}
Provided that Phase $N$ is absent, i.e., $\phi_N \equiv -1$, we have $\tilde{\nabla} \phi_N=\mathbf{0}$ and $\tilde{\nabla} \cdot (\tilde{\nabla} \phi_N)=0$. From Eq.(\ref{Eq g1 and g2}), we have $g_1(\phi)=g_2(\phi-1)$, $g'_1(\phi)=g'_2(\phi-1)$, and $g_1(-1)=0$. 

The balanced-force method Eq.(\ref{Eq The balanced-force method}) becomes
\begin{eqnarray}\nonumber
\mathbf{f}_s
=
\frac{1}{2} \sum_{p=1}^N \overline{\xi_p} \tilde{\nabla} \phi_p
=
\frac{1}{2} \sum_{p=1}^{N-1} \overline{\xi_p} \tilde{\nabla} \phi_p,
\end{eqnarray}
where
\begin{eqnarray}\nonumber
\xi_{p}
=
\sum_{q=1}^{N}
\lambda_{p,q} \left[
\frac{1}{\eta^2} \left( g'_1(\phi_p)-g'_2(\phi_p+\phi_q) \right)
+
\tilde{\nabla} \cdot (\tilde{\nabla} \phi_q)
\right]
=
\sum_{q=1}^{N-1}
\lambda_{p,q} \left[
\frac{1}{\eta^2} \left( g'_1(\phi_p)-g'_2(\phi_p+\phi_q) \right)
+
\tilde{\nabla} \cdot (\tilde{\nabla} \phi_q)
\right].
\end{eqnarray}

The conservative method Eq.(\ref{Eq The conservative method}) becomes
\begin{eqnarray}\nonumber
\mathbf{f}_s
=
\frac{1}{2} \sum_{p,q=1}^N \lambda_{p,q} \left[ 
\frac{1}{\eta^2} \left(  
\tilde{\nabla} g_1(\phi_p) 
-
\frac{1}{2} \tilde{\nabla} g_2(\phi_p+\phi_q)  \right) 
+  
\overline{\tilde{\nabla} \cdot (\tilde{\nabla} \phi_q)} \tilde{\nabla} \phi_p
\right]\\
\nonumber
=
\frac{1}{2} \sum_{p,q=1}^{N-1} \lambda_{p,q} \left[ 
\frac{1}{\eta^2} \left(  
\tilde{\nabla} g_1(\phi_p)  
-
\frac{1}{2} \tilde{\nabla} g_2(\phi_p+\phi_q)  \right)
+  
\overline{\tilde{\nabla} \cdot (\tilde{\nabla} \phi_q)} \tilde{\nabla} \phi_p
\right].
\end{eqnarray}
As a result, both the balanced-force method Eq.(\ref{Eq The balanced-force method}) and the conservative method Eq.(\ref{Eq The conservative method}) recovers the corresponding $(N-1)$-phase formulations excluding the contribution from absent Phase $N$, and they therefore are reduction consistent.
\end{proof}

\textit{\textbf{Remark}:
\begin{itemize}
    \item 
    Theorem \ref{Theorem Consistency of mass conservation discrete}, Theorem \ref{Theorem consistency of mass and momentum transport discrete}, and Theorem \ref{Theorem consistency of reduction mass flux discrete} illustrate correspondences of numerical operations in the fully discrete Phase-Field and momentum equations. These correspondences discretely reproduce the physical connection among the Phase-Field equation, the mass conservation, and the momentum transport.
    \item
    The issue of the viscous force in the \textit{consistency of redaction} is totally numerical, and does not appear on the continuous level. An alternative way to meet the requirement in Theorem \ref{Theorem consistency of reduction momentum discrete} is to first transform $\nabla \cdot (\mu \nabla \mathbf{u}^T)$ to $\nabla \mu \cdot \nabla \mathbf{u}^T$, and then discretize $\nabla \mu \cdot \nabla \mathbf{u}^T$. Although this helps to achieve the \textit{consistency of reduction}, it unfortunately destroys the discrete momentum conservation. On the other hand, the ``DGT'' operator developed in \citep{Huangetal2020} retains the conservative form of the viscous force and helps to achieve the \textit{consistency of reduction} and at the same time momentum conservation on the discrete level.
\end{itemize}
}

\section{The consistent and conservative scheme for the present multiphase flow model}\label{Sec Scheme}
In this section, we develop a consistent and conservative scheme specific to the present multiphase flow model including the Phase-Field equation Eq.(\ref{Eq Volume fraction contrast equation conservative}) with the diffusion flux in Eq.(\ref{Eq Volume fraction contrast equation}) and the momentum equation Eq.(\ref{Eq Momentum}) from the \textit{consistency of mass conservation} and the \textit{consistency of mass and momentum transport}. For brevity, in this section, the Phase-Field equation refers to the one with the diffusion flux in Eq.(\ref{Eq Volume fraction contrast equation}).

The present scheme is based on the second-order decoupled scheme in \citep{Dong2018} for the Phase-Field equation and the second-order projection scheme in \citep{Huangetal2020} for the momentum equation, but implements the novel numerical techniques developed in Section \ref{Sec Discretization} to satisfy all the consistency conditions, to achieve mass and momentum conservation, and to assure the summation of the volume fractions to be unity (or Eq.(\ref{Eq Summation of volume fraction contrasts})), on the discrete level. Analyses of those properties of the scheme are performed, which were not reported in \citep{Dong2018,Huangetal2020}. Hereafter, the time discretization scheme is the second-order backward difference, i.e., $\gamma_t=1.5$ and $\hat{f}=2f^n-0.5f^{n-1}$, and the second-order $f^{*,n+1}$ is $(2f^n-f^{n-1})$. The discrete differential operators defined in Section \ref{Sec Notations of discritization} are used. This ensures that the same time discretization scheme and discrete divergence operator will be used to discretize the Phase-Field and momentum equations, which is critical to achieve the \textit{consistency of mass conservation} and the \textit{consistency of mass and momentum transport} on the discrete level, see Theorem~\ref{Theorem Consistency of mass conservation discrete} and Theorem~\ref{Theorem consistency of mass and momentum transport discrete}. The interpolation denoted by $\tilde{f}$ for example is specific to the 5th-order WENO scheme \citep{JiangShu1996} here.

\subsection{Scheme for the Phase-Field equation}\label{Sec The volume fraction contrast equation}
The Phase-Field equation is solved by the following two steps.

\textbf{Step 1:} Solve for $\{\psi_p^{n+1}\}_{p=1}^N$ from
\begin{eqnarray} \label{Eq Volume fraction contrast Step1}
\tilde{\nabla} \cdot (\tilde{\nabla}\psi_p^{n+1})
-
(\alpha + S) \psi_p^{n+1}
=
\left[
\tilde{\nabla} \cdot (\tilde{\nabla} \psi_p^{*,n+1})
-
(\alpha+S) \tilde{\nabla} \cdot (\tilde{\nabla} \phi_p^{*,n+1} )
\right] \\
\nonumber
+
\frac{1}{\gamma_0} \left[
\frac{\widehat{\phi_p}}{\Delta t}
-
\tilde{\nabla} \cdot \left( \mathbf{u}^{*,n+1} \widetilde{\phi_p^{*,n+1}} \right)
+
\sum_{q=1}^{N} \tilde{\nabla} \cdot \left( \overline{M_{p,q}^{*,n+1}} \tilde{\nabla} \xi_q^{*,n+1} \right)
\right],
\quad 
1\leqslant p \leqslant N.
\end{eqnarray}

\textbf{Step 2:} Solve for $\{\phi_p^{n+1}\}_{p=1}^N$ from
\begin{equation} \label{Eq Volume fraction contrast Step2}
\tilde{\nabla} \cdot (\tilde{\nabla} \phi_p^{n+1} ) + \alpha \phi_p^{n+1}=\psi_p^{n+1},
\quad
1 \leqslant p \leqslant N.
\end{equation}
Here, $\alpha$ satisfies
\begin{equation} \label{Eq alpha}
\alpha^2+S  \alpha+\frac{\gamma_t }{\gamma_0 \Delta t}  =0,
\end{equation}
or explicitly, $\alpha=\frac{1}{2} \left[-S + \sqrt{S^2-\frac{4\gamma_t }{\gamma_0 \Delta t} } \right]$ with $S \geqslant \sqrt{\frac{4\gamma_t }{\gamma_0 \Delta t}}$, and $\gamma_0=N M_0 \sum_{p,q=1}^N \lambda_{p,q}$.

The boundary conditions, unless otherwise specified, for $\{\phi_p^{n+1}\}_{p=1}^N$ and $\{\psi_p^{n+1}\}_{p=1}^N$ are, respectively,
\begin{equation} \label{Eq Boundary condition Step2}
\mathbf{n} \cdot \tilde{\nabla} \phi_p^{n+1}
=
\sum_{q=1}^N 
\zeta_{p,q} \frac{1+\phi_p^{*,n+1}}{2} \frac{1+\phi_q^{*,n+1}}{2},
\quad
1 \leqslant p \leqslant N,
\end{equation}
\begin{equation} \label{Eq Boundary condition Step1}
\mathbf{n} \cdot \tilde{\nabla} \psi_p^{n+1}=
\mathbf{n} \cdot \tilde{\nabla} \left[
\psi_p^{*,n+1}
+
(\alpha+S)(\phi_p^{n+1}-\phi_p^{*,n+1}) \right]
+
\frac{1}{\gamma_0}\sum_{q=1}^{N} \mathbf{n} \cdot \left( \overline{M_{p,q}^{*,n+1}} \tilde{\nabla} \xi_q^{*,n+1} \right),
\quad
1 \leqslant p \leqslant N,
\end{equation}
where $\mathbf{n}$ is the unit outward normal at the domain boundary, and $\zeta_{p,q}$ is defined in Eq.(\ref{Eq Zeta Phi}).

This scheme for the Phase-Field equation was originated in \citep{Dong2018} and several modifications have been performed in the present study to eliminate unphysical results.
(i) To conserve the order parameters and therefore the mass of each phase, the convection term is computed from its conservative form, i.e., $\nabla \cdot(\mathbf{u}\phi)$, while in \citep{Dong2018} it is computed from its non-conservative form, i.e., $ \mathbf{u} \cdot \nabla \phi$. 
(ii) To satisfy the \textit{consistency of reduction}, all the order parameters are solved from the scheme, i.e., Eq.(\ref{Eq Volume fraction contrast Step1}) and Eq.(\ref{Eq Volume fraction contrast Step2}), while in \citep{Dong2018} only the first $(N-1)$ order parameters are solved from the scheme and the last one is obtained algebraically from their summation, i.e., Eq.(\ref{Eq Summation of volume fraction contrasts}). As discussed in Section \ref{Sec Phase selection}, such an implementation in \citep{Dong2018} can easily violate the \textit{consistency of reduction} and produce fictitious phases numerically.
(iii) To additionally satisfy the summation of the order parameters, the gradient-based phase selection procedure, developed in Section~\ref{Sec Phase selection}, has been implemented to the convection terms $\left\{ \tilde{\nabla} \cdot \left(\mathbf{u}^{*,n+1} \widetilde{\phi_p^{*,n+1}}\right)\right\}_{p=1}^N$ in Eq.(\ref{Eq Volume fraction contrast Step1}).

In summary, the present scheme for the Phase-Field equation including the newly proposed gradient-based phase selection procedure in Section \ref{Sec Phase selection} has the following provable properties on the discrete level:
\begin{itemize}
    \item 
    It satisfies the \textit{consistency of reduction}, see Theorem \ref{Theorem consistency of reduction Phase-Field discrete}, and therefore no fictitious phases can be numerically produced.
    \item 
    It satisfies the summation of the order parameters, see Theorem \ref{Theorem Summation Phi}, and therefore no local voids or overfilling can be numerically produced.
    \item 
    It conserves the order parameters and therefore the mass of each phase and the mass of the fluid mixture, see Theorem \ref{Theorem Mass conservation p}.
\end{itemize}

\subsubsection{Properties of the scheme for the Phase-Field equation}\label{Sec Scheme Phase-Field properties}
Here, we analyze the physical properties of the scheme for the Phase-Field equation and illustrate the significance of including the gradient-based phase selection procedure in Section \ref{Sec Phase selection}. It should be noted that the following analyses were not available in \citep{Dong2018} where the scheme was originated. 

The corresponding fully discrete Phase-Field equation of the scheme is
\begin{eqnarray} \label{Eq Fully-discretized volume fraction contrast equation}
\frac{\gamma_t \phi_p^{n+1} - \widehat{\phi_p}}{\Delta t}
+
\tilde{\nabla} \cdot \left( \mathbf{u}^{*,n+1} \widetilde{\phi_p^{*,n+1}} \right)
=
\tilde{\nabla} \cdot \langle \mathbf{J}_p \rangle,\\
\nonumber
\langle \mathbf{J}_p \rangle
=
\sum_{q=1}^{N} \overline{M_{p,q}^{*,n+1}} \tilde{\nabla} \xi_q^{*,n+1}
-
\gamma_0 \tilde{\nabla} \left[
(\psi_p^{n+1}-\psi_p^{*,n+1})
-
(\alpha+S)(\phi_p^{n+1}-\phi_p^{*,n+1})
\right].
\end{eqnarray}
Compared to its continuous correspondence Eq.(\ref{Eq Volume fraction contrast equation}), the last term in the discrete diffusion flux $\langle \mathbf{J}_p \rangle$ is newly introduced. It comes from decoupling the Phase-Field equation among the fluid phases and is zero on the continuous level. Since the time derivative is approximated by the second-order backward difference and the difference between $(\cdot)^{*,n+1}$ and $(\cdot)^{n+1}$ is $O(\Delta t^2)$, the scheme is formally second-order accurate in time. As one shall see in Section \ref{Sec The momentum equation and the divergence-free condition}, the cell-face velocity is discretely divergence-free at all the time level (Eq.(\ref{Eq Discrete Divergence-Free condition})), and therefore $\tilde{\nabla} \cdot \mathbf{u}^{*,n+1}=\tilde{\nabla} \cdot (2\mathbf{u}^{n}-\mathbf{u}^{n-1})= (2\tilde{\nabla} \cdot\mathbf{u}^{n}-\tilde{\nabla} \cdot\mathbf{u}^{n-1})=(\tilde{\nabla} \cdot \mathbf{u})^{*,n+1}=0$. This property will be used in the following analyses.

\begin{theorem}\label{Theorem consistency of reduction Phase-Field discrete}
The scheme for the Phase-Field equation in Section \ref{Sec The volume fraction contrast equation} including the gradient-based phase selection procedure in Section \ref{Sec Phase selection} satisfies the \textit{consistency of reduction}.
\end{theorem}
\begin{proof}
Provided that Phase $N$ is absent up time level $n$, i.e., $\phi_N^{n}, \phi_N^{n-1}, ... ,\phi_N^{0} \equiv -1$, we first consider the scheme for Phase $N$. The provided condition implies $\phi_N^{*,n+1} = -1$, $\widehat{\phi_N}=-\gamma_t$, $M_{N,q}^{*,n+1}=M_{q,N}^{*,n+1}=0$ from Eq.(\ref{Eq Mobility Phi}), $\psi_N^{*,n+1}=-\alpha$ from Eq.(\ref{Eq Volume fraction contrast Step2}), $\tilde{\nabla} \cdot \left(\mathbf{u}^{*,n+1}  \widetilde{\phi_N^{*,n+1}}\right)=-\tilde{\nabla} \cdot \mathbf{u}^{*,n+1}=0$ thanks to the gradient-based phase selection procedure, see Section \ref{Sec Phase selection}, and to satisfying the divergence-free condition discretely, see Section \ref{Sec The momentum equation and the divergence-free condition}. As a result, Eq.(\ref{Eq Volume fraction contrast Step1}) for Phase $N$ is
\begin{equation} \label{Proof Consistency of reduction Phi}\nonumber
\tilde{\nabla} \cdot (\tilde{\nabla}\psi_N^{n+1})
-
(\alpha + S) \psi_N^{n+1}
=
-\frac{\gamma_t}{\gamma_0\Delta t},
\end{equation}
and its solution is $\psi_N^{n+1}=-\alpha$ after noticing that $\alpha$ satisfies Eq.(\ref{Eq alpha}). Using $\psi_N^{n+1}=-\alpha$ into Eq.(\ref{Eq Volume fraction contrast Step2}), it is obvious that $\phi_N^{n+1}=-1$ is the solution. Therefore, Phase $N$, which is absent initially, won't be generated numerically by the scheme.

Then, we consider the scheme for Phase $p$ $(p \neq N)$. From Eq.(\ref{Eq Volume fraction contrast Step1}) and Eq.(\ref{Eq Volume fraction contrast Step2}), the only term including other phases is $\sum_{q=1}^N \tilde{\nabla} \cdot \left( \overline{M_{p,q}^{*,n+1}} \tilde{\nabla} \xi_q^{*,n+1} \right)$, and it becomes $\sum_{q=1}^{N-1} \tilde{\nabla} \cdot \left( \overline{M_{p,q}^{*,n+1}} \tilde{\nabla} \xi_q^{*,n+1} \right)$ since $M_{p,N}^{*,n+1}=0$ from Eq.(\ref{Eq Mobility Phi}). The contribution of absent Phase $N$ to the chemical potential $\{\xi_p\}_{p=1}^N$ is removed due to $g'_2(\phi-1)=g'_1(\phi)$ and $\tilde{\nabla} \cdot (\tilde{\nabla} \phi_N)=0$, see the definition of the chemical potential in Eq.(\ref{Eq Chemical potential Phi}). As a result, the scheme for the present phases recovers the corresponding $(N-1)$-phase formulation excluding the contribution from absent Phase $N$. 

Combining the analyses for the absent and present phases, the scheme for the Phase-Field equation including the gradient-based phase selection procedure is reduction consistent. Correspondingly, the fully discrete Phase-Field equation Eq.(\ref{Eq Fully-discretized volume fraction contrast equation}) becomes
\begin{eqnarray}\nonumber
\frac{\gamma_t \phi_N^{n+1} - \widehat{\phi_N}}{\Delta t}
+
\underbrace{\tilde{\nabla} \cdot \left( \mathbf{u}^{*,n+1} \widetilde{\phi_N^{*,n+1}} \right)}_{0}
=
\tilde{\nabla} \cdot \underbrace{\langle \mathbf{J}_N\rangle}_{\mathbf{0}},\\
\nonumber
\frac{\gamma_t \phi_p^{n+1} - \widehat{\phi_p}}{\Delta t}
+
\tilde{\nabla} \cdot \left( \mathbf{u}^{*,n+1} \widetilde{\phi_p^{*,n+1}} \right)
=
\tilde{\nabla} \cdot \langle \mathbf{J}_p \rangle, \quad 1 \leqslant p \leqslant (N-1),\\
\nonumber
\langle \mathbf{J}_p \rangle
=
\sum_{q=1}^{N-1} \overline{M_{p,q}^{*,n+1}} \tilde{\nabla} \xi_q^{*,n+1}
-
\gamma_0 \tilde{\nabla} \left[
(\psi_p^{n+1}-\psi_p^{*,n+1})
-
(\alpha+S)(\phi_p^{n+1}-\phi_p^{*,n+1})
\right],
\end{eqnarray}
which discretely reproduces Theorem \ref{Theorem RC Phase-Field} and is reduction consistent.
\end{proof}

\begin{theorem}\label{Theorem Summation Phi}
The scheme for the Phase-Field equation in Section \ref{Sec The volume fraction contrast equation} including the gradient-based phase selection procedure in Section \ref{Sec Phase selection} satisfies the summation of the order parameters in Eq.(\ref{Eq Summation of volume fraction contrasts}).
\end{theorem}
\begin{proof}\label{Proof Summation Phi}
For a clear presentation, we define $\Phi=\sum_{p=1}^N \phi_p$ and $\Psi=\sum_{p=1}^N \psi_p$. Provided that $\Phi=(2-N)$ is true up to time level $n$, i.e., $\Phi^n=\Phi^{n-1}=...=\Phi^0=(2-N)$, we have $\Phi^{*,n+1}=(2-N)$, $\hat{\Phi}=\gamma_t (2-N)$, $\Psi^{*,n+1}=\alpha(2-N)$ from Eq.(\ref{Eq Volume fraction contrast Step2}), $\sum_{p=1}^N \overline{M_{p,q}^{*,n+1}}=\overline{\sum_{p=1}^N M_{p,q}^{*,n+1}}=0$ from Eq.(\ref{Eq Mobility Phi}), and $\sum_{p=1}^N \tilde{\nabla} \cdot \left(\mathbf{u}^{*,n+1}\widetilde{\phi_p^{*,n+1}}\right)=(2-N) \tilde{\nabla} \cdot \mathbf{u}^{*,n+1}=0$ thanks to the gradient-based phase selection procedure, see Section \ref{Sec Phase selection}, and to satisfying the divergence-free condition discretely, see Section \ref{Sec The momentum equation and the divergence-free condition}. Summing Eq.(\ref{Eq Volume fraction contrast Step1}) over $p$, we reach the equation for $\Psi^{n+1}$, i.e.,
\begin{equation} \label{Eq Psi}\nonumber
\tilde{\nabla} \cdot (\tilde{\nabla} \Psi^{n+1})-(\alpha+S) \Psi^{n+1}=\frac{\gamma_t (2-N)}{\gamma_0 \Delta t},
\end{equation}
and $\Psi^{n+1}=\alpha(2-N)$ is the solution after noticing that $\alpha$ satisfies Eq.(\ref{Eq alpha}). The equation for $\Phi$ is obtained by summing Eq.(\ref{Eq Volume fraction contrast Step2}) over $p$, i.e.,
\begin{equation} \label{Eq Phi}\nonumber
\tilde{\nabla} \cdot (\tilde{\nabla} \Phi^{n+1} )+\alpha \Phi^{n+1}=\Psi^{n+1},
\end{equation}
and the solution of $\Phi^{n+1}$ is $(2-N)$ after using $\Psi^{n+1}=\alpha(2-N)$. 

As a result, we have $\Phi^{n+1}=\sum_{p=1}^N \phi_p=(2-N)$, as well as the summation of the volume fractions to be unity, i.e., $\sum_{p=1}^N C_p^{n+1}=\sum_{p=1}^N \frac{1+\phi_p^{n+1}}{2}=1$, at every cell center.
\end{proof}

\textit{\textbf{Remark}:
The novel gradient-based phase selection procedure in Section \ref{Sec Phase selection} is essential in the proofs of Theorem \ref{Theorem consistency of reduction Phase-Field discrete} and Theorem \ref{Theorem Summation Phi}. Without performing this procedure, the convection term will not become zero for the absent phases or after summing over all the phases, but acts as a numerical source to generate fictitious phases, local voids, or overfilling. These unphysical behaviors are observed in the numerical tests in Section \ref{Sec Horizontal shear layer} when the gradient-based phase selection procedure is deactivated.
}

\begin{theorem}\label{Theorem Mass conservation p}
The scheme for the Phase-Field equation in Section \ref{Sec The volume fraction contrast equation} including the gradient-based phase selection procedure in Section \ref{Sec Phase selection} conserves the order parameters, and therefore the mass of each phase and the mass of the fluid mixture, i.e.,
\begin{equation}\nonumber
\sum_{i,j} [\phi_p^n]_{i,j} \Delta \Omega=\sum_{i,j} [\phi_p^0]_{i,j} \Delta \Omega,
\quad
[\mathrm{Mass}]_p^n=[\mathrm{Mass}]_{p}^0,
\quad
1 \leqslant p \leqslant N,
\quad \mathrm{and} \quad
[\mathrm{Mass}]^n=[\mathrm{Mass}]^0,
\end{equation}
where $[\mathrm{Mass}]_p^n=\sum_{i,j} \left[\rho_p \frac{1+\phi_p^n}{2} \right]_{i,j} \Delta \Omega$ and $[\mathrm{Mass}]^n=\sum_{i,j} [\rho^n]_{i,j} \Delta \Omega$.
\end{theorem}
\begin{proof}\label{Proof Mass conservation p}
Given $\sum_{i,j} [\phi_p^n]_{i,j}=\sum_{i,j} [\phi_p^{n-1}]_{i,j}=...=\sum_{i,j} [\phi_p^0]_{i,j}$, we have $\sum_{i,j} [\widehat{\phi_p}]_{i,j}= \widehat{\sum_{i,j} [\phi_p]_{i,j}}=\gamma_t \sum_{i,j} [\phi_p^0]_{i,j}$. Summing the fully discrete Phase-Field equation Eq.(\ref{Eq Fully-discretized volume fraction contrast equation}) over all the cells and applying Lemma \ref{Lemma Conservation discrete gradient face}, we reach
\begin{equation}\nonumber
\sum_{i,j} [\phi_p^{n+1}]_{i,j}=\frac{1}{\gamma_t} \sum_{i,j} [\widehat{\phi_p}]_{i,j}= \sum_{i,j} [\phi_p^0]_{i,j},
\quad
1 \leqslant p \leqslant N.    
\end{equation}
As a result, $\sum_{i,j} [\phi_p]_{i,j} \Delta \Omega$ and $[\mathrm{Mass}]_p=\sum_{i,j} \left[\rho_p \frac{1+\phi_p}{2} \right]_{i,j} \Delta \Omega$ ($1 \leqslant p \leqslant N$) will not change as time advances. From Eq.(\ref{Eq Mixture density}), we have $[\mathrm{Mass}]=\sum_{p=1}^N [\mathrm{Mass}]_p$, and therefore $[\mathrm{Mass}]$ will not change as time advances.
\end{proof}

\subsection{The discrete consistent mass flux}\label{Sec The discrete mass flux}
To obtain the discrete consistent mass flux after solving the Phase-Field equation with the scheme in Section~\ref{Sec The volume fraction contrast equation}, we follow Theorem \ref{Theorem Consistency of mass conservation discrete} and Theorem \ref{Theorem consistency of reduction mass flux discrete} developed in Section \ref{Sec Consistency on the discrete level}. First from Theorem \ref{Theorem Consistency of mass conservation discrete}, we need to obtain the discrete Phase-Field flux, which includes both the discrete convection and diffusion fluxes, from the fully discrete Phase-Field equation Eq.(\ref{Eq Fully-discretized volume fraction contrast equation}), i.e.,
\begin{eqnarray}\label{Eq Discrete Phase-Field flux}
\langle \mathbf{m}_{\phi_p} \rangle
=
\left( \mathbf{u}^{*,n+1} \widetilde{\phi_p^{*,n+1}} \right)
-
\langle \mathbf{J}_p \rangle\\
\nonumber
=
\left( \mathbf{u}^{*,n+1} \widetilde{\phi_p^{*,n+1}} \right)
-
\sum_{q=1}^{N} \overline{M_{p,q}^{*,n+1}} \tilde{\nabla} \xi_q^{*,n+1}\\
\nonumber
+
\gamma_0 \tilde{\nabla} \left[
(\psi_p^{n+1}-\psi_p^{*,n+1})
-
(\alpha+S)(\phi_p^{n+1}-\phi_p^{*,n+1})
\right],\\
\nonumber
\quad 1 \leqslant p \leqslant N.
\end{eqnarray}
Then, we need to choose a discretely divergence-free cell-face velocity, and Theorem \ref{Theorem consistency of reduction mass flux discrete} tells that in order to achieve the \textit{consistency of reduction} on the discrete level, this velocity should be the same as the one in the convection term of the fully discrete Phase-Field equation Eq.(\ref{Eq Fully-discretized volume fraction contrast equation}), which is $\mathbf{u}^{*,n+1}$ in the present case. Notice that $\mathbf{u}^{*,n+1}$ is discretely divergence-free, i.e., $\tilde{\nabla} \cdot \mathbf{u}^{*,n+1}=0$, see Section \ref{Sec The momentum equation and the divergence-free condition}. Finally, using the formulation in Theorem \ref{Theorem Consistency of mass conservation discrete} we obtain the discrete consistent mass flux in the present scheme:
\begin{equation} \label{Eq Discrete mass flux}
\langle \mathbf{m} \rangle=\sum_{p=1}^N \frac{\rho_p}{2} (\mathbf{u}^{*,n+1}+\langle \mathbf{m}_{\phi_p} \rangle).
\end{equation}

In summary, the discrete consistent mass flux in the present scheme, i.e., the one in Eq.(\ref{Eq Discrete mass flux}) with the discrete Phase-Field flux in Eq.(\ref{Eq Discrete Phase-Field flux}), has the following properties on the discrete level:
\begin{itemize}
    \item 
    It satisfies the \textit{consistency of mass conservation}, see Theorem \ref{Theorem Consistency of mass conservation discrete}.
    \item 
    It satisfies the \textit{consistency of reduction}, see Theorem \ref{Theorem consistency of reduction mass flux discrete}, since the same consistency condition has been satisfied by the fully discrete Phase-Field equation, see Theorem \ref{Theorem consistency of reduction Phase-Field discrete}. 
\end{itemize}

\textit{\textbf{Remark}:
It should be noted that starting from the continuous formulation of the consistent mass flux in Section \ref{Sec Definitions and governing equations} and then performing discretization usually fails to obtain the discrete consistent mass flux that satisfies the \textit{consistency of mass conservation} on the discrete level. The present scheme provides an example that some extra terms, in the order of the truncation error, may appear in the fully discrete Phase-Field equation and therefore in the discrete Phase-Field flux, like the last term in Eq.(\ref{Eq Discrete Phase-Field flux}). As a result, directly discretizing the continuous formulation will miss those terms, since they are zero on the continuous level. Moreover, if the numerical operations chosen to discretize the continuous formulation are different from those used in the fully discrete Phase-Field equation, the resulting mass flux will also fail to satisfy the \textit{consistency of mass conservation}. Therefore, one should follow Theorem \ref{Theorem Consistency of mass conservation discrete} and Theorem \ref{Theorem consistency of reduction mass flux discrete} developed in Section \ref{Sec Consistency on the discrete level}, like we did here, to obtain the discrete consistent mass flux. 
}

\subsection{Scheme for the momentum equation} \label{Sec The momentum equation and the divergence-free condition}
The momentum equation Eq.(\ref{Eq Momentum}) is solved by the following 7 steps.

\textbf{Step 1:} Solve for $\mathbf{u}^*$ at the cell centers from
\begin{equation} \label{Eq Momentum Step1}
\frac{\gamma_t \rho^{n+1} \mathbf{u}^*-\widehat{(\rho \mathbf{u})}}{\Delta t}
+
\tilde{\nabla} \cdot ( \langle \mathbf{m} \rangle \otimes \widetilde{\mathbf{u}^{*,n+1}} )
=
\rho^{n+1} \overline{\mathbf{G}^n}
+
\tilde{\nabla} \cdot ( \overline{\mu^{n+1}} \tilde{\nabla} \mathbf{u}^* )
+
\tilde{\nabla} \cdot ( \overline{\mu^{n+1}} (\tilde{\nabla} \widetilde{\mathbf{u}^{*,n+1}})^T ).
\end{equation}

\textbf{Step 2:} Solve for $\mathbf{u}^{**}$ at the cell centers from
\begin{equation} \label{Eq Momentum Step2}
\frac{\gamma_t \mathbf{u}^{**}-\gamma_t \mathbf{u}^*}{\Delta t}
=
-\overline{\mathbf{G}^n}.
\end{equation}

\textbf{Step 3:} Solve for $\mathbf{u}^{*}$ at the cell faces from
\begin{equation} \label{Eq Momentum Step3}
\frac{\gamma_t \mathbf{u}^*-\gamma_t \overline{\mathbf{u}^{**}}}{\Delta t}
=
-\frac{1}{\overline{\rho^{n+1}}} \tilde{\nabla} P^n
+
\mathbf{G}_s^{n+1}.
\end{equation}

\textbf{Step 4:} Solve for $P'$ at the cell centers from
\begin{equation} \label{Eq Momentum Step4}
\frac{\gamma_t}{\Delta t} ( \tilde{\nabla} \cdot \mathbf{u}^{n+1}- \tilde{\nabla} \cdot \mathbf{u}^*)
=
-\tilde{\nabla} \cdot \left( \frac{1}{\overline{\rho^{n+1}}} \tilde{\nabla} P' \right).
\end{equation}

\textbf{Step 5:} Solve for $P^{n+1}$ at the cell centers from
\begin{equation} \label{Eq Momentum Step5}
P^{n+1}=P^n+P'.
\end{equation}

\textbf{Step 6:} Solve for $\mathbf{u}^{n+1}$ at the cell faces from
\begin{equation} \label{Eq Momentum Step6}
\frac{\gamma_t \mathbf{u}^{n+1}-\gamma_t \mathbf{u}^*}{\Delta t}
=
-\frac{1}{\overline{\rho^{n+1}}} \tilde{\nabla} P'.
\end{equation}

\textbf{Step 7:} Solve for $\mathbf{u}^{n+1}$ at the cell centers from
\begin{equation} \label{Eq Momentum Step7}
\frac{\gamma_t \mathbf{u}^{n+1}-\gamma_t \mathbf{u}^{**}}{\Delta t}
=
\overline{\mathbf{G}^{n+1}}.
\end{equation}
The cell-face velocity is required to be discretely divergence-free at all the time level, i.e.,
\begin{equation} \label{Eq Discrete Divergence-Free condition}
\tilde{\nabla} \cdot \mathbf{u}^n=0,
\quad n=0,1,2,...
\end{equation}
Therefore, $\tilde{\nabla} \cdot \mathbf{u}^{n+1}=0$ is applied in \textbf{Step 4}. In the above steps, $\rho$ and $\mu$ are computed from Eq.(\ref{Eq Mixture density}) and Eq.(\ref{Eq Mixture viscosity}), respectively, at the cell centers. Recall that $\mathbf{G}$ and $\mathbf{G}_s$, defined in Eq.(\ref{Eq G}), are the net force per unit mass and the net force per unite mass excluding the pressure gradient, respectively, when $\mathbf{u} \equiv \mathbf{0}$. To address the odd-even decoupling of the pressure in the collocated grid arrangement, the ``Rhie-Chow'' interpolation \citep{RhieChow1983,FerzigerPeric2002,Francoisetal2006,Popinet2009} has been applied (\textbf{Step 2} and \textbf{Step 3}). The boundary conditions for velocity and pressure are problem-dependent and will be specified in individual cases in Section \ref{Sec Validations and applications}.

This scheme for the momentum equation Eq.(\ref{Eq Momentum}) is originated in \citep{Huangetal2020} for the two-phase flows. Here, it is applied to the multiphase flows, incorporating with the novel numerical techniques developed in Section~\ref{Sec Discretization}.
Following Theorem \ref{Theorem consistency of mass and momentum transport discrete} in Section \ref{Sec Consistency on the discrete level}, the discrete consistent mass flux $\langle \mathbf{m} \rangle$ obtained from Theorem \ref{Theorem Consistency of mass conservation discrete} and Theorem \ref{Theorem consistency of reduction mass flux discrete} has been used in the inertia term, in order to achieve the \textit{consistency of mass and momentum transport} on the discrete level.
Following Theorem \ref{Theorem consistency of reduction momentum discrete} in Section~\ref{Sec Consistency on the discrete level}, we retain the ``DGT'' operator in \citep{Huangetal2020}, denoted as $\tilde{\nabla} \cdot ( \overline{\mu^{n+1}} (\tilde{\nabla} \widetilde{\mathbf{u}^{*,n+1}})^T )$ here, so that, along with the discretely divergence-free cell-face velocity, $\tilde{\nabla} \cdot (\tilde{\nabla} \widetilde{\mathbf{u}^{*,n+1}})^T=\overline{\tilde{\nabla} (\tilde{\nabla} \cdot \mathbf{u}^{*,n+1})}=\mathbf{0}$ is true, which contributes to the \textit{consistency of reduction} and momentum conservation on the discrete level.
Moreover, the balanced-force method Eq.(\ref{Eq The balanced-force method}) or the conservative method Eq.(\ref{Eq The conservative method}), developed in Section \ref{Sec Discreteization of the surface force}, is implemented to compute the surface force in $\mathbf{G}_s$ to model multiphase interfacial tensions. 

In summary, the present scheme for the momentum equation including the novel numerical techniques in Section \ref{Sec Discretization} has the following properties on the discrete level:
\begin{itemize}
    \item 
    It satisfies the \textit{consistency of mass and momentum transport}, see the fully discrete momentum equation Eq.(\ref{Eq Fully-discretized momentum equation}) and Theorem \ref{Theorem consistency of mass and momentum transport discrete}.
    \item 
    It satisfies the \textit{consistency of reduction}, see Theorem \ref{Theorem consistency of reduction momentum discrete}, Theorem \ref{Theorem consistency of reduction mass flux discrete}, and Theorem \ref{Theorem consistency of reduction surface force discrete}, since the same consistency condition has been satisfied by the fully discrete Phase-Field equation, see Theorem~\ref{Theorem consistency of reduction Phase-Field discrete}, and $\tilde{\nabla} \cdot (\tilde{\nabla} \widetilde{\mathbf{u}^{*,n+1}})^T$ is zero from the property of the ``DGT'' operator in \citep{Huangetal2020}. As a result, the fully discrete momentum equation inside the bulk-phase region of Phase $p$ ($1 \leqslant p \leqslant N$) becomes the corresponding fully discrete Navier-Stokes equation
    \begin{equation}\nonumber
    \frac{\gamma_t \rho_p \mathbf{u}^{n+1}-\rho_p \hat{\mathbf{u}}}{\Delta t}
    +
    \tilde{\nabla} \cdot ( \rho_p \mathbf{u}^{*,n+1} \otimes \widetilde{\mathbf{u}^{*,n+1}} )
    =
    -\overline{\tilde{\nabla} P^{n+1}}
    +
    \mu_p \tilde{\nabla} \cdot (\tilde{\nabla} \mathbf{u}^*)
    +
    \rho_p \mathbf{g},
    \end{equation}
    see the fully discrete momentum equation Eq.(\ref{Eq Fully-discretized momentum equation}) and Corollary~\ref{Corollary consistency of reduction momentum discrete  bulk-phase}.
    \item 
    It solves advection (or translation) problems exactly, independent of the material properties of the fluid phases or the initial shapes of the interfaces, see Theorem \ref{Theorem constant velocity discrete}.
    \item 
    It satisfies the momentum conservation if the surface force is either neglected or computed conservatively, e.g., using the conservative method developed in Section \ref{Sec Discreteization of the surface force}, see Theorem \ref{Theorem Momentum conservation} and its corollaries.
    \item
    It preserves the kinetic energy transport on the semi-discrete level, i.e., the space is discretized but the time is continuous, with the linear interpolation, see Theorem \ref{Theorem Kinetic energy conservation semi-discrete}.
\end{itemize}

\subsubsection{Properties of the scheme for the momentum equation}\label{Sec Scheme momentum properties}
Here, we analyze the physical properties of the scheme for the momentum equation Eq.(\ref{Eq Momentum}) including the novel numerical techniques developed in Section \ref{Sec Discretization}. It should be noted that the following analyses were not available in \citep{Huangetal2020} where the scheme was originated. The corresponding fully discrete momentum equation of the scheme is
\begin{eqnarray} \label{Eq Fully-discretized momentum equation}
\frac{\gamma_t \rho^{n+1} \mathbf{u}^{n+1}-\widehat{(\rho \mathbf{u})}}{\Delta t}
+
\tilde{\nabla} \cdot ( \langle \mathbf{m} \rangle \otimes \widetilde{\mathbf{u}^{*,n+1}} )
=
\rho^{n+1} \overline{\left(-\frac{1}{\overline{\rho^{n+1}}} \tilde{\nabla} P^{n+1} + \frac{1}{\overline{\rho^{n+1}}} \mathbf{f}_s^{n+1} + \mathbf{g} \right)}\\
\nonumber
+
\tilde{\nabla} \cdot ( \overline{\mu^{n+1}} \tilde{\nabla} \mathbf{u}^* )
+
\tilde{\nabla} \cdot ( \overline{\mu^{n+1}} (\tilde{\nabla} \widetilde{\mathbf{u}^{*,n+1}})^T ).
\end{eqnarray}
The formal order of accuracy of the scheme is second-order, as analyzed in \citep{Huangetal2020}. 

\begin{theorem}\label{Theorem constant velocity discrete}
Provided that all the forces except the pressure gradient are neglected, the scheme for the momentum equation in Section \ref{Sec The momentum equation and the divergence-free condition} admits any homogeneous velocity and pressure as its solution. 
\end{theorem}
\begin{proof}\label{Proof constant velocity discrete}
Given an initial homogeneous velocity $\mathbf{u}_0$ and pressure $P_0$, and suppose they are still the solution of the scheme up to time level $n$, we seek for the solution of the scheme at $t^{n+1}$.

From \textbf{Step 1}, the right-hand side of Eq.(\ref{Eq Momentum Step1}) is zero, since the pressure gradient at $t^n$ is zero and the other forces are neglected. As a result, we have
\begin{equation} \label{Eq constant velocity Discrete}\nonumber
\mathbf{u}^*
=
\frac{1}{\rho^{n+1}} \underbrace{\left( \frac{\hat{\rho}  -\Delta t \tilde{\nabla} \cdot \langle \mathbf{m} \rangle}{\gamma_t } \right) }_{\rho^{n+1}} \mathbf{u}_0
=
\mathbf{u}_0.
\end{equation}
Notice that the terms inside the parentheses are equal to $\rho^{n+1}$ due to satisfying the \textit{consistency of mass conservation} by the discrete consistent mass flux. From \textbf{Step 2}, we obtain $\mathbf{u}^{**}=\mathbf{u}^*=\mathbf{u}_0$ and further $\overline{\mathbf{u}^{**}}=\mathbf{u}_0$ due to $\overline{\mathbf{G}^{n}}=\mathbf{0}$. 
From \textbf{Step 3}, we have $\mathbf{u}^*$ at the cell faces is again $\mathbf{u}_0$, by noticing that both $\tilde{\nabla} P^n$ and $\mathbf{G}_s^{n+1}$ are zero. From \textbf{Step 4}, we have $\tilde{\nabla} \cdot \mathbf{u}^*=\tilde{\nabla} \cdot \mathbf{u}_0=0$, and therefore $P'$ is a constant. Without loss of generality, we set $P'=0$. 
After that, from \textbf{Step 5}, we have $P^{n+1}=P_0$, and from \textbf{Step 6}, $\mathbf{u}^{n+1}$ at the cell faces is $\mathbf{u}_0$. 
Finally from \textbf{Step 7}, $\overline{\mathbf{G}^{n+1}}$ is zero and $\mathbf{u}^{n+1}$ at the cell centers is $\mathbf{u}_0$. 
\end{proof}
Theorem \ref{Theorem constant velocity discrete} is rooted in satisfying the \textit{consistency of mass conservation} and the \textit{consistency of mass and momentum transport}. Otherwise, the terms inside the parentheses from \textbf{Step 1} are not equal to $\rho^{n+1}$ anymore, resulting in $\mathbf{u}^* \neq \mathbf{u}_0$. This error will be transferred into the upcoming steps of the scheme, leading to a non-homogeneous $P'$. As a result, $\mathbf{u}^{n+1}$ is not equal to $\mathbf{u}_0$ either. Such unphysical velocity and pressure fluctuations depend on the densities of the phases and probably become a source of numerical instability. A simplified two-phase 1D analysis only considering \textbf{Step 1} of the scheme are available in our previous work \citep{Huangetal2020}, where even the pressure gradient is neglected. Theorem \ref{Theorem constant velocity discrete} here is more general since it considers the entire scheme including both the velocity and pressure, and works in mulitphase and multidimensional problems. A comparison study will be performed in Section \ref{Sec Advection} to demonstrate Theorem \ref{Theorem constant velocity discrete} and therefore the significance of satisfying the consistency conditions.

\textit{\textbf{Remark}:
\begin{itemize}
    \item 
    Theorem \ref{Theorem constant velocity discrete} corresponds to solving advection (or translation) problems, and it holds independent of the number, the material properties, or the initial shapes of the phases. This is consistent with the physical configuration of the advection problems.
    \item
    Theorem \ref{Theorem constant velocity discrete} works when including the viscous force because the viscous force is identically zero with a homogeneous velocity. 
    \item 
    Theorem \ref{Theorem constant velocity discrete} can be extended to the case where the hydrostatic mechanical equilibrium is always satisfied numerically, i.e., $\mathbf{G} \equiv 0$. Recall that $\mathbf{G}$ is the net force per unit mass under $\mathbf{u} \equiv 0$, see its definition in Eq.(\ref{Eq G}). The initial homogeneous velocity is still the solution of the scheme in this case, and the proof is similar. The only difference is that $P'$ can be non-homogeneous. Nonetheless, after combining \textbf{Step~3} and \textbf{Step 6}, we still have $\mathbf{u}^{n+1}=\overline{\mathbf{u}^{**}}=\mathbf{u}_0$ at the cell faces, since $\mathbf{G}^{n+1}$, which is zero, is recovered. This case corresponds to the one in Corollary \ref{Corollary Discrete momentum transport}. It should be noted that $\mathbf{G} \equiv \mathbf{0}$ is seldom achievable in numerical practice.
\end{itemize}
}

\begin{theorem}\label{Theorem Momentum conservation}
If the discrete surface force is conservative, i.e.,
\begin{equation}\nonumber
\sum_{i,j} [f_s^x]_{i+1/2,j}=\sum_{i,j} [f_s^y]_{i,j+1/2}=0,
\end{equation}
the scheme for the momentum equation in Section \ref{Sec The momentum equation and the divergence-free condition} satisfies the momentum conservation on the discrete level, i.e.,
\begin{equation}\nonumber
[\mathrm{\mathbf{Momentum}}]^n=[\mathrm{\mathbf{Momentum}}]^0,
\end{equation}
where $[\mathrm{\mathbf{Momentum}}]^n=\sum_{i,j}[\rho^n \mathbf{u}^n]_{i,j}\Delta \Omega$.
\end{theorem}
\begin{proof}\label{Proof Momentum conservation}
Given $\sum_{i,j} [\rho^n \mathbf{u}^n]_{i,j} =\sum_{i,j} [\rho^{n-1} \mathbf{u}^{n-1}]_{i,j}=...=\sum_{i,j} [\rho^0 \mathbf{u}^0]_{i,j}$, we have $\sum_{i,j} [\widehat{(\rho \mathbf{u})}]_{i,j}= \widehat{\sum_{i,j} [\rho \mathbf{u}]_{i,j}}=\gamma_t \sum_{i,j} [\rho^0 \mathbf{u}^0]_{i,j}$. Then, summing the fully discrete momentum equation Eq.(\ref{Eq Fully-discretized momentum equation}) over all the cell centers, and using Lemma \ref{Lemma Conservation discrete gradient face}, we obtain
\begin{equation}\nonumber
\frac{\gamma_t \sum_{i,j}[\rho^{n+1} \mathbf{u}^{n+1}]_{i,j}-\gamma_t \sum_{i,j} [\rho^0 \mathbf{u}^0]_{i,j}}{\Delta t}
=
\sum_{i,j} \left\{
\begin{array}{ll}
&[f_s^x]_{i+1/2,j}^{n+1}, \quad x-\mathrm{component}\\
&[f_s^y]_{i,j+1/2}^{n+1}, \quad y-\mathrm{component}\\
\end{array}
\right.
=\mathbf{0}.
\end{equation}
Notice that the gravity is neglected. As a result, the momentum, i.e., $[\mathrm{\mathbf{Momentum}}]=\sum_{i,j}[\rho \mathbf{u}]_{i,j}\Delta \Omega$, is conserved.
\end{proof}
\begin{corollary}\label{Corollary Momentum conservation}
If the surface force in Eq.(\ref{Eq Surface force Phi}) is computed by the conservative method Eq.(\ref{Eq The conservative method}), the scheme for the momentum equation in Section \ref{Sec The momentum equation and the divergence-free condition} satisfies the momentum conservation on the discrete level.
\end{corollary}
\begin{corollary}\label{Corollary Momentum conservation no fs}
If the surface force is neglected, the scheme for the momentum equation in Section \ref{Sec The momentum equation and the divergence-free condition} satisfies the momentum conservation on the discrete level.
\end{corollary}

Corollary \ref{Corollary Momentum conservation} is obvious after noticing Theorem \ref{Theorem conservative surface force}. On the other hand, when the balanced-force method Eq.(\ref{Eq The balanced-force method}) is used to compute the surface force in Eq.(\ref{Eq Surface force Phi}), the momentum is not guaranteed to be conserved on the discrete level. Since the surface force is a local force, which is zero except close to interfaces, and the difference between the balanced-force and conservative methods is in the order of the truncation error, see Section \ref{Sec Discreteization of the surface force}, it is reasonable to expect that the momentum change introduced by the balanced-force method is small. Effects of the balanced-force method and the conservative method on the momentum conservation will be studied with a numerical experiment in Section \ref{Sec Horizontal shear layer}.

\begin{theorem}\label{Theorem Kinetic energy conservation semi-discrete}
The semi-discrete momentum transport of the scheme for the momentum equation in Section~\ref{Sec The momentum equation and the divergence-free condition}, i.e.,
\begin{equation}\nonumber
\frac{\partial (\rho \mathbf{u}) }{\partial t}
+
\tilde{\nabla} \cdot ( \langle \mathbf{m} \rangle \otimes \widetilde{\mathbf{u}} )
=
\mathbf{0},
\end{equation}
implies the following semi-discrete kinetic energy transport, i.e.,
\begin{eqnarray}\nonumber
\frac{\partial e_K}{\partial t}
+
\tilde{\nabla} \cdot \left(\langle \mathbf{m} \rangle \frac{1}{2} \widetilde{(\mathbf{u}\cdot \mathbf{u})} \right)
=
0,
\quad
e_K=\frac{1}{2} \rho \mathbf{u} \cdot \mathbf{u},\\
\nonumber
[\widetilde{(\mathbf{u} \cdot \mathbf{u})}]_{i+1/2,j}=\mathbf{u}_{i,j} \cdot \mathbf{u}_{i+1,j},
\quad
[\widetilde{(\mathbf{u}\cdot \mathbf{u})}]_{i,j+1/2}=\mathbf{u}_{i,j} \cdot \mathbf{u}_{i,j+1},
\end{eqnarray}
if $\tilde{\mathbf{u}}$ is computed by the linear interpolation, i.e., $\tilde{\mathbf{u}}=\overline{\mathbf{u}}$.
\end{theorem}
\begin{proof}\label{Proof Kinetic energy conservation semi-discrete}
Performing the dot product between $\mathbf{u}$ and the semi-discrete momentum transport, we obtain
\begin{equation}\nonumber
\frac{\partial e_K}{\partial t}
+
\tilde{\nabla} \cdot \left( \langle \mathbf{m} \rangle \frac{1}{2} \widetilde{(\mathbf{u}\cdot\mathbf{u})} \right)
+
\frac{\mathbf{u} \cdot \mathbf{u}}{2} \underbrace{\left( \frac{\partial \rho}{\partial t}+\nabla \cdot \langle \mathbf{m} \rangle \right)}_{0}
=
0.
\end{equation}
Notice that the semi-discrete mass conservation equation, which is zero, is recovered, thanks to satisfying the \textit{consistency of mass conservation} by $\langle \mathbf{m} \rangle$.
\end{proof}

The second-order backward difference in time, the WENO scheme for the interpolation, and, as demonstrated in \citep{Morinishietal1998,Huangetal2020}, the linearly interpolated pressure gradient, used in the present scheme to solve the momentum equation, are dissipative. After considering Theorem \ref{Theorem Kinetic energy conservation semi-discrete}, it is reasonable to infer that the kinetic energy will decay on the fully discrete level. Analyses of the energy law Eq.(\ref{Eq Energy law Phi}) on the fully discrete level, which are related to the property of the secondary conservation of the scheme, are very complicated. Thus, we will perform numerical experiments to examine Eq.(\ref{Eq Energy law Phi}) in Section \ref{Sec Horizontal shear layer}.

\textit{\textbf{Remark}:
It is worth mentioning that Theorem \ref{Theorem constant velocity discrete}, Theorem \ref{Theorem Momentum conservation}, and Theorem \ref{Theorem Kinetic energy conservation semi-discrete} in this section do not require a specific form of the discrete mass flux $\langle \mathbf{m} \rangle$. The only requirement on $\langle \mathbf{m} \rangle$ in Theorem \ref{Theorem constant velocity discrete} and Theorem \ref{Theorem Kinetic energy conservation semi-discrete} is that it satisfies the \textit{consistency of mass conservation}.
}

\subsection{Summary of the present consistent and conservative scheme}\label{Sec Summary scheme}
The present consistent and conservative scheme for the multiphase flow model, including the Phase-Field equation Eq.(\ref{Eq Volume fraction contrast equation conservative}) and Eq.(\ref{Eq Volume fraction contrast equation}) and the momentum equation Eq.(\ref{Eq Momentum}), is performed following the steps below:
\begin{enumerate}
    \item 
    Implement the gradient-based phase selection procedure in Section \ref{Sec Phase selection} to compute the convection term in Eq.(\ref{Eq Volume fraction contrast Step1}) and then solve for $\{\psi_p^{n+1}\}_{p=1}^N$ and $\{\phi_p^{n+1}\}_{p=1}^N$ from Eq.(\ref{Eq Volume fraction contrast Step1}) and Eq.(\ref{Eq Volume fraction contrast Step2}), respectively.
    \item
    Compute the discrete consistent Phase-Field fluxes $\{\langle \mathbf{m}_{\phi_p} \rangle\}_{p=1}^N$ from Eq.(\ref{Eq Discrete Phase-Field flux}) and then the discrete consistent mass flux from Eq.(\ref{Eq Mass flux discrete}).
    \item
    Compute $\rho^{n+1}$ and $\mu^{n+1}$ from Eq.(\ref{Eq Mixture density}) and Eq.(\ref{Eq Mixture viscosity}), respectively, using $\{\phi_p^{n+1}\}_{p=1}^N$.
    \item
    Compute $\mathbf{f}_s^{n+1}$ from either the balanced-force method Eq.(\ref{Eq The balanced-force method}) or the conservative method Eq.(\ref{Eq The conservative method}).
    \item
    Solve for $\mathbf{u}^{n+1}$ at both the cell centers and cell faces and $P^{n+1}$ from Eq.(\ref{Eq Momentum Step1}) to Eq.(\ref{Eq Momentum Step7}), provided $\tilde{\nabla} \cdot \mathbf{u}^{n+1}=0$, see Eq.(\ref{Eq Discrete Divergence-Free condition}).
\end{enumerate}
Then, we can proceed to the next time step.

\section{Validation and application} \label{Sec Validations and applications}
In the first part of this section, numerical experiments are performed to validate the present multiphase flow model and scheme. In the second part, realistic multiphase flow problems are performed to demonstrate the capability of the model and scheme.

\subsection{Validation} \label{Sec Validations}
In this section, we investigate the formal order of accuracy, the numerical force balance, and the convergence of the numerical solution of the present multiphase flow model to the sharp-interface solution, and validate the analyses and discussions of the present scheme in Section \ref{Sec Scheme}, through numerical experiments. All the setups and results are reported in their dimensionless forms.

\subsubsection{Manufactured solution} \label{Sec Manufactured solution}
We perform a manufactured solution problem to validate the formal order of accuracy in both time and space of the present scheme in Section \ref{Sec Scheme}. The following source terms
\begin{eqnarray} \label{Eq Manufactured solution Source Phase-Field}
S_{\phi_p}
=
\frac{\partial \phi_p^E}{\partial t}
+
\nabla \cdot (\mathbf{u}^E \phi_p^E)
-
\sum_{q=1}^N \nabla \cdot ( M_{p,q}^E \nabla \xi_q^E ),
\quad 1 \leqslant p \leqslant N,\\
\nonumber
\mathbf{S_u}
=
\frac{\partial (\rho^E \mathbf{u}^E)}{\partial t} 
+ 
\nabla \cdot (\mathbf{m}^E \otimes \mathbf{u}^E)
+
\nabla p^E
-
\nabla \cdot \left[ \mu^E (\nabla \mathbf{u}^E + (\nabla \mathbf{u}^E)^T ) \right]
-
\rho^E \mathbf{g}
-
\mathbf{f}_s^E,
\end{eqnarray}
are added to the Phase-Field and momentum equations, respectively. Here,
\begin{eqnarray} \label{Eq Manufactured solution Exact}
\phi_1^E= \frac{1}{3}  \cos(x)  \cos(y)  \sin(t)-\frac{2}{3},\\
\nonumber
\phi_2^E= \frac{1}{3}  \cos(x)  \cos(y)  \sin(2t)-\frac{2}{3},\\
\nonumber
\phi_3^E= \frac{1}{3}  \cos(x)  \cos(y)  \sin \left(\frac{1}{2} t\right)-\frac{2}{3},\\
\nonumber
\phi_4^E= -\frac{1}{3}  \cos(x)  \cos(y) \left[\sin(t)+\sin(2t)+\sin \left(\frac{1}{2} t\right) \right],\\
\nonumber
u^E= \sin(x)  \cos(y) \cos(t),\\
\nonumber
v^E= -\cos(x)  \sin(y) \cos(t),\\
\nonumber
P^E=\cos(x) \cos(y) \sin(t),
\end{eqnarray}
are the exact solution of the problem, and $M_{p,q}^E$, $\xi_p^E$, $\rho^E$, $\mu^E$, $\mathbf{m}^E$ and $\mathbf{f}_s^E$ are computed from their definitions in Section \ref{Sec Definitions and governing equations} using Eq.(\ref{Eq Manufactured solution Exact}). Notice that Eq.(\ref{Eq Manufactured solution Exact}) has already satisfied $\nabla \cdot \mathbf{u}^E=0$ and $\sum_{p=1}^{N} \phi_p^E=(2-N)$.

The domain considered is $[-\pi,\pi]\times[-\pi,\pi]$, and its boundaries are all free-slip, which is consistent with Eq.(\ref{Eq Manufactured solution Exact}). The material properties and parameters are $\rho_1=1$, $\rho_2=5$, $\rho_3=10$, $\rho_4=15$, $\mu_1=0.01$, $\mu_2=0.02$, $\mu_3=0.05$, $\mu_4=0.10$, $\sigma_{1,2}=0.01$, $\sigma_{1,3}=0.02$, $\sigma_{1,4}=0.03$, $\sigma_{2,3}=0.04$, $\sigma_{2,4}=0.05$, $\sigma_{3,4}=0.06$, $\mathbf{g}=\{1,-2\}$, $\eta=0.1$, and $M_0=0.001$. The initial conditions are obtained from Eq.(\ref{Eq Manufactured solution Exact}) with $t=0$. We first fix the time step to be $\Delta t=10^{-3}$, and successively refine the cell size $h$ from $2\pi/8$ to $2\pi/128$. The computations are stopped at $t=1$ and we compute the $L_2$ error as the root mean square of $f-f^E$, and the $L_\infty$ error as $\mathrm{max}|f-f^E|$, where $f$ is the variable of interest and $f^E$ is its exact value. 

Table \ref{Table Manufactured solution balaned-force space} lists results obtained by using the balanced-force method Eq.(\ref{Eq The balanced-force method}). We observe second-order convergences of both the $L_2$ and $L_\infty$ errors for all the unknowns. We can conclude that the formal order of accuracy in space of the present scheme is second-order. In addition, Table \ref{Table Manufactured solution balaned-force space} also lists the $L_\infty$ norms, i.e, the maximum absolute values, of $\tilde{\nabla} \cdot \mathbf{u}$, $\tilde{\nabla} \cdot (\tilde{\nabla}\tilde{\mathbf{u}})^T$, and $\mathrm{Res}$ which is the residue of the fully discrete mass conservation equation computed with the discrete consistent mass flux, and all of them reach the machine zero. This implies that the cell-face velocity is discretely divergence-free, that the viscous term $\tilde{\nabla} \cdot (\mu \tilde{\nabla}\tilde{\mathbf{u}}^T)$ vanishes inside each bulk-phase region, and that the \textit{consistency of mass conservation} is achieved on the discrete level. Table \ref{Table Manufactured solution conservative space} lists results obtained by using the conservative method Eq.(\ref{Eq The conservative method}) and the same conclusion can be drawn. Next, we set the time step proportional to the cell size, i.e., $\Delta t=\frac{h}{2\pi}$ and run the simulations again. Results obtained by using the balanced-force method and the conservative method are listed in Table~\ref{Table Manufactured solution balaned-force time} and Table \ref{Table Manufactured solution conservative time}, respectively. We again observe second-order convergences of both the $L_2$ and $L_\infty$ errors, no matter which method is used. This implies that the formal order of accuracy of the present scheme in time is second-order as well.
\begin{table*}[!t]
	\centering
	\caption{Results of the manufactured solution problem with fixed $\Delta t$ and with the balanced-force method.} \label{Table Manufactured solution balaned-force space}
	\includegraphics[scale=0.45]{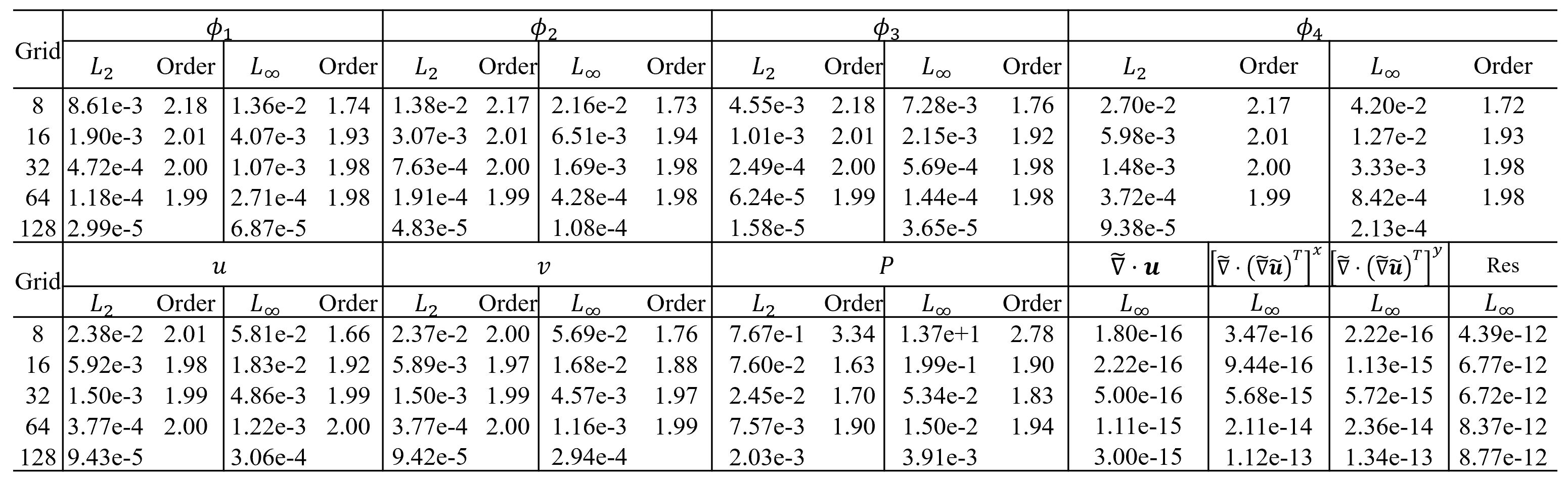}
\end{table*}
\begin{table*}[!t]
	\centering
	\caption{Results of the manufactured solution problem with fixed $\Delta t$ and with the conservative method.} \label{Table Manufactured solution conservative space}
	\includegraphics[scale=0.45]{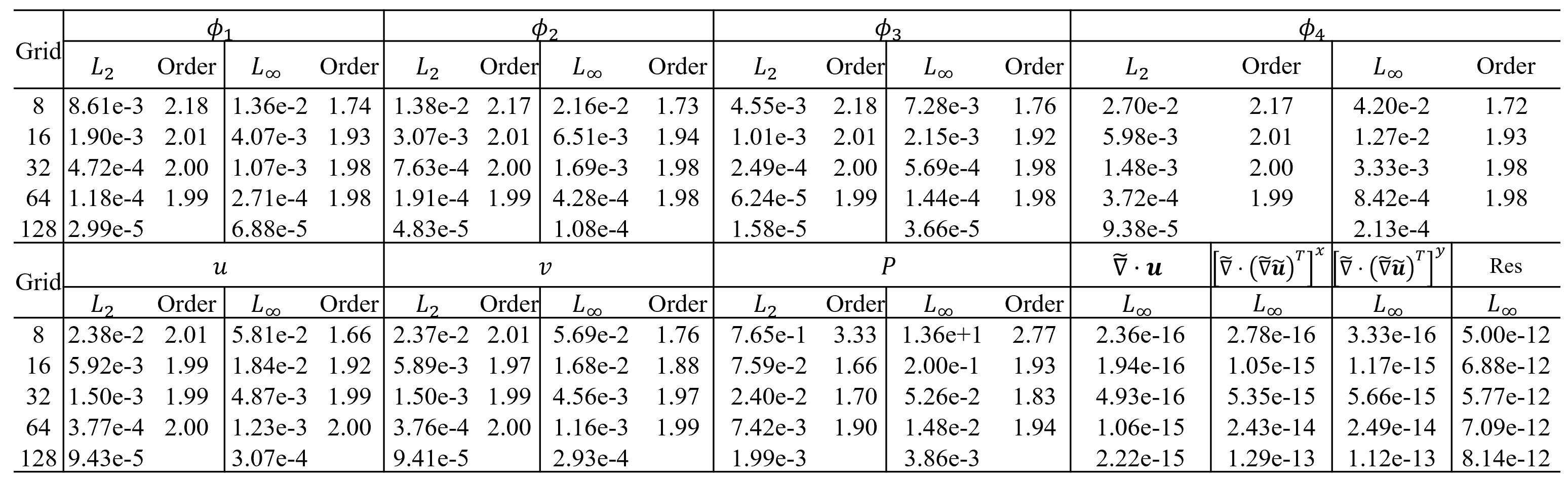}
\end{table*}
\begin{table*}[!t]
	\centering
	\caption{Results of the manufactured solution problem with $\Delta t=\frac{h}{2\pi}$ and with the balanced-force method.} \label{Table Manufactured solution balaned-force time}
	\includegraphics[scale=0.45]{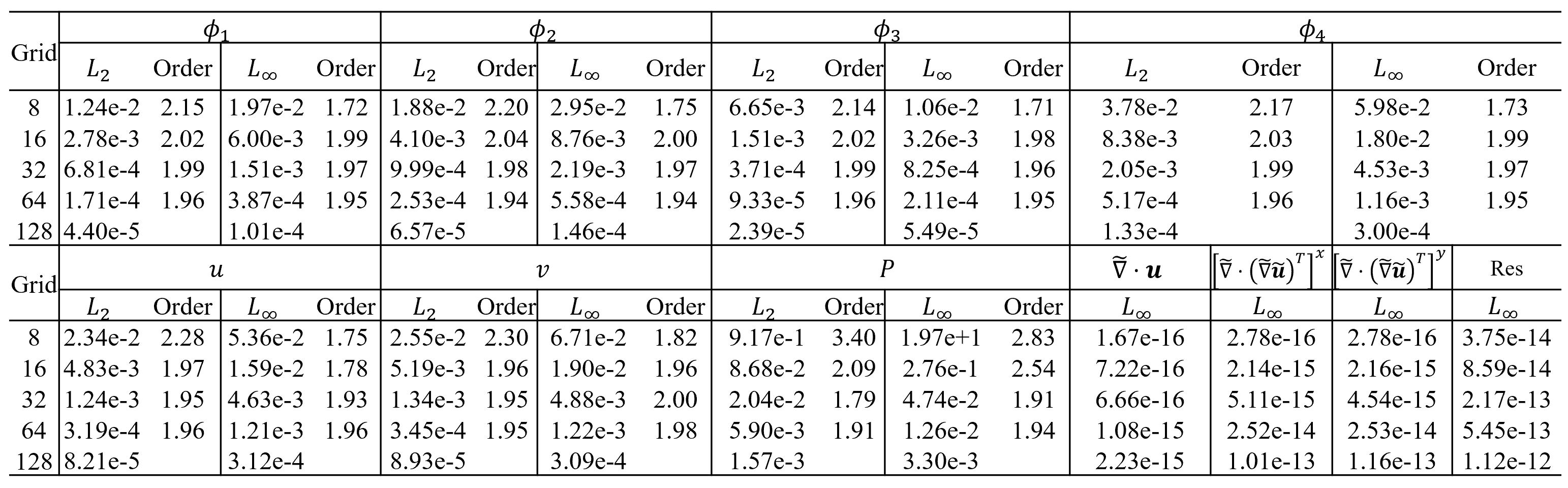}
\end{table*}
\begin{table*}[!t]
	\centering
	\caption{Results of the manufactured solution problem with $\Delta t=\frac{h}{2\pi}$ and with the conservative method.} \label{Table Manufactured solution conservative time}
	\includegraphics[scale=0.45]{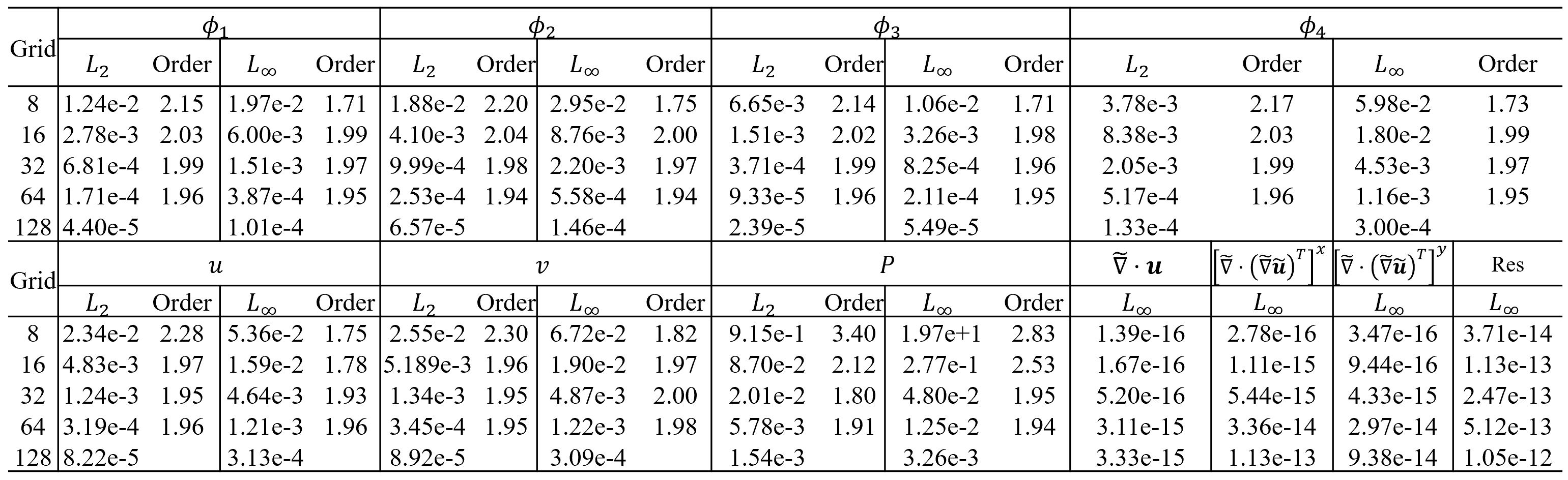}
\end{table*}

It should be noted that, in this case, the residue of the fully discrete mass conservation equation is $\mathrm{Res}=
\left(
\frac{\gamma_t \rho^{n+1}-\hat{\rho}}{\Delta t}
+
\tilde{\nabla} \cdot \langle \mathbf{m} \rangle
-
S_m
\right)
$, 
where $S_m=\sum_{p=1}^N \frac{\rho_p}{2} S_{\phi_p}$, because $S_{\phi_p}$ is added to the Phase-Field equation. Therefore, the \textit{consistency of mass conservation} is satisfied in a more general sense. In the rest of the cases, we have $S_{\phi_p} \equiv 0$, resulting in $S_m \equiv 0$, and the fully discrete mass conservation equation in Eq.(\ref{Eq Mass discrete}) is recovered. 

In summary, the proposed scheme is formally second-order accurate in both time and space, and it satisfies the \textit{consistency of mass conservation}, $\tilde{\nabla} \cdot \mathbf{u}=0$, and $\tilde{\nabla}\cdot(\tilde{\nabla}\tilde{\mathbf{u}}^T)=\mathbf{0}$, on the discrete level.

\subsubsection{Steady drops} \label{Sec Steady drops}
We perform the steady drop problem to quantify the numerical force balance of the present scheme. 
Ample studies of the two-phase spurious current have been performed, e.g., in \citep{Huangetal2019,Huangetal2020,Mirjalilietal2019,Mirjalili2019,AkhlaghiAmiriHamouda2013,Jametetal2002,Kim2005,Popinet2018,Brackbilletal1992,Francoisetal2006,Bogeretal2010,Yokoi2013}. Due to the \textit{consistency of reduction} analyzed in Section \ref{Sec Momentum reduction} and Section \ref{Sec Consistency of reduction}, the two-phase behavior of the spurious current will be automatically recovered by the multiphase flow model and scheme, and, therefore, it is not to be repeated in the present study. Instead, we consider the three-phase steady drops in the present section, which is unable to be easily modeled by previous two-phase methods, to illustrate the properties of the two proposed methods on numerical force balance in the multiphase flows.

\begin{figure}[!t]
	\centering
	\includegraphics[scale=0.4]{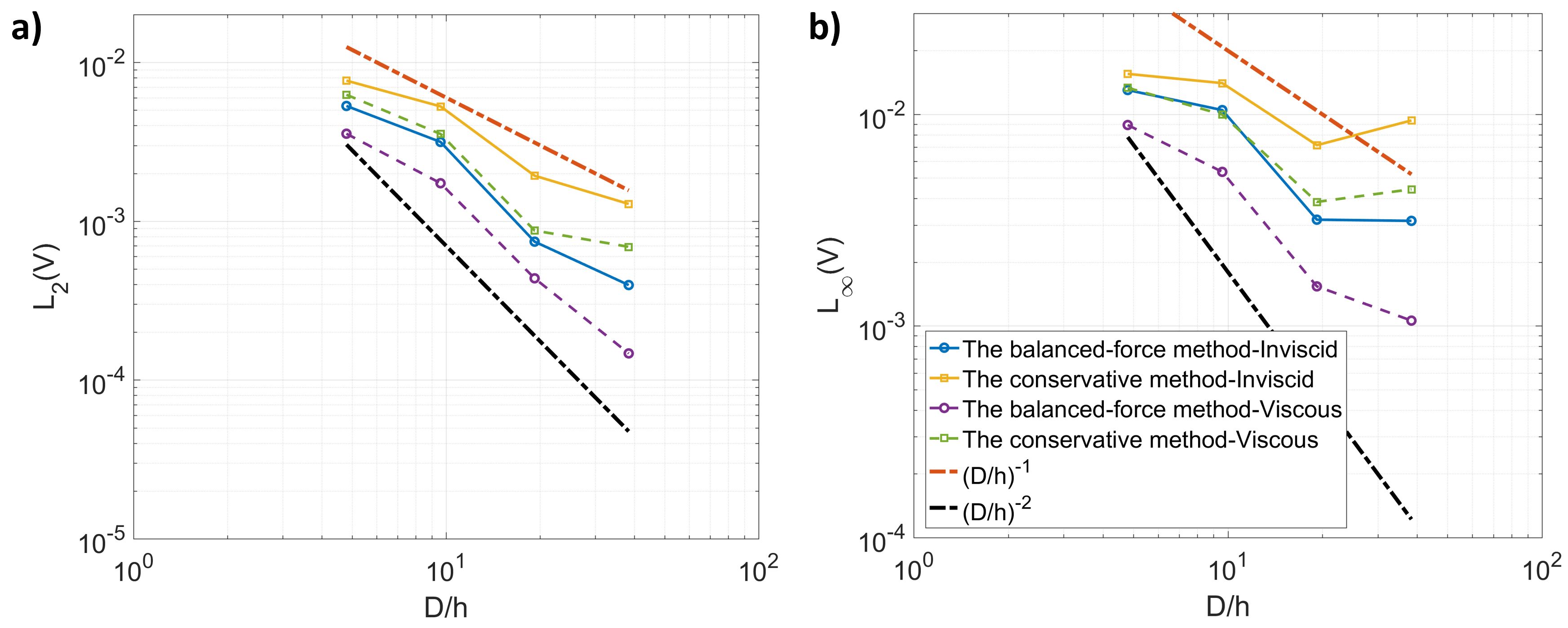}
	\caption{Results of the steady drops. \textbf{a)} $L_2$ and \textbf{b)} $L_{\infty}$ norms of the spurious current $V$($=|\mathbf{u}|$). \textbf{a)} and \textbf{b)} share the same legend.
	$(D/h)$ represents the spatial resolution where $D$ is the diameter of the Phase 1 drop and $h$ denotes the grid size.
	} \label{Fig SD}
\end{figure}

The domain considered is $[1\times1]$ and its boundaries are all free-slip. The material properties are $\rho_1=1000$, $\rho_2=100$, $\rho_3=1$, $\sigma_{1,2}=0.5$, $\sigma_{1,3}=0.1$, $\sigma_{2,3}=1$. The cell size $h$ decreases from $\frac{1}{16}$ to $\frac{1}{128}$ and the time step is $\Delta t=10^{-3}$. $\eta$ is $\eta_0 \left(\frac{h}{h_0}\right)^{2/3}$, where $\eta_0=h_0=\frac{1}{32}$, and $M_0$ is $10^{-7} \left( \frac{\eta}{\eta_0}\right)^{3/2}$, like those in \citep{Mirjalilietal2019,Huangetal2020}. The spurious current is measured at $t=10$. Initially, the drop (Phase 1) at $(0.5,0.5)$ with a radius 0.15 is surrounded by a circular ring (Phase 2) whose outer radius is 0.3. Since all the interfaces are circular, they are in the equilibrium state. The interfacial tensions should be exactly balanced by the pressure gradient, and the interfaces should not move or deform. However, such an exact force balance is seldom achievable in numerical implementation. As a result, the numerical error introduced by the force imbalance drives the fluids to move and generates the so-called spurious current. The strength of the spurious current is quantified by computing the $L_2$ and $L_{\infty}$ norms of the total velocity $V$, where $V$ is $\sqrt{u^2+v^2}$. We consider both the inviscid and viscous cases and compare the performance of the balanced-force method to that of the conservative method. In the viscous case, $\mu_1=0.1$, $\mu_2=0.005$, and $\mu_3=0.0001$. The results are shown in Fig.\ref{Fig SD}. 

It can be observed that the balanced-force method performs better than the conservative method in this case. The balanced-force method generates a smaller spurious current and the norms of the spurious current reduce faster during the grid refinement. Its $L_2$ norm reduces with a rate close to second order, while the convergence rate of its $L_{\infty}$ norm is about 1st order. In the cases using the conservative method, both the $L_2$ and $L_{\infty}$ norms converge at a rate close to 1st order. In addition, the spurious currents in the viscous cases are smaller than those in the inviscid cases. However, the appearance of the viscosity does not change the convergence rate. 

In summary, the spurious currents generated by the balanced-force method and by the conservative method are in an acceptable range. However, the balanced-force method achieves better numerical force balance than the conservative method.

\subsubsection{Rising bubble: a convergence test}
We perform a convergence test to quantify the convergence behavior of the numerical solution of the present mulitphase flow model to the sharp-interface solution. There are two convergence behaviors to consider when we numerically apply a Phase-Field based multiphase flow model. The first one is the convergence of the numerical solution to the exact solution of the multiphase flow model where the interface has a fixed thickness, as the cell size decreases. The second one is the convergence of the exact solution of the multiphase flow model to the sharp-interface solution, as the interface thickness decreases. In numerical implementations, it is more common and practical to consider those two together by decreasing both the cell size and the interface thickness at the same time, and then to evaluate how fast the numerical solution of the multiphase flow model converges to the sharp-interface solution. Hereafter in this section, we call the numerical solution of the present multiphase flow model from the present scheme in Section \ref{Sec Scheme} as the Phase-Field solution.

The domain considered is $[1\times2]$ and it has no-slip boundaries at the top and bottom but free-slip at the lateral. A bubble (Phase 1), whose density is 1 and viscosity is 0.1, is surrounded by Phase 2, whose density is 1000 and viscosity is 10. The density and viscosity ratios are 1000 and 100, respectively. The interfacial tension between them is 1.96, and the gravity is $\mathbf{g}=\{0,-0.98\}$. The cell size $h$ is ranging from $\frac{1}{16}$ to $\frac{1}{256}$, and the time step is $\Delta t= \mathrm{CFL} h$, where $\mathrm{CFL}=0.128$. We set $\eta=\eta_0 \frac{h}{h_0}$, where $\eta_0$=$h_0=\frac{1}{32}$, and $M_0=10^{-7}\frac{\eta}{\eta_0}$, so that both of them decrease as the cell is refined. Initially, the bubble is at $(0.5,0.5)$ and its radius is 0.25. The computation is stopped at $t=1$. We use the circularity $\psi_c$ to quantify the shape of the bubble, while the dynamics of the bubble is quantified by the center of mass $y_c$ and the rising velocity $v_c$. These three benchmark quantities are defined as
\begin{eqnarray} \label{Eq Circularity}	
\psi_c=\frac{P_a}{P_b}=\frac{2\sqrt{\int_{\phi_1>0} \pi d\Omega }}{P_b},\\
\nonumber
y_c=\frac{ \int_{\Omega} y \frac{1+\phi_1}{2} d \Omega}{ \int_{\Omega} \frac{1+\phi_1}{2} d \Omega },\\
\nonumber
v_c=\frac{ \int_{\Omega} v \frac{1+\phi_1}{2} d \Omega}{ \int_{\Omega} \frac{1+\phi_1}{2} d \Omega },
\end{eqnarray}
where $P_a$ is the perimeter of a circle whose area is identical to the bubble, and $P_b$ is the perimeter of the bubble. If the bubble is circular, $\psi_c$ is 1. Otherwise, it is less than 1. The same case was performed in \citep{Hysingetal2009} using the sharp-interface models, where the Level-Set and the Arbitrary-Lagrange-Euler (ALE) methods were implemented and excellent agreements were reached in the time zone considered. The same benchmark quantities from the sharp-interface models are available in \citep{Hysingetal2009} and they are considered as the references. 

\begin{figure}[!t]
	\centering
	\includegraphics[scale=0.5]{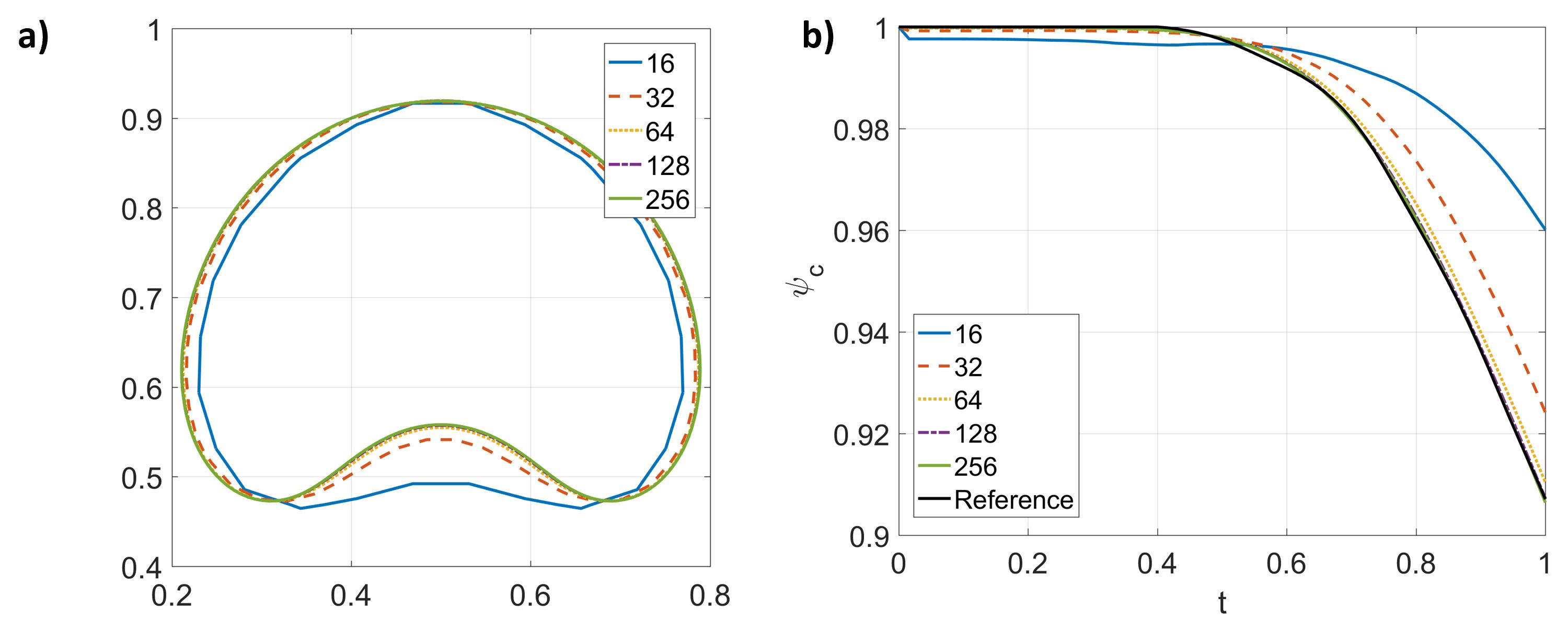}\\
	\includegraphics[scale=0.5]{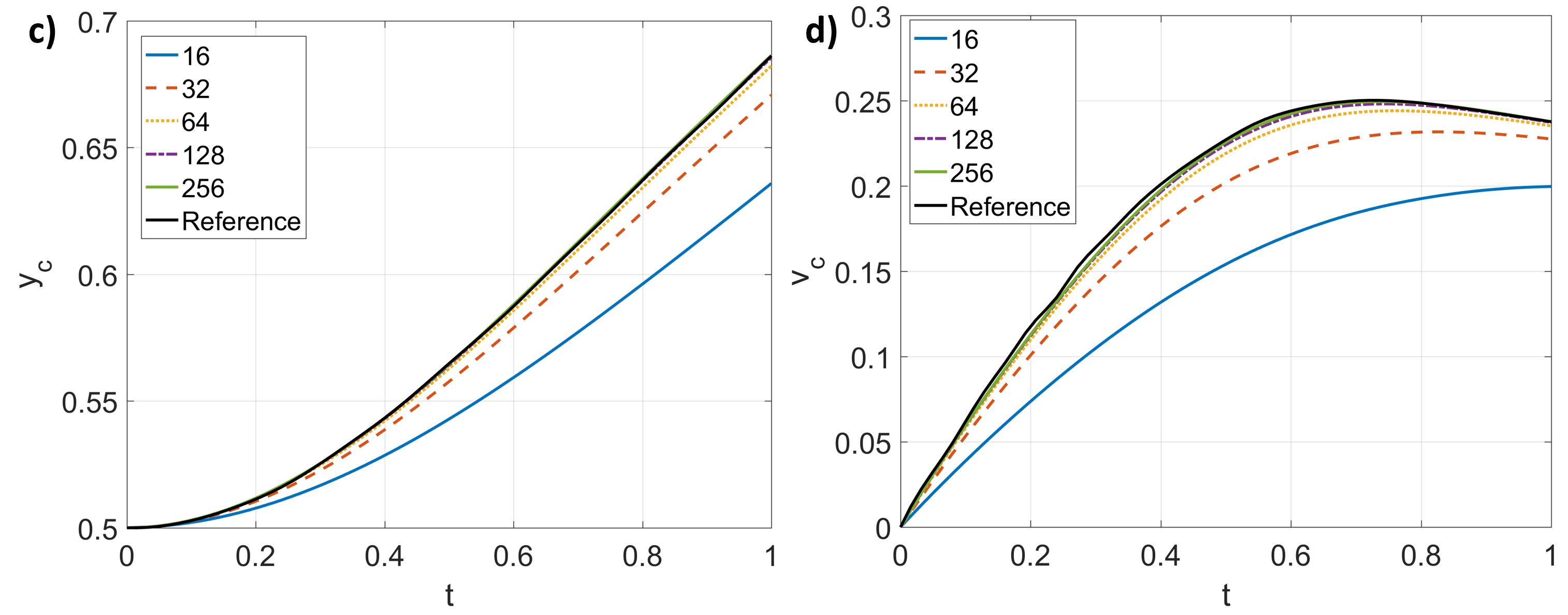}
	\caption{Results of the convergence test using the balanced-force method. \textbf{a)} Bubble shape at $t=1$. \textbf{b)} Circularity $\psi_c$ vs. $t$. \textbf{c)} Center of mass $y_c$ vs. $t$. \textbf{d)} Rising velocity $v_c$ vs. $t$. The legend shows the number of grid cells per unit length, and ``Reference'' is the sharp-interface solution from \citep{Hysingetal2009}.} \label{Fig RBC-FB}
\end{figure}

\begin{figure}[!t]
	\centering
	\includegraphics[scale=0.5]{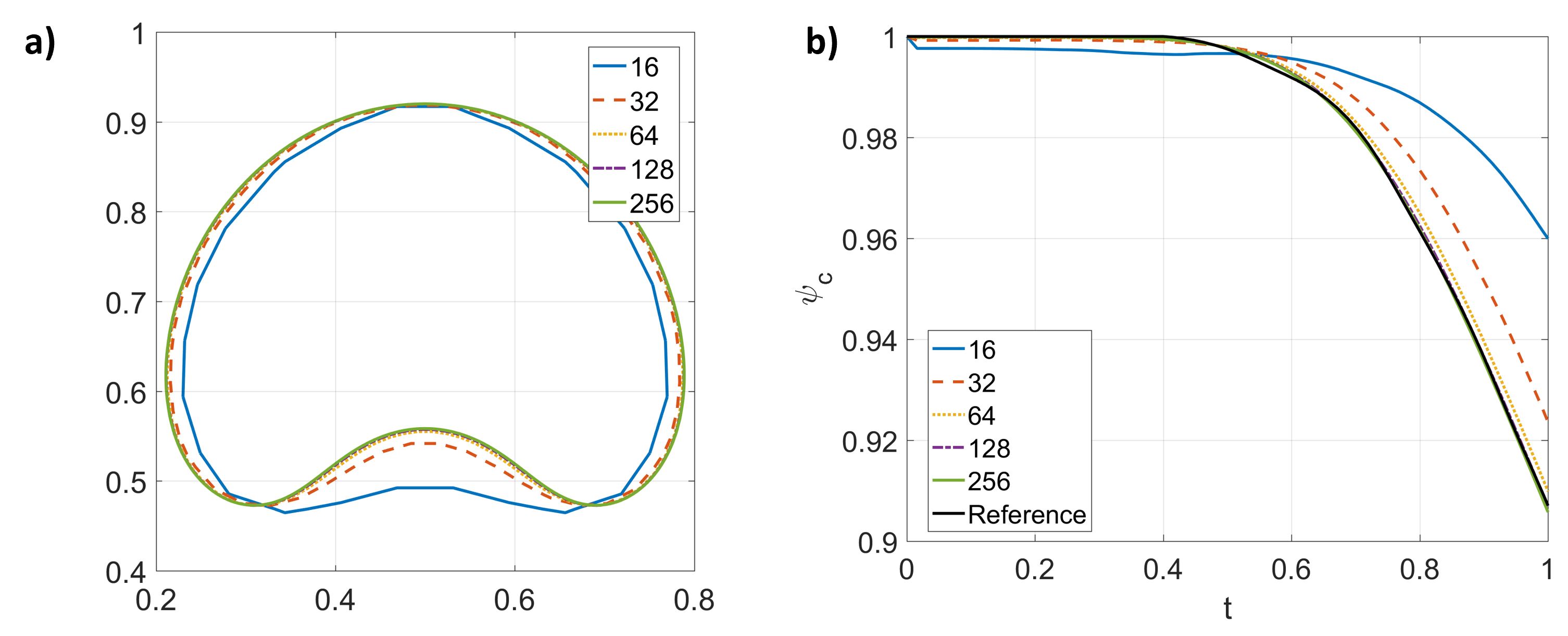}\\
	\includegraphics[scale=0.5]{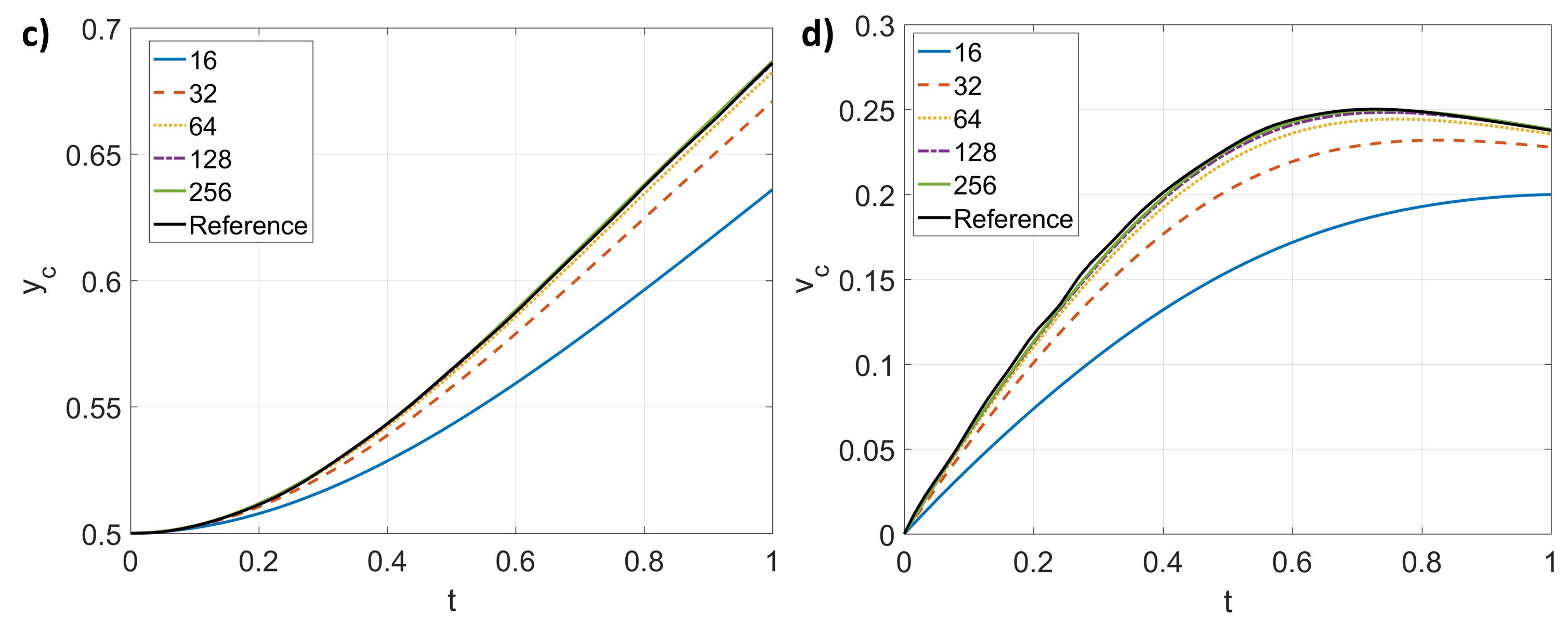}
	\caption{Results of the convergence test using the conservative method. \textbf{a)} Bubble shape at $t=1$. \textbf{b)} Circularity $\psi_c$ vs. $t$. \textbf{c)} Center of mass $y_c$ vs. $t$. \textbf{d)} Rising velocity $v_c$ vs. $t$. The legend shows the number of grid cells per unit length, and ``Reference'' is the sharp-interface solution from \citep{Hysingetal2009}.} \label{Fig RBC-CS}
\end{figure}

Fig.\ref{Fig RBC-FB} and Fig.\ref{Fig RBC-CS} show results using the balanced-force method and the conservative method, respectively. It is clear that the Phase-Field solution converges to the sharp-interface solution (labeled as ``Reference'' in the figures) as the cell size $h$, as well as the interface thickness $\eta$, reduces. To quantify the convergence rate, we compute the $L_2$ errors, i.e, the root mean square of the difference between the Phase-Field and the sharp-interface solutions, of the three benchmark quantities. The results are listed in Table \ref{Table RBC}, where ``B-method'' refers to the balanced-force method and ``C-method'' represents the conservative method. There is no significant difference between the results of the two methods, although the conservative method performs slightly better, giving smaller errors and faster convergence rates. Overall, both the circularity $\psi_c$ and the center of mass converge to their sharp-interface solutions with a rate close to second order, while the convergence rate of the rising velocity is slightly slower about 1.5th order.

In summary, the numerical solution of the present multiphase flow model from the present scheme in Section \ref{Sec Scheme} converges to the sharp-interface solution, even in a case including large density and viscosity ratios. The convergence rate is between 1.5th- and 2nd-order. 

\begin{table*}[!t]
	\centering
	\caption{Results of the convergence test. B-method: the balanced-force method; C-method: the conservative method.} \label{Table RBC}
	\includegraphics[scale=0.6]{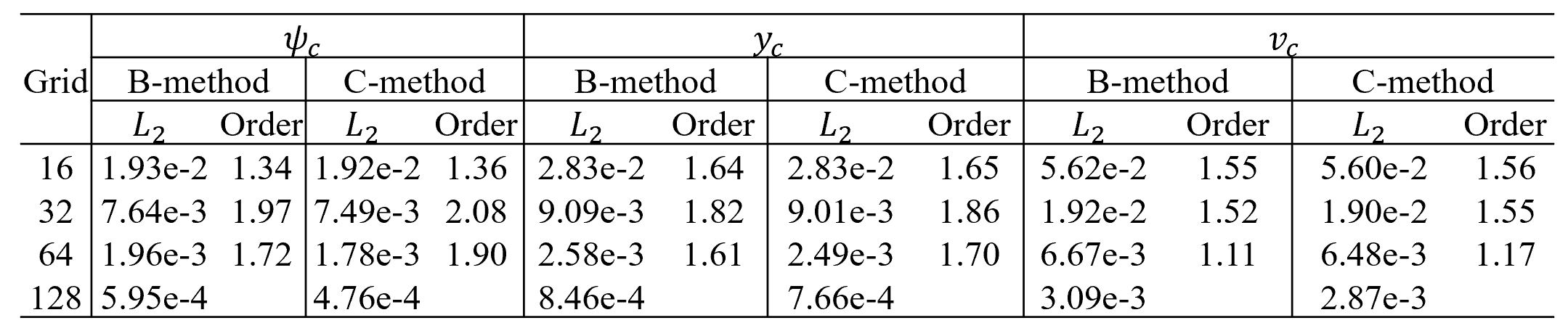}
\end{table*}

\subsubsection{Large-density-ratio advection}\label{Sec Advection}
We perform the large-density-ratio advection problem to demonstrate that the \textit{consistency of mass conservation} and the \textit{consistency of mass and momentum transport} are satisfied on the discrete level by the present scheme , and to illustrate the significance of satisfying these two consistency conditions. The domain considered is $[1\times1]$ with doubly periodic boundaries. The material properties, unless otherwise specified, are $\rho_1=10^{9}$, $\rho_2=10^{6}$, and $\rho_1=1$ and there is neither viscosity nor interfacial tension. The cell size $h$ is $\frac{1}{128}$, and the time step is $\Delta t=\mathrm{CFL} h$, where $\mathrm{CFL}=0.1$. All the computations are stopped at $t=1$. We set $\eta$ to be $3h$ and $M_0$ to be $10^{-7}$. Initially, a circular drop (Phase 1) with a radius $0.1$ is at $(0.5,0.5)$, surrounded by Phase 2, whose outer interface is an ellipse with its semi-major axis 0.3 along the $y$ axis and with its semi-minor axis 0.25 along the $x$ axis. Both $u$ and $v$ are unity everywhere. Based on Theorem~\ref{Theorem constant velocity discrete} in Section~\ref{Sec Scheme momentum properties}, the interfaces should be translated by the homogeneous velocity and therefore return to their original location at $t=1$ without any deformation. Moreover, the initial homogeneous velocity should not be changed, thanks to satisfying the \textit{consistency of mass conservation} and the \textit{consistency of mass and momentum transport} on the discrete level. 

\begin{figure}[!t]
	\centering
	\includegraphics[scale=0.5]{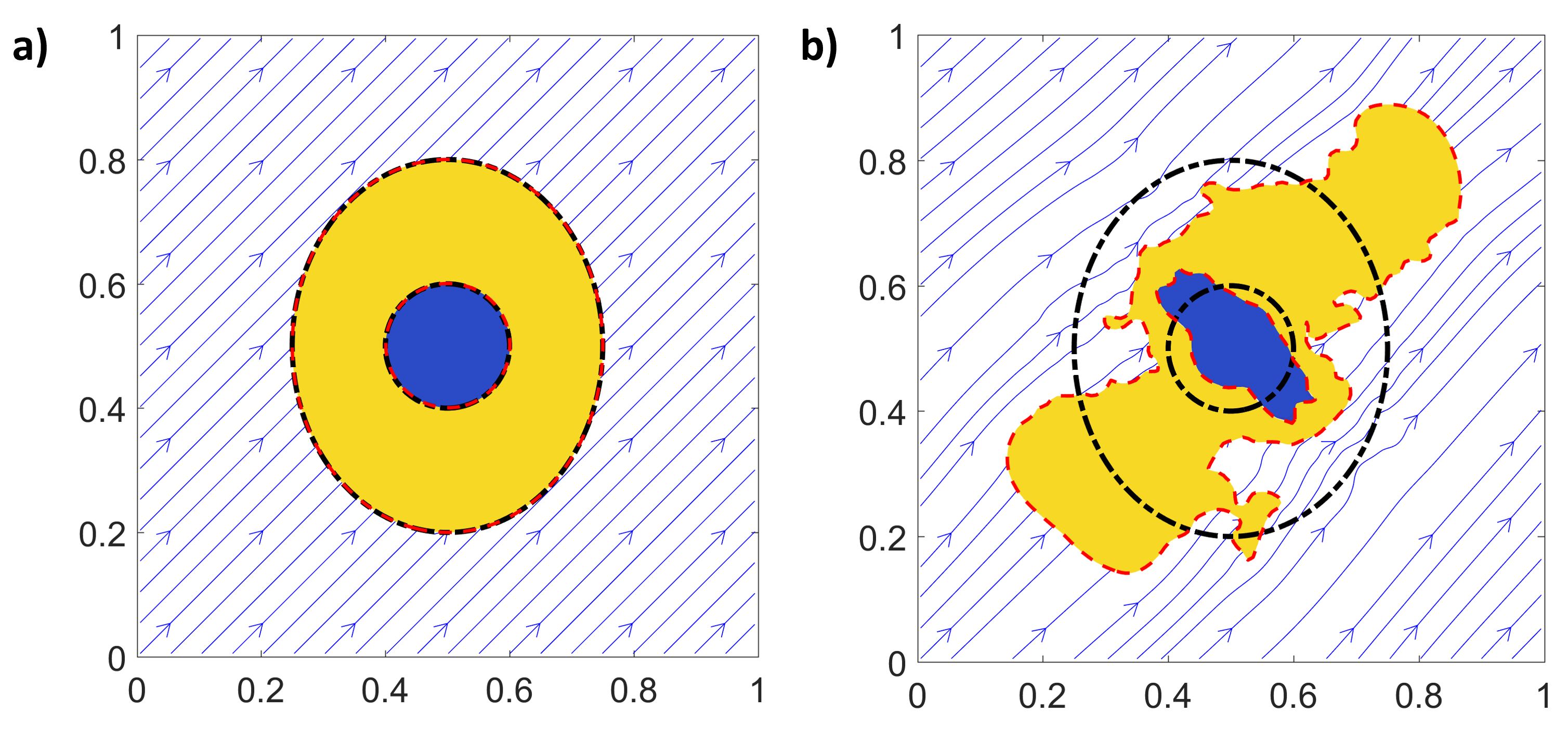}
	\includegraphics[scale=0.5]{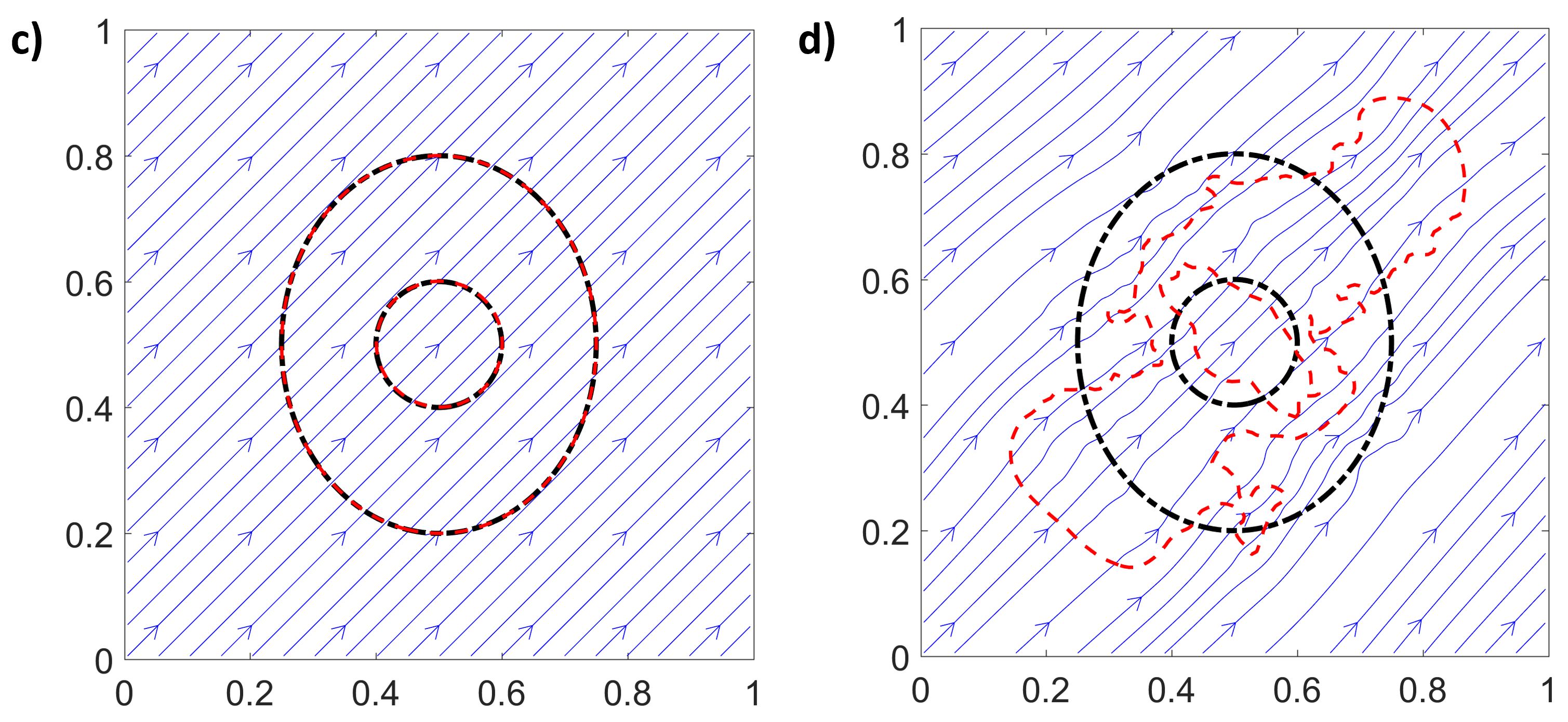}
	\caption{Results of Large-density-ratio advection. \textbf{a)} and \textbf{c)} Results from the present scheme with densities $\rho_1=10^9$, $\rho_2=10^6$, and $\rho_1=1$. \textbf{b)} and \textbf{d)} Results of using the inconsistent mass flux with densities $\rho_1=10^3$, $\rho_2=10$, and $\rho_1=1$. Blue arrow lines: Streamlines at $t=1$, Black dash-dotted lines: Interfaces at $t=0$, Red dashed lines: Interfaces at $t=1$, Blue: Phase 1, Yellow: Phase 2, Background: Phase 3.} \label{Fig AB}
\end{figure}

Fig.\ref{Fig AB} \textbf{a)} and \textbf{c)} show the result from the present scheme. It can be observed that the interfaces, as expected, return to their initial location without any deformation, and all the streamlines are straightly pointing towards $45^0$, as what they were at $t=0$. Fig.\ref{Fig AB} \textbf{b)} and \textbf{d)} show the result from an inconsistent mass flux, i.e., using $\langle \mathbf{m} \rangle=\overline{\rho} \mathbf{u}$. Due to violating the consistency conditions, the interfaces experience significant unphysical deformations and the streamlines are fluctuating. It should be noted that the densities of the phases are $1000$, $10$, and $1$, respectively, in the case using the inconsistent mass flux, in order to have a stable solution up to $t=1$.

In summary, the present scheme satisfies the \textit{consistency of mass conservation} and the \textit{consistency of mass and momentum transport} on the discrete level, and Theorem \ref{Theorem constant velocity discrete} in Section \ref{Sec Scheme momentum properties} is validated.

\subsubsection{Horizontal shear layer}\label{Sec Horizontal shear layer}
We perform the horizontal shear layer problem to validate the \textit{consistency of reduction}, summation of the order parameters, and mass and momentum conservation of the present scheme, and further to discuss the energy law on the discrete level. Different layers of fluids are moving horizontally at different speeds. A vertical velocity perturbation is applied and the fluids begin to interact with each other. The domain considered is $[1\times1]$ and all its boundaries are periodic. We use $[128\times128]$ cells to discretize the domain in default, and the time step is $\Delta t=\mathrm{CFL} h$, where $h$ is the cell size and $\mathrm{CFL}$ is $0.1$. The material properties, unless otherwise specified, are $\rho_1=50$, $\rho_2=10$, $\rho_3=1$, $\rho_4=0.5$, $\mu_1=0.01$, $\mu_2=0.1$, $\mu_3=0.05$, $\mu_4=0.08$, $\sigma_{1,2}=0.05$, $\sigma_{1,3}=0.01$, $\sigma_{1,4}=0.08$, $\sigma_{2,3}=0.1$, $\sigma_{2,4}=0.02$, $\sigma_{3,4}=0.2$. $\eta$ and $M_0$ are $\frac{1}{128}$ and $10^{-7}$, respectively. Initially, Phase 1 is stationary in $y_0<y<y_2$, Phase 2 is in $y_1<y<y_0$, moving from the left to right at a unity speed, and Phase 3 moves in the opposite direction, filling the rest of the domain. Phase~4 is absent, i.e., $\phi_4|_{t=0} = -1$. We set $y_0=0.5$, $y_1=0.25$, and $y_2=0.75$. This configuration is perturbed by a vertical velocity $v|_{t=0}=\delta \sin(k_0 x)$, where $\delta=0.05$ and $k_0=2\pi$.

\begin{figure}[!t]
	\centering
	\includegraphics[scale=0.5]{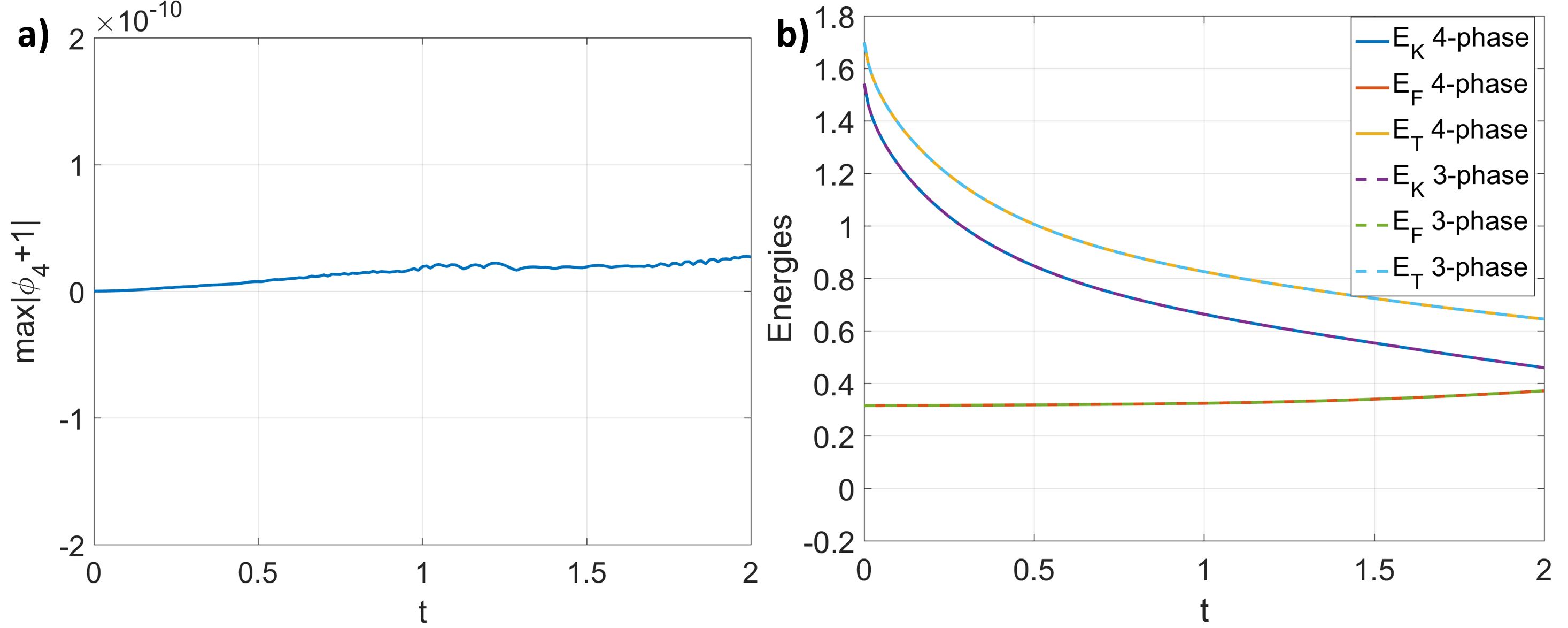}\\
	\includegraphics[scale=0.5]{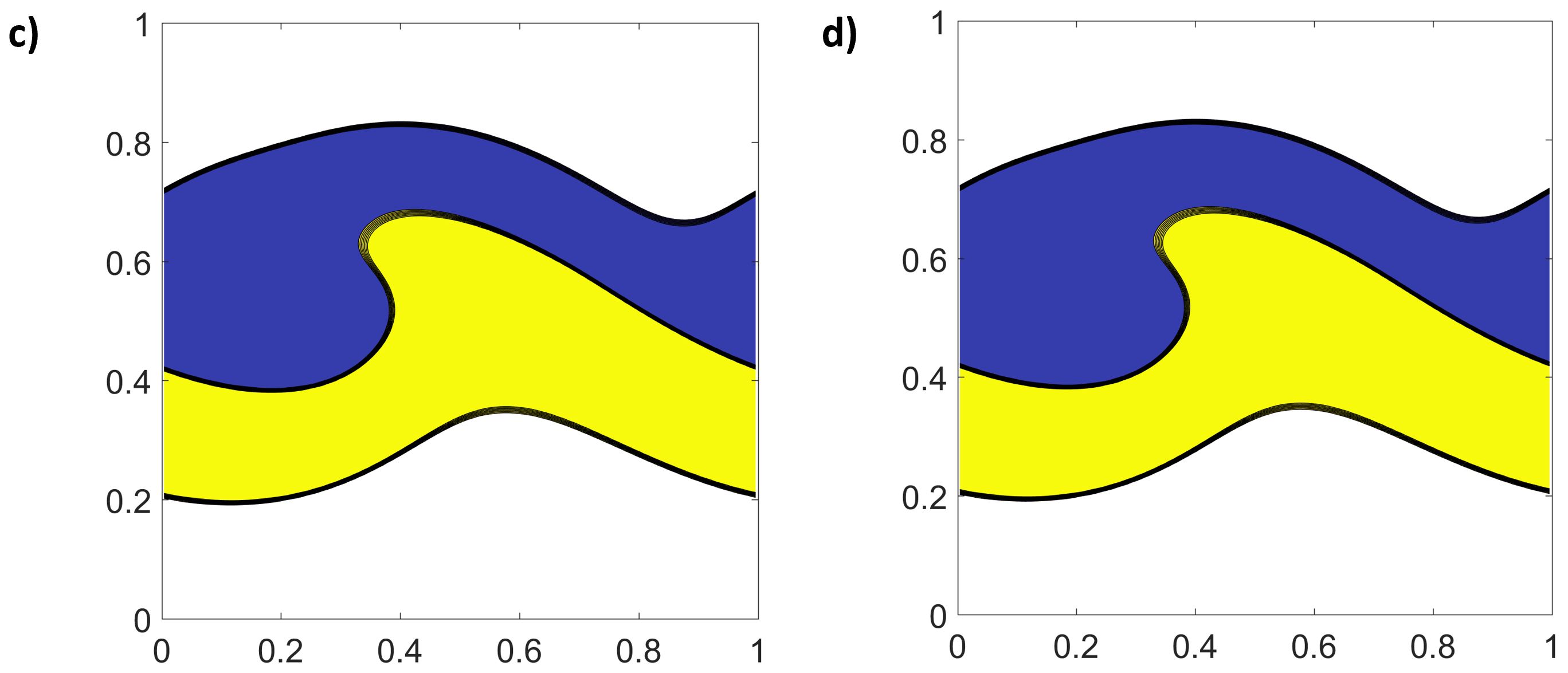}
	\caption{Validation of the \textit{consistency of reduction}. \textbf{a)} Time history of $\mathrm{max}|\phi_4+1|$. \textbf{b)} Time histories of the kinetic, free and total energies from the 4-phase and 3-phase solutions. \textbf{c)} Snapshot of the phases at $t=2$ from the 4-phase solution. \textbf{d)} Snapshot of the phases at $t=2$ from the 3-phase solution. Blue: Phase~1, Yellow: Phase 2, White: Phase 3.}\label{Fig HSL-RC}
\end{figure}

\begin{figure}[!t]
	\centering
	\includegraphics[scale=0.4]{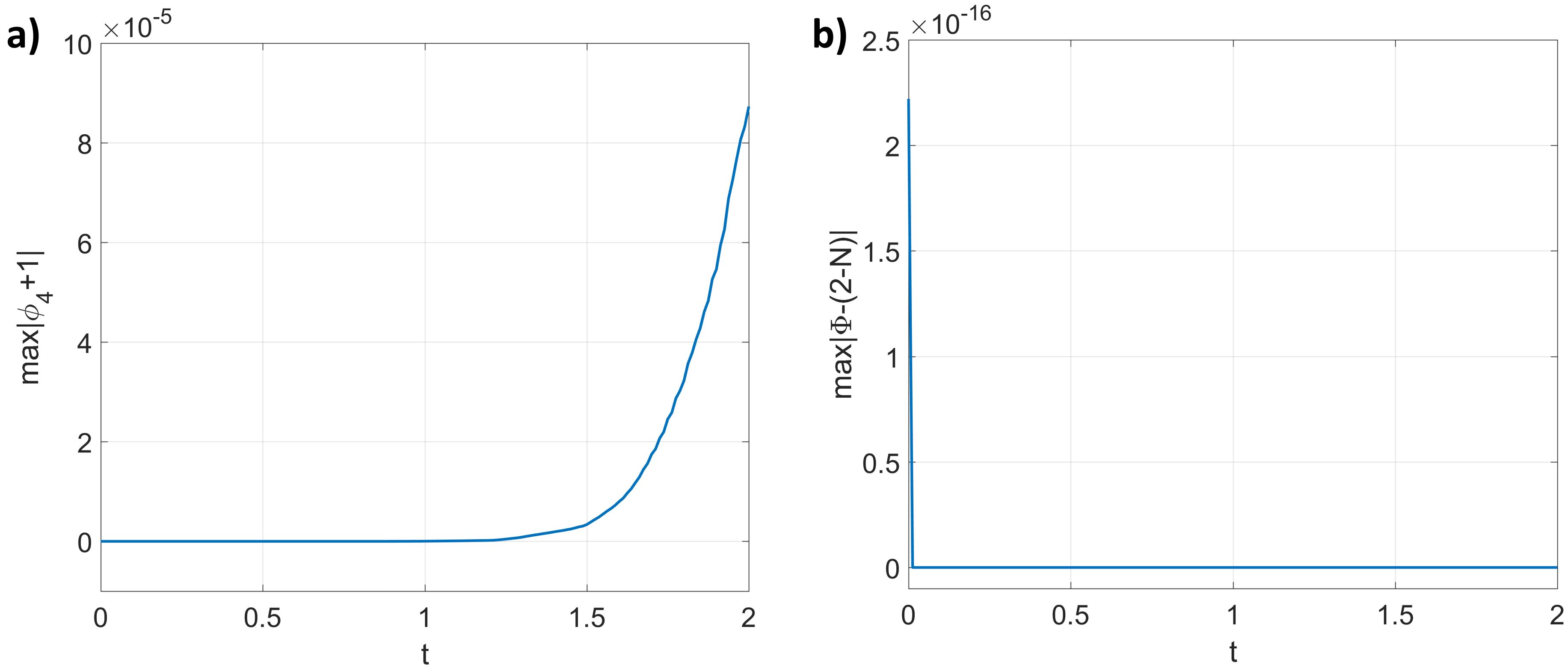}
	\caption{Generation of fictitious phases from a reduction inconsistent scheme: solving $\phi_1$, $\phi_2$, and $\phi_3$ numerically from their Phase-Field equation and obtaining $\phi_4$ algebraically from $\phi_4=-2-(\phi_1+\phi_2+\phi_3)$. \textbf{a)} Time history of $\mathrm{max}|\phi_4+1|$. \textbf{b)} Time history of $\mathrm{max}|\Phi-(2-N)|$ ($\Phi=\sum_{p=1}^N \phi_p$).}\label{Fig HSL-RIC}
\end{figure}

We first validate the \textit{consistency of reduction} on the discrete level. From Theorem \ref{Theorem consistency of reduction Phase-Field discrete} in Section~\ref{Sec Scheme Phase-Field properties}, $\phi_4$, which is absent at the beginning of the computation, should not appear during the computation, i.e., $\phi_4 \equiv -1$. In addition, since Phase 4 is absent, the problem considered is actually a three-phase problem. The result from setting $N=4$ and $\phi_4|_{t=0}=-1$ should be the same as that from setting $N=3$. Fig.\ref{Fig HSL-RC} \textbf{a)} shows the time history of the maximum value of $|\phi_4+1|$. It is clear that $\phi_4$ equals to $-1$ up to the round off error. Fig.\ref{Fig HSL-RC} \textbf{b)} shows the time histories of the kinetic energy $E_K$ ($=\sum_{i,j} [e_K]_{i,j}\Delta \Omega$), the free energy $E_F$ ($=\sum_{i,j} [e_F]_{i,j}\Delta \Omega$), and the total energy $E_T$ ($=E_K+\frac{1}{2}E_F$) from the 4-phase and 3-phase solutions, and they are indistinguishable from each other. Fig.\ref{Fig HSL-RC} \textbf{c)} and \textbf{d)} show the snapshots of the three phases at $t=2$ from the 4-phase and 3-phase solutions, respectively, and they are identical. The results validate that the \textit{consistency of reduction} is satisfied by the present scheme on the discrete level. 
Instead of applying the present scheme, where all the order parameters are numerically solved from their Phase-Field equation, we repeat the 4-phase setup and follow the practice in \citep{Dong2018}, where only $\phi_1$, $\phi_2$, and $\phi_3$ are numerically solved from their Phase-Field equation and $\phi_4$ is algebraically obtained from their summation, i.e., $\phi_4=-2-(\phi_1+\phi_2+\phi_3)$. The result is shown in Fig.\ref{Fig HSL-RIC}. It can be observed that Phase 4, which is absent at the beginning, is being generated numerically, and, therefore, the \textit{consistency of reduction} is violated on the discrete level, although the summation of the order parameters, i.e., Eq.(\ref{Eq Summation of volume fraction contrasts}), is exactly enforced. As discussed in Section \ref{Sec Phase selection}, $\{\widetilde{\phi_p}\}_{p=1}^3$ in the convection term of the Phase-Field equation do not follow $\sum_{p=1}^3 \widetilde{\phi_p}=-1$ because they are computed independently from the WENO scheme. This error contributes to $\sum_{p=1}^3 \phi_p \neq -1$, and consequently, $\phi_4=-2-(\phi_1+\phi_2+\phi_3)$ can be larger than $-1$. In other words, $\phi_4$ is generated numerically. This numerical test highlights that even though the Phase-Field equation satisfies the \textit{consistency of reduction}, its numerical scheme still needs to be carefully designed in order to preserve this property and, consequently, to avoid generating fictitious phases numerically.
Since we have validated that the present scheme satisfies the \textit{consistency of reduction}, in the rest of this section, we only use the 3-phase solution.

\begin{figure}[!t]
	\centering
	\includegraphics[scale=0.5]{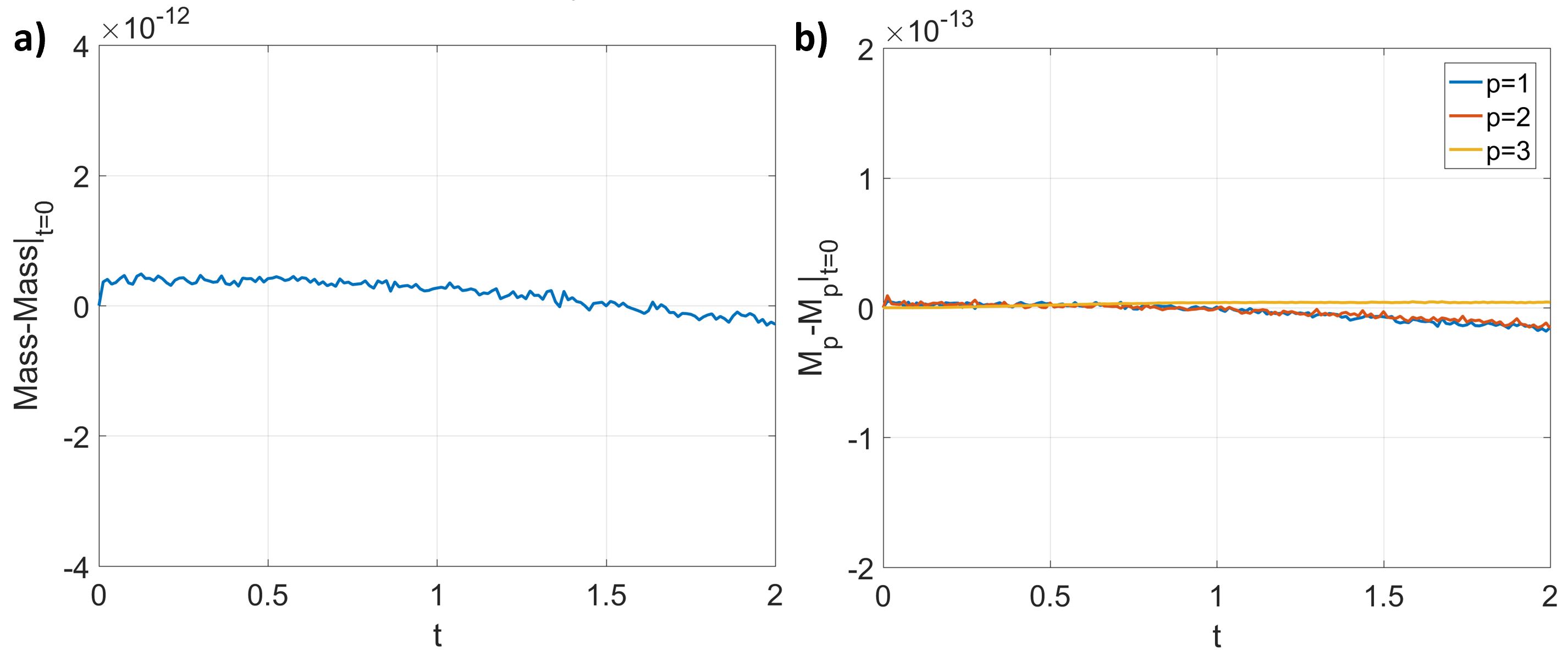}\\
	\includegraphics[scale=0.5]{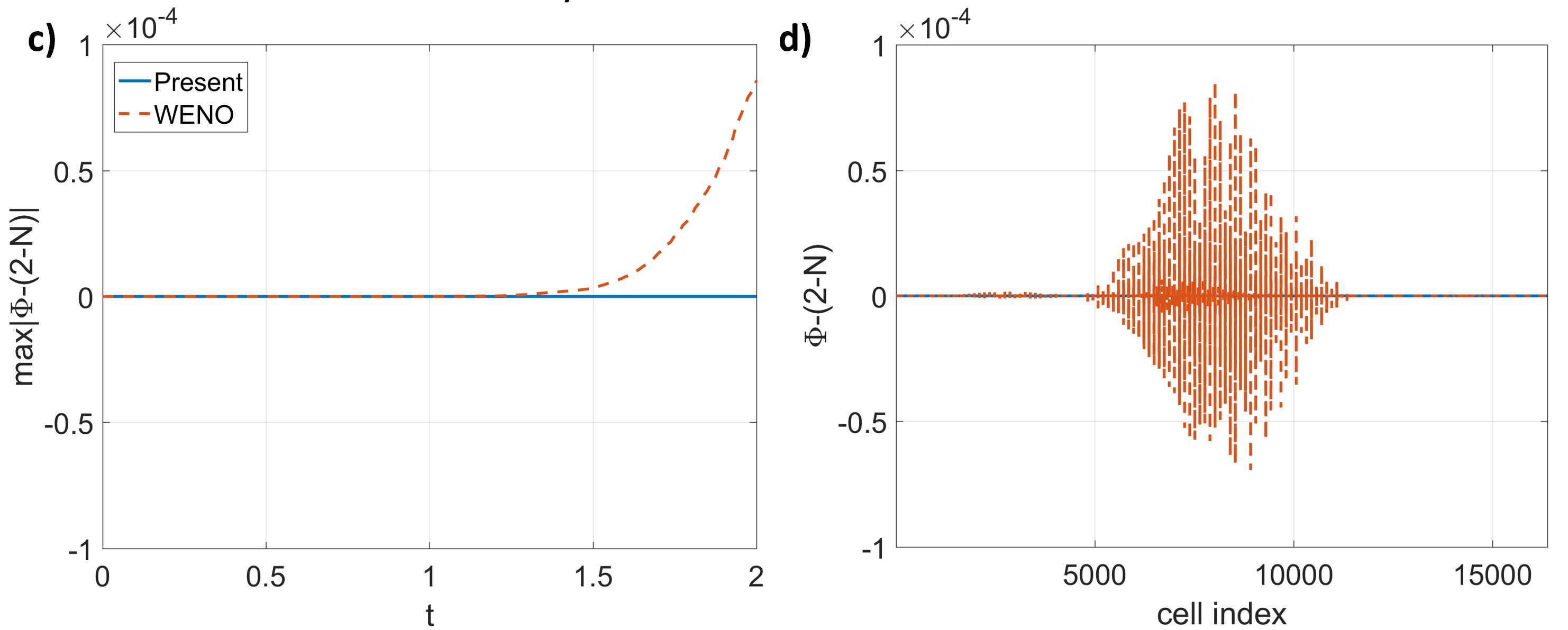}
	\caption{Validation of the mass conservation and summation of the order parameters. \textbf{a)} Time history of the change of $\mathrm{Mass}$($=\sum_{i,j} [\rho]_{i,j} \Delta \Omega$). \textbf{b)} Time histories of the change of $\mathrm{M}_p$($=\sum_{i,j} [\phi_p]_{i,j} \Delta \Omega$). \textbf{c)} Time histories of $\mathrm{max}|\Phi-(2-N)|$ ($\Phi=\sum_{p=1}^N \phi_p$) from the present scheme including the gradient-based phase selection procedure (Present) and the scheme using WENO only (WENO). \textbf{d)} Spatial distributions of $\Phi-(2-N)$ at $t=2$ from the present scheme and the scheme using WENO only. \textbf{c)} and \textbf{d)} share the same legend.
	}\label{Fig HSL-MassC}
\end{figure}

We next consider the mass conservation and the summation of the order parameters on the discrete level. From Theorem \ref{Theorem Mass conservation p} in Section \ref{Sec Scheme Phase-Field properties}, the order parameters and therefore the mass of fluid mixture are conserved, i.e., $\mathrm{Mass}=\sum_{i,j} [\rho]_{i,j} \Delta \Omega$ and $\mathrm{M}_p=\sum_{i,j} [\phi_p]_{i,j} \Delta \Omega$ do not change with time. In addition, from Theorem \ref{Theorem Summation Phi}, $\Phi=\sum_{p=1}^N \phi_p=(2-N)$ is true at every cell center, which avoids generating local voids or overfilling. Fig.\ref{Fig HSL-MassC} \textbf{a)} and \textbf{b)} show that the changes of $\mathrm{Mass}$ and $\mathrm{M}_p$, respectively, are the machine zero. These results confirm that the mass conservation is satisfied by the present scheme on the discrete level. Fig.\ref{Fig HSL-MassC} \textbf{c)} and \textbf{d)} show the results including the gradient-based phase selection procedure (Present) and using the WENO scheme only (WENO), respectively. Because of the non-linearity of the WENO scheme, the summation and the interpolation are not commutable. As a result, $\sum_{p=1}^N \tilde{\nabla} \cdot (\mathbf{u} \widetilde{\phi_p})$ is not zero, which leads to $\Phi \neq (2-N)$. Besides, the difference between $\Phi$ and $(2-N)$ increases with time, as shown in Fig.\ref{Fig HSL-MassC} \textbf{c)}, although the difference between $\Phi^{n+1}$ and $\Phi^{n}$ at the beginning is tiny. The error is accumulating and finally becomes noticeable in long-time simulations. Physically speaking, this error corresponds to generating local voids or overfilling. On the contrary, the present scheme, which implements the gradient-based phase selection procedure, guarantees $\Phi=(2-N)$ all the time. Fig.\ref{Fig HSL-MassC} \textbf{d)} shows the spatial distribution of $\Phi-(2-N)$ at $t=2$. The present scheme guarantees $\Phi=(2-N)$ at every cell center, while only using the WENO scheme generates local voids or overfilling at about one third of the total discrete cells.

\begin{figure}[!t]
	\centering
	\includegraphics[scale=0.5]{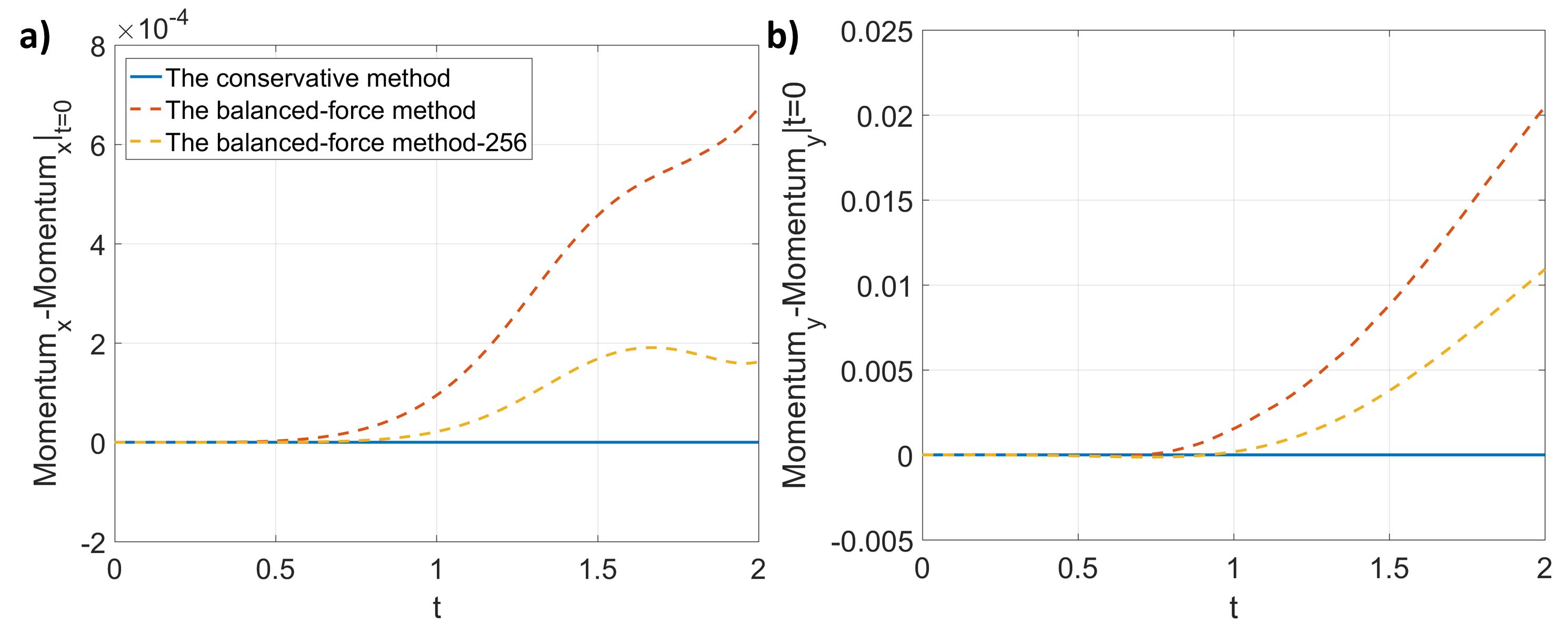}\\
	\includegraphics[scale=0.5]{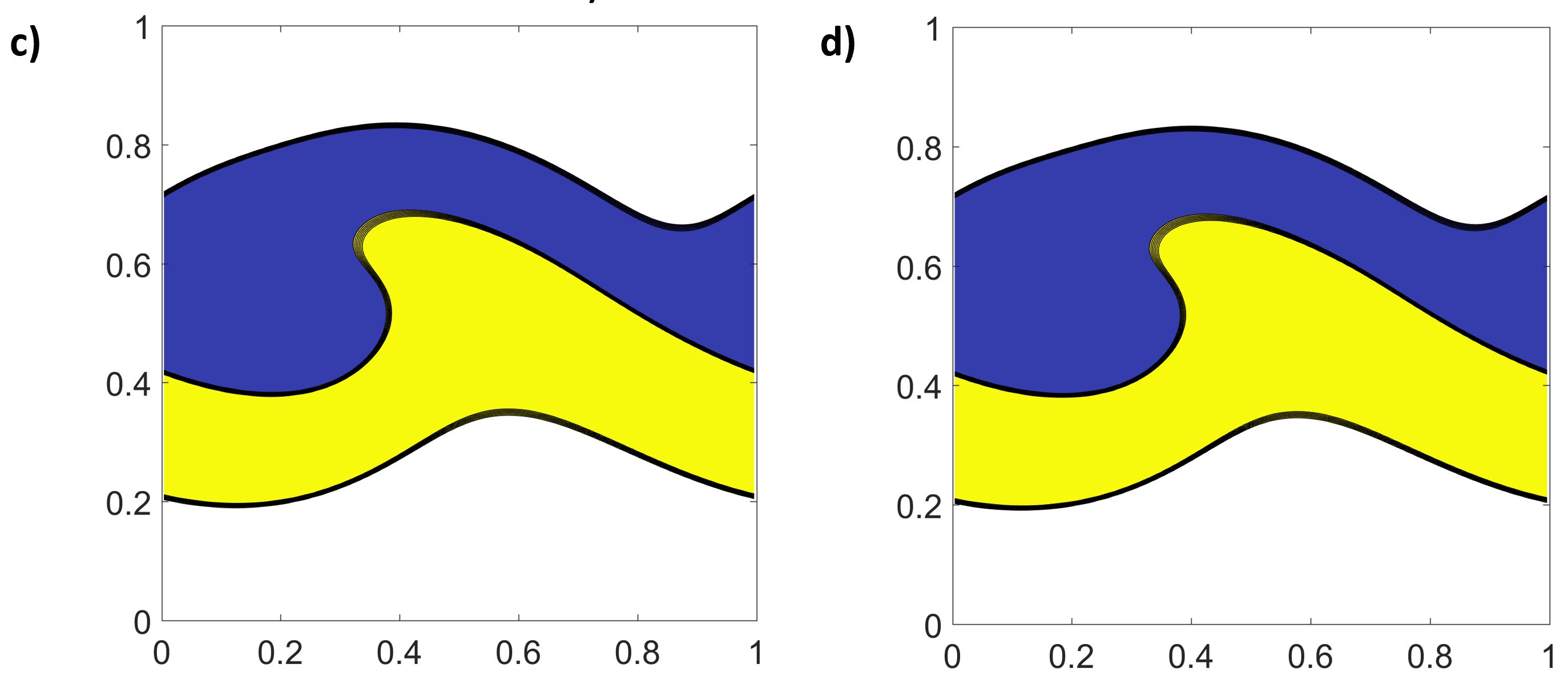}
	\caption{Validation of the momentum conservation. \textbf{a)} Time history of the change of $\mathrm{Momentum}_x$(=$\sum_{i,j} [\rho u]_{i,j} \Delta \Omega$). \textbf{b)} Time histories of the change of $\mathrm{Momentum}_y$(=$\sum_{i,j} [\rho v]_{i,j} \Delta \Omega$). \textbf{a)} and \textbf{b)} share the same legend, and ``256'' in the legend represents the results obtained from a finer cell size $h=\frac{1}{256}$.
	\textbf{c)} Snapshot of the phases at $t=2$ from the balanced-force method. \textbf{d)} Snapshot of the phases at $t=2$ from the conservative method. Blue: Phase 1, Yellow: Phase 2, White: Phase 3.
	}\label{Fig HSL-MomC}
\end{figure}

We now consider the momentum conservation on the discrete level. From Theorem \ref{Theorem Momentum conservation} and Corollary~\ref{Corollary Momentum conservation} in Section \ref{Sec Scheme momentum properties}, the momentum is conserved, i.e., $\mathrm{\mathbf{Momentum}}=\sum_{i,j} [\rho \mathbf{u}]_{i,j} \Delta \Omega$ does not change with time, if the conservative method is used to compute the surface force in Eq.(\ref{Eq Surface force Phi}). However, this is not necessarily true when the balanced-force method is implemented. Results in Fig.\ref{Fig HSL-MomC} \textbf{a)} and \textbf{b)} confirm our analysis. The momentum from the balanced-force method increases with time in this case, although it is not significant. Quantitatively, the change of $\mathrm{Momentum}_x$ is less than $0.04\%$ of its initial value. On the other hand, the momentum from the conservative method does not change with time. Since the difference between the balanced-force method and the conservative method is in the order of the truncation error, the non-conservation of the momentum from the balanced-force method should be reduced under grid refinement, and this is also shown in Fig.\ref{Fig HSL-MomC} \textbf{a)} and \textbf{b)}. Fig.\ref{Fig HSL-MomC} \textbf{c)} and \textbf{d)} show the snapshots of the three phases from the balanced-force method and the conservative method, respectively. The difference is very small. We summarize that, on the discrete level, the momentum is conserved by the present scheme if the conservative method is used, while it is essentially conserved if the balanced-force method is used. 

We finally consider the energy law of the present multiphase flow model, i.e., Eq.(\ref{Eq Energy law Phi}), on the discrete level. We will show through the following numerical experiments that the present scheme reproduces the behaviors of the energy law. 

\begin{figure}[!t]
	\centering
	\includegraphics[scale=0.5]{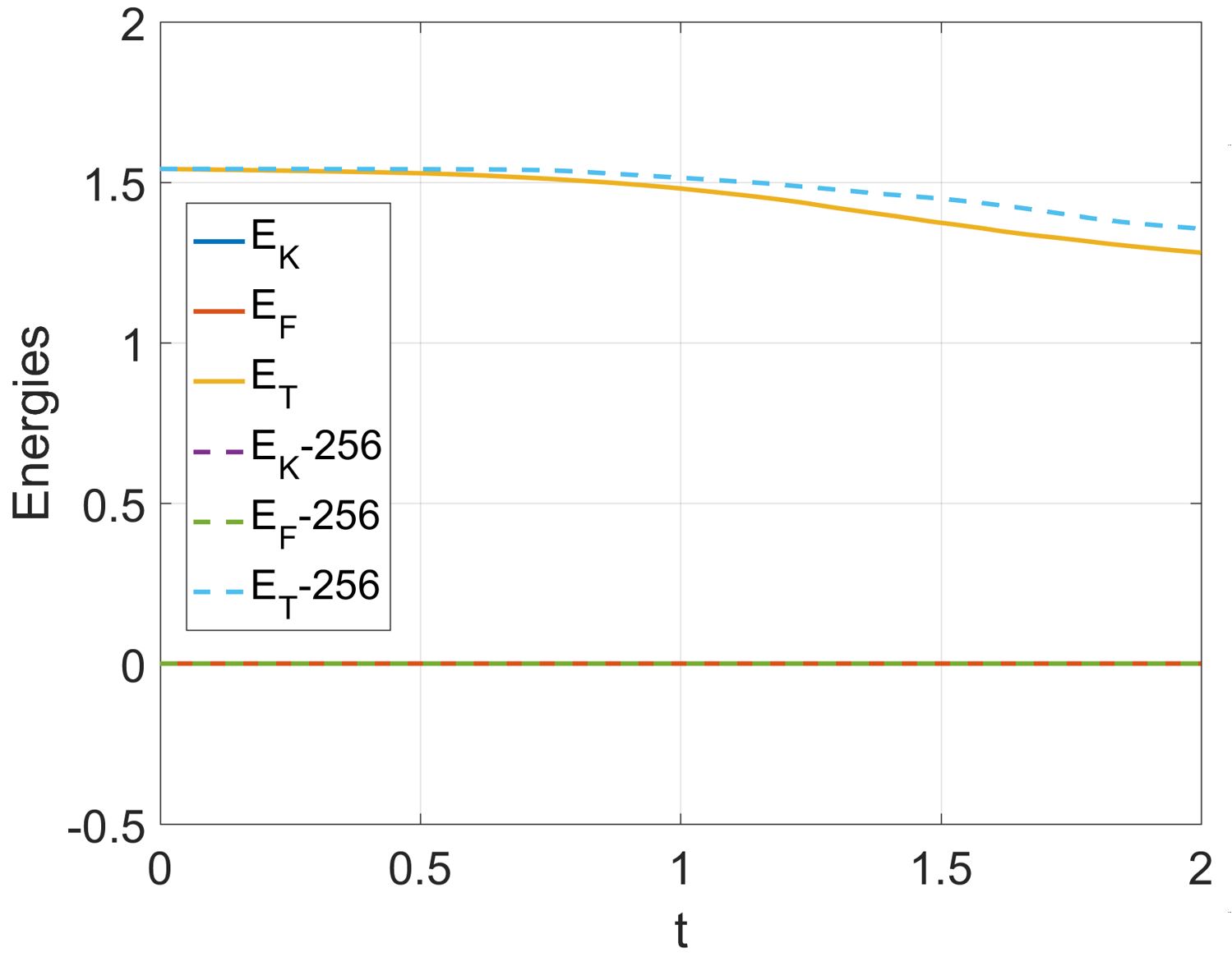}
	\caption{Time histories of the kinetic ($E_K$), free ($E_F$), and total ($E_T$) energies in the case without viscosity and interfacial tensions. ``256'' in the legend: Results obtained from a finer cell size $h=\frac{1}{256}$.}\label{Fig HSL-Eng-Inviscid-NoFs}
\end{figure}

In the first case, where there is neither viscosity nor interfacial tension, the free energy is zero so the total energy is the same as the kinetic energy. From the energy law in Eq.(\ref{Eq Energy law Phi}), the total energy, the same as the kinetic energy in this case, does not change with time. However, in practice, the total energy is decreased due to numerical dissipation. Fig.\ref{Fig HSL-Eng-Inviscid-NoFs} shows the time histories of the kinetic, free, and total energies. The free energy is identically zero, while the kinetic energy and the total energy are decaying. When we refine the cell, the change of the kinetic energy is reduced, implying that the numerical dissipation is the source of changing the kinetic energy. The numerical dissipation comes from the backward difference in time, the WENO scheme for the interpolation, and the linearly interpolated pressure gradient in Eq.(\ref{Eq Fully-discretized momentum equation}). 

\begin{figure}[!t]
	\centering
	\includegraphics[scale=0.4]{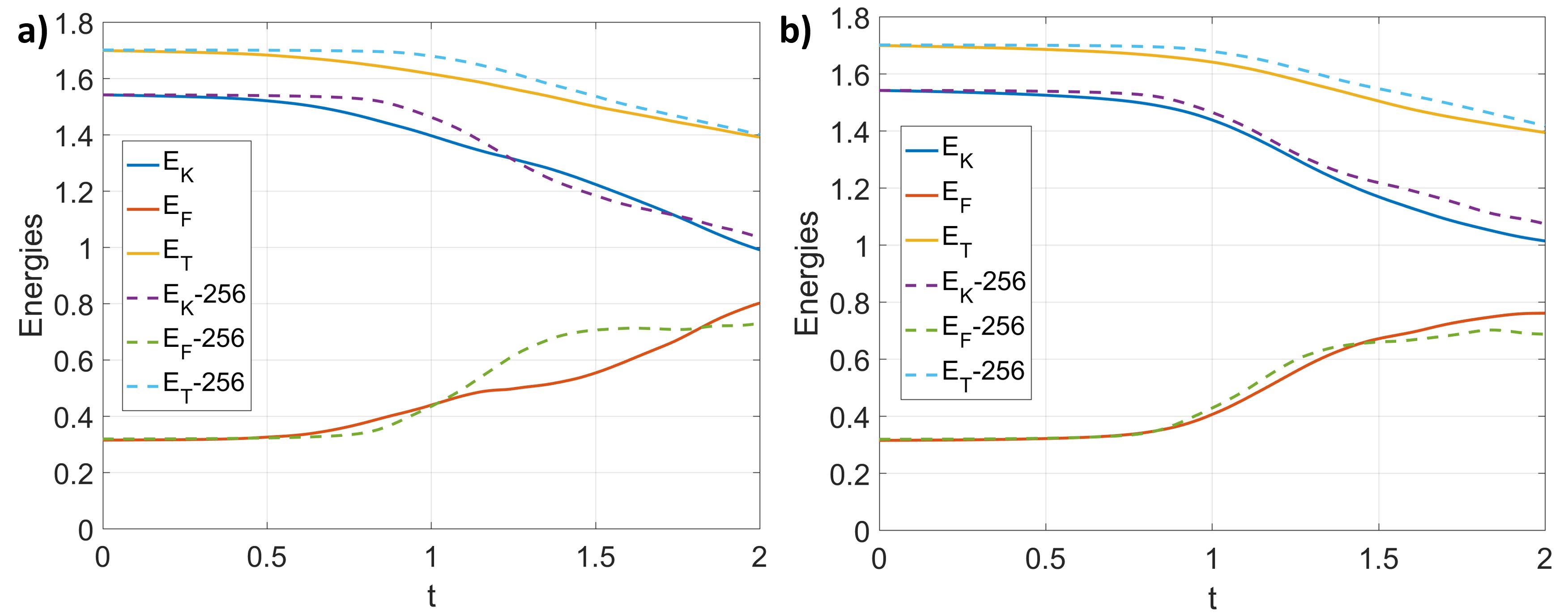}\\
	\includegraphics[scale=0.4]{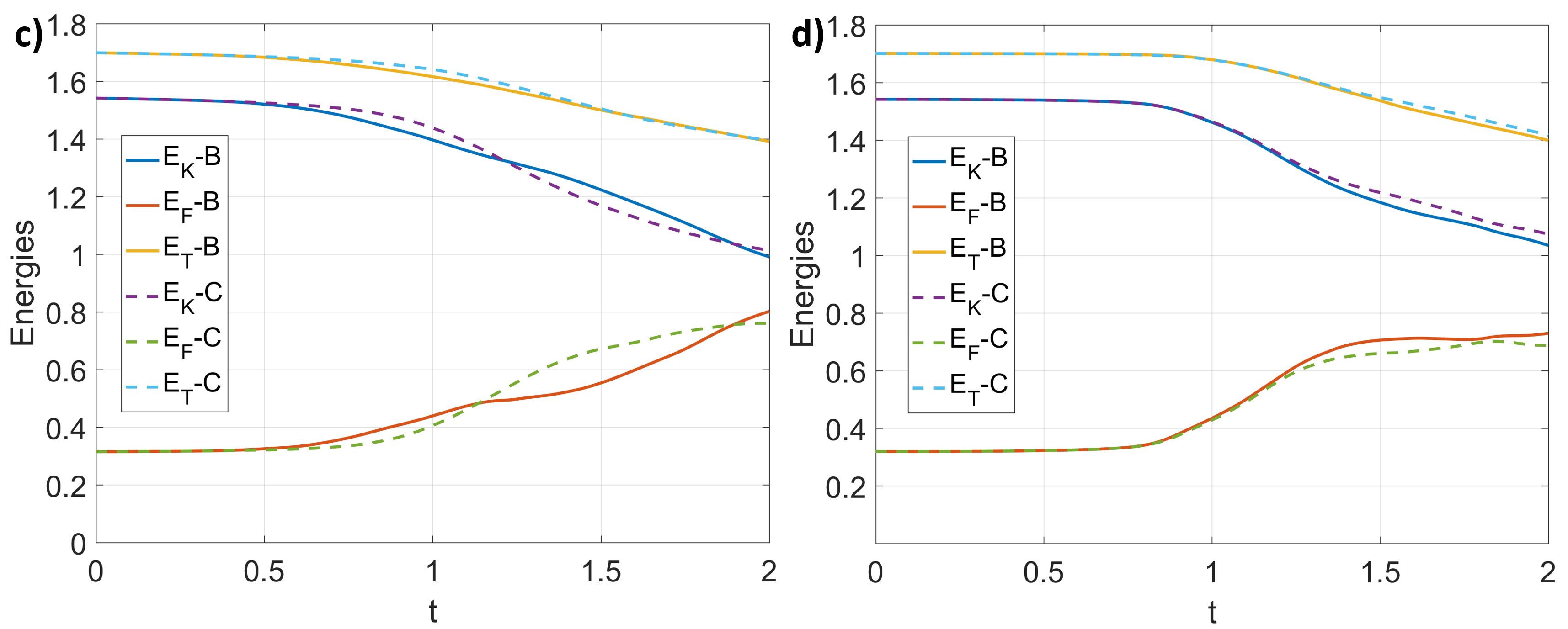}
	\caption{Time histories of the kinetic ($E_K$), free ($E_F$), and total ($E_T$) energies in the case without viscosity but with interfacial tensions. \textbf{a)} from the balanced-force method. \textbf{b)} from the conservative method. \textbf{c)} from the cell size $h=\frac{1}{128}$. \textbf{d)} from the cell size $h=\frac{1}{256}$. B: The balanced-force method, C: The conservative method, ``256'': Results obtained from a finer cell size $h=\frac{1}{256}$.}\label{Fig HSL-Eng-Inviscid}
\end{figure}

The second case has no viscosity but interfacial tensions. From Eq.(\ref{Eq Energy law Phi}), there is energy transfer between the kinetic energy and the free energy in this case, while the total energy should decrease. Fig.\ref{Fig HSL-Eng-Inviscid} shows the results from the balanced-force method and the conservative method. Both methods produce results that respect the energy law, i.e., the total energy decreases with time, as shown in Fig.\ref{Fig HSL-Eng-Inviscid} \textbf{a)} and \textbf{b)}. We can also observe the energy transfer from the kinetic energy to the free energy due to the deformation of the initial flat interfaces. In Fig.\ref{Fig HSL-Eng-Inviscid} \textbf{c)}, although the total energies from the two methods are very close to each other, the kinetic energy from the conservative method decreases more than the one from the balanced-force method at a later time in this case. As a result, the free energy from the conservative method increases faster than the one from the balanced-force method. The difference between the two methods is much less after we refine the cell size, as shown in Fig.\ref{Fig HSL-Eng-Inviscid} \textbf{d)}. 

\begin{figure}[!t]
	\centering
	\includegraphics[scale=0.4]{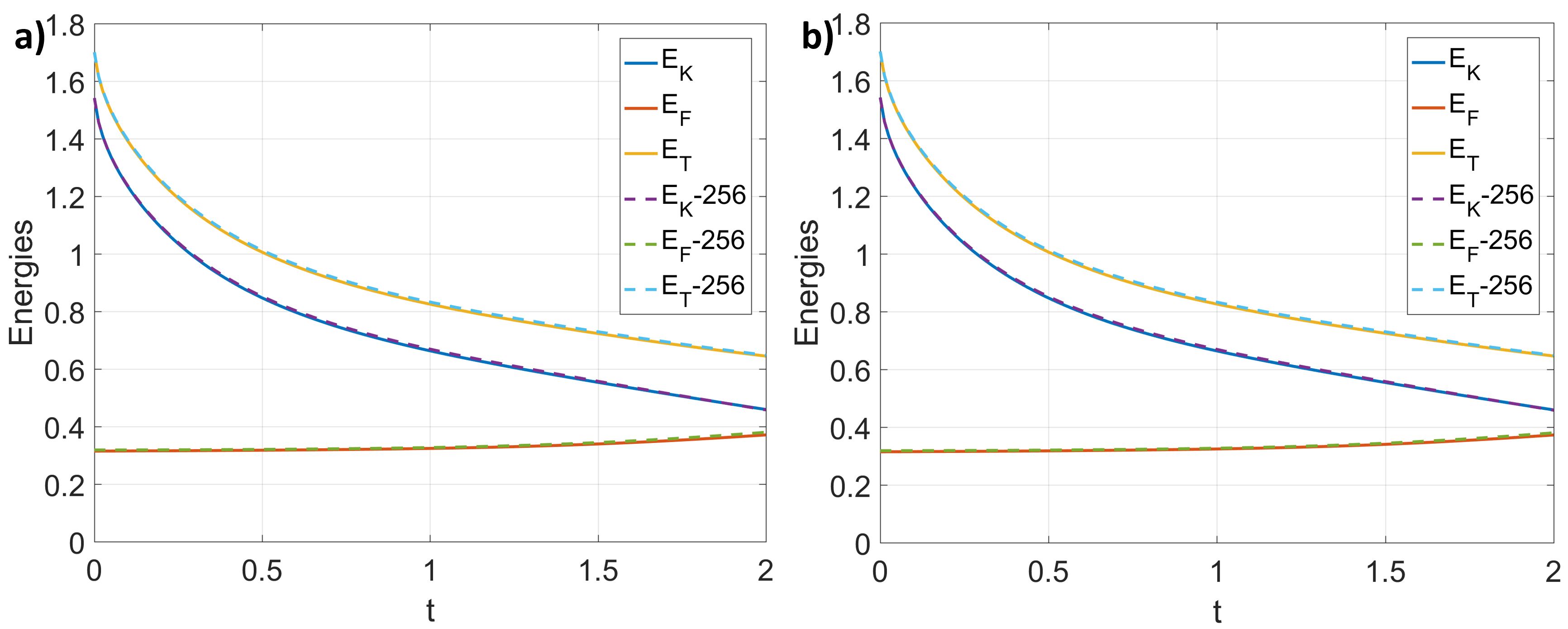}\\
	\includegraphics[scale=0.4]{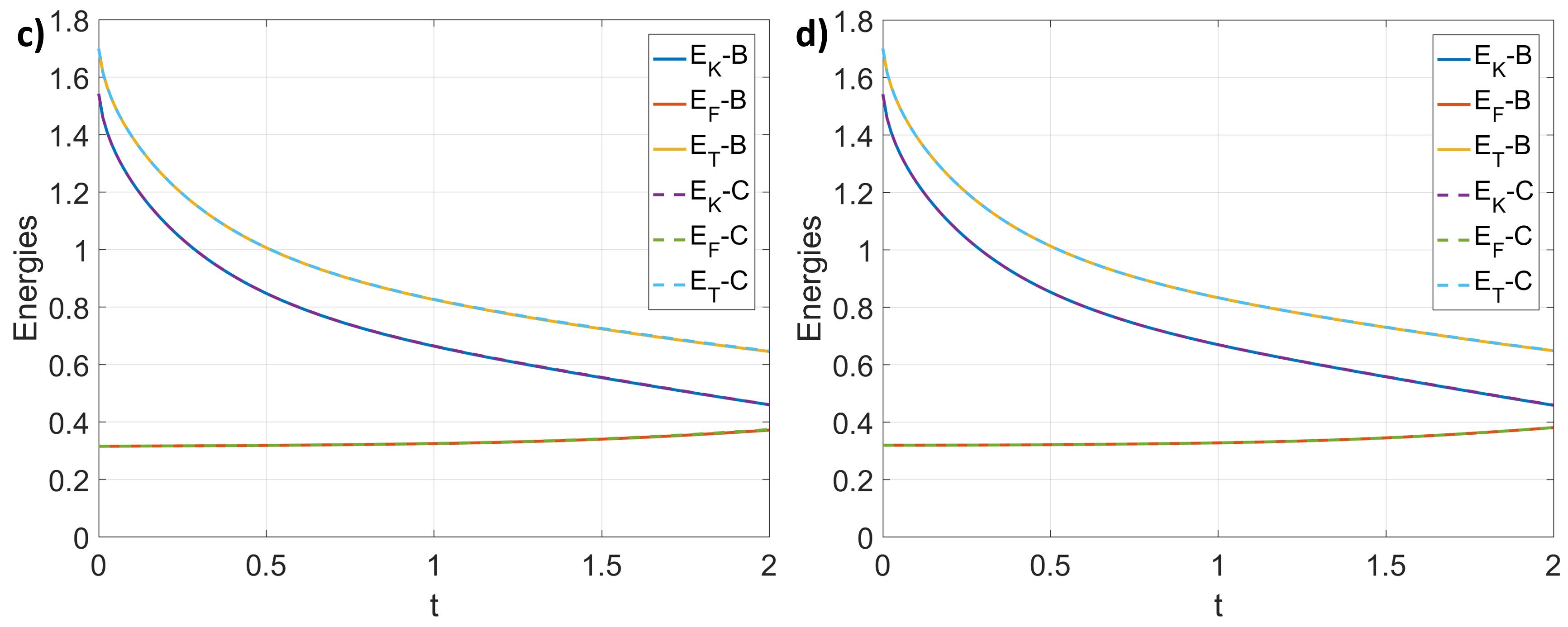}
	\caption{Time histories of the kinetic ($E_K$), free ($E_F$), and total ($E_T$) energies in the case with viscosity and interfacial tensions. \textbf{a)} from the balanced-force method. \textbf{b)} from the conservative method. \textbf{c)} from the cell size $h=\frac{1}{128}$. \textbf{d)} from the cell size $h=\frac{1}{256}$. B: The balanced-force method, C: The conservative method, ``256'': Results obtained from a finer cell size $h=\frac{1}{256}$.}\label{Fig HSL-Eng}
\end{figure}

The last case considered includes both viscosity and interfacial tensions, corresponding to the complete form of Eq.(\ref{Eq Energy law Phi}). The results are shown in Fig.\ref{Fig HSL-Eng}, and both the balanced-force method and the conservative method produce results that follow Eq.(\ref{Eq Energy law Phi}), as shown in Fig.\ref{Fig HSL-Eng} \textbf{a)} and \textbf{b)}. There is little difference between the default and the fine grid solutions, implying that the decay of the total energy results from the physical interaction in the multiphase flow model, instead of numerical dissipation. In addition, the difference between the two methods is hardly observable in both the default and the find grid solutions, as shown in Fig.\ref{Fig HSL-Eng} \textbf{c)} and \textbf{d)}.

In summary, the following properties of the present scheme analyzed in Section \ref{Sec Scheme} are validated. The present scheme satisfies the \textit{consistency of reduction}, mass conservation, and summation of the order parameters. The momentum, on the discrete level, is conserved when the conservative method is used to compute the surface force, while it is essentially conserved when the balanced-force method is used. Moreover, the behaviors of the energy law of the present multiphase flow model is reproduced.

\subsection{Application} \label{Applications}
In this section, we apply the present mutliphase flow model and scheme to various realistic multiphase flow problems to further verify their capability. The water, oil, and air have the material properties in Table~\ref{Table WaterAirOil}, and the contact angles, if needed, are all $\frac{\pi}{2}$, unless otherwise specified. Results are reported in their dimensionless forms under $\eta=0.01$ and $M_0=10^{-7}$. 
\begin{table*}[!t]
	\centering
	\caption{Material properties of the water, air, and oil.} \label{Table WaterAirOil}
	\includegraphics[scale=0.6]{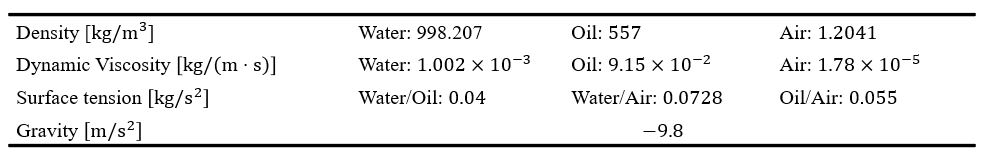}
\end{table*}

\subsubsection{Three-phase Rayleigh-Taylor instability} \label{Sec Three-phase Rayleigh-Taylor instability}
We consider a three-phase Rayleigh-Taylor instability problem, where the heaviest phase (Phase 1) stays on the top of Phase 2, whose density is in the middle. At the bottom, there is the lightest phase (Phase~3). The Atwood number between Phases $p$ and $q$ is defined as $At_{p,q}=\frac{\rho_p-\rho_q}{\rho_p+\rho_q}$. The domain of interest is $1\times6$. Its left and right boundaries are periodic while the top and bottom boundaries are free-slip. We use $128 \times 768$ cells to discretize the domain and the time step is $\Delta t=10^{-3}$. The densities of the fluids are $\rho_1=4.5, \rho_2=3, \rho_3=1$, and their dynamic viscosities are all $10^{-3}$. As a result, we have $At_{1,2}=0.2$ and $At_{2,3}=0.5$. The pairwise interfacial tensions are $\sigma_{p,q} = 10^{-12}, p \neq q$. The gravity is pointing downward with a magnitude of unity, i.e., $\mathbf{g}=\{0,-1\}$. Here, the chosen material properties follow the setups in \citep{Tryggvason1988,GuermondQuartapelle2000,Dingetal2007,Huangetal2019,Huangetal2020}, so that we can compare interface locations from the present model to existing numerical data. Initially, the flow is stationary, and the interface between Phases 1 and 2 is at $y=4$ and that between 2 and 3 is at $y=2$. A horizontal sinusoidal perturbation is applied to both of the interfaces, whose amplitude is 0.1 and wavelength is $2\pi$. 

\begin{figure}[!t]
	\centering
	\includegraphics[scale=0.5]{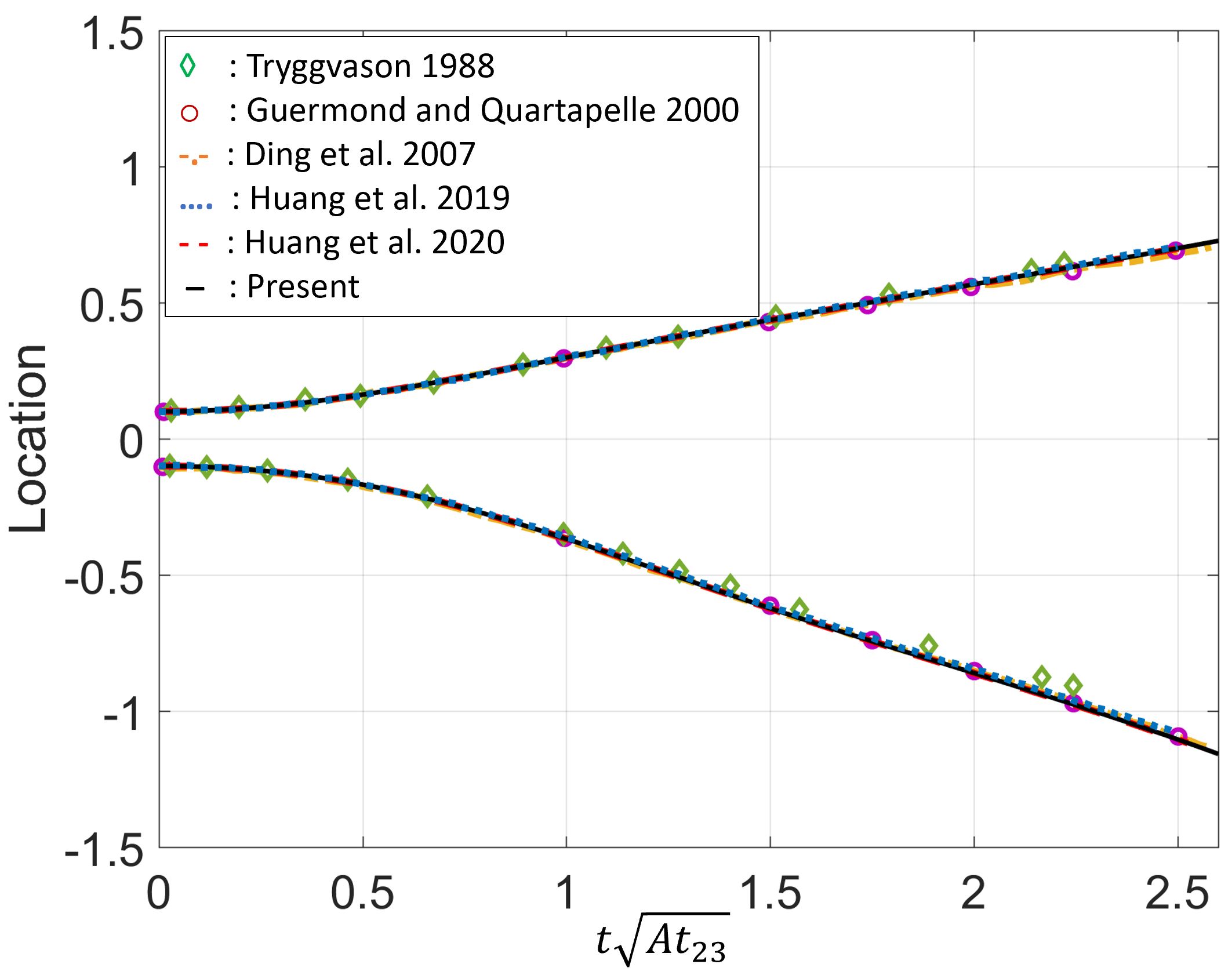}
	\caption{Locations of the interface between Phases 2 and 3 at the center (bottom curves) and at the lateral (top curves) of the domain in the three-phase Rayleigh-Taylor instability problem.} \label{Fig RT-Location}
\end{figure}

Since the interfaces are initially far away separated, there is no interaction between them, and the three-phase Rayleigh-Taylor instability problem is actually a combination of two independent two-phase Rayleigh-Taylor instability problems at the beginning of the simulation. As a result, we can compare the locations of the interface between Phases 2 and 3 at the center and at the lateral edge of the domain to the two-phase results in \citep{Tryggvason1988,GuermondQuartapelle2000,Dingetal2007,Huangetal2019,Huangetal2020}, where the Atwood number $At$ is 0.5. The comparisons are shown in Fig.\ref{Fig RT-Location}, and the present results under a three-phase setup are hardly distinguished from the published ones under a two-phase setup.

\begin{figure}[!t]
	\centering
	\includegraphics[scale=0.35]{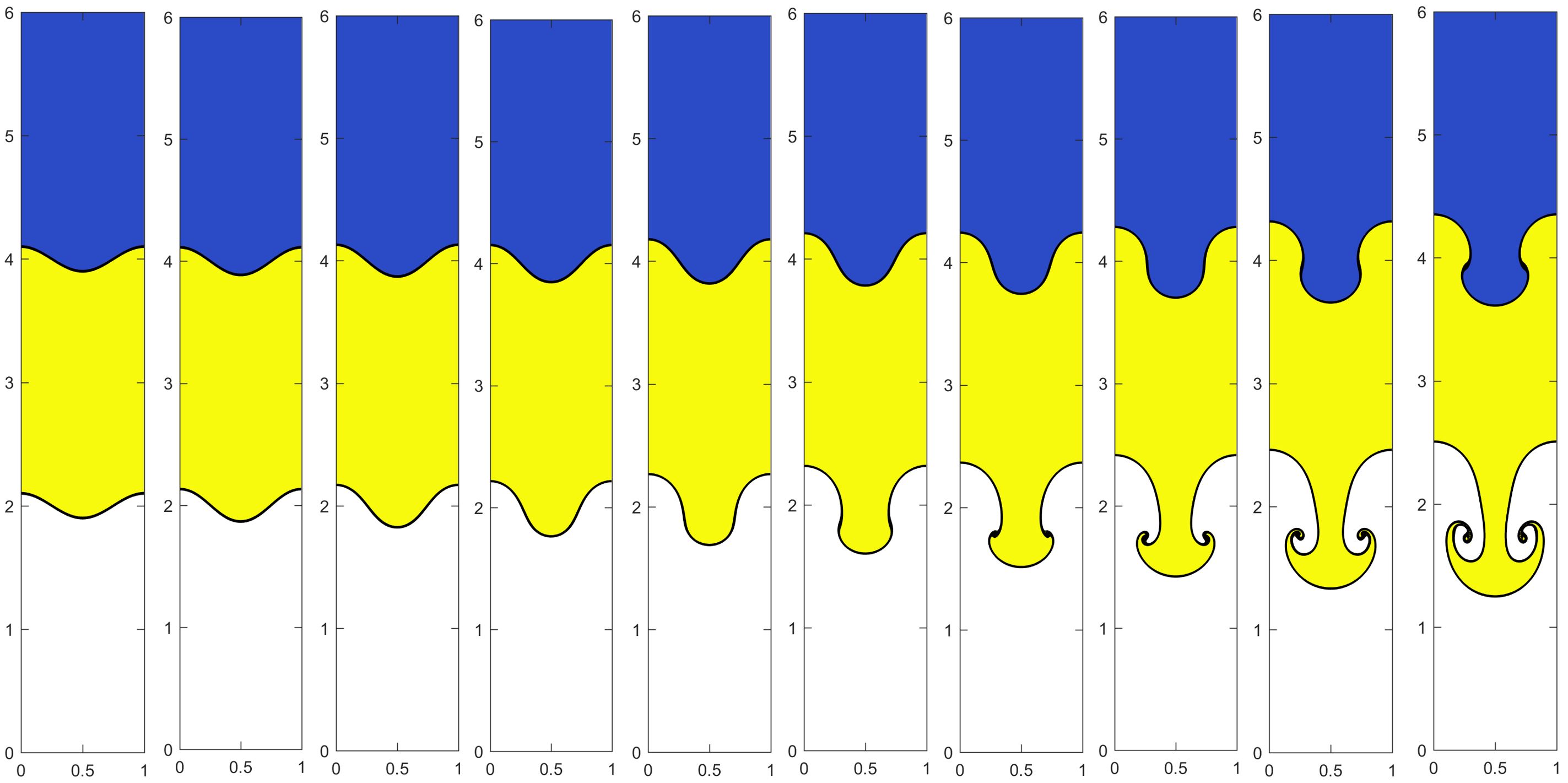}\\
	\includegraphics[scale=0.35]{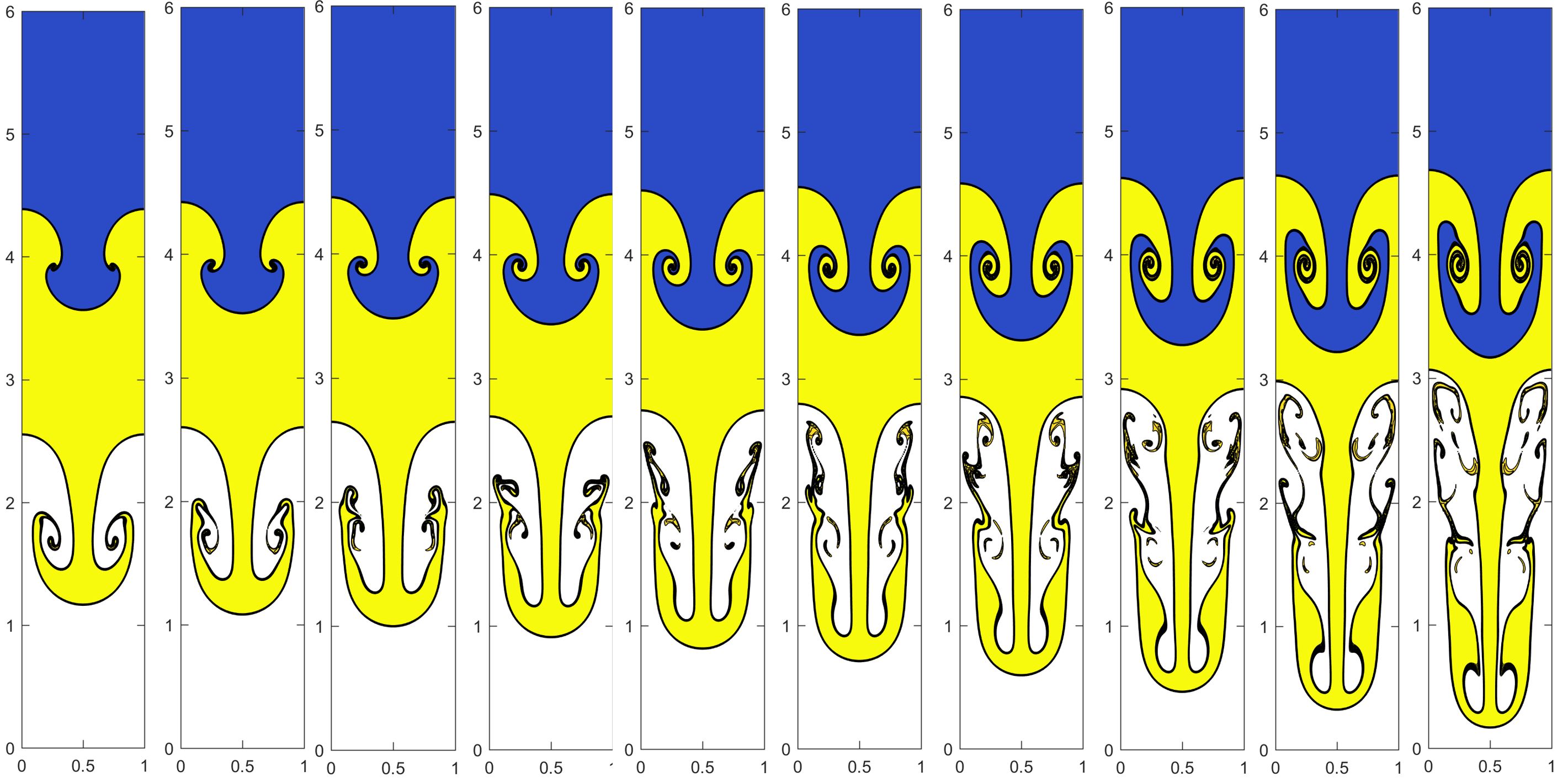}\\
	\includegraphics[scale=0.35]{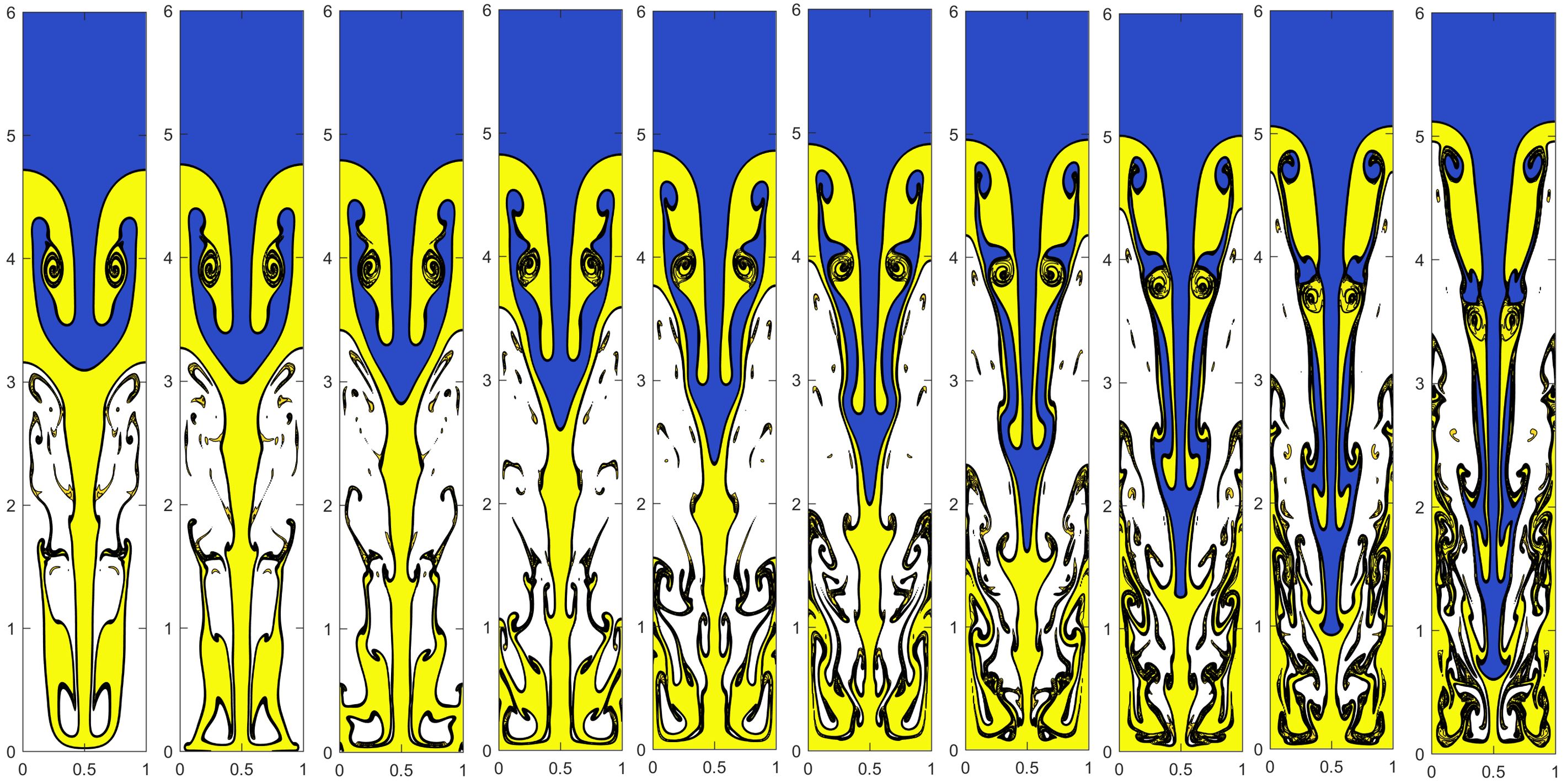}\\
	\includegraphics[scale=0.35]{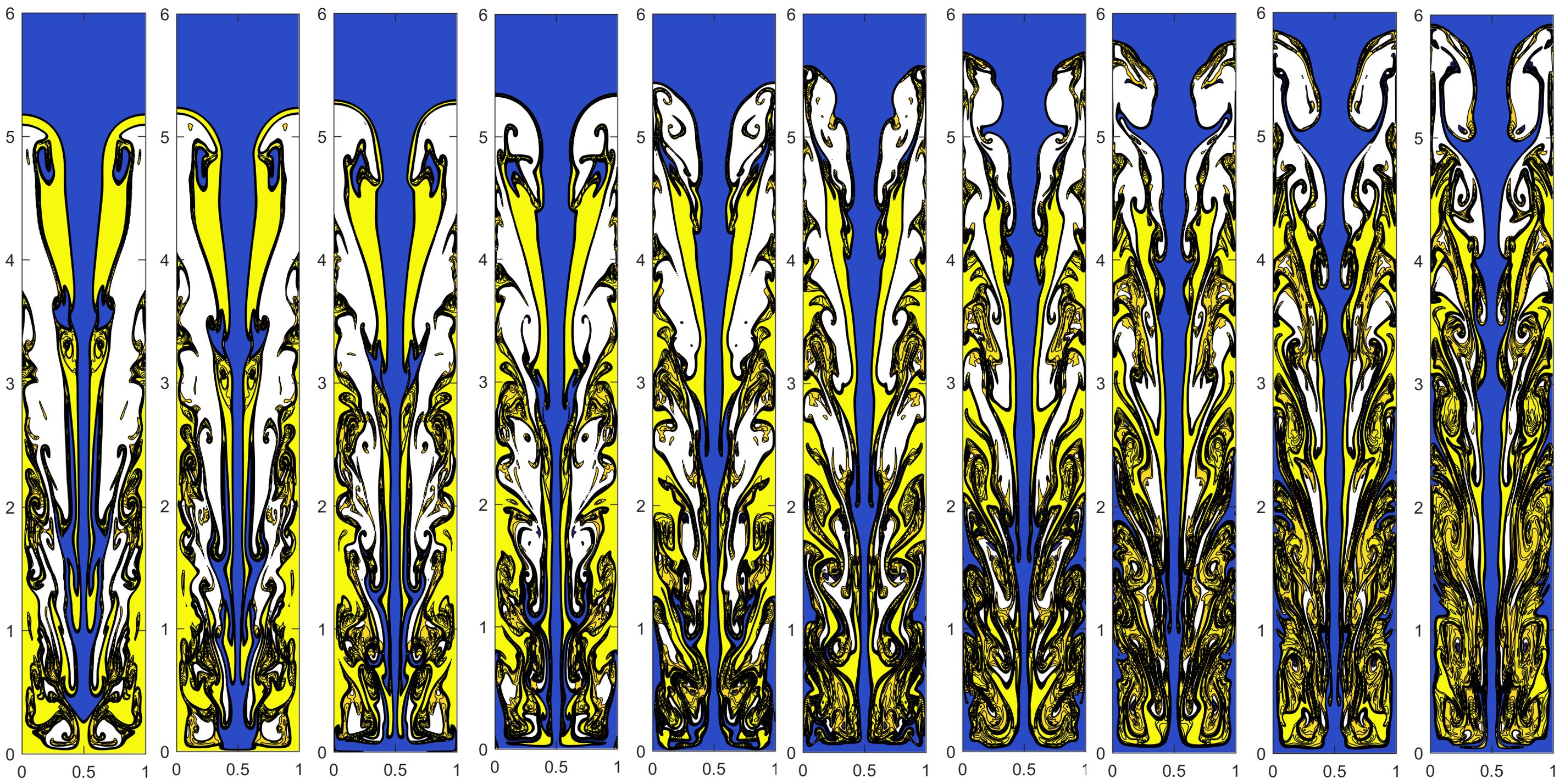}\\
	\caption{Evolution of the three phases in the three-phase Rayleigh-Taylor instability problem. From left to right, top to bottom, $t$ is 0.00, 0.50, 0.75, 1.00, ..., 10.00. Blue: Phase1, Yellow: Phase2, and White: Phase3.} \label{Fig RT-Interface}
\end{figure}

The evolution of the three phases is presented in Fig.\ref{Fig RT-Interface} up to $t=10$. Before $t=5$, the three-phase dynamics is not observed, and the interfaces between Phases 1 and 2 and between Phases 2 and 3 evolve independently, without noticing the existence of an additional phase. The configurations of the individual interfaces are almost identical to those reported in \citep{Dingetal2007, Huangetal2019, Huangetal2020} where only two phases were considered. Phase~2 moves downward faster than Phase 1 because the Atwood number between Phases 2 and 3 is larger than that between Phases 1 and 2. We can observe multiple droplets and filaments being generated from Phase 2 before it reaches the bottom of the domain. After $t=5$, the three-phase dynamics begins. Phase 2 accumulates at the bottom of the domain, and then moves upward along the lateral edge of the domain. A channel of Phase 2 is built up in the middle of the domain and Phase 1 tries to move downward along this channel. As a result, the configuration of the interface between Phases 1 and 2 is shrunk compared to the two-phase one in \citep{Huangetal2019}. Phase 1 gradually fills up the middle channel, producing complex patterns resulting from its strong interactions with the other phases. Since Phase 1 is the heaviest, it takes the place of Phase~2 at the bottom and the lateral edge of the domain at the end of the simulation. As a result, Phase 2 is pushed to aggregate between the middle and lateral columns of Phase 1. The long-time dynamics of this problem is highly sophisticated because of the strong interactions between the three phases. 

\subsubsection{Floating liquid lens} \label{Sec Floating liquid lens}
We perform the floating liquid lens problem to investigate the equilibrium states of the multiphase flows. An oil drop (Phase 2) is floating on the interface between the water (Phase 1) at the bottom and the air (Phase 3) on the top. Because the forces are imbalanced at the triple points of the three phases, the drop will be stretched horizontally while compressed vertically and finally reaches its equilibrium state. If there is no gravity, we can derive an exact solution based on the force balance at the triple points, the normal stress balance at the interfaces, and the mass conservation. Details of the derivation are provided in Appendix. If the gravity is dominant, asymptotic analyses from Langmuir and de Gennes \citep{Langmuir1933,deGennesetal2003} give the final thickness of the drop as
\begin{equation}\label{Eq Floating liquid lens Thickness}
e_d=\sqrt{ \frac{2(\sigma_{2,3}+\sigma_{1,2}-\sigma_{1,3})\rho_1}{\rho_2 (\rho_1-\rho_2) |\mathbf{g}|} }.
\end{equation}
The material properties are listed in Table \ref{Table WaterAirOil}. Non-dimensionalization is performed based on a density scale $1.2041\mathrm{kg/m^3}$, a length scale $0.04\mathrm{m}$, and an acceleration scale $1\mathrm{m/s^2}$. After that, the domain considered is $[-1,1]\times[0,0.8]$ with periodic boundaries at the lateral and no-slip walls at the top and bottom. The domain is discretized by $[201 \times 81]$ cells and the time step is $\Delta t=10^{-5}$. Initially, a circular oil drop (Phase 2) of a radius 0.2 is at $(0,0.4)$, floating on the horizontal interface between the water (Phase 1) and the air (Phase~3) at $y=0.4$. We consider three different magnitudes of the gravity, i.e., $|\mathbf{g}|=0,5,9.8$, and the direction of the gravity is pointing downward.

\begin{figure}[!t]
	\centering
	\includegraphics[scale=0.5]{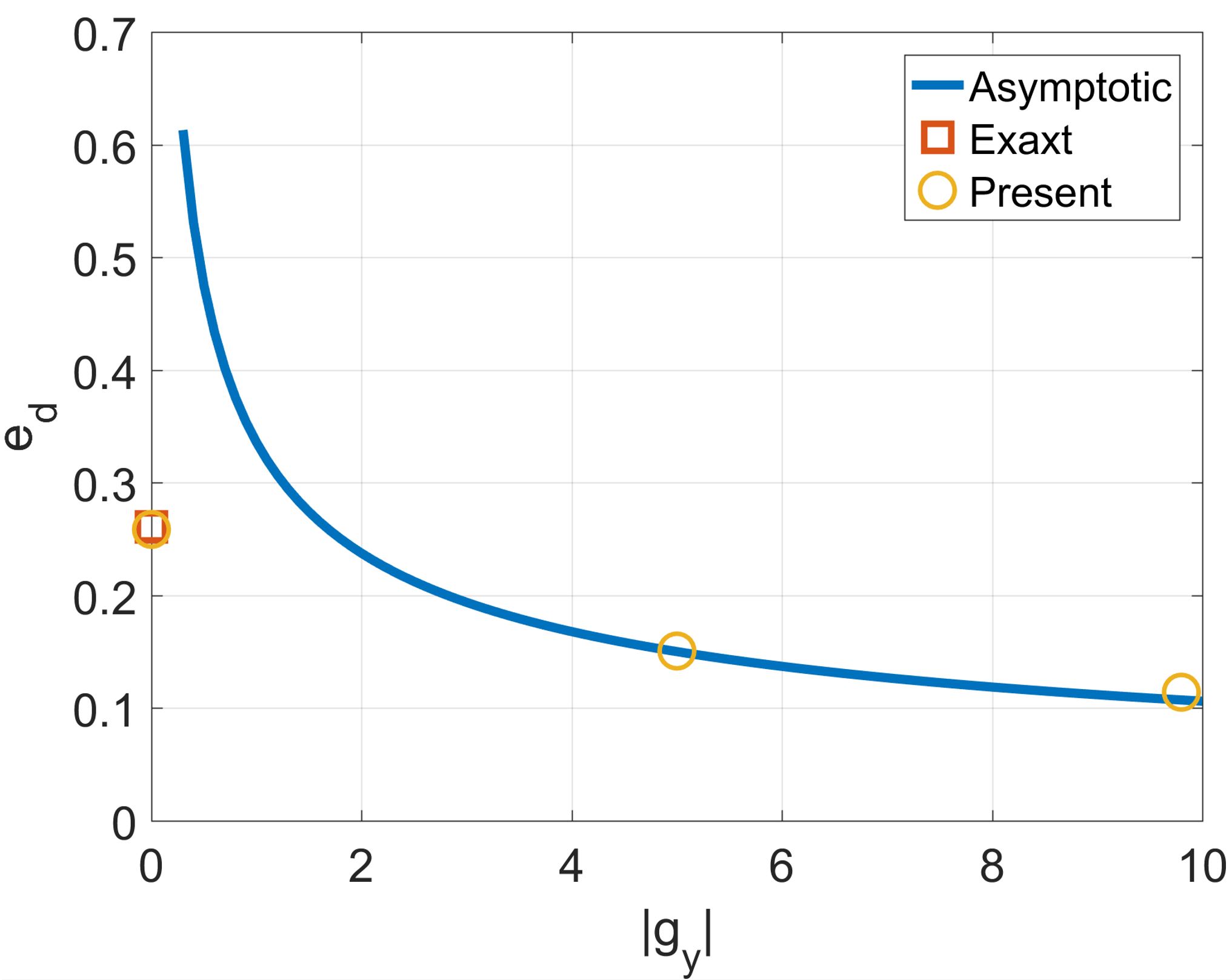}
	\caption{Final thickness of the drop vs. magnitude of the gravity in the floating liquid lens problem.} \label{Fig LL-Location}
\end{figure}

\begin{figure}[!t]
	\centering
	\includegraphics[scale=0.36]{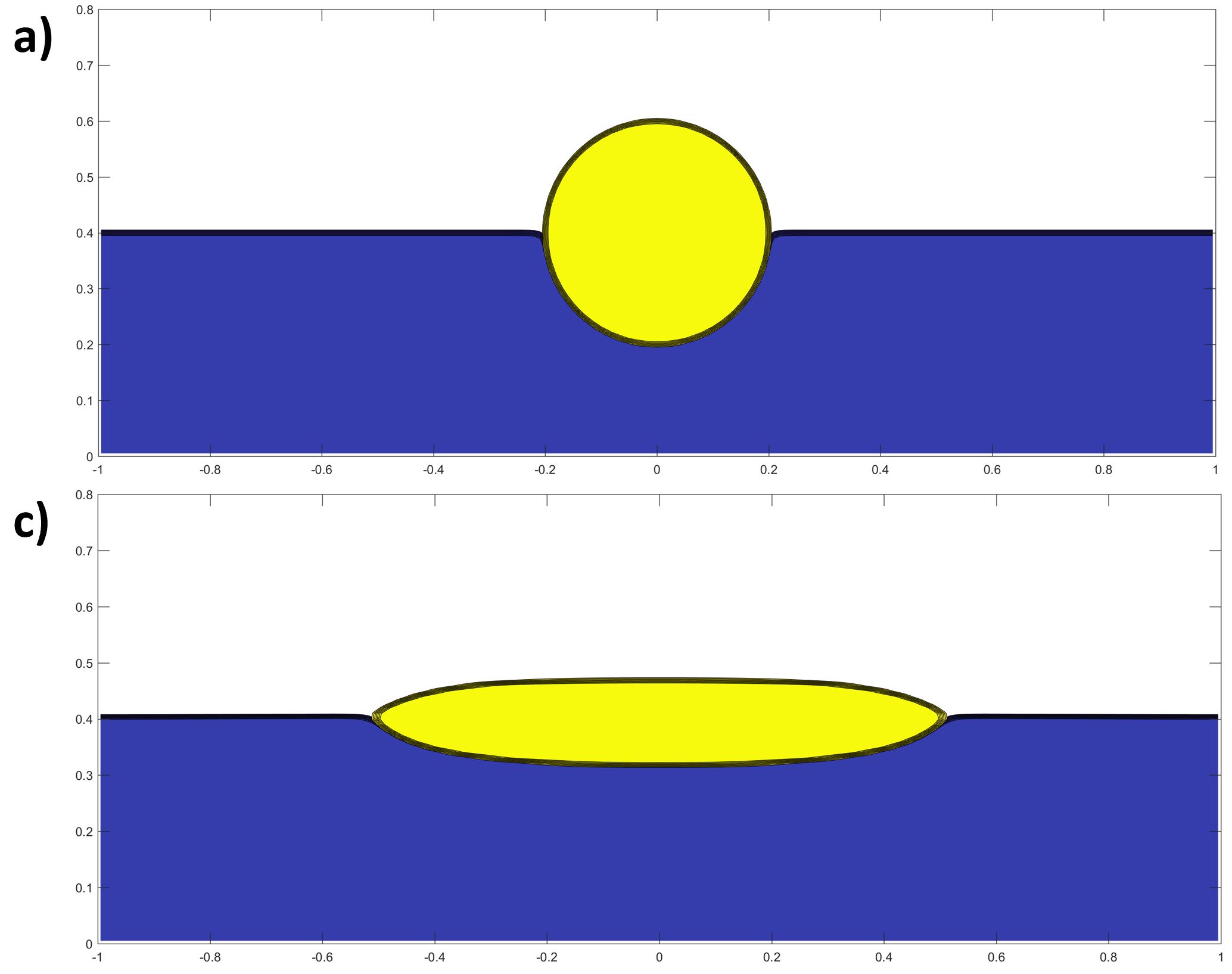}
	\includegraphics[scale=0.36]{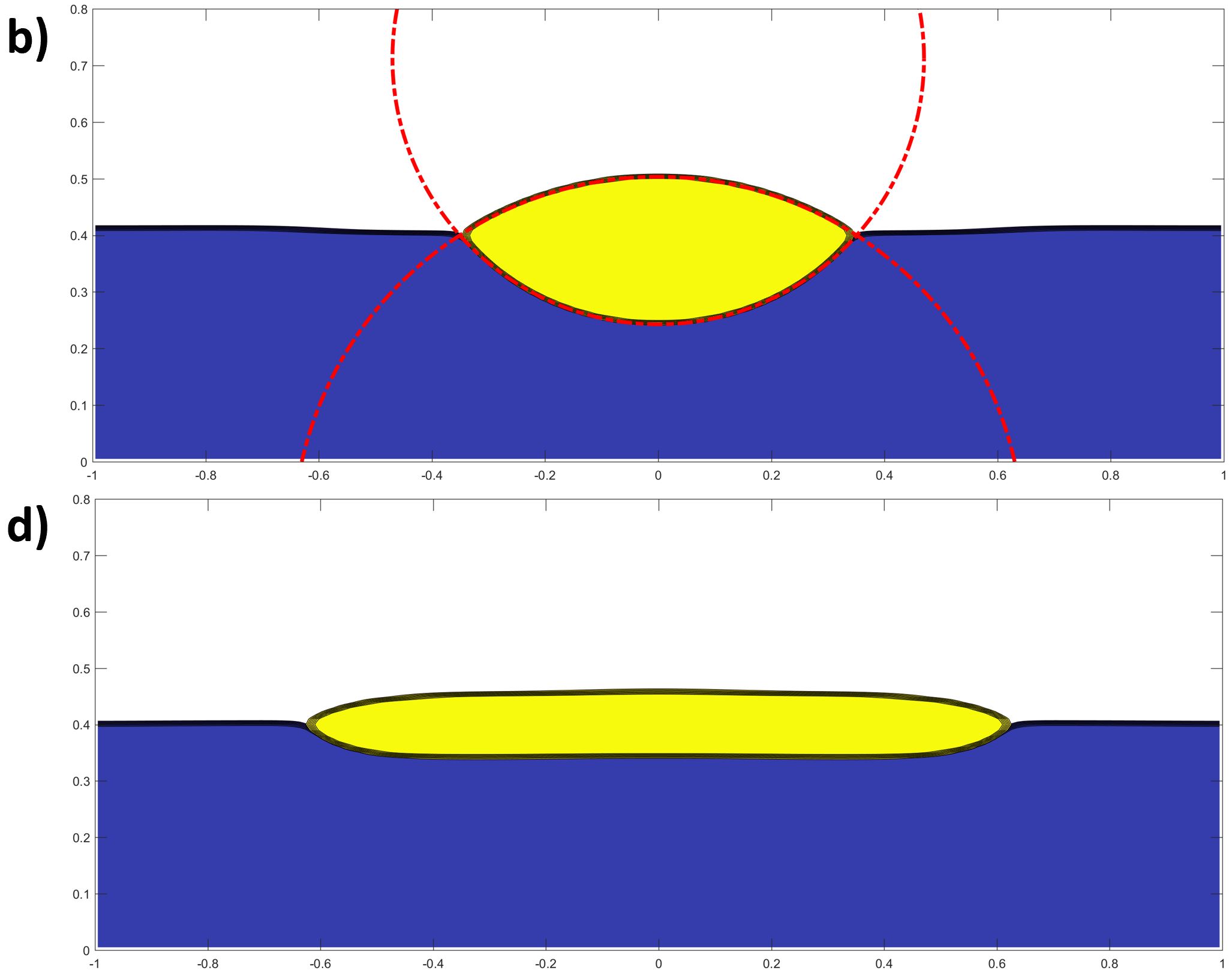}
	\caption{Initial and final shapes of the three phases in the floating liquid lens problem. \textbf{a)} Initial shapes. \textbf{b)} Final shapes with $|\mathbf{g}|=0$. \textbf{c)} Final shapes with $|\mathbf{g}|=5$. \textbf{d)} Final shapes with $|\mathbf{g}|=9.8$. Blue: Phase 1, Yellow: Phase 2, White: Phase 3. Red dash-dotted lines in \textbf{b)}: Exact solution for $|\mathbf{g}|=0$.} \label{Fig LL-Interface}
\end{figure}

Fig.\ref{Fig LL-Location} shows the final thickness of the oil drop (Phase 2) with respect to the magnitude of the gravity, along with the exact solution for the zero gravity and with the asymptotic solution Eq.(\ref{Eq Floating liquid lens Thickness}). Our numerical results match both the exact and asymptotic solutions in the two limits very well. 
Fig.\ref{Fig LL-Interface} shows the initial and final shapes of the three phases. When there is no gravity, the drop becomes a combination of two circular segments, which agrees well with the exact solution. As the gravity increases, the drop is horizontally elongated and vertically compressed.

\subsubsection{Drops on a surface} \label{Sec Equilibrium drop}
In this case, we include the effect of the contact angles, and investigate the equilibrium shapes of the drops on a flat surface. Without considering the gravity, any semi-circular drop, whose radius is $R_0$, on a homogeneous solid wall will finally reach an equilibrium state that it becomes a circular segment with a radius $R$, height $H_d$, and spreading length $L_d$ as
\begin{eqnarray} \label{Eq Equilibrium drop}
R&=&R_0 \sqrt{\frac{\pi/2}{\theta_s-\cos(\theta_s)\sin(\theta_s)}},\\
\nonumber
H_d&=&R\left(1-\cos(\theta_s)\right),\\
\nonumber
L_d&=&2R\sin(\theta_s),
\end{eqnarray}
where $\theta_s$, measured by radian, is the contact angle between the drop and the wall. The derivation of Eq.(\ref{Eq Equilibrium drop}) is provided in Appendix. We consider three phases appearing in the domain, where the water (Phase 1) is in contact with the bottom wall, the oil (Phase 2) is in contact with the top wall, and the air (Phase~3) fills the rest of the domain. The material properties of the phases are the same as those in Table \ref{Table WaterAirOil} except $\rho_2=870\mathrm{kg/m^3}$ and $\mu_3=1.78\times 10^{-4}\mathrm{Pa \cdot s}$ in this case. Non-dimensionalization is performed based on a density scale $1.2041\mathrm{kg/m^3}$, a length scale $0.04\mathrm{m}$, and an acceleration scale $1\mathrm{m/s^2}$. After that, the domain considered is $[-1,1]\times[0,0.5]$ with periodic boundaries at the lateral and no-slip walls at the top and bottom. The domain is discretized by $[200 \times 50]$ cells and the time step is $\Delta t=10^{-4}$. Three sets of the contact angles between Phases 1 and 3 at the bottom wall and between Phases 2 and 3 at the top wall are $(45^0,135^0)$, $(120^0,60^0)$, and $(75^0,105^0)$. Initially, the centers of Phases 1 and 2 are at $(-0.5,0)$ and $(0.5,0.5)$, respectively, and both of them are semi-circular with a radius $R_0=0.2$. 

\begin{figure}[!t]
	\centering
	\includegraphics[scale=0.5]{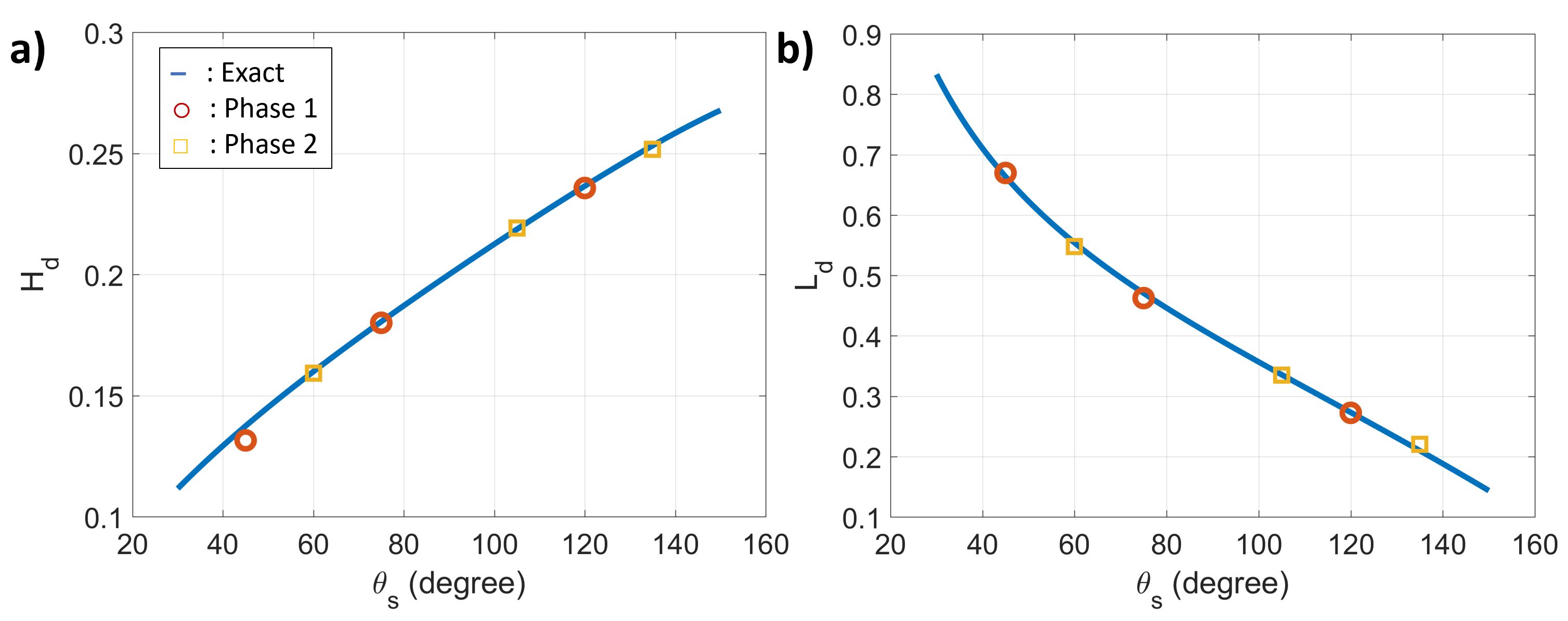}
	\caption{Heights and spreading lengths of the drops on a surface. \textbf{a)} Height ($H_d$) vs. contact angle ($\theta_s$). \textbf{b)} Spreading length ($L_d$) vs. contact angle ($\theta_s$). \textbf{a)} and \textbf{b)} share the same legend.} \label{Fig ED-Location}
\end{figure}

\begin{figure}[!t]
	\centering
	\includegraphics[scale=0.25]{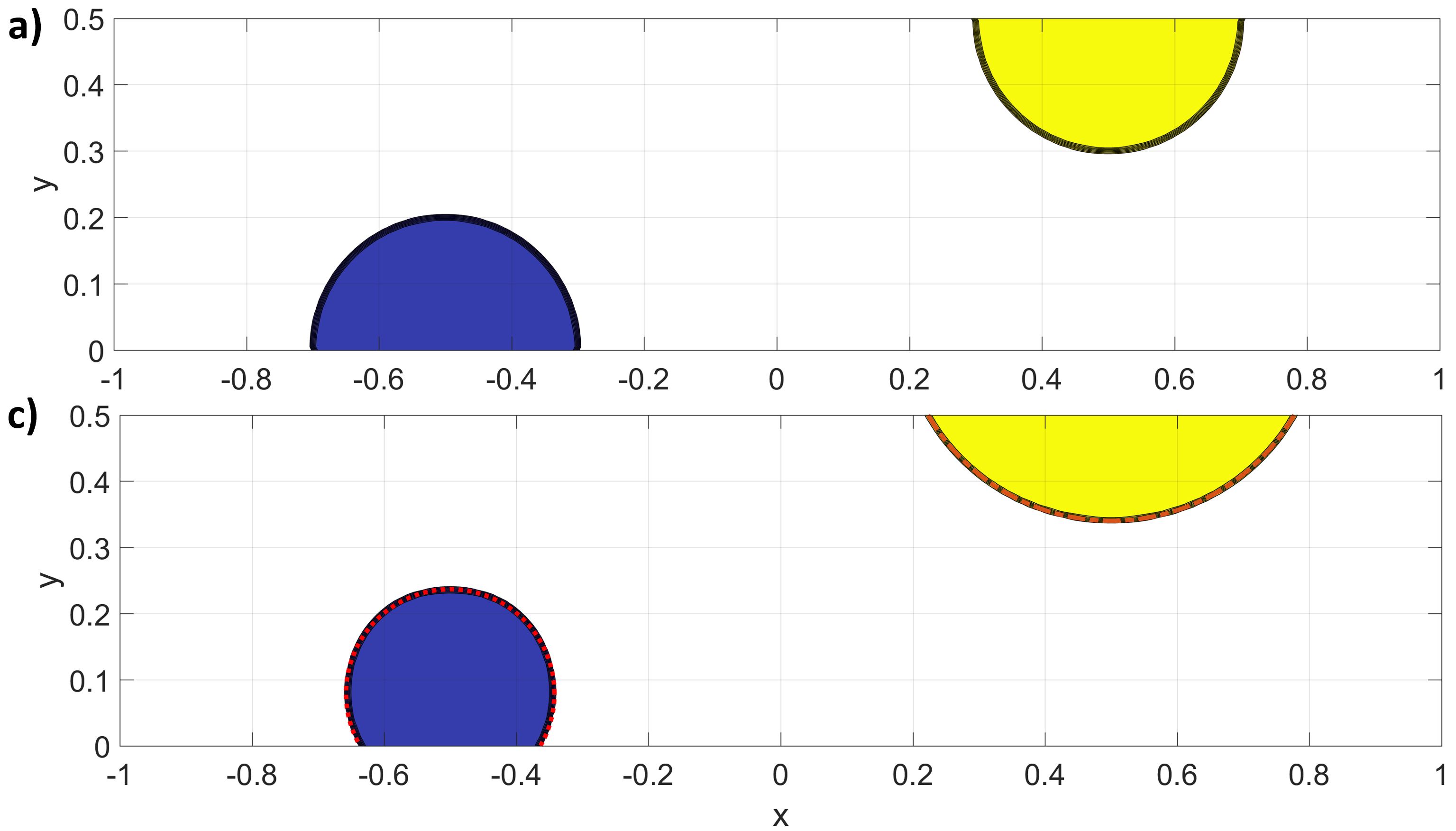}
	\includegraphics[scale=0.25]{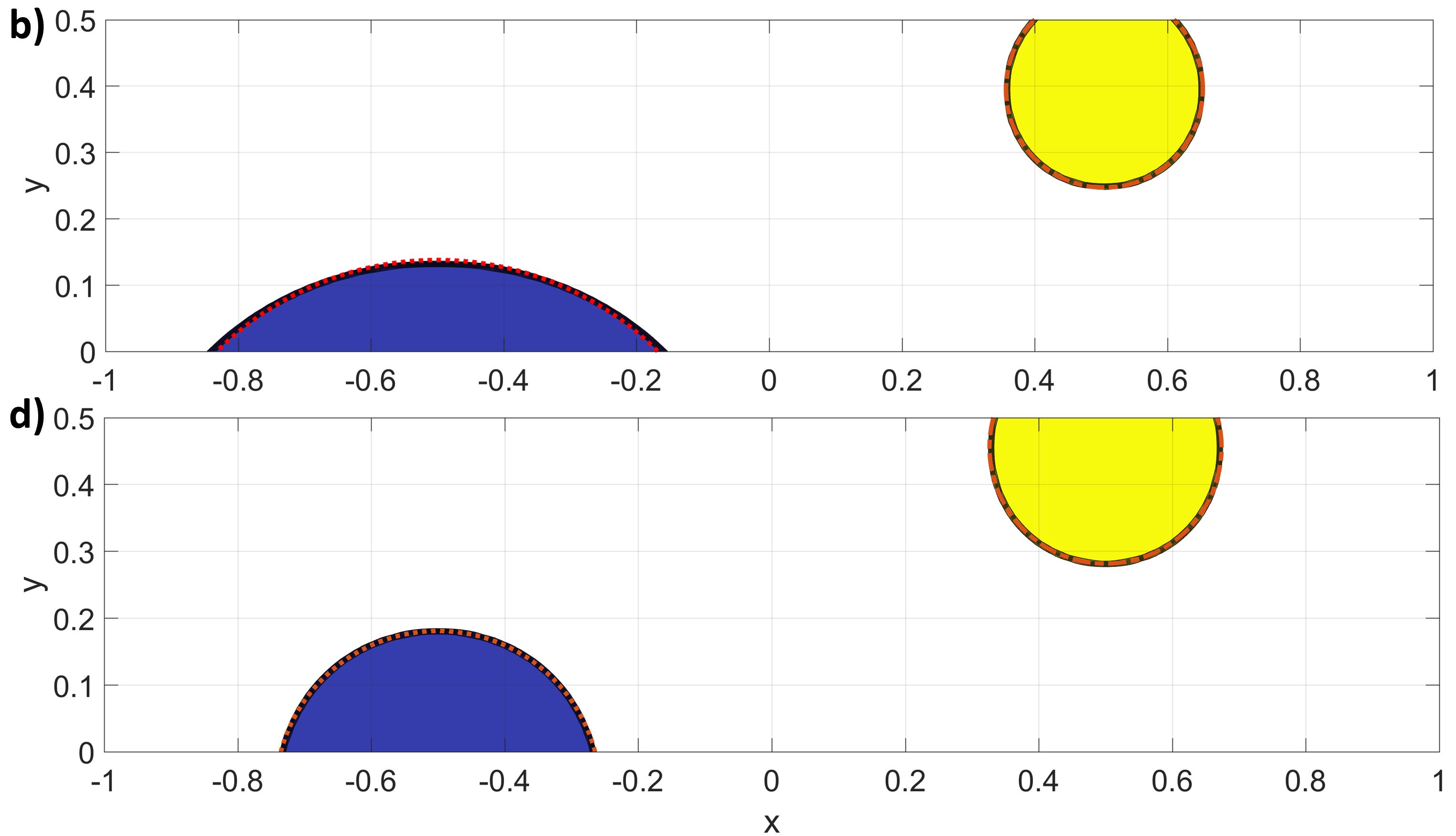}
	\caption{Initial and final shapes of the drops on a surface. \textbf{a)} Initial shapes. \textbf{b)} Final shapes with contact angles $(45^0,135^0)$. \textbf{c)} Final shapes with contact angles $(120^0,60^0)$. \textbf{d)} Final shapes with contact angles $(75^0,105^0)$. Blue: Phase 1, Yellow: Phase 2, White: Phase 3. Red dotted line: Exact equilibrium shape of Phase 1, Orange dash-dotted line: Exact equilibrium shape of Phase 2.} \label{Fig ED-Interface}
\end{figure}

Fig.\ref{Fig ED-Location} shows the heights and the spreading lengths obtained from the numerical solutions under different contact angles, along with the exact solution from Eq.(\ref{Eq Equilibrium drop}). Our numerical results agree well with the exact solution. 
Fig.\ref{Fig ED-Interface} shows the initial and final shapes of the drops, along with the circular segments determined from Eq.(\ref{Eq Equilibrium drop}). The interfaces between Phases 1 and 3 and between Phases 2 and 3 overlap onto those circular segments, which are the exact equilibrium shapes of the drops.

\subsubsection{Three-phase dam break} \label{Sec Three-phase dam break}
We consider a three-phase dam break problem, where the water (Phase 1), oil (Phase 2), and air (Phase 3) interact with each other. Their material properties are listed in Table \ref{Table WaterAirOil}. Non-dimensionalization is performed based on a density scale $1.2041 \mathrm{kg/m^3}$, a length scale $0.05715\mathrm{m}$, and an acceleration scale $9.8\mathrm{m/s^2}$. After that, the domain considered is $[8 \times 2]$ and all the domain boundaries are no-slip. The domain is discretized by $[512\times128]$ cells and the time step is $\Delta t=5\times10^{-4}$. Initially, the water column has a unit height and is at the left of the domain, while the oil column is at the right, as high as the water column, and the fluids are stationary. 

\begin{figure}[!t]
	\centering
	\includegraphics[scale=0.5]{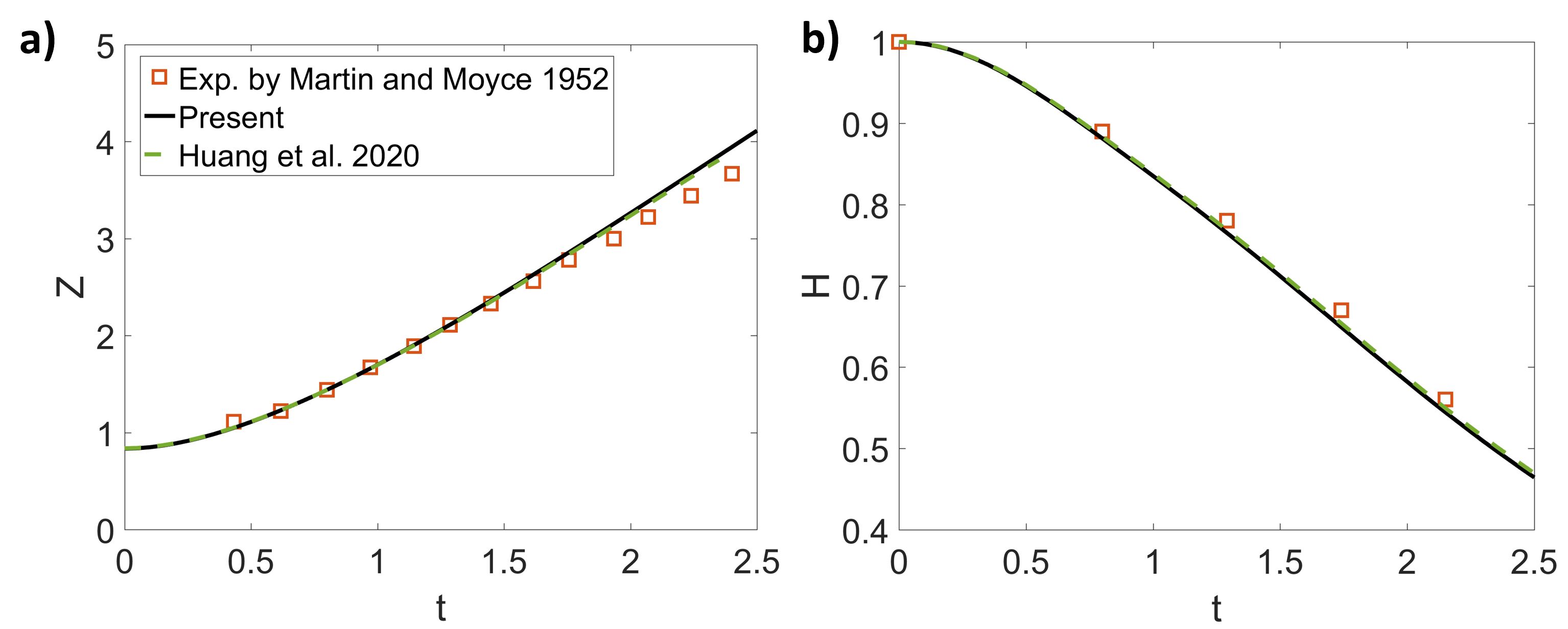}
	\caption{Front and height of the water column in the three-phase dam break problem. \textbf{a)} Front ($Z$) vs. $t$. \textbf{b)} Height ($H$) vs. $t$. \textbf{a)} and \textbf{b)} share the same legend.} \label{Fig DB-Location}
\end{figure}

Because of the gravity, both the water and oil columns will collapse. Since the water and oil are initially separated far enough, the motion of the water should not be influenced by the oil at the beginning of the dynamics. Thus, we can compare the locations of the front $Z$ and the height $H$ of the water column to the experimental measurements by Martin and Moyce \citep{MartinMoyce1952} and to the two-phase (water-air) results by Huang et al. \citep{Huangetal2020}. We calibrated the data by setting $Z=1.44$ at $t=0.8$ and $H=1$ at $t=0$, like those in \citep{MartinMoyce1952}. The comparisons are shown in Fig.\ref{Fig DB-Location}, and the present results not only agree well with the experimental measurements but also are almost the same as the two-phase solution. 

\begin{figure}[!t]
	\centering
	\includegraphics[scale=0.3]{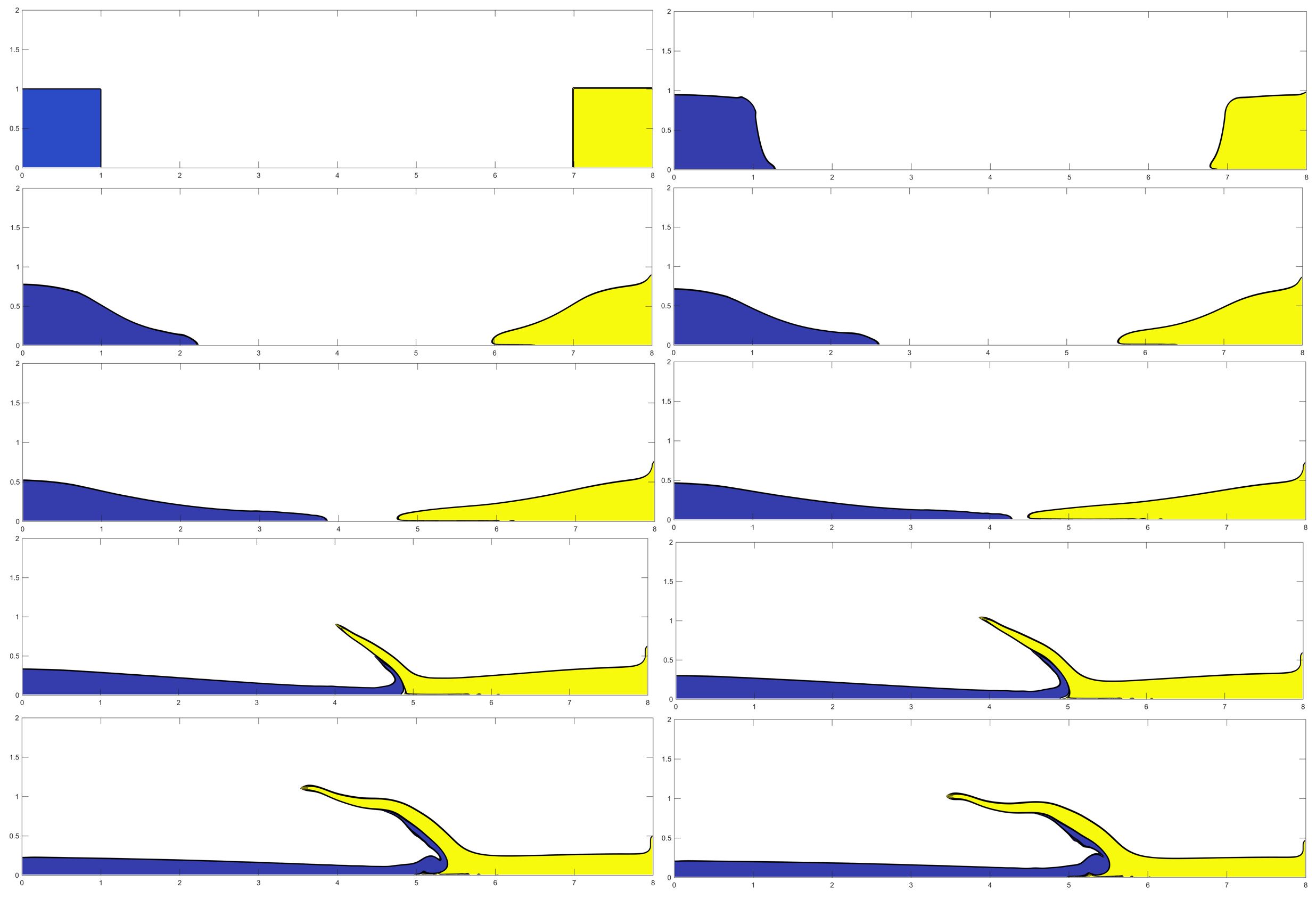}
	\includegraphics[scale=0.3]{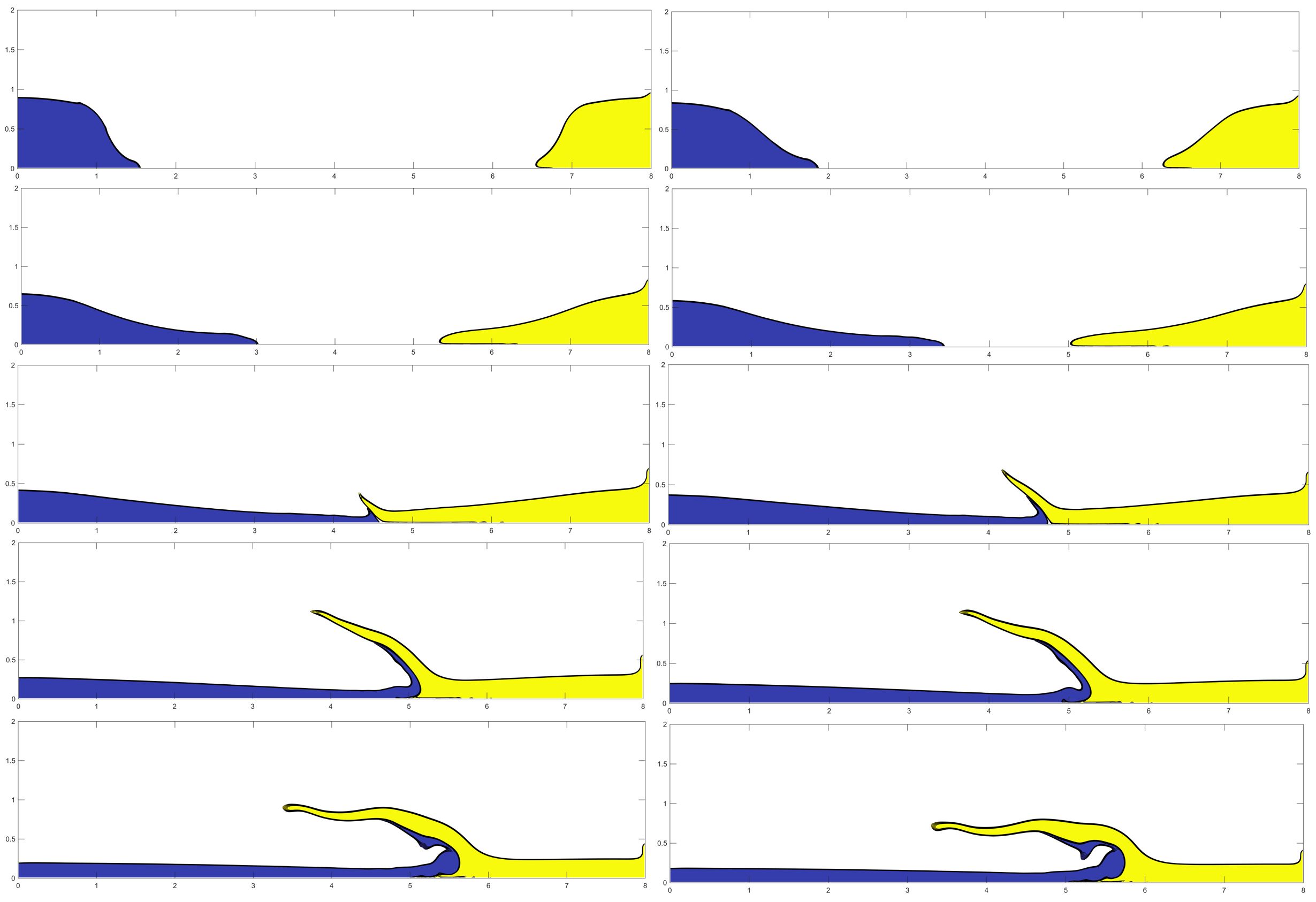}\\
	\includegraphics[scale=0.3]{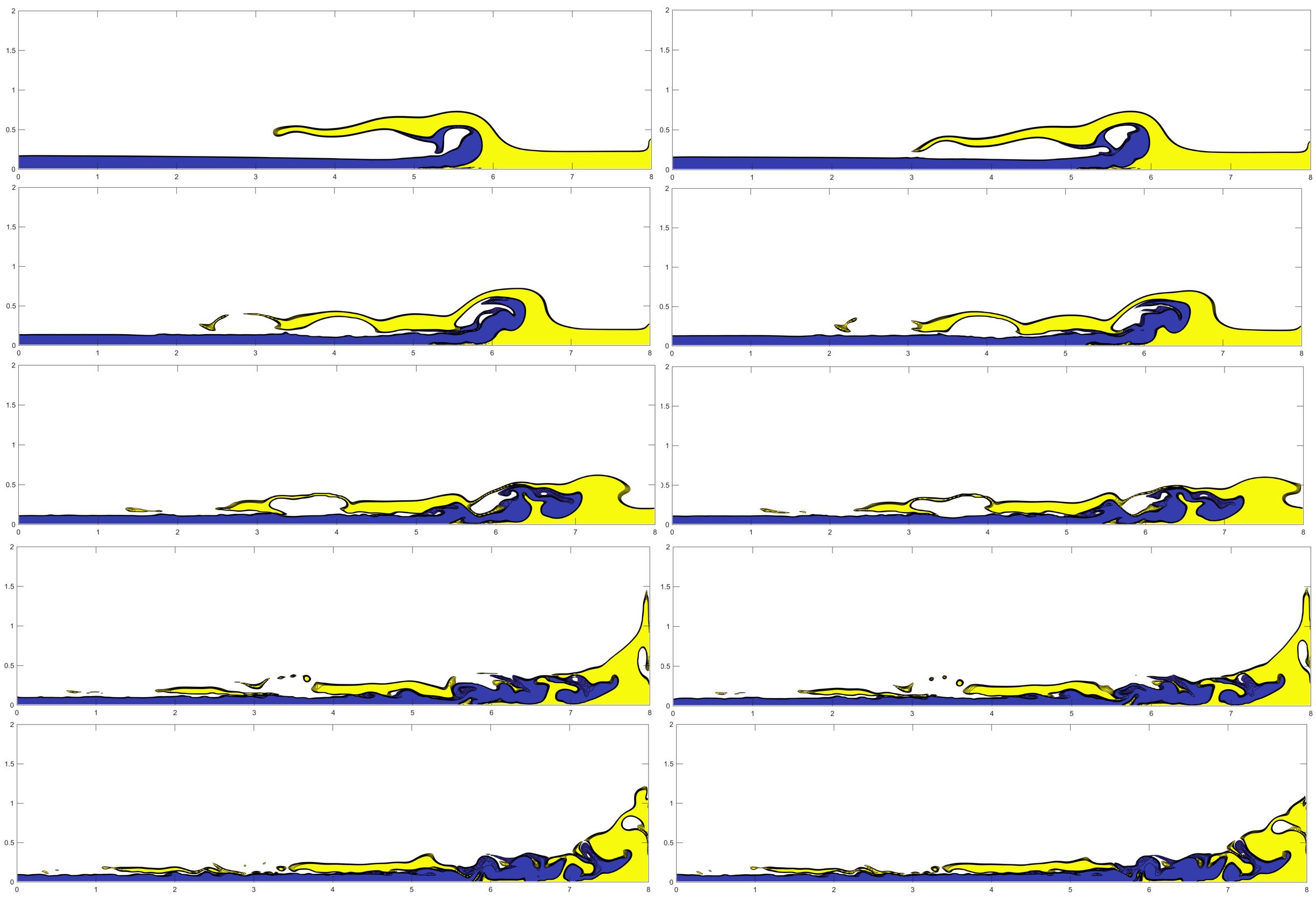}
	\includegraphics[scale=0.3]{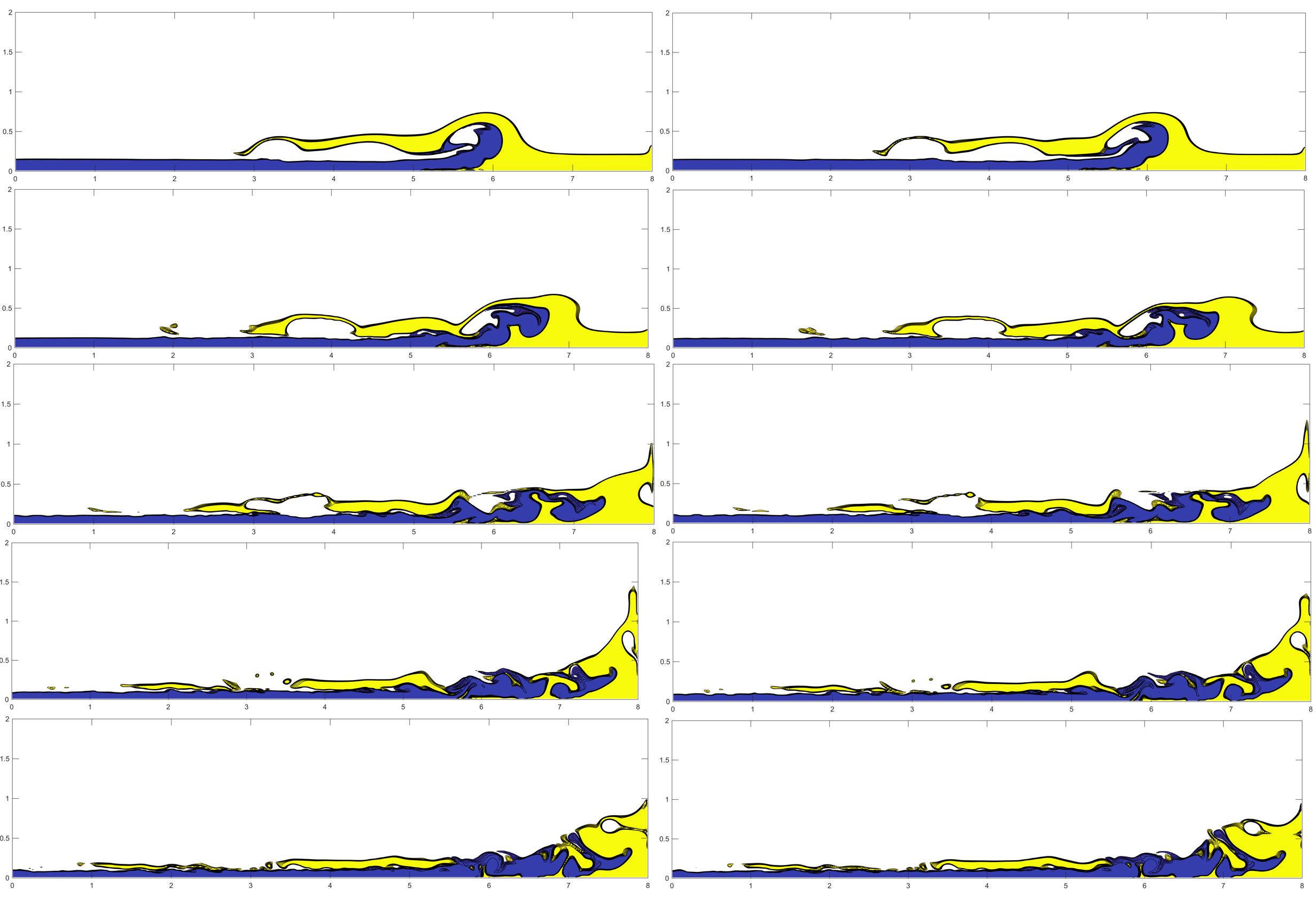}
	\caption{Evolution of the three phases in the three-phase dam break problem. From left to right, top to bottom, $t$ is 0.00, 0.50, 0.75, 1.00, ..., 10.00. Blue: Water, Yellow: Oil, and White: Air.} \label{Fig DB-Interface}
\end{figure}

The evolution of the three phases is shown in Fig.\ref{Fig DB-Interface} up to $t=10$. At the beginning, both the water (Phase 1) and oil (Phase 2) columns collapse towards the center of the domain, like two independent two-phase problems. Since the water is about a hundred times less viscous than the oil, the water front moves faster than the oil front. When the two fronts meet together, the oil front is squeezed upward, and the water moves below the oil, with some amount of air trapped by the water and oil. The oil is then pushed backward to the right wall. In the meantime, the squeezed oil front collapses again onto the water, breaking up into small droplets and filaments, and as a result, the trapped air is released. The interactions between the water, oil, and air are very strong and sophisticated in the long-time dynamics of the problem.

\subsubsection{Three-phase rising bubbles with moving contact lines} \label{Sec Three-phase rising bubble with moving contact lines}
\begin{figure}[!t]
	\centering
	\includegraphics[scale=0.4]{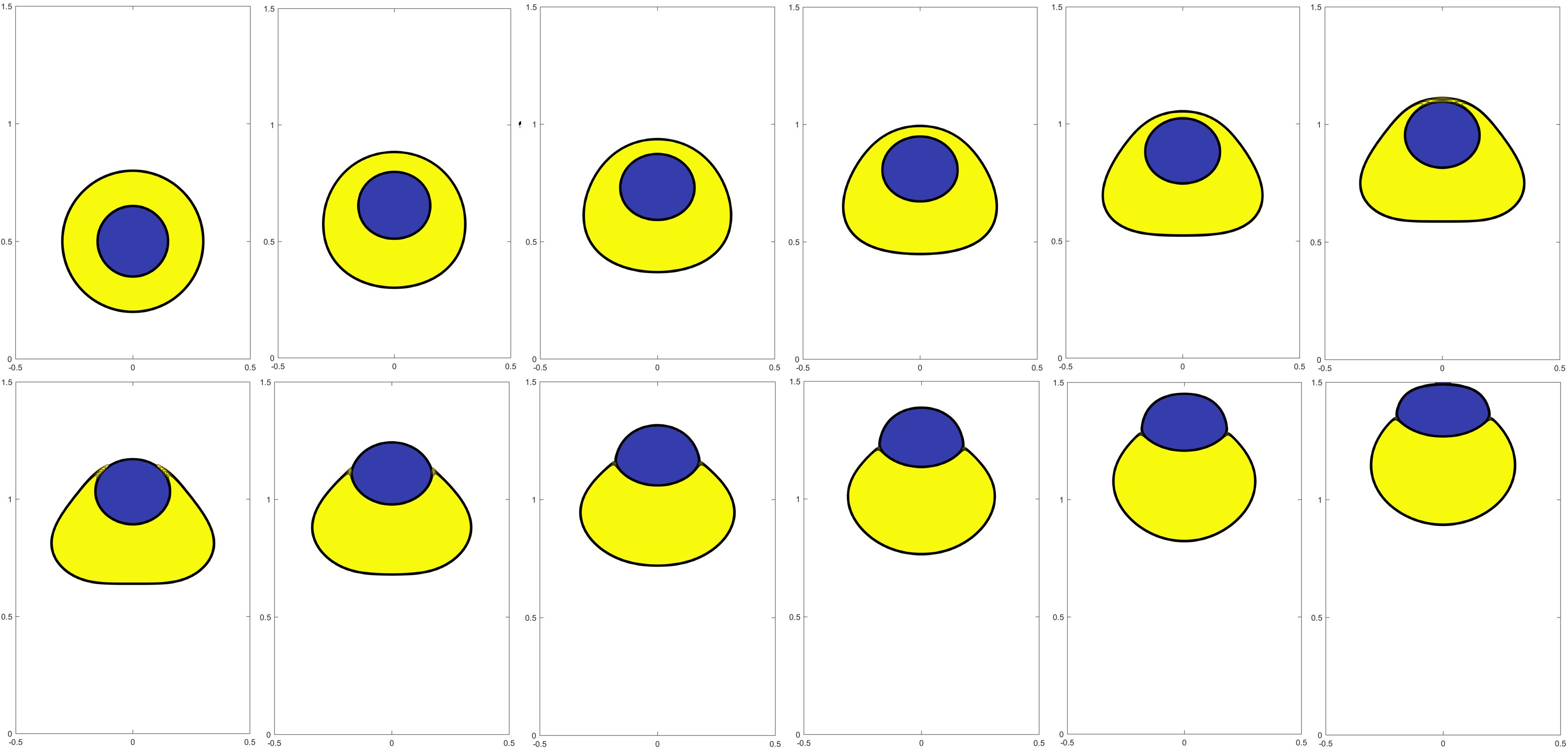}
	\includegraphics[scale=0.4]{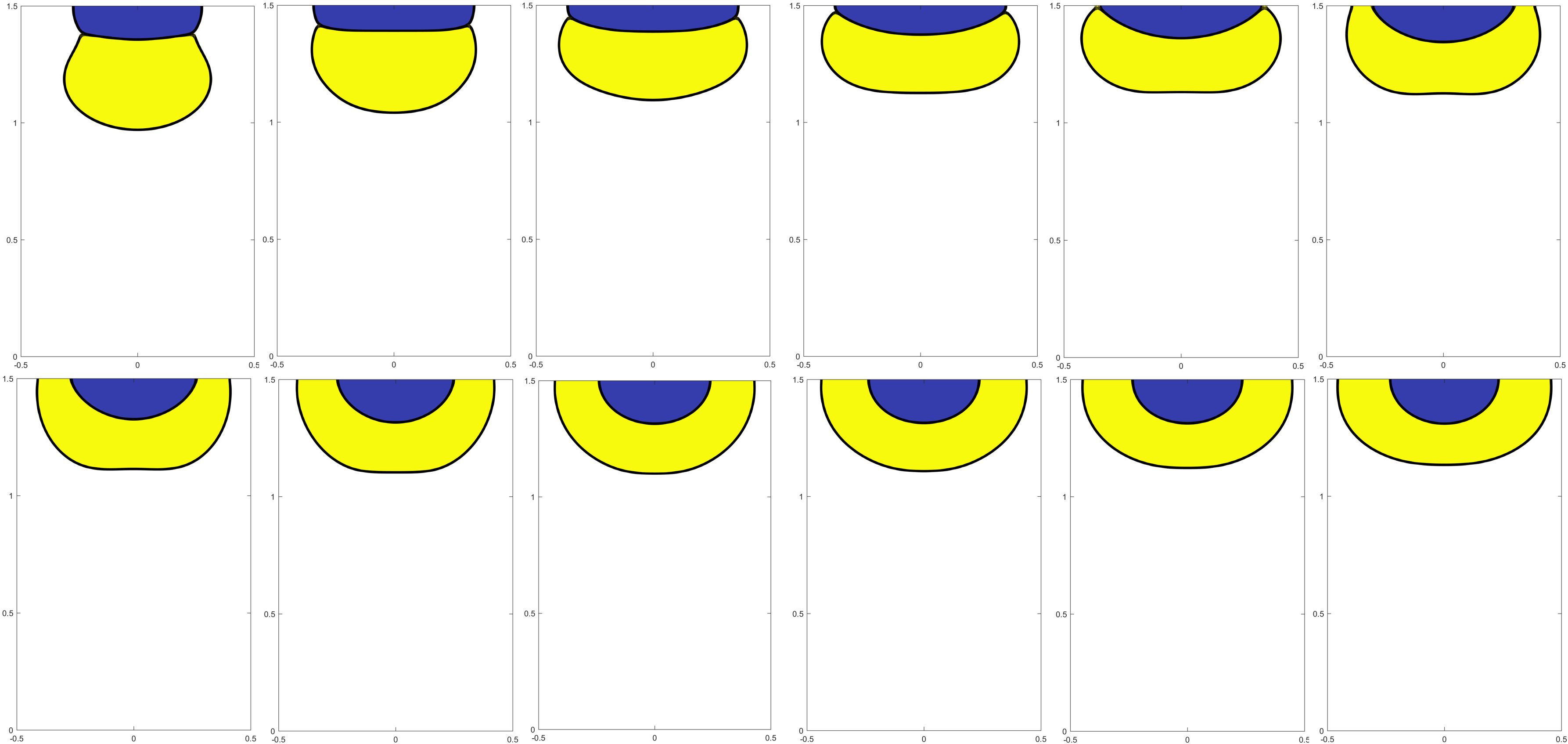}
	\caption{Evolution of the three phases in the three-phase rising bubble with moving contact lines, where $\theta_{1,2}^T=\frac{\pi}{2}$ and $\theta_{2,3}^T=\frac{\pi}{2}$. From left to right, top to bottom, $t$ is 0.00, 0.3, 0.4, 0.5, ..., 2.5. Blue: Air (Phase~1), Yellow: Oil (Phase 2), and White: Water (Phase 3).} \label{Fig RB9090-Interface}
\end{figure}

\begin{figure}[!t]
	\centering
	\includegraphics[scale=0.4]{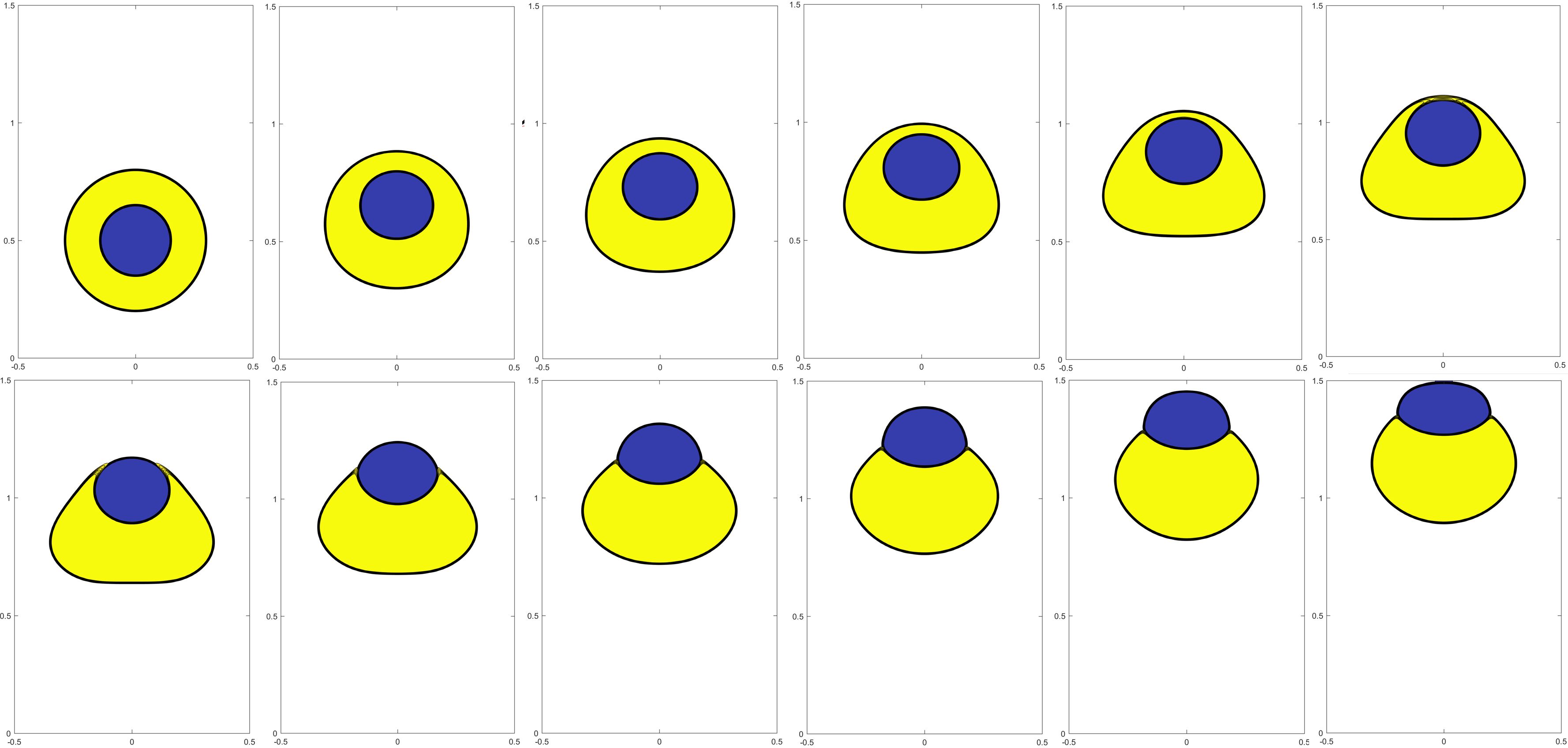}
	\includegraphics[scale=0.4]{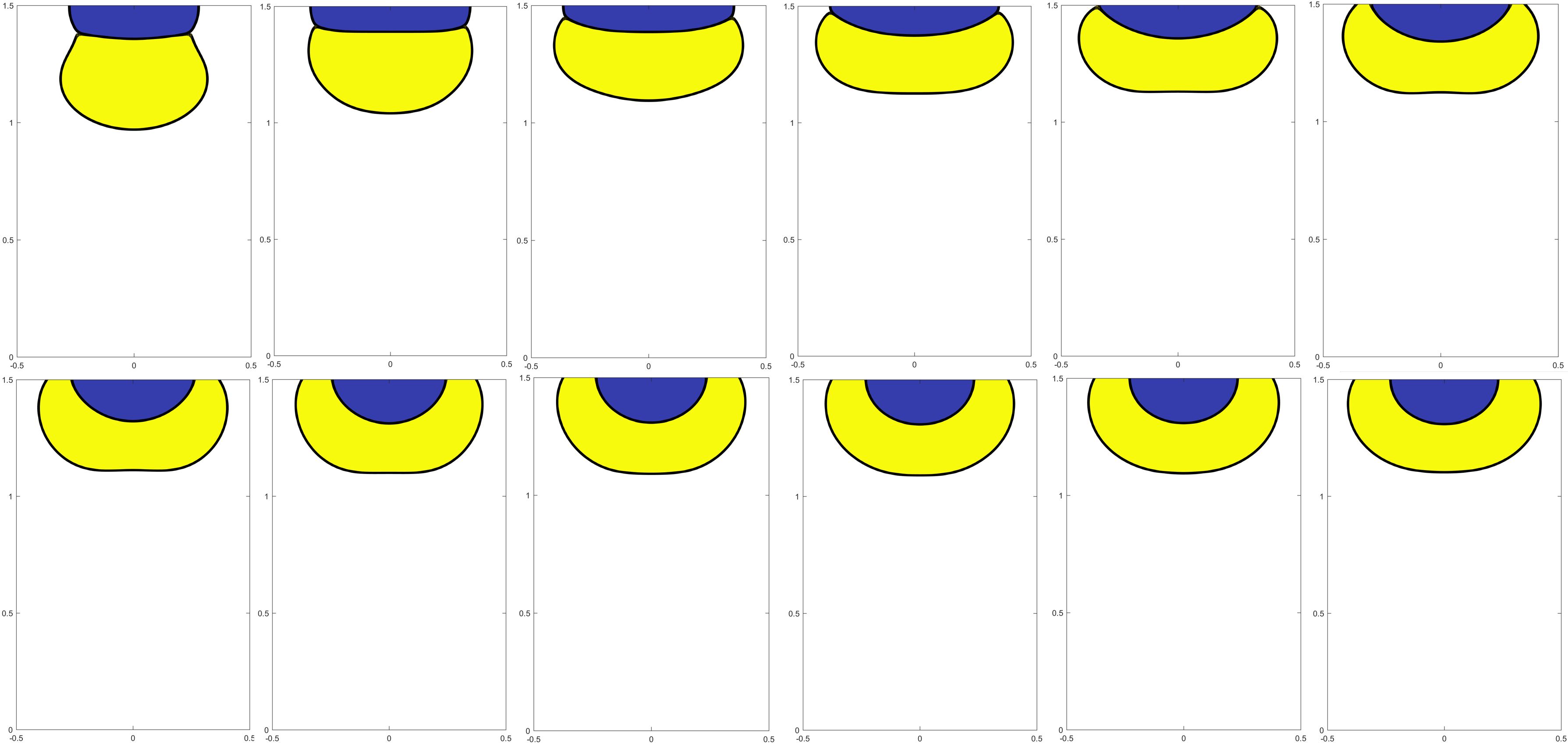}
	\caption{Evolution of the three phases in the three-phase rising bubble with moving contact lines, where $\theta_{1,2}^T=\frac{\pi}{2}$ and $\theta_{2,3}^T=\frac{2\pi}{3}$. From left to right, top to bottom, $t$ is 0.00, 0.3, 0.4, 0.5, ..., 2.5. Blue: Air (Phase~1), Yellow: Oil (Phase 2), and White: Water (Phase 3).} \label{Fig RB12090-Interface}
\end{figure}

\begin{figure}[!t]
	\centering
	\includegraphics[scale=0.4]{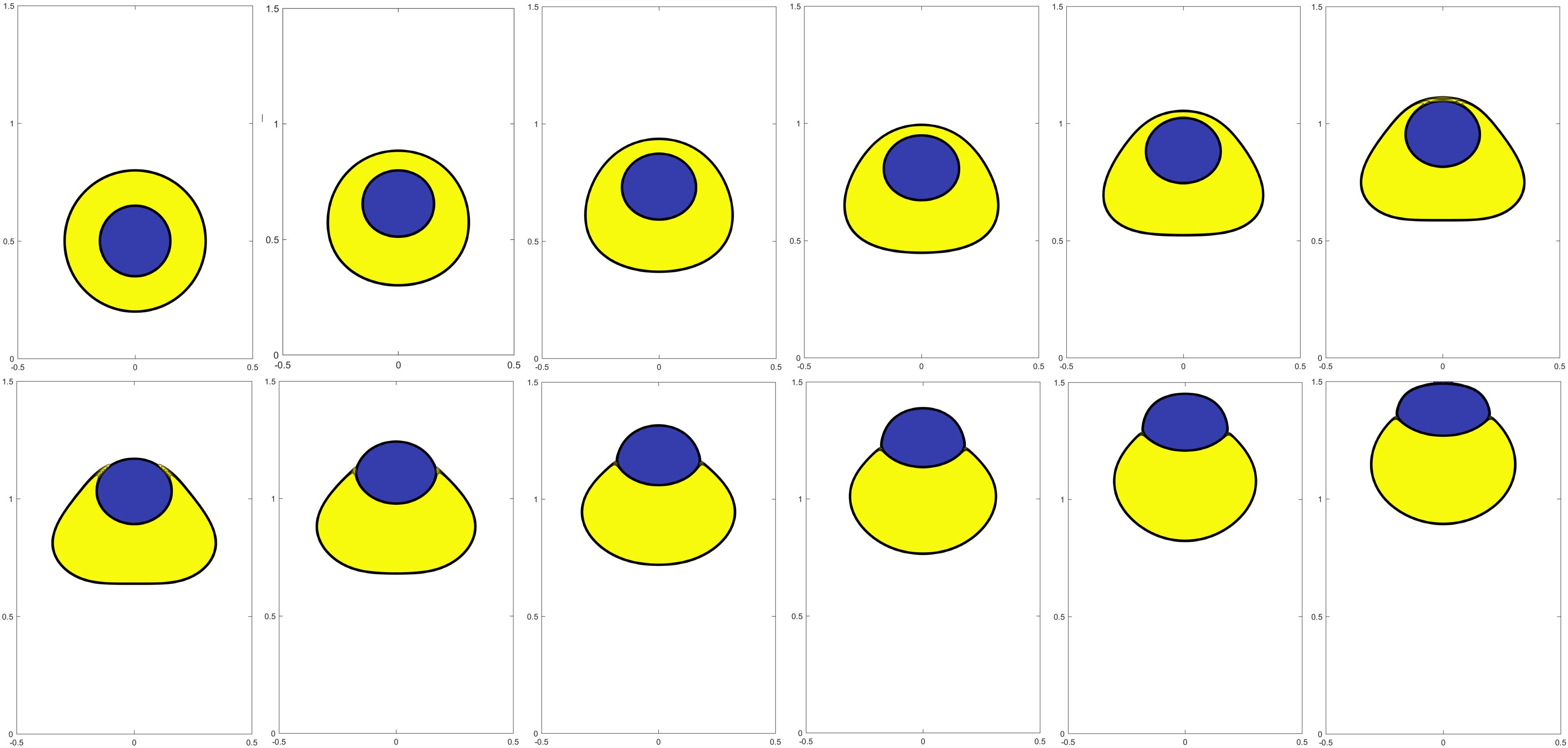}
	\includegraphics[scale=0.4]{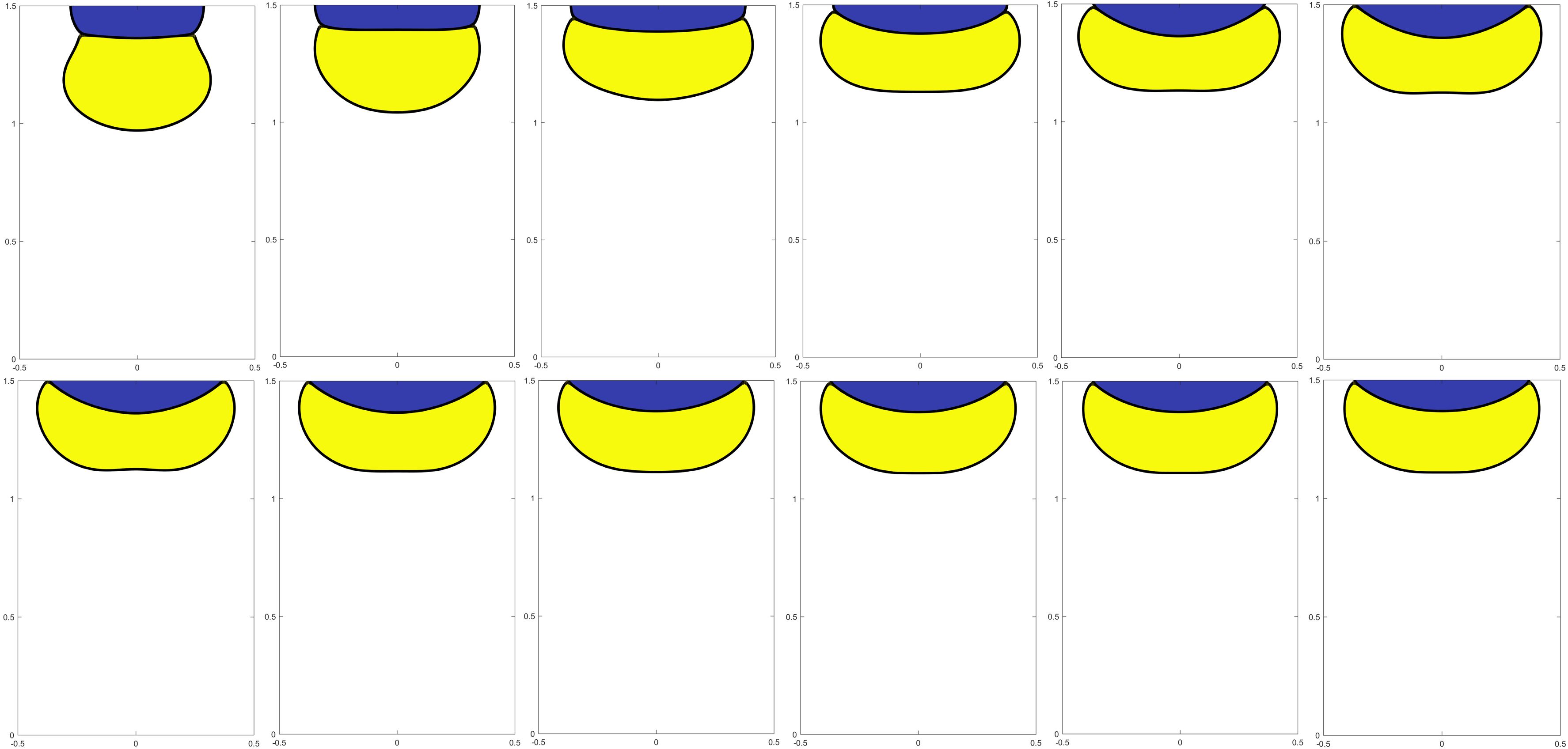}
	\caption{Evolution of the three phases in the three-phase rising bubble with moving contact lines, where $\theta_{1,2}^T=\frac{\pi}{6}$ and $\theta_{2,3}^T=\frac{2\pi}{3}$. From left to right, top to bottom, $t$ is 0.00, 0.3, 0.4, 0.5, ..., 2.5. Blue: Air (Phase~1), Yellow: Oil (Phase 2), and White: Water (Phase 3).} \label{Fig RB12030-Interface}
\end{figure}

We consider a three-phase rising bubble problem with moving contact lines, where an air (Phase 1) bubble is surrounded by an oil (Phase 2) drop, inside a water (Phase 3) tank. The buoyancy effect drives both the air bubble and the oil drop to move upward. After they touch the top wall of the water tank, contact lines between the fluids and the wall are formed and they begin to slide along the wall. The material properties of the fluids are listed in Table \ref{Table WaterAirOil}. Non-dimensionalization is performed based on a density scale $1.2041 \mathrm{kg/m^3}$, a length scale $0.01 \mathrm{m}$, and an acceleration scale $1 \mathrm{m/s^2}$. After that, the domain considered is $[-0.5,0.5]\times[0,1.5]$ and its boundaries are all no-slip. We use $[100\times150]$ cells to discretize the domain and the time step is $\Delta t=2.5\times10^{-5}$. Initially, the flow is stationary. Both the air bubble and the oil drop are circular, whose centers are at $(0,0.5)$. Their radii are $0.15$ and $0.3$, respectively. We consider three different sets of the contact angles at the top wall, which are $\theta_{1,2}^T=\theta_{2,3}^T=\frac{\pi}{2}$ in the first case, $\theta_{1,2}^T=\frac{\pi}{2}$ and $\theta_{2,3}^T=\frac{2\pi}{3}$ in the second case, and $\theta_{1,2}^T=\frac{\pi}{6}$ and $\theta_{2,3}^T=\frac{2\pi}{3}$ in the last case. The evolution of the three phases in the three cases are shown in Fig.\ref{Fig RB9090-Interface}, Fig.\ref{Fig RB12090-Interface}, and Fig.\ref{Fig RB12030-Interface}, respectively, up to $t=2.5$. 

There is no significant difference among Fig.\ref{Fig RB9090-Interface}, Fig.\ref{Fig RB12090-Interface}, and Fig.\ref{Fig RB12030-Interface} before $t=1.8$. This is expected since the contact angle boundary condition should only influence the dynamics close to the top wall. At the beginning, the air bubble moves upward, relative to the oil drop, driven by the buoyancy effect. The drag introduced by the relative motion between the air bubble and the oil drop, along with the buoyancy effect between the oil and the water, drives the oil drop moving upward. Since the density contrast between the air and the oil is much larger than that between the water and the oil, the motion of the air bubble is more pronounced, and as a result, the air bubble crosses the water-oil interface. After that, the triple points between the air, oil, and water is formed, and the three phases gradually reach an equilibrium configuration. This equilibrium configuration keeps moving upward without observable deformation until the air bubble touches the top wall. After touching the top wall, the air bubble quickly spreads along the top wall and is further compressed along the vertical direction because of the rising motion of the oil drop. The difference among the three cases begins with the oil drop touching the top wall, i.e., right after $t=1.8$, since the contact angle between the water and oil at the top wall varies from case to case. In Fig.\ref{Fig RB9090-Interface} and Fig.\ref{Fig RB12090-Interface}, the spreading length of the oil drop is larger than that of the air bubble, and the compressed air bubble is relaxed to a shape close to a semi circle. In Fig.\ref{Fig RB12030-Interface}, the spreading length of the oil drop is very close to that of the air bubble, and the relaxation of the air bubble is not obvious. Finally, both the air bubble and the oil drop stay at the top wall with the assigned contact angles.

\section{Conclusions} \label{Sec Conclusions and future works}
In the present study, we develop a consistent and conservative Phase-Field method, including both the model and scheme, for the multiphase flows where there can be an arbitrary number of incompressible and immiscible phases.

The first important question answered is how to physically couple a mass conservative Phase-Field equation to the hydrodynamics. The \textit{consistency of mass conservation} and the \textit{consistency of mass and momentum transport} are implemented, based on a generic multiphase Phase-Field equation, to obtain the ``actural'' mass conservation equation and the consistent mass flux of the multiphase flow model, which can be different from the sharp-interface one particularly with a divergence-free velocity, and then the momentum equation that is compatible with the ``actual'' mass conservation equation and in general different from the Navier-Stokes equation. The direct consequence of satisfying these two consistency conditions is that the resulting momentum equation satisfies the Galilean invariance and implies the conservation of kinetic energy, which is independent of the Phase-Field equation. We further illustrate that the \textit{consistency of mass conservation} and the \textit{consistency of mass and momentum transport} produce the ``optimal'' coupling in the sense that failures of satisfying the second law of thermodynamics or the \textit{consistency of reduction} of the multiphase flow model only result from the same failures of the Phase-Field equation but are not contributed by the newly obtained momentum equation from those two consistency conditions. 
The consistency conditions as well as the formulations derived from them can be physically explained using the control volume analysis and mixture theory. Moreover, those consistency conditions are valid to alternative mutliphase flow models whose velocity can be non-divergence-free, and the general formulations independent of the interpretation of the velocity are summarized.
Specifically in the present study, we choose the Phase-Field equation in \citep{Dong2018} to complete the present consistent and conservative multiphase flow model because it is fully reduction consistent, although the developed theory and formulations are directly applicable to other existing Phase-Field equations. The present multiphase flow model conserves the mass and momentum and assures the summation of the volume fractions to be unity, in addition to satisfying the consistency conditions mentioned.

Several novel numerical techniques are developed, which helps to preserve the physical properties of the multiphase flow model and therefore avoids unphysical behaviors due to numerical approximation. The gradient-based phase selection procedure is developed to address the issue of producing fictitious phases, local voids, or overfilling in the multiphase flows by the discretized convection term, and this procedure admits any numerical method to compute the convection flux. The balanced-force method and the conservative method are proposed to discretize the surface force, and the conservative method is a general mulitphase interfacial tension model that contributes to the discrete conservation of momentum. The general theorems are developed to address the difficulty of preserving the consistency conditions on the discrete level, and therefore discretely reproduce the physical coupling between the Phase-Field and momentum equations. Those theorems have no restrictions on the number of phases, the form of the Phase-Field equation, or the numerical scheme to solve the Phase-Field equation. They illustrate that the numerical operations in the fully discrete Phase-Field and momentum equations are not independent.

A consistent and conservative scheme is developed for the present multiphase flow model, based on the schemes in \citep{Dong2018,Huangetal2020} for the Phase-Field and momentum equations, respectively, and with the help of the novel numerical techniques. The physical properties of the scheme are carefully analyzed, which were not performed in \citep{Dong2018,Huangetal2020} or other works. The scheme is consistent because it satisfies the \textit{consistency of mass conservation}, \textit{consistency of mass and momentum transport}, and \textit{consistency of reduction}. As a result, it solves advection (or translation) problems exactly regardless of the number, the material properties, or the interface shapes of the phases, eliminates the production of fictitious phases, and discretely recovers the single-phase Navier-Stokes dynamics in each bulk-phase region. The scheme is conservative, because it conserves the order parameters and therefore the mass of each phase and the mass of the fluid mixture, and the momentum if the surface force is either neglected or computed conservatively, e.g., using the newly developed conservative method. Besides, the scheme guarantees the summation of the volume fractions to be unity at every grid point so that no voids or overfilling are produced. 

A series of numerical experiments are performed to validate all the aforementioned properties of the scheme, and the numerical results are consistent with our analyses. 
In addition, we observe the convergence of the numerical solution of the present multiphase flow model from the developed scheme to the sharp-interface solution with a rate between 1.5th- and 2nd-order, no matter whether the balanced-force or conservative method for the surface force is used. 
In the steady drop problem, the balanced-force method gives a smaller spurious current and faster convergence rate than the conservative method.
As illustrated in the large-density-ratio advection problem, violating the consistency conditions results in unphysical velocity fluctuations and interface deformations and finally triggers numerical instability, while the physical result is well reproduced by our scheme even with an extremely large density ratio.
In the horizontal shear layer problem, we demonstrate the generation of fictitious phases, local voids, or overfilling when the gradient-based phase selection procedure does not implemented. These unphysical behaviors disappear after the procedure is activated.
Different from using the conservative method, the momentum is only essentially conserved with the balanced-force method, and the non-conservation is very small and is reduced after grid refinement.
Although it is not shown in the analyses, our numerical results reproduce the behaviors of the energy law of the present multiphase flow model, i.e., Eq.(\ref{Eq Energy law Phi}). The conservative method performs better in the inviscid case when the grid is coarse, while the difference between the two methods is not significant in the fine-grid solution or in the case including viscosity. Therefore, both the balanced-force method and the conservative method are practical and effective for the multiphase flows.
Various realistic multiphase flow problems are performed to verify and demonstrate the capability of the present multiphase flow model and scheme. The numerical results agree well with the exact/asymptotic solutions, other numerical results, and experimental measurements, in problems having significant density and/or viscosity ratios. The multiphase interface configuration is far more complicated than the two-phase one due to strong interactions among different phases in the long-time dynamics.

It is worth mentioning that all the analyses, theorems, and formulations in the present study are valid for an arbitrary number of phases and dimensions, although some of them are reported in two dimensions for a clear presentation. 
In summary, strong interactions among different phases produce complicated dynamics.
Implementing the consistency conditions provides a practical and convenient way to directly obtain physical ``mixture-level'' models for the multiphase problems.
The proposed scheme for the present multiphase flow model preserves many physical properties of the multiphase flows in a discrete sense. It is robust, accurate, and applicable to study various multiphase flows even when there are large differences of material properties.

\section*{Acknowledgments}
A.M. Ardekani would like to acknowledge the financial support from National Science Foundation (CBET-1705371) and the Extreme Science and Engineering Discovery Environment (XSEDE) \citep{Townsetal2014}, which is supported by the National Science Foundation grant number ACI-1548562 through allocation TG-CTS180066 and TG-CTS190041. G. Lin gratefully acknowledge the support from the National Science Foundation (DMS-1555072, DMS-1736364, CMMI-1634832, and CMMI-1560834), and Brookhaven National Laboratory Subcontract 382247, ARO/MURI grant W911NF-15-1-0562, and U.S. Department of Energy (DOE) Office of Science Advanced Scientific Computing Research program DE-SC0021142.

\section*{Appendix}

\subsection*{Equilibrium solution of the floating liquid lens}
\begin{figure}[!t]
	\centering
	\includegraphics[scale=0.5]{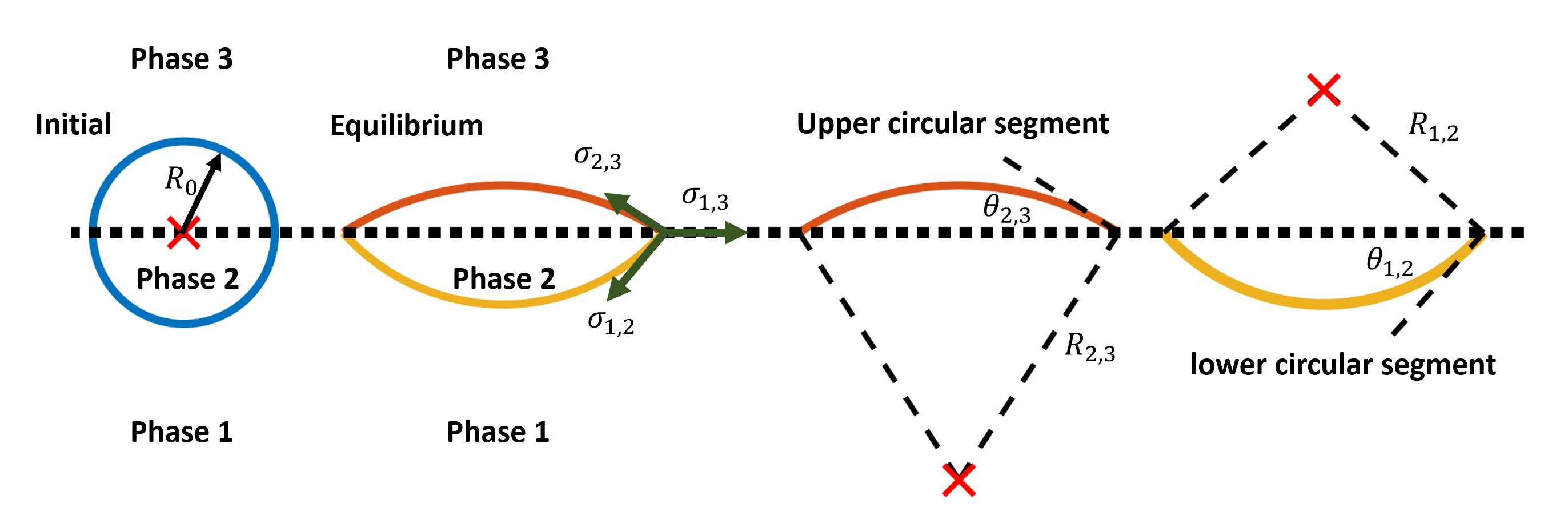}
	\caption{Schematic of a floating liquid lens.} \label{Fig Appendix LL}
\end{figure}

At the equilibrium state, the free energy of the floating lens is minimized. In other words, the area (or the length in 2D) of the interface of the drop  should be minimized under the mass conservation (or the area constraint in 2D), the force balance at the triple points, and the normal stress balance at the interfaces. As a result, in 2D, the equilibrium shape of the drop is a combination of two circular segments, which satisfies the aforementioned constraints. As illustrated in Fig.\ref{Fig Appendix LL}, the drop, called Phase 2, has an initial radius $R_0$. The interfacial tensions between Phases 1 and 2, Phases 1 and 3, and Phases 2 and 3 are $\sigma_{1,2}$, $\sigma_{1,3}$, and $\sigma_{2,3}$, respectively. When the drop reaches its equilibrium shape, the radius of the upper circular segment is $R_{2,3}$ and the angle between the circular segment and the horizontal line is $\theta_{2,3}$, while they are denoted by $R_{1,2}$ and $\theta_{1,2}$ for the lower circular segment.

First, we consider the force balance at the triple points, which are
\begin{eqnarray} \label{Eq Force balance at the triple points}
\sigma_{2,3} \cos(\theta_{2,3}) + \sigma_{1,2} \cos(\theta_{1,2})=\sigma_{1,3} \\
\nonumber
\sigma_{2,3} \sin(\theta_{2,3})=\sigma_{1,2} \sin(\theta_{1,2}),
\end{eqnarray}
and its solutions for $\theta_{1,2}$ and $\theta_{2,3}$ are
\begin{eqnarray} \label{Eq Solutions of theta}
\theta_{1,2}=\cos^{-1} \left( \frac{ \sigma_{1,3}^2+\sigma_{1,2}^2-\sigma_{2,3}^2 }{2\sigma_{1,2} \sigma_{1,3}} \right),\\
\nonumber
\theta_{2,3}= \sin^{-1} \left( \frac{\sigma_{1,2}}{\sigma_{2,3}} \sin(\theta_{1,2}) \right).
\end{eqnarray}
Second, we can compute the areas of the two circular segments as 
\begin{eqnarray} \label{Eq Areas of circular segments}
S_{1,2}=R_{1,2}^2 \left(\theta_{1,2}-\sin(\theta_{1,2}) \cos(\theta_{1,2} ) \right)\\
\nonumber
S_{2,3}=R_{2,3}^2 \left(\theta_{2,3}-\sin(\theta_{2,3}) \cos(\theta_{2,3} ) \right).
\end{eqnarray}	
The area constraint requires that
\begin{equation}
S_{1,2}+S_{2,3}=\pi R_0^2.
\end{equation}
Third, the normal stress balances at the upper and lower circular segments are considered. By applying the Young-Laplace Law at the two circular segments, we have
\begin{equation}
\frac{\sigma_{1,2}}{R_{1,2}}=p_{1,2}=p_{drop}=p_{2,3}=\frac{\sigma_{2,3}}{R_{2,3}},
\end{equation}
where $p_{1,2}$ is the pressure inside the drop and right beside the lower circular segment, $p_{2,3}$ is the pressure inside the drop and right beside the upper circular segment, and $p_{drop}$ is the pressure inside the drop.
By combining the area constraint and the normal stress balances, $R_{1,2}$ and $R_{2,3}$ are obtained, i.e.,
\begin{eqnarray}
R_{2,3}=R_0 \sqrt{\frac{\pi}{(\theta_{2,3}-\sin(\theta_{2,3}) \cos(\theta_{2,3}) ) + (\frac{\sigma_{1,2}}{\sigma_{2,3}})^2(\theta_{1,2}-\sin(\theta_{1,2}) \cos(\theta_{1,2}))}} \\
\nonumber
R_{1,2}=R_{2,3}\frac{\sigma_{1,2}}{\sigma_{2,3}}.
\end{eqnarray}
Finally, it's straightforward to obtain the equilibrium thickness of the drop, 
\begin{equation}
e_d = R_{2,3} \left( 1-\cos(\theta_{2,3}) \right)
+R_{1,2} \left( 1-\cos(\theta_{1,2}) \right)
\end{equation}

\subsection*{Equilibrium solution of a drop on a homogeneous flat surface}
\begin{figure}[!t]
	\centering
	\includegraphics[scale=0.5]{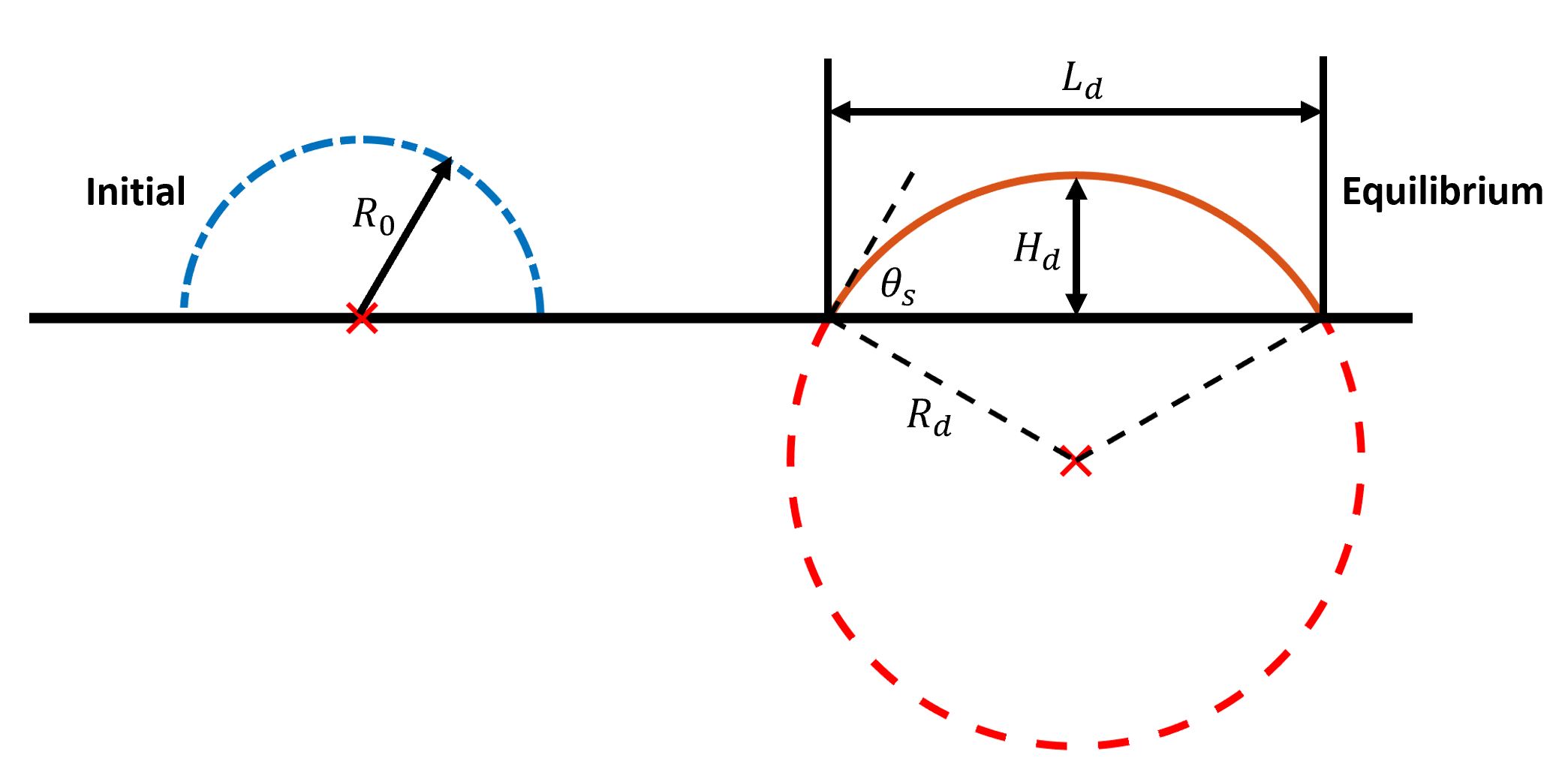}
	\caption{Schematic of a drop on a homogeneous flat surface.} \label{Fig Appendix ED}
\end{figure}

At the equilibrium state, the free energy of a drop placed on a homogeneous surface is minimized. In other words, the area (or the length in 2D) of the interface of the drop should be minimized under the mass conservation (or the area constraint in 2D) and the contact angle constraint. As a result, in 2D, the equilibrium shape of the drop is a circular segment, which intersects the homogeneous flat surface with the given contact angle, and the enclosed area is the same as the initial one. As illustrated in Fig.\ref{Fig Appendix ED}, the area enclosed by the circular segment is 
\begin{equation}
S_d=R_d^2 \left( \theta_s-\sin(\theta_s)\cos(\theta_s) \right),
\end{equation}
where $R_d$ is the radius of the circular segment and $\theta_s$ is the contact angle between the drop and the surface. The area constraint requires that $S_d=\frac{\pi}{2} R_0^2$, where $R_0$ is the radius of the initial semi-circular drop. Thus, we can obtain
\begin{equation}
R_d=R_0 \sqrt{\frac{\pi/2}{\theta_s-\sin(\theta_s)\cos(\theta_s)}}.
\end{equation}
Once $R_d$ is found by given $R_0$ and $\theta_s$, the height and the spreading length of the equilibrium drop can be computed as
\begin{equation}
H_d=R_d \left( 1-\cos(\theta_s) \right),
\end{equation}
\begin{equation}
L_d= 2R_d \sin(\theta_s).
\end{equation}

\bibliographystyle{plain}
\bibliography{refs.bib}

\end{document}